\documentclass[11pt]{article}
\AtBeginDocument{\fontsize{11.5pt}{13.8pt}\selectfont}
\usepackage[utf8]{inputenc}
\usepackage[margin=1in]{geometry}
\usepackage[round]{natbib}
\usepackage{amsmath,amsthm,amssymb}
\usepackage{graphicx}
\usepackage{float}

\usepackage{microtype}
\usepackage{setspace}
\usepackage{tabularx}
\usepackage{longtable}
\usepackage{booktabs}
\usepackage{subcaption}
\usepackage{nicefrac}
\usepackage{etoolbox}
\usepackage{xcolor}
\makeatletter
\renewenvironment{proof}[1][\proofname]{\par
  \pushQED{\qed}%
  \normalfont
  \topsep6\p@\@plus6\p@\relax
  \trivlist
  \item[\hskip\labelsep
        \bfseries #1\@addpunct{.}]\ignorespaces
}{%
  \popQED\endtrivlist\@endpefalse
}
\makeatother
\usepackage{hyperref}
\hypersetup{
  hidelinks,
  pdfborder={0 0 0}
}
\usepackage{url}
\usepackage[english]{isodate}
\usepackage{enumitem}
\usepackage{titlesec}
\titleformat{\paragraph}[block]
  {\normalfont\normalsize\bfseries}{\theparagraph}{1em}{}
\titlespacing*{\paragraph}
  {0pt}{1.5ex plus .5ex minus .2ex}{0.8ex plus .2ex}
\IfFileExists{newpxtext.sty}{\usepackage{newpxtext,newpxmath}}{\usepackage{mathpazo}}
\graphicspath{{./Images/}}
\newcommand{\R}{\mathbb{R}}
\newcommand{\E}{\mathbb{E}}
\newcommand{\wgt}[2]{w_{#1}({\scriptstyle #2})}
\DeclareMathOperator{\Var}{Var}
\DeclareMathOperator{\Cov}{Cov}
\theoremstyle{plain}
\newtheorem{theorem}{Theorem}
\newtheorem{proposition}{Proposition}
\newtheorem{lemma}{Lemma}

\newtheorem{assumption}{Assumption}

\theoremstyle{plain}
\newtheorem{definition}{Definition}

\usepackage{mathrsfs}                       
\theoremstyle{plain}
\newtheorem{oatheorem}{Theorem}[section]

\newtheorem{oalemma}[oatheorem]{Lemma}
\newtheorem{oacorollary}[oatheorem]{Corollary}

\theoremstyle{definition}
\newtheorem{oadefinition}[oatheorem]{Definition}
\newtheorem{oaassumption}[oatheorem]{Assumption}
\theoremstyle{remark}

\theoremstyle{plain}

\usepackage{etoc}
\etocstandardlines
\makeatletter
\patchcmd{\l@section}{1.5em}{3.4em}{}{}
\renewcommand*\l@subsection{\@dottedtocline{2}{3.4em}{3.2em}}
\makeatother
\newcommand{\onlineappendix}[2]{%
  \clearpage
  \newgeometry{margin=0.9in}%
  \setcounter{section}{0}\setcounter{figure}{0}\setcounter{table}{0}\setcounter{equation}{0}%
  \renewcommand{\thesection}{#1.\arabic{section}}%
  \renewcommand{\theHsection}{oa#1.\arabic{section}}%
  \renewcommand{\thefigure}{#1.\arabic{figure}}%
  \renewcommand{\theHfigure}{oa#1.\arabic{figure}}%
  \renewcommand{\thetable}{#1.\arabic{table}}%
  \renewcommand{\theHtable}{oa#1.\arabic{table}}%
  \counterwithin*{equation}{section}%
  \renewcommand{\theequation}{\thesection.\arabic{equation}}%
  \renewcommand{\theHequation}{\theHsection.\arabic{equation}}%
  \normalsize\setstretch{1.3}%
  \begin{center}
    {\LARGE Online Appendix #1}\\[2mm]
    {\large to ``Inflation and the Distortion of Relative Prices''}\\[2mm]
    {\large #2}
  \end{center}
  \vspace{2mm}
  \phantomsection
  \addcontentsline{toc}{part}{Online Appendix #1: #2}%
  \etocsettocstyle{\subsection*{Contents}}{}%
  \etocsetnexttocdepth{subsection}%
  \localtableofcontents
}
\title{Inflation and the Distortion of Relative Prices\thanks{Vipin P. Veetil acknowledges the support of Large Research Grant Project of IIMK.}}

\author{Vipin P. Veetil\thanks{Economics Area, Indian Institute of Management Kozhikode, Kerala 673 570, India.}}

\date{\small\printdayoff\today}
\begin{document}

\maketitle
\begin{abstract}
\setstretch{1.3}
This paper presents a production network model of inflationary dynamics in which inflation can have near-zero correlation with the size of price change yet generate significant distortions in relative prices. New money enters unevenly across firms and percolates through buyer--seller links, thereby displacing relative prices even when every price is free to adjust and every market clears. Under state-dependent price stickiness, firms absorb higher inflation through more frequent rather than larger price changes, leaving the distortion intact. The distortion depends critically on the spectral gap of the production network, which sets the rate at which the economy converges to equilibrium. We quantify the mechanism on a reconstructed production network with the universe of firms in the United States. At a moderate rate of inflation, the typical firm's relative-price distortion is of the same magnitude as one quarter's inflation. The size of price change remains essentially uncorrelated with the magnitude of relative-price distortion.
\end{abstract}
\vspace{2mm}

\setstretch{1.3}
\newpage
\section{Introduction}
Economists have long believed that inflation is capable of disturbing relative prices. Data, however, cannot be brought to bear upon this hypothesis without a model of exactly how inflation affects prices. In most empirical work that model is either Calvo or menu-cost. While the two differ in significant respects, one of their common predictions is that the extent to which inflation disturbs relative prices can be measured by the size of price changes. As \citet[p. 75]{woodford2003} put it, in so far as the old price ``has drifted away from the current optimal relative price by the cumulate effects of inflation... the size of the reset adjustment exactly reflects that accumulated deviation''. Therefore, within Calvo, menu-cost, or even standard sticky-information settings, if inflation disturbs relative prices, then higher rates of inflation must be associated with larger price changes. But such a relation is seldom found in the data.\footnote{See \citet{dhyne2006price}, \citet{gagnon2009price}, \citet{klenow2008state}, \citet{alvarez2019price}, and \citet{berardi2015more}. Some older studies do find a positive correlation during periods of high inflation \citep{lach1992behavior,ball1994asymmetric}.} This has led some economists to conclude that the cost of inflation is elusive \citep{nakamura2018elusive}.

There is, however, reason to doubt this inference: the size of the price change may not be a good measure of relative-price distortion. In fact, placing inflationary dynamics on a production network proves sufficient to generate relative-price distortions that the size of price change does not identify. Our model of inflationary dynamics builds on \citet{gualdi2016emergence}, who generalize the network economy of \citet{acemoglu2012network} to incorporate transient dynamics. Firms are related to each other as buyers and sellers of intermediate inputs. They face cash-in-advance constraints. Each firm sets its price using information about local demand and supply. Put differently, we have distributed price-setting in segmented markets. The probability of price change is governed by a state-dependent hazard function embedded within a production network, wherein the `state' of a firm can be neatly decomposed into the firm's position within the production network and aggregate inflation.

We introduce inflation within the afore-described network economy through cash injections. The cash injections are heterogeneous, reflecting differential access of firms to the banking and financial sector. Since firms face cash-in-advance constraints, an increase in cash balance generates an increase in demand for intermediate inputs. This in turn triggers a sequence of price changes as the money percolates through the economy. Each monetary injection generates a wave of nominal demand change that travels through the economy via the production network. The waves from the different rounds of monetary shocks interfere with each other, dampening one another insofar as they are out of phase. The `phase tilt' between successive waves depends on how the heterogeneity of monetary injection couples with the assortativity of the production network. Ultimately, the price set by a firm depends on the sum of the nominal demand waves faced by it, with the sum itself depending on the number of waves, their decay, and phase profiles. For sufficiently general bounds on the spectral gap of the production network, these sums of waves of nominal demands, and therefore the prices of all firms, converge to a constant growth rate. During the transient, however, the growth of prices varies by firm size.

Our model generates the standard result that higher rates of inflation do not generate larger price changes. Instead, they generate more frequent price changes. This, however, does not mean that inflation does not distort relative prices. The network economy is capable of exhibiting a near-zero correlation between inflation and the size of price change on the one hand, and sizable relative-price distortions on the other. The ultimate cause of relative-price distortions is the fact that different firms experience the waves of nominal demand generated by monetary shocks at different points in time and with different amplitudes. This, in turn, produces differential price changes. Consider the nominal demand wave from a single monetary shock exhibiting a distinct crest and trough soon after its origin, only to spread and tend toward infinite dilution as it travels through the production network. Early on, firms that find themselves at the crest of the monetary wave increase their prices more than inflation, while those at the trough increase by less. In the later periods, as the monetary wave flattens out to a neutral vector, firms raise prices by roughly the rate of inflation. Under sustained injection at a constant rate, divided across firms by a fixed rule, each new monetary wave adds to the cumulative relative-price distortion. This process of increasing relative-price distortions reaches a steady state once the sum of past monetary waves is sufficiently in phase with new injections to create a `standing wave' pattern. The resulting envelope grows steadily, its growth outweighing the fluctuations around its seam. Once such a pattern is reached, all prices grow at the rate of inflation, and there are no new relative-price distortions. 

Naturally, if new money injections are brought to an end, the economy relaxes to an equilibrium with a price level corresponding to the new monetary mass and relative prices corresponding to the real primitives of the economy. But insofar as money injections persist, so will the relative-price distortions. And the extent of this distortion increases with the heterogeneity in the initial impact of monetary shocks and the degree of disassortativity of the production network. In short, in a network setting relative-price distortions emerge from structural features of the system that have little to do with the flexibility of prices. This is not all. Price stickiness and price flexibility can trade places along the inflationary path. Early on, while the monetary waves are still building, an economy with fully flexible prices can distort relative prices more than a sticky one. Flexible prices pass every displacement straight through, whereas staggered prices still answer to the calmer vintages of the recent past. Once the standing wave has formed, that attenuation disappears, and the sticky-price regime is at least as distorted as the flexible-price regime, for staggered resets can only add a dispersion of price vintages to the distortion that the network creates. These relative-price distortions do not fade away through a central-limit mechanism. Large economies are not immune to them.

We characterize the mechanisms by which the forces that drive an economy toward equilibrium are periodically thwarted by the injection of new money. In fact, it is at the horizon of tension between these opposing forces that we see the emergence of relative-price distortions that are essentially uncorrelated with the size of price changes. Two objects suffice to characterize the entire adjustment path: the spectral gap of the production network, and certain moments of the degree distribution that depend on the degree-assortativity and injection-heterogeneity parameters. This closed-form characterization of transient dynamics takes us beyond the usual comparative statics into the domain of comparative dynamics. That is, we are in earnest able to take derivatives of statistics that capture the size of price change and relative-price distortions with respect to model parameters without having to assume the system is always in equilibrium. We can, for instance, ask what would happen to the relative-price distortion if the rate of inflation increased well before the economy had reached the equilibrium pertaining to the old rate. These closed-form solutions, however, depend on certain strict assumptions about the size of the monetary shocks and output responses to price changes. We relax these assumptions through a battery of computational experiments, which include experiments on a reconstructed network of the universe of firms in the United States. Our analytical results prove robust. And on the reconstructed United States network the distortion is of the same magnitude as the inflation rate itself: at one percent inflation per period, the `typical' firm's relative price stands about one percent from its equilibrium value.

\subsection{Related literature}

There has, in recent years, been a perceptible interest in the role of production networks in transmitting monetary shocks, see for instance \citet{pasten2020propagation}, \citet{DiGiovanniHale2022}, and \citet{PelletTahbazSalehi2023}.\footnote{The Banque de France, together with the Toulouse School of Economics, organized a conference on the theme, `Unbalanced Sectoral Inflation Dynamics: Lessons for Monetary Policy', in Paris in December 2023.} Much, if not all, of this literature implicitly assumes that monetary shocks do not generate disequilibrium dynamics. Our concern in this paper is precisely the price distortions generated in monetary disequilibrium.\footnote{An older generation of monetary economists, Patinkin and others, had a serious interest in monetary disequilibrium \citep{patinkin1965,clower1965keynesian,leijonhufvud1968keynesian}, an interest which, for reasons known and unknown, faded away after the 1970s. The known bit of the reason is the discovery of rational expectations. The unknown bit perhaps has to do with the peculiarities of personalities, which in science, as elsewhere, have a bearing on all that is human.} We have in previous papers studied how a single monetary shock influences price dynamics on a production network \citep{mandel2019price,mandel2021monetary}. Here we extend that line of thinking by studying constant-cadence monetary shocks, with a particular eye to ascertaining the relative-price distortions they generate. The singular reason why the extension has proved so difficult is that a sequence of monetary shocks necessarily means an interaction between them. And such interactions within a network setting are capable of generating certain subtle---dare I say beautiful---phenomena, whose appeal is matched by the analytical difficulties in dissecting them.

Naturally, our paper is reminiscent of the rich literature on heterogeneity in the impact and propagation of monetary shocks, the oldest formal account of which is perhaps the Islands Model \citep{lucas1972expectations}. It is but a special case of segmented markets in which frictions impede the flow of money and monetary conditions across markets \citep{Rotemberg1984JPE,AlvarezAtkesonKehoe2002}.\footnote{Several papers in this literature use cash-in-advance constraints to generate some heterogeneity in the impact of monetary shocks across markets. This heterogeneity in turn influences decisions to trade, the formation of relative prices, and the allocation of resources \citep{cooley1991welfare,gillman1993welfare,silva2012rebalancing}.} The choice is merely one of where to locate the heterogeneity and how to design the friction that keeps markets segmented. The heterogeneous-agent New Keynesian literature, for instance, locates heterogeneity on the consumption side \citep{Auclert2019}, while other models locate it in participation in the asset market \citep{AlvarezAtkesonKehoe2002}. In our model, heterogeneity is on the production side, driven by the size distribution of firms. The literature also offers a wide variety of frictions that keep markets `segmented', including staggered access to financial markets \citep{Grossman1983}, restricted trading relationships in goods markets \citep{williamson2008monetary}, and incomplete insurance combined with portfolio illiquidity \citep{KaplanMollViolante2018}. Our friction is the production network itself.

Much of what is said in this paper relates rather directly to the extensive literature that models price setting through generalized-hazard functions.\footnote{These encompass both state-dependent and time-dependent specifications \citep{ChristianoEichenbaumEvans2005,GopinathItskhoki2010,CostainNakov2011,AlvarezLippiOskolkov2020}.} Under flexible prices, the nominal demand waves left by past shocks synchronize with one another, so much so that in steady state the size of price change carries almost no trace of the network. Staggered price adjustment breaks that alignment. Firms reset their prices on different dates. It is as if each firm were living in a different nominal year. How many vintages exist, and how they accumulate to generate relative-price distortions, depends on the topology of the production network. The `local state-dependent pricing' of this paper differs from the pricing of \citet{nakamura2010multisector} and \citet{pasten2020propagation} in one respect. Here the price vintages are endogenous to the production network. This endogeneity allows higher rates of inflation to be absorbed by more frequent price changes, while the spread of vintages keeps relative prices away from their equilibrium values. With more frequent price changes, the size of price changes can be small even when the dispersion of relative prices is large. 

\subsection{Organization of the paper}
Section~\ref{model} presents the model, a production network in which firms pay for inputs out of their money balances. Section~\ref{sec:framework} develops the analytical framework. Section~\ref{sec:flexible} studies inflation under flexible prices. Section~\ref{sec:sticky} studies inflation under sticky prices and establishes the central result, that the size of price change does not measure relative-price distortion. Section~\ref{sec:abm} relaxes the analytical assumptions with computational experiments. It also measures the size of relative-price distortion with a computational experiment calibrated to the United States production network. Section~\ref{sec:conclusion} presents concluding thoughts. Appendix~\ref{appendix:mathematics} contains the proofs of the paper's central formal claims. Appendix~B contains supplementary formal claims and their proofs. Among them is the result that the size of price change and the two measures of relative-price distortion are asymptotically Gaussian across random draws of the production network (Appendix~\ref{oa:gaussian_concentration}). And Appendix~C documents the calibration and collects additional computational results.

\section{Model}
\label{model}

We now build a production network in which firms pay for their inputs out of the money they hold and new money enters unevenly. The economy has two blocks: a nominal block, in which money balances move from buyers to suppliers, and a real block, in which firms turn delivered inputs into output. When no new money enters, the economy has a unique equilibrium and returns to it from every positive starting point. With sustained injection the money stock grows without bound, and the economy converges to a balanced-growth path, on which the distribution of money across firms, output, and prices relative to the money stock stay constant.

\subsection{The production network and general equilibrium}
\label{subsec:production_network}

Firms buy intermediate inputs from one another and pay for them out of their money balances. Money balances stand for internal liquidity or committed short-term credit. There are $n$ firms, indexed by $N=\{1,\dots,n\}$.\footnote{We omit a representative household because the object of interest is propagation through heterogeneous local trading relations. A household that earns from every firm and spends at every firm would spread money across the whole economy each period and speed up the convergence of balances by construction.} Firm $j$ spends the fraction $a_{ij}\in\mathbb R_+$ of its input expenditure on the good of firm $i$. These expenditure shares sum to one, $\sum_{i\in N}a_{ij}=1$, and $a_{ij}>0$ exactly when firm $i$ supplies firm $j$. The matrix $\mathbf A=(a_{ij})_{i,j\in N}$ carries each firm's spending to its suppliers, and we call it the `propagation matrix'. Its columns sum to one, so it is column-stochastic.

We characterize how money circulates through the network with Assumption~\ref{assu:primitivity}, maintained throughout.

\begin{assumption}[Primitivity]\label{assu:primitivity}
The propagation matrix $\mathbf A$ is primitive, that is, irreducible and aperiodic.
\end{assumption}

\noindent Assumption~\ref{assu:primitivity} guarantees that money, spent and respent at the expenditure shares, settles into a unique positive distribution $\mathbf m^{*}$, the right Perron eigenvector of $\mathbf A$, from every starting distribution with the same total. We normalize the money stock to one, $\mathbf 1^\top\mathbf m^{*}=1$, and call $\mathbf m^{*}$ the `stationary balances'.

Each firm
$j\in N$ has a Cobb--Douglas production function with common returns to scale $\varsigma\in(0,1)$, mapping inputs $(x_{1j},\dots,x_{nj})$ into output
\[
\prod_{i\in N} x_{ij}^{\varsigma\, a_{ij}}.
\]
Because $\sum_{i\in N} a_{ij}=1$, the input elasticities sum to $\varsigma$, so production has decreasing returns.\footnote{Decreasing returns make the equilibrium determinate in levels. At $\varsigma=1$ the matrix $\mathbf I-\varsigma\mathbf A^\top$ is singular, because the Perron root of $\mathbf A^\top$ is one, and output levels are indeterminate, although the economy may still grow along a balanced-growth path. The Hawkins--Simon viability condition and the von Neumann growth model draw a related distinction \citep{hawkins1949,vonneumann1945,sraffa1960}.} A firm that minimizes cost keeps its expenditure shares $a_{ij}$ whatever the prices. Let $\mathcal D_i^{(t)}$ denote the nominal demand firm $i$ faces at date $t$, and $p_i^{(t)}$ and $q_i^{(t)}$ its price and output.

The zero-injection equilibrium is the benchmark against which every displacement of relative prices is measured. At the zero-injection equilibrium, firms allocate their balances according to the expenditure shares, transform the resulting inputs into output, clear their markets, and reproduce the stationary balances:
\[
x_{ij}^{*}=a_{ij}\,\frac{m_j^{*}}{p_i^{*}},
\qquad
q_j^{*}=\prod_{i\in N}(x_{ij}^{*})^{\varsigma a_{ij}},
\qquad
p_i^{*}\,q_i^{*}=\mathcal D_i^{*}=\sum_{j\in N}a_{ij}m_j^{*}=m_i^{*}.
\]
The vectors $(p_i^{*})_{i\in N}$ and $(q_i^{*})_{i\in N}$ and the matrix $(x_{ij}^{*})_{i,j\in N}$ form the fixed point of this system. We write $(b_A)_j:=\sum_{i\in N}a_{ij}\log a_{ij}$ for the technological constant of firm $j$, and collect these constants in the vector $\mathbf b_A$.

\begin{proposition}[Existence and uniqueness of equilibrium]\label{prop:equilibrium_existence}
Let the returns to scale satisfy $\varsigma\in(0,1)$, and fix the nominal scale by the normalization $\mathbf 1^\top\mathbf m^{*}=1$. Then the zero-injection economy admits a unique positive stationary equilibrium $(\mathbf m^{*},\mathbf q^{*},\mathbf p^{*})$, given by
\[
\mathbf A\mathbf m^{*}=\mathbf m^{*},\qquad
\log\mathbf q^{*}
=(\mathbf I-\varsigma\mathbf A^\top)^{-1}
\varsigma\!\left[\mathbf b_A+(\mathbf I-\mathbf A^\top)\log\mathbf m^{*}\right],
\qquad
p_i^{*}=m_i^{*}/q_i^{*}\ \ (i\in N).
\]
\end{proposition}

\noindent\textit{Proof in Appendix~\ref{app:proof_equilibrium_existence}.}

Money is neutral in this equilibrium, for doubling the money stock doubles every price and leaves every firm's output unchanged.

\subsection{Zero-injection adjustment dynamics}
\label{subsec:transient_dynamics}

Each firm $j$ begins period $t$ with the money balance $m_j^{(t)}\in\mathbb R_+$, which finances its input expenditure, and spends $a_{ij}m_j^{(t)}$ on the good of firm $i$. Money moves one step ahead of goods: an order placed at date $t$ is settled at date $t+1$, at the price its seller then posts, and the revenue it carries becomes the seller's date-$(t+1)$ balance. We index demand by the date of the revenue balance it generates, so under zero injection the nominal demand faced by seller $i$ is
\[
\mathcal D_i^{(t+1)}\;:=\;\sum_{j\in N} a_{ij}\, m_j^{(t)}.
\]
This settlement convention is maintained throughout. Prices are flexible in the baseline. Each seller posts the price $p_i^{(t)}:=\mathcal D_i^{(t)}/q_i^{(t)}$, at which the nominal demand it faces buys exactly its current output. Settlement at this price delivers the physical inputs $x_{ij}^{(t)}=a_{ij}m_j^{(t-1)}/p_i^{(t)}$, which sum to $q_i^{(t)}$. The inputs a firm receives at date $t$ produce its output at date $t+1$, $q_j^{(t+1)}=\prod_{i\in N}(x_{ij}^{(t)})^{\varsigma a_{ij}}$. Revenue flows then imply $m_i^{(t+1)}=\sum_{j\in N}a_{ij}m_j^{(t)}$, or, in vector form, $\mathbf m_{t+1}=\mathbf A\,\mathbf m_t$. The nominal block is therefore autonomous, for balances evolve whatever happens to prices and output. But the real block is not. This is because output depends on the balances that finance input orders. Using the technological constant $\mathbf b_A$ and $p_i^{(t)}=m_i^{(t)}/q_i^{(t)}$, logarithms of the production relation give
\begin{equation}\label{eq:zero_injection_real_dynamics}
\log\mathbf q_{t+1}\;=\;\varsigma\mathbf A^\top\log\mathbf q_t\;+\;\varsigma\,\mathbf b_A\;+\;\varsigma\bigl(\log\mathbf m_{t-1}-\mathbf A^\top\log\mathbf m_t\bigr),
\qquad t\ge1,
\end{equation}
a stable recursion forced by the evolving distribution of balances, whose fixed point is the equilibrium output of Proposition~\ref{prop:equilibrium_existence}. From every positive initial configuration with $\mathbf 1^\top\mathbf m_0=1$ the economy converges to that equilibrium.

We characterize the speed of this convergence with Assumption~\ref{assu:spectral_gap}, maintained throughout.

\begin{assumption}[Second eigenvalue]\label{assu:spectral_gap}
The second eigenvalue $\lambda_2$ of $\mathbf A$ is real, simple, and strictly positive, $0<\lambda_2<1$, and is separated in modulus from the remainder of the spectrum, $\lambda_2>|\lambda_3|\ge\cdots\ge|\lambda_n|$.
\end{assumption}

\noindent We call $1-\lambda_2$ the `spectral gap'. Because $\lambda_2>|\lambda_3|$, deviations of balances from $\mathbf m^{*}$ die out most slowly along the second eigenvector of $\mathbf A$, so they come to line up with it. And because $\lambda_2$ is real and positive, a deviation along that eigenvector shrinks without changing sign, so successive disturbances to balances add up. Output and prices converge at the slower of two rates, the rate $\lambda_2$ at which balances mix and the rate $\varsigma$ at which decreasing returns damp output (Lemma~\ref{lemma:monotone_convergence}).

\subsection{Monetary injection and balance dynamics} \label{subsec:monetary_process}

At every date, new money equal to the fraction $\pi\in[0,1]$ of the money stock $\mathbf 1^\top\mathbf m_t=\sum_{i\in N}m_i^{(t)}$ is injected before trade. The injection exponent $\theta$, the central monetary primitive of the paper, governs the incidence of the new money across firms. Under the injection rule, firm $i$'s share of the new money is proportional to $m_i^{\theta}$, so $\theta$ is the elasticity of its intake of new money with respect to its balance. At $\theta=1$ the injection reproduces the existing distribution, for each firm receives in proportion to what it already holds, and relative balances are undisturbed. At $\theta=0$ every firm receives the same absolute amount. We maintain the intermediate range $\theta\in(0,1)$, in which new money favors small-balance firms relative to their holdings. The evidence on the firm-size gradient of credit supports this Cantillon configuration (Appendix~\ref{oa:theta_calibration}). New money is added before trade and then spent:
\begin{equation}\label{eq:demand_dynamics}
\boldsymbol{\gamma}_t
:=
\frac{\mathbf m_t^{\circ \theta}}{\mathbf 1^\top \mathbf m_t^{\circ \theta}}
\in\bigl\{x\in\mathbb R^n_{\ge0}:\mathbf 1^\top x=1\bigr\},
\qquad
\widetilde{\mathbf m}_t
:=
\mathbf m_t+\pi(\mathbf 1^\top\mathbf m_t)\,\boldsymbol{\gamma}_t,
\qquad
\mathbf m_{t+1}
=
\mathbf A\,\widetilde{\mathbf m}_t,
\end{equation}
where $\circ$ denotes the Hadamard (elementwise) power. We call $\boldsymbol\gamma_t$, which gives each firm's share of the new money, the `injection profile'. The vector $\widetilde{\mathbf m}_t$ collects the post-injection balances that finance the orders placed at date $t$. Trade conserves the money stock, $\mathbf 1^\top\mathbf A=\mathbf 1^\top$, and the injection profile sums to one, so the money stock grows by the factor $1+\pi$ each period. Under the normalization $\mathbf 1^\top\mathbf m_0=1$, the money stock is $\mathbf 1^\top\mathbf m_t=(1+\pi)^{t}$. The settlement convention and demand indexing of Section~\ref{subsec:transient_dynamics} carry over: orders placed at date $t$ are financed by $\widetilde{\mathbf m}_t$, settle at date $t+1$, and $\boldsymbol{\mathcal D}_{t+1}=\mathbf m_{t+1}$. Throughout, we call $\pi$ the `rate of inflation'. It is the growth rate of the money stock, and every price grows at $\log(1+\pi)$ on the balanced-growth path of Proposition~\ref{prop:balanced_growth_convergence}.

We write $M_t:=\mathbf 1^\top\mathbf m_t=(1+\pi)^{t}$ for the money stock and $\widehat{\mathbf m}_t:=\mathbf m_t/M_t$ for normalized balances, the shares of the money stock that firms hold. Dividing the balance recursion by $M_{t+1}=(1+\pi)M_t$ gives
\begin{equation}\label{eq:normalised_balance_dynamics}
\widehat{\mathbf m}_{t+1}
=\frac{1}{1+\pi}\,\mathbf A\bigl(\widehat{\mathbf m}_t+\pi\,\boldsymbol\gamma_t\bigr).
\end{equation}
The money stock $M_t$ carries the common trend, and normalized balances $\widehat{\mathbf m}_t$ carry the distribution of money across firms, on which every relative price depends.

The injection profile is set by current balances, which reflect past injections, so the profile is endogenous. The concavity of the injection rule keeps this feedback in check. The injection rule $\mathbf m_t\mapsto\boldsymbol\gamma_t$ contracts Hilbert's projective metric by exactly $\theta$, and mixing through the network contracts it further, so for $\theta<1$ the stationary injection profile $\boldsymbol\gamma_{\mathrm{ub}}$ is unique \citep{bhattathiripad2026}. At $\theta=1$ the injection rule is a projective isometry, but convergence follows directly, since $\boldsymbol\gamma_t=\widehat{\mathbf m}_t$ and primitivity gives $\widehat{\mathbf m}_t\to\mathbf m^*$. For $\theta\in(0,1)$, convergence also requires that forcing by new injections not outrun dissipation through network mixing, and we call this requirement the `mixing--forcing condition'. The mixing--forcing condition holds on every network once the rate of inflation $\pi$ is below a threshold that depends on the propagation matrix $\mathbf A$ and on the injection exponent $\theta$. Below this threshold the injection profile converges to $\boldsymbol\gamma_{\mathrm{ub}}$ from every initial condition \citep[Proposition~5 and Supplementary Material, Corollary~S1.3]{bhattathiripad2026}.\footnote{The threshold is specific to each network. The explicit sufficient conditions of \citet[Supplementary Material, Theorems~S1.2 and~S1.4]{bhattathiripad2026} divide by the smallest entry of a column-stochastic matrix built from $\mathbf A$, which is at most $1/n$, so they are restrictive on large sparse networks. The reconstructed networks of our computational experiments do not meet them at the calibrated rates of inflation. Doubling the simulation horizon in those experiments leaves the long-run averages of the simulated measures of relative-price distortion essentially unchanged.} We maintain $\pi$ below this threshold throughout. A result that needs a smaller $\pi$ states its own restriction.

The convergence of the injection profile carries the distribution of balances with it. The distribution of balances in turn brings output and prices with it.

\begin{proposition}[Balanced-growth convergence under sustained injection]\label{prop:balanced_growth_convergence}
Let $\pi>0$, $\theta\in(0,1)$, and $\varsigma\in(0,1)$, and let the injection profile converge, $\boldsymbol\gamma_t\to\boldsymbol\gamma_{\mathrm{ub}}$. Then normalized balances converge to the unique solution of
\begin{equation}\label{eq:balanced_growth_money_fixed_point}
\widehat{\mathbf m}_{\mathrm{ub}}
=\frac{1}{1+\pi}\,\mathbf A\bigl(\widehat{\mathbf m}_{\mathrm{ub}}+\pi\,\boldsymbol\gamma_{\mathrm{ub}}\bigr),
\qquad\text{equivalently}\qquad
\widehat{\mathbf m}_{\mathrm{ub}}
=\pi\bigl((1+\pi)\mathbf I-\mathbf A\bigr)^{-1}\mathbf A\,\boldsymbol\gamma_{\mathrm{ub}},
\end{equation}
which is strictly positive with $\mathbf 1^\top\widehat{\mathbf m}_{\mathrm{ub}}=1$. Under flexible prices, output converges to the unique vector $\mathbf q_{\mathrm{ub}}$ with
\[
\log\mathbf q_{\mathrm{ub}}
=(\mathbf I-\varsigma\mathbf A^\top)^{-1}\bigl[\varsigma\,\mathbf b_A+\varsigma\bigl(\log\widehat{\mathbf n}_{\mathrm{ub}}-\mathbf A^\top\log\widehat{\mathbf m}_{\mathrm{ub}}\bigr)\bigr],
\qquad
\widehat{\mathbf n}_{\mathrm{ub}}:=\frac{\widehat{\mathbf m}_{\mathrm{ub}}+\pi\boldsymbol\gamma_{\mathrm{ub}}}{1+\pi},
\]
where $\widehat{\mathbf n}_{\mathrm{ub}}$ collects the normalized post-injection balances that finance orders at the fixed point, $\mathbf A\widehat{\mathbf n}_{\mathrm{ub}}=\widehat{\mathbf m}_{\mathrm{ub}}$. Detrended prices $\widehat{\mathbf p}_t:=\mathbf p_t/M_t$ converge to $\widehat p_{i,\mathrm{ub}}=\widehat m_{i,\mathrm{ub}}/q_{i,\mathrm{ub}}$, so every price ultimately grows at the rate of inflation, $\Delta\log p_i^{(t)}\to\log(1+\pi)$ for every $i$.
\end{proposition}

\noindent\textit{Proof in Appendix~\ref{oa:proof_balanced_growth}.}

\noindent The growth of the money stock does not move output. It raises the orders a firm places and the prices at which they settle by the same factor, so this factor cancels from the inputs the firm receives. Output therefore responds only to the distribution of balances across firms. Relative prices converge to $\widehat p_{i,\mathrm{ub}}/\widehat p_{k,\mathrm{ub}}$, which need not equal the equilibrium relative prices $p_i^{*}/p_k^{*}$. Sustained injection can therefore hold relative prices away from equilibrium for as long as it lasts. We call this displacement of relative prices from their equilibrium values the `standing network distortion'.

\section{The Analytical Framework}
\label{sec:framework}

We now derive the relation between the rate of inflation and the displacement of relative prices under sustained injection. We also trace how that displacement builds up over time, starting from the zero-injection equilibrium. The exact economy is nonlinear, because prices and output respond to each other as money moves. We make three analytical reductions, the first of which is that we reduce price dynamics to the dynamics of nominal demand by assuming quantities are fixed. The second, and equally important, reduction is that we approximate the network dynamics by its first two eigenmodes. And we represent these eigenmodes by appropriate functions of the degree vector of the production network. Which in turn gives us each firm's nominal demand as the sum of a permanent trend of money growth proportional to its size and a transitory monetary wave that depends on its position in the network. The three reductions together mean that we are able to write the relative price dynamics of each firm as the ratio of the transitory monetary wave to the permanent monetary trend. What is most insightful of all is that this ratio proves to be a rather simple object: a product of the rate of inflation, the firm's position in the network, and a factor common to all firms.

\subsection{The nominal-demand projection}
\label{subsec:nominal_projection}

A firm's price is the nominal demand it faces divided by its output. Price growth is therefore the growth of nominal demand less the growth of output,
\begin{equation}
\label{eq:price_is_demand}
\Delta\log p_i^{(t)}:=\log p_i^{(t)}-\log p_i^{(t-1)}=\Delta\log\mathcal D_i^{(t)}-\Delta\log q_i^{(t)}.
\end{equation}
The nominal block determines nominal demand on its own, while output responds to balances only with a production lag, as in the recursion~\eqref{eq:zero_injection_real_dynamics}. We therefore assume that quantities are fixed at their equilibrium values, so that prices move one for one with nominal demand. We call this reduction the `nominal-demand projection'.

\begin{assumption}[Nominal-demand projection]\label{assu:small_returns}
Output stays at its zero-injection equilibrium value, $q_i^{(t)}=q_i^{*}$, for every firm $i$ and every date $t$.
\end{assumption}

\noindent Under Assumption~\ref{assu:small_returns} a firm's detrended price moves only with its share of the money stock.\footnote{On the balanced-growth path of Proposition~\ref{prop:balanced_growth_convergence} output is stationary, so there price growth equals the growth of nominal demand even when output responds. Off that path output responds to new money, and the response of output modifies the response of balances without removing its differences across firms, except at isolated values of the returns to scale (Lemmas~\ref{lem:first_order_price} and~\ref{lem:non_degeneracy}). The response of output is small beside the response of balances when the projection ratio $\varkappa:=\varsigma(1-\lambda_2)/(1-\varsigma)$ is small. Contrary to the natural presumption, $\varkappa$ is small when money spreads slowly through the network, so the regime in which relative prices are most displaced is the regime in which Assumption~\ref{assu:small_returns} is most accurate (Appendix~\ref{oa:sec:projection_ratio}).}

We write
\[
\boldsymbol\gamma_0:=\frac{(\mathbf m^{*})^{\circ\theta}}{\mathbf 1^\top(\mathbf m^{*})^{\circ\theta}}
\]
for the injection profile evaluated at the zero-injection equilibrium, and $\boldsymbol\mu:=\partial_\pi\log\widehat{\mathbf m}_{\mathrm{ub}}\big|_{\pi=0}$ for the response of normalized balances on the balanced-growth path to the rate of inflation. We call $\boldsymbol\mu$ the `balance response'. Under Assumption~\ref{assu:small_returns} it is also the first-order response of detrended prices. The balance response vanishes at $\theta=1$, where new money is distributed like the stationary balances. It differs across firms unless one round of trade wipes out the tilt of the injection, that is, unless $\mathbf A(\boldsymbol\gamma_0-\mathbf m^{*})=\mathbf 0$ (Lemma~\ref{lem:first_order_price}).

\subsection{The Perron--Frobenius decomposition and two-eigenmode approximation}
\label{subsec:pf_decomposition}

Under the nominal-demand projection, prices follow nominal demand, and nominal demand obeys the balance recursion~\eqref{eq:demand_dynamics}. A monetary injection traveling through a production network can be traced firm by firm, as money spent by its recipients on their suppliers and by the suppliers on their own suppliers. The injection can also be traced in the eigenmodes of the propagation matrix $\mathbf A$. An eigenmode is a pattern of balances that a round of trade reproduces in the same proportions across firms, and its eigenvalue records how the eigenmode changes with each round. The Perron vector is reproduced exactly by trade and equals the stationary balances $\mathbf m^{*}$, for it is the eigenvector of $\lambda_1=1$. Every other eigenvalue has modulus below one, so its eigenmode is transitory. Our question is one of propagation: what remains of uneven monetary incidence after repeated rounds of trade, and what prices do while it decays. For such questions the modal representation is the natural one \citep{LevinPeres2017,golubjackson2012}. The modal representation takes its simplest form when the propagation matrix is diagonalizable.

\begin{assumption}[Diagonalizable propagation]
\label{assu:diagonalisable}
The propagation matrix $\mathbf A$ is diagonalizable, with a complete biorthogonal system of right and left eigenvectors $\{(\mathbf v_j,\mathbf u_j)\}_{j=1}^{n}$ normalized so that
\[
\mathbf u_j^\top\mathbf v_k=\mathbf 1\{j=k\}.
\]
\end{assumption}

\noindent Under Assumption~\ref{assu:diagonalisable} every distribution of balances splits into the $n$ eigenmodes, and trade scales each component of an injection separately.

Trade acts linearly on balances, so the effects of injections from different dates add up. A period-$t$ injection
\[
\boldsymbol\epsilon_t
:=\pi(\mathbf 1^\top\mathbf m_{t-1})\,\boldsymbol\gamma_{t-1}
\]
reaches horizon $T$ as $\mathbf A^{T-t+1}\boldsymbol\epsilon_t$: it is propagated once on entry and $T-t$ times thereafter. We start the economy at the stationary balances, $\mathbf m_0=\mathbf m^{*}$, so every later departure from them comes from these injections.

Under Assumptions~\ref{assu:spectral_gap} and~\ref{assu:diagonalisable}, $h$ rounds of trade have the exact spectral representation
\begin{equation}
\label{eq:spectral_decomposition}
\mathbf A^h
=\mathbf v_1\mathbf u_1^\top
+\lambda_2^h\mathbf v_2\mathbf u_2^\top
+\sum_{j\ge3}\lambda_j^h\mathbf v_j\mathbf u_j^\top,
\qquad h\ge1,
\end{equation}
where $|\lambda_j|\le|\lambda_3|<\lambda_2<1$ for $j\ge3$. The two-eigenmode approximation omits the final sum in~\eqref{eq:spectral_decomposition} and retains one permanent component and the slowest transitory component, the second eigenmode:
\begin{equation}
\label{eq:pf_decomposition}
\mathbf A^{T-t+1}\boldsymbol\epsilon_t
\;\approx\;
\underbrace{\mathbf v_1
\bigl(\mathbf u_1^\top\boldsymbol\epsilon_t\bigr)}
_{\text{permanent}}
+
\underbrace{\lambda_2^{T-t+1}\mathbf v_2
\bigl(\mathbf u_2^\top\boldsymbol\epsilon_t\bigr)}
_{\text{transitory}}.
\end{equation}
The spectral separation $|\lambda_3|<\lambda_2$ makes the omitted eigenmodes decay faster than the second eigenmode, injection vintage by injection vintage.\footnote{After $h$ rounds of trade the omitted eigenmodes contribute at most $\operatorname{cond}(\mathbf V)\,|\lambda_3|^{h}\lVert\boldsymbol\epsilon_t\rVert$, with $\mathbf V$ the eigenvector matrix. Appendix~\ref{oa:sec:perron_degree_bound} bounds their accumulated contribution relative to that of the second eigenmode. The ratio of the two accumulated contributions is small when the spectral gap $1-\lambda_2$ is small beside $1-|\lambda_3|$, the injection loads substantially on the second eigenmode, and the eigenbasis is well conditioned.} Because $\lambda_2>0$, the contribution of each injection vintage to the second eigenmode keeps its sign as it ages.

Summing~\eqref{eq:pf_decomposition} over injection vintages through horizon $T$ gives, within the two-eigenmode approximation,
\begin{equation}
\label{eq:demand_split}
\begin{aligned}
\mathcal D_i^{(T)}
&\approx P_i^{(T)}+S_i^{(T)},\\[2pt]
P_i^{(T)}
&=(\mathbf v_1)_i\,\mathbf 1^\top\mathbf m_T,\\[2pt]
S_i^{(T)}
&=\sum_{t=1}^{T}
\lambda_2^{T-t+1}(\mathbf v_2)_i\,
\mathbf u_2^\top\boldsymbol\epsilon_t.
\end{aligned}
\end{equation}
We call $P_i^{(T)}$ the `permanent trend' and $S_i^{(T)}$ the `transitory wave'. The permanent trend carries the growth of the money stock along the Perron vector $\mathbf v_1=\mathbf m^{*}$. The transitory wave records what survives of past injections in the second eigenmode, one injection vintage per date, each decaying by the factor $\lambda_2$ per period of age.

\subsection{The degree representation}
\label{subsec:degree_closure}

The eigenvectors $\mathbf v_1$, $\mathbf u_2$, and $\mathbf v_2$ decide how an injection moves each firm's nominal demand, and each is an implicit function of the whole propagation matrix $\mathbf A$. The degree representation replaces them by functions of degree, so nominal demand comes to depend on the distribution of degrees and the pattern of who trades with whom. Normalized degree stands in for the Perron vector, and centered functions of degree stand in for the second eigenmode. The degree representation also replaces empirical degree moments by their population counterparts, which ties the distortion of relative prices to the network primitives. The three approximations of the degree representation are distinct, and Appendix~\ref{appendix:mathematics} and Appendix~\ref{oa:sec:perron_degree_bound} bound or record each separately.

We characterize the degrees of firms and the division of their spending with Assumptions~\ref{assu:degree_structure} and~\ref{assu:degree_distribution}, the first of which ties each firm's spending to its degree and the second fixes how unequal firms are in degree.

\begin{assumption}[Degree symmetry and even expenditure shares]\label{assu:degree_structure}
Each firm has as many customers as suppliers and divides its input expenditure evenly across its suppliers. We write $d_i$ for the common number of customers and suppliers of firm $i$, its degree, so that $a_{ij}=1/d_j$ for every supplier $i$ of firm $j$.
\end{assumption}

\begin{assumption}[Degree distribution]\label{assu:degree_distribution}
Degrees follow a power law truncated above, with survival function $\Pr(D>d)\propto d^{-\alpha}$ for $d\in[1,d_{\max}]$ and $\alpha>1$.
\end{assumption}

\noindent The two assumptions together guarantee that firm sizes, measured by stationary balances, follow a power law. This is because under Assumption~\ref{assu:degree_structure} the degree vector reproduces itself under trade, $\mathbf A\mathbf d=\mathbf d$, so the stationary balances, which make up the Perron vector, are the normalized degrees,
\[
m_i^{*}=\frac{d_i}{\sum_jd_j}=c_m d_i,
\qquad
c_m:=\frac1{\sum_jd_j}.
\]
We call $\alpha$ the `tail exponent', for it sets how unequal firms are in size as well as in degree. The truncation at $d_{\max}$ keeps every moment of degree finite. Realized networks meet degree symmetry and even expenditure shares only approximately, and Appendix~\ref{oa:sec:perron_degree_bound} bounds the resulting Perron--degree error.\footnote{Under a neutral-network benchmark and bounded degree-preserving perturbations, a Stewart--Sun bound holds on the angle between degree and the leading eigenvector, which depends on the perturbation size, spectral separation, and eigenbasis conditioning \citep{puravankara2025}.}

For a realized degree sequence, we write $\E_n[H(d)]:=n^{-1}\sum_{i=1}^{n}H(d_i)$ for the equally weighted empirical moment and $\E_{\boldsymbol\gamma_t}[H(d)]:=\sum_{i=1}^{n}\gamma_i^{(t)}H(d_i)$ for the moment that weights firms by their share of the date-$t$ injection. We write $\E[H(d)]$ for population moments under the degree law of Assumption~\ref{assu:degree_distribution}.

We characterize who trades with whom with Assumption~\ref{assu:assortativity}.

\begin{assumption}[Assortativity]\label{assu:assortativity}
The mean degree $\bar d_i$ of firm $i$'s suppliers scales as $\bar d_i\propto d_i^{\nu}$, with $\nu\in(-1,0)$. Under degree symmetry, the mean degree of firm $i$'s customers scales in the same way.
\end{assumption}

\noindent We call $\nu$ the `assortativity exponent'. A negative $\nu$ makes the network disassortative: high-degree firms trade disproportionately with low-degree firms. A positive $\nu$ makes the network assortative: high-degree firms trade disproportionately with other high-degree firms. At $\nu=0$ the network is degree-neutral. The range $\nu\in(-1,0)$ covers the disassortativity documented for production networks \citep{bachilieri2023topology}.

Unlike the Perron vector, the second eigenmode has no exact degree benchmark. We instead use the assortativity relation of Assumption~\ref{assu:assortativity}. In the mean-field approximation, one round of trade maps a degree tilt into the reading shape $d^{\nu}$. A second round of trade maps the reading shape into the receiving shape $d^{\nu^2}$. Centering the receiving shape gives firm $i$'s excess degree,
\[
\delta_i
:=d_i^{\nu^2}-\E_n[d^{\nu^2}],
\]
while the reading shape is centered as $d_i^{\nu}-\E_n[d^{1+\nu}]/\E_n[d]$. Both centerings hold exactly on the realized degree sequence, $\sum_i\delta_i=0$ and $\sum_i d_i\bigl(d_i^{\nu}-\E_n[d^{1+\nu}]/\E_n[d]\bigr)=0$. Define the threshold degree as $d^{*}:=(\E_n[d^{\nu^2}])^{1/\nu^2}$, so that $\delta_i=d_i^{\nu^2}-(d^{*})^{\nu^2}$. A firm's excess degree is positive when its degree exceeds the threshold degree and negative when it falls short. The second eigenmode reaches firm $i$ in proportion to its excess degree, and its response to an injection profile is represented as
\begin{equation}
\label{eq:degree_response_proxy}
\bigl(\mathbf v_2\mathbf u_2^\top\boldsymbol\gamma_t\bigr)_i
\;\approx\;
\kappa\,\delta_i
\left(
\E_{\boldsymbol\gamma_t}[d^{\nu}]
-\frac{\E_n[d^{1+\nu}]}{\E_n[d]}
\right),
\end{equation}
where $\kappa>0$ is the amplitude of the second-eigenmode response. This amplitude scales every magnitude of the distortion of relative prices. The centered reading shape and the excess degree describe who reads the injection and who receives it. The sign of the second-eigenmode response is a result of the model, for the first round of trade orients it.\footnote{The projector $\mathbf v_2\mathbf u_2^\top$ is unchanged when both eigenvectors switch sign, so only propagation can orient the second-eigenmode response.} New money lands disproportionately on small firms, which spend it on their suppliers, and under disassortative mixing those suppliers are larger, so the second-eigenmode response tilts positively along the degree coordinate (Lemma~\ref{lemma:newest_dominates}). Note that the degree representation of the second eigenmode is a mean-field approximation (Lemma~\ref{oa:lem:eigenvector_proxies}).

\subsection{The master equation}
\label{subsec:master_equation}

The effects of injections from different dates add up, so a firm's nominal demand is its stationary share of the growing money stock plus the second-eigenmode responses~\eqref{eq:degree_response_proxy} of all earlier injection vintages.

\begin{lemma}[Nominal demand separates into a permanent trend and a transitory wave]
\label{lemma:exposure}
Within the two-eigenmode degree representation, and starting from the stationary balances, $\mathbf m_0=\mathbf m^{*}$, nominal demand at horizon $T$ satisfies
\begin{equation}
\label{eq:master_demand}
\mathcal D_i^{(T)}
\;\approx\;
P_i^{(T)}+S_i^{(T)},
\qquad
P_i^{(T)}=m_i^{*}(1+\pi)^T,
\qquad
S_i^{(T)}=\kappa\,\pi(1+\pi)^T\delta_i\mathcal X_T.
\end{equation}
The excess degree $\delta_i$ is firm specific and constant over time. The propagation kernel
\begin{equation}
\label{eq:master_kernel}
\mathcal X_T
:=\sum_{t=0}^{T-1}
\xi_t\,\lambda_2^{T-t}(1+\pi)^{-(T-t)}
\end{equation}
is common to all firms and accumulates the injection misalignments
\begin{equation}
\label{eq:Ct_def}
\xi_t
:=\E_{\boldsymbol\gamma_t}[d^{\nu}]
-\frac{\E_n[d^{1+\nu}]}{\E_n[d]}.
\end{equation}
\end{lemma}

\noindent\textit{Proof in Appendix~\ref{app:money_injection_and_transient}.}

Because excess degrees sum to zero, $\sum_i\delta_i=0$, the transitory wave redistributes nominal demand across firms and leaves the money stock unchanged.

The injection rule enters through the injection misalignment $\xi_t$. The second term of $\xi_t$ in~\eqref{eq:Ct_def} is the balance-weighted mean of $d^{\nu}$, so $\xi_t$ measures how far the date-$t$ injection departs from the stationary flow along the degree coordinate of the second eigenmode. Under the small-firm tilt of the injection rule, $\xi_t>0$ at every date, so successive injection vintages load on the second eigenmode with a common sign and reinforce one another. And since normalized balances stay close to $\mathbf m^{*}$ at every date, the injection misalignment barely moves along the transition, and the propagation kernel is close to a geometric sum that rises with the horizon (Lemma~\ref{lem:Ct_positive}).

In the steady state the injection profile is the stationary injection profile $\boldsymbol\gamma_{\mathrm{ub}}$ of Section~\ref{subsec:monetary_process}. At a given positive spectral gap, the fixed point~\eqref{eq:balanced_growth_money_fixed_point} places the limit $\widehat{\mathbf m}_{\mathrm{ub}}$ of normalized balances within $O(\pi)$ of the stationary balances $\mathbf m^{*}$, which are proportional to degree, so $\boldsymbol\gamma_{\mathrm{ub}}\propto d^{\theta}+O(\pi)$. The stationary misalignment is then
\begin{equation}
\label{eq:Cub_empirical}
\xi_{\mathrm{ub}}
=
\frac{\E_n[d^{\theta+\nu}]}{\E_n[d^\theta]}
-
\frac{\E_n[d^{1+\nu}]}{\E_n[d]}
+O(\pi),
\end{equation}
and the propagation kernel converges to
\begin{equation}
\label{eq:X_infty_exact}
\mathcal X_\infty
=\xi_{\mathrm{ub}}\frac{\lambda_2}{1+\pi-\lambda_2}
=\xi_{\mathrm{ub}}\frac{\lambda_2}{1-\lambda_2}+O(\pi).
\end{equation}

Population moments make the anatomy of the stationary misalignment explicit. By the law of large numbers, empirical moments converge to their population counterparts under the degree law, and at a large cutoff the truncated Pareto moment approaches $\E[d^{k}]=\alpha/(\alpha-k)$ for $k<\alpha$. Under that substitution the stationary misalignment is
\begin{equation}
\label{eq:Cub_pareto_exact}
\xi_{\mathrm{ub}}
=
\frac{-\nu(1-\theta)}
{(\alpha-\theta-\nu)(\alpha-1-\nu)}
+O(\pi).
\end{equation}
Expanding the leading term of~\eqref{eq:Cub_pareto_exact} in $\nu$ gives $-\nu(1-\theta)/[(\alpha-1)(\alpha-\theta)]+O(\nu^2)$, which isolates the two ingredients of the stationary misalignment. The factor $1-\theta$ sets how far new money departs from proportionality on arrival, and the assortativity exponent $\nu$ sets how strongly that departure loads on the second eigenmode.

There are two cases in which the stationary misalignment is zero, and they differ in status. At $\theta=1$ injection is proportional to balances and creates no relative nominal displacement, an exact property of the balance recursion. At $\nu=0$ both centered degree shapes vanish, so the degree representation predicts no displacement of relative prices. An exact degree-neutral network can still carry a second-eigenmode response, so this zero is a property of the degree representation (Appendix~\ref{oa:subsec:eigenvector_proxies}).

Dividing the transitory wave by the permanent trend gives the paper's master equation, in which a firm's place in the network enters through a single coefficient. Define firm $i$'s normalized excess degree as its excess degree per unit of degree, $\widehat\delta_i:=\delta_i/d_i$, and its Cantillon exposure as
\begin{equation}
\label{eq:cantillon_exposure}
\mathcal C_i
:=\kappa\,\frac{\delta_i}{m_i^{*}}
=\frac{\kappa}{c_m}\,\widehat\delta_i.
\end{equation}
The exponent $\nu^{2}$ is the only place the pattern of who trades with whom enters the Cantillon exposure. The assortativity exponent $\nu$ enters squared, for the second-eigenmode response reads the assortativity relation once on the way in and once on the way out. Injections that depart from proportionality are the Cantillon impulse of this economy. The injection misalignment $\xi_t$ measures that impulse date by date, and the propagation kernel $\mathcal X_T$ accumulates what survives of past impulses. The Cantillon exposure $\mathcal C_i$ is the rate at which the accumulated impulse displaces firm $i$'s nominal demand and, under the nominal-demand projection, its price. This rate differs from the firm's share of the new money, $\gamma_i^{(t)}$. And it is here that a Cantillon economy parts from the classical one \citep{Cantillon1755}. What matters for a firm is not how much of the new money it receives first, but how much of it the network brings to its door through its customers.

We call $\kappa/c_m$ the `intensive response scale'.\footnote{Under the scaling maintained in Lemma~\ref{oa:lem:eigenvector_proxies}, $\kappa/c_m$ stays bounded away from zero and infinity as the network grows, so the Cantillon exposure is intensive.} Being positive, $\kappa/c_m$ determines no sign, so every sign and threshold in the analysis is a property of the normalized excess degree $\widehat\delta_i$, which is negative below the threshold degree $d^{*}$ and positive above it. The transient and limiting level distortions are
\begin{equation}
\label{eq:level_distortion}
\ell_i^{(T)}
:=\frac{S_i^{(T)}}{P_i^{(T)}}
\;\approx\;
\pi\,\mathcal C_i\mathcal X_T,
\qquad
\ell_i^{(\infty)}
\;\approx\;
\pi\,\mathcal C_i\mathcal X_\infty.
\end{equation}
The level distortion removes the common trend $(1+\pi)^T$ of nominal demand and moves with the propagation kernel during the transition. Once the propagation kernel has converged, every price grows at the rate of inflation. But the limiting level distortions $\ell_i^{(\infty)}$ are the standing network distortion, a displacement of relative prices that remains for as long as new money continues to arrive unevenly.

\section{Inflation under Flexible Prices}
\label{sec:flexible}

We now trace the relative-price distortion that inflation generates when prices are fully flexible. We find that during the transition prices grow faster than inflation at firms above a threshold size and more slowly below it. We also find that in the steady state sustained injection holds relative prices away from equilibrium, by an amount proportional to the rate of inflation. The distortion is larger the more persistent the network, the more uneven the injection, and the more large firms trade with small ones.

\subsection{Firm size and transient price growth}
\label{subsec:transient_growth}

Every price grows at the rate of inflation on the balanced-growth path (Proposition~\ref{prop:balanced_growth_convergence}). But along the transition path prices grow at different rates. This is because new money enters unevenly and takes several rounds of trade to spread. Firm size decides which prices run ahead of inflation. A firm's size is its stationary balance, which is proportional to its degree. New money lands disproportionately on small firms. They spend it on their suppliers, and under disassortative mixing those suppliers are larger. While the propagation kernel builds, nominal demand therefore grows faster than the money stock at large firms and more slowly at small ones. But the very largest firms sell to a representative cross-section of the economy. The tilt an injection carries is diluted across their many links, so their nominal demand grows close to the rate of inflation. Under flexible prices each price follows its nominal demand. The Cantillon exposure~\eqref{eq:cantillon_exposure} records this pattern. It is negative below the threshold degree $d^{*}$ and positive above it, and it falls back toward zero among the very largest firms (Corollary~\ref{oa:cor:departure_peak}).

\begin{proposition}[Transient price growth varies with firm size]
\label{prop:size_price_change}
Let prices be flexible and let Assumptions~\mbox{\ref{assu:primitivity}--\ref{assu:degree_structure}} and~\ref{assu:assortativity} hold. Within the two-eigenmode degree representation, each firm's price grows at
\[
\Delta\log p_i^{(T)}
\;=\;
\pi
+\pi\,\bigl(\mathcal X_T-\mathcal X_{T-1}\bigr)\mathcal C_i
+o(\pi).
\]
If the propagation kernel is building, $\mathcal X_T>\mathcal X_{T-1}$, as it is early in the transition, price growth exceeds inflation at firms above the threshold degree $d^{*}$ and falls short of it below.
\end{proposition}

\noindent\textit{Proof in Appendix~\ref{app:proof_thm1}.}

The transient departure from inflation falls on the many small and mid-sized firms and not on the giants. Large firms carry much of the transitory wave in absolute terms, but little of it relative to their size. Below the threshold degree the shortfall of price growth from inflation is deepest among the smallest firms. This is because their customers are the large firms that receive the least new money per unit of size.

\subsection{Measuring relative-price distortion}
\label{subsec:distortion_flex}

Scarcities alone set relative prices at the zero-injection equilibrium, so a firm's relative price there is the benchmark for its relative price under inflation. A relative price needs a numeraire, and a firm at the threshold degree $d^{*}$ is the natural one. It has zero excess degree, so the transitory wave passes it by and its price keeps pace with the money stock, $p_{d^{*}}^{(t)}=p_{d^{*}}^{*}M_t$. We assume that the network contains such a firm, and we call it the `numeraire firm'. Define firm $i$'s log displacement as
\[
g_i^{(t)}:=\log\frac{p_i^{(t)}/M_t}{p_i^{*}}=\log\frac{p_i^{(t)}/p_{d^{*}}^{(t)}}{p_i^{*}/p_{d^{*}}^{*}},
\]
the log of its relative price in terms of the numeraire firm, measured from its equilibrium value. Under Assumption~\ref{assu:small_returns} the master equation~\eqref{eq:level_distortion} gives $g_i^{(T)}=\log\bigl(1+\ell_i^{(T)}\bigr)=\pi\,\mathcal C_i\,\mathcal X_T+O(\pi^{2})$. A distortion of relative prices is a dispersion of these displacements. We weight firms by their stationary balances $m_i^{*}$, their shares of the money stock at the zero-injection equilibrium and, since $m_i^{*}=p_i^{*}q_i^{*}$, of revenue. Under these weights the displacements average to zero (Lemma~\ref{oa:lem:balance_centering}). This is because a firm's displacement is its transitory wave relative to its permanent trend, which is proportional to its stationary balance, and the transitory wave only redistributes nominal demand across firms. Which means that the numeraire firm sits at the center of the distribution of relative prices. We measure the distortion in two ways.

\begin{definition}[Relative price gap]
\label{def:relative_price_gap}
The date-$t$ relative price gap is the balance-weighted standard deviation of the log displacements,
\[
\omega_t
\;:=\;
\min_{c\in\R}\Bigl(\sum_{i=1}^{n}m_i^{*}\bigl(g_i^{(t)}-c\bigr)^{2}\Bigr)^{1/2}
\;=\;\Bigl(\Var_{\mathbf m^{*}}\bigl(\mathbf g^{(t)}\bigr)\Bigr)^{1/2},
\qquad
\omega
\;:=\;
\lim_{T\to\infty}\frac1T\sum_{t=1}^{T}\omega_t.
\]
\end{definition}

\begin{definition}[Relative-price entropy]
\label{def:relative_price_entropy}
We write $\widehat r_i^{(t)}:=p_i^{(t)}/\sum_j p_j^{(t)}$ and $\widehat r_i^{*}:=p_i^{*}/\sum_j p_j^{*}$ for the current and equilibrium price shares. Date-$t$ relative-price entropy is the Kullback--Leibler divergence of the current price shares from the equilibrium price shares,
\[
\psi_t
\;:=\;
\sum_{i=1}^n\widehat r_i^{(t)}
\log\frac{\widehat r_i^{(t)}}{\widehat r_i^{*}},
\qquad
\psi
\;:=\;
\lim_{T\to\infty}\frac1T\sum_{t=1}^{T}\psi_t.
\]
\end{definition}

The relative price gap measures how far relative prices have moved from equilibrium, as a standard deviation of the log displacements. Relative-price entropy measures how far the distribution of price shares has been reshaped, for it is nothing but the \citeauthor{theil1967economics} index applied to price shares.\footnote{Neither measure depends on which price serves as the numeraire. The minimization over a common rescaling $c$ removes whatever all displacements share, and the normalization into shares does the same for relative-price entropy. Relative-price entropy is nonnegative and is zero only when the current and equilibrium price shares coincide (Gibbs' inequality).} Both measures of relative-price distortion record how far sustained injection holds relative prices from the zero-injection equilibrium of Proposition~\ref{prop:equilibrium_existence}.\footnote{Because the degree sequence is a draw from the degree law, the relative price gap and relative-price entropy are random. Conditional on the spectral statistics, they concentrate around their population values in large economies (Lemma~\ref{oa:lem:gaussian_omega_psi}).}

\subsection{Relative-price distortion under flexible prices}
\label{subsec:flexible_distortion}

Once the propagation kernel has converged, every price grows at the rate of inflation (Proposition~\ref{prop:size_price_change}). The question is whether relative prices then settle at their equilibrium values, as they do under proportional injection. They do not. New money keeps entering unevenly and travels through the network at a finite speed, so each firm's price settles where the uneven flow of money leaves its nominal demand, away from where scarcities alone would place it. The displacement of each price is proportional to the rate of inflation, and so is the relative price gap, for it is nothing but the standard deviation of these displacements. Relative-price entropy is a divergence, and a divergence is flat where the current and equilibrium price shares coincide. Which means that relative-price entropy is proportional to the square of the rate of inflation.

\begin{theorem}[Sustained inflation distorts relative prices under flexible prices]
\label{theorem:price_deviation}
Let prices be flexible, let Assumptions~\ref{assu:primitivity}--\ref{assu:small_returns} hold, let $\theta\in(0,1)$, and let $\mathbf A(\boldsymbol\gamma_0-\mathbf m^{*})\neq\mathbf 0$. The steady-state relative price gap and relative-price entropy are then
\[
\omega\;=\;\pi\,\digamma_\omega\;+\;o(\pi),
\qquad
\psi\;=\;\pi^{2}\,\digamma_\psi\;+\;o(\pi^{2}),
\]
with strictly positive coefficients
\[
\digamma_\omega
\;=\;
\Bigl(\Var_{\mathbf m^{*}}\bigl(\boldsymbol\mu\bigr)\Bigr)^{1/2},
\qquad
\digamma_\psi
\;=\;
\frac12\Var_{\widehat{\mathbf r}^{*}}\bigl(\boldsymbol\mu\bigr).
\]
If Assumptions~\ref{assu:diagonalisable}, \ref{assu:degree_structure}, and~\ref{assu:assortativity} also hold, the two coefficients $\digamma_\omega$ and $\digamma_\psi$ are represented within the two-eigenmode degree representation by
\[
\digamma_\omega
\;=\;
\frac{\kappa}{c_m}\,\xi_{\mathrm{ub}}\,\frac{\lambda_2}{1-\lambda_2}
\Bigl(\Var_{\mathbf m^{*}}\bigl(\widehat{\boldsymbol\delta}\bigr)\Bigr)^{1/2},
\qquad
\digamma_\psi
\;=\;
\frac12\left(\xi_{\mathrm{ub}}\,\frac{\lambda_2}{1-\lambda_2}\right)^{\!2}
\Var_{\widehat{\mathbf r}^{*}}\bigl(\boldsymbol{\mathcal C}\bigr).
\]
\end{theorem}

\noindent\textit{Proof in Appendix~\ref{app:proof_thm2}.}

The distortion arises even though every price is free to adjust and every market clears.\footnote{In the full model, where output also responds to new money, the two coefficients carry the balance response net of the response of output (Lemma~\ref{lem:first_order_price}). They stay strictly positive except at isolated values of the returns to scale (Corollary~\ref{coro:full_model_distortion}).} Halving the rate of inflation halves the relative price gap and quarters relative-price entropy (Corollary~\ref{coro:inflation_elasticity_distortion}). No positive rate of sustained non-proportional injection is neutral for relative prices. Stopping injections lets the distortion dissipate, and resuming them rebuilds it.

The degree representation in Theorem~\ref{theorem:price_deviation} shows what the distortion is made of. The two coefficients $\digamma_\omega$ and $\digamma_\psi$ each multiply three terms. The stationary misalignment $\xi_{\mathrm{ub}}$ measures how unevenly new money arrives relative to the stationary flow of money. Network persistence $\lambda_2/(1-\lambda_2)$ measures how much of an uneven injection the network carries forward, for it is nothing but the sum $\lambda_2+\lambda_2^{2}+\cdots$ of what remains of the injection's transitory part after each later round of trade. The dispersion of the Cantillon exposure measures how differently firms are placed to receive what survives. The first two terms are common to all firms, and their product is the limit of the propagation kernel. The relative price gap weights the dispersion of the Cantillon exposure by stationary balances, and relative-price entropy weights it by equilibrium price shares. The equilibrium price shares depend in turn on the elasticity $\vartheta$ of equilibrium output with respect to degree, which we call the `output--degree exponent'.\footnote{Within the degree representation the output--degree exponent is set by the returns to scale and the assortativity exponent, and it lies below one half (Lemma~\ref{oa:lem:output_degree}). Firm-level revenue data cannot identify it, for a firm's revenue is its stationary balance, which is proportional to its degree whatever the value of $\vartheta$.}

\subsection{Comparative statics of the distortion}
\label{subsec:comparative_statics}

The three terms in the coefficients of Theorem~\ref{theorem:price_deviation} depend on how new money is injected and on how the network relays it. We now ask which economies are the most distorted. The distortion is largest where new money is injected most unevenly, where the network absorbs it most slowly, and where large firms trade most with small ones. We measure these three features by the injection exponent $\theta$, the second eigenvalue $\lambda_2$, and the assortativity exponent $\nu$. There are two ways in which the network shapes the dispersion of the Cantillon exposure, through how unequal firms are in size and through who trades with whom. Under the degree law of Assumption~\ref{assu:degree_distribution}, the tail exponent $\alpha$ summarizes the first and the assortativity exponent $\nu$ the second (Appendix~\ref{app:proof_thm2}). We hold the tail exponent fixed.\footnote{The tail exponent enters both the dispersion of the Cantillon exposure and, for relative-price entropy, the equilibrium price shares, so the sign of its net effect depends on the calibration (Appendix~\ref{app:proof_comparative_statics}).}

\begin{proposition}[Network comparative statics]
\label{prop:comparative_statics}
Let prices be flexible, let Assumptions~\ref{assu:primitivity}--\ref{assu:assortativity} hold, and let $\theta\in(0,1)$. Hold the intensive response scale $\kappa/c_m$, the tail exponent $\alpha$, and the output--degree exponent $\vartheta$ fixed. Within the two-eigenmode degree representation, the leading terms of the steady-state relative price gap $\omega$ and relative-price entropy $\psi$ then satisfy
\[
\begin{aligned}
\partial_{\nu}\omega&<0, &\qquad \partial_{\theta}\omega&<0, &\qquad \partial_{\lambda_2}\omega&>0,\\[2pt]
\partial_{\nu}\psi&<0, &\qquad \partial_{\theta}\psi&<0, &\qquad \partial_{\lambda_2}\psi&>0.
\end{aligned}
\]
\end{proposition}

\noindent\textit{Proof in Appendix~\ref{app:proof_comparative_statics}.}

An injection proportional to balances is distributed like the stationary balances and leaves nothing uneven for trade to propagate, so the stationary misalignment vanishes at $\theta=1$. Lowering $\theta$ directs a larger share of each injection toward firms with smaller balances, and the network propagates the part that arrives out of proportion. The injection exponent acts only through the stationary misalignment, since the dispersion of the Cantillon exposure does not depend on it.

A smaller spectral gap $1-\lambda_2$ lets each uneven injection decay more slowly, so successive injections accumulate before they dissipate. Note that nothing forbids the persistence of a network, $\lambda_2/(1-\lambda_2)$, from varying with its size.\footnote{Many growth processes change how a network is knit together as it grows. Some produce communities of firms that trade far more within than across one another, and money then leaves each community slowly. Others lengthen supply chains as the division of labor deepens, and money then takes more rounds of trade to reach every firm. The degree law of a realized network can also change with its size, for in a power-law network the largest hub grows with the number of firms, and with it the heterogeneity of firms (Appendix~\ref{subsec:sim_size}).} But the stationary misalignment and the dispersion of the Cantillon exposure are set by the injection and by the degree law, and not by the number of firms. Which means that it is not the number of firms as such that distorts relative prices, but the slowness with which the network absorbs new money.

Disassortative mixing raises both measures of relative-price distortion on the whole range of the assortativity exponent. A more negative $\nu$ makes links cross size classes more sharply, so an injection tilted toward small firms is relayed more strongly toward large firms in the next round of trade. Near degree neutrality this raises the stationary misalignment. And since stronger disassortativity also widens the spread of exposures across firms, the dispersion of the Cantillon exposure rises throughout the range (Appendix~\ref{app:proof_comparative_statics}). The effect of disassortative mixing fades as injection approaches proportionality and vanishes at $\theta=1$.

\section{Inflation under Sticky Prices}
\label{sec:sticky}

We now derive the distortion of relative prices when prices are sticky. We use a state-dependent rule to update prices. More specifically, we use a variant of state-dependent pricing where the state of each firm depends on its network position, and therefore, we call it `local state-dependent pricing'. A firm keeps its posted price until it resets it, and the probability that it resets rises with the gap between the price it posts and the price that would clear its market. Naturally, any pricing rule apportions the response to an increase in inflation to a change in the size of price change at every reset and a change in the frequency of resets. Local state-dependent pricing is not different in this respect. But what it does is generate considerable relative-price distortion with very weak to no relation between the size of price change and the rate of inflation. Which means that the size of price change reveals almost nothing about the distortion that inflation generates.

\subsection{State-dependent price adjustment}
\label{subsec:sticky_prices}

Under flexible prices every firm trades at the price that clears its market. But under sticky prices a firm trades at the price it last posted, which in general differs from the price that would clear its market. We write $\bar p_i^{(t)}$ for the price firm $i$ posts at date $t$, and $p_i^{(t)}$ for the price that would clear its market, which we define exactly as under flexible prices and call the `market-clearing price'. From here on we mark a posted value with a bar and leave its market-clearing counterpart unadorned. In the sticky-price regime we compute the measures of relative-price distortion at posted prices, so the log displacement of Definition~\ref{def:relative_price_gap} becomes $\bar g_i^{(t)}:=\log[(\bar p_i^{(t)}/M_t)/p_i^{*}]$.\footnote{Here we divide by the market-clearing price of the numeraire firm, which keeps pace with the money stock, and not by its posted price. Otherwise the relative price of every firm would move whenever the numeraire firm resets its own price.}

The timing is as follows. At the start of date $t$, before trade, firm $i$ decides whether to reset its price. We record this decision in the `price reset indicator' $I_i^{(t)}$, which equals one if the firm resets and zero if it does not. And we measure the age of a price after this decision, so $u_i^{(t)}$ is the age of the price at which the firm trades at date $t$. If the firm resets, the age of its price falls to zero, and if it does not, the age rises by one period. We call the dates over which a firm keeps one posted price a `price spell', and so a firm that resets at date $t$ ends a price spell of length $u_i^{(t-1)}+1$. When a firm resets, it posts the price that clears its market that day. A posted price is therefore always the market-clearing price of the date at which it was last set, and we call that date the `vintage' of the posted price,
\[
\bar p_i^{(t)}=p_i^{(t-u_i^{(t)})}.
\]

Money is settled exactly as under flexible prices, and the posted price decides only how much a seller delivers. Buyer $j$ commits $a_{ij}\widetilde m_j^{(t-1)}$ to seller $i$, so seller $i$ faces nominal demand $\mathcal D_i^{(t)}:=\sum_j a_{ij}\widetilde m_j^{(t-1)}$. The price that would clear seller $i$'s current output is $p_i^{(t)}:=\mathcal D_i^{(t)}/q_i^{(t)}$, and the seller delivers $a_{ij}\widetilde m_j^{(t-1)}/\max\{\bar p_i^{(t)},p_i^{(t)}\}$ to buyer $j$. If the posted price is at or below the market-clearing price, buyers ask for more than the seller has, and the seller divides its whole output among them in proportion to what they committed. If the posted price is above the market-clearing price, each buyer receives what its commitment buys at the posted price, and the seller discards the rest of its output.\footnote{The analytical economy has no storage. The computational economy carries unsold output forward as inventory (Appendix~\ref{oa:proof_sticky_convergence}).} In both cases the buyer pays what it committed, so seller $i$ receives $\mathcal D_i^{(t)}$, and money still moves by
\begin{equation}
\label{eq:sticky_money_recursion}
\mathbf m_{t+1}=\mathbf A\,\widetilde{\mathbf m}_t .
\end{equation}

The nominal side of the economy is therefore exactly what it was under flexible prices, and sticky prices act only on deliveries and, through deliveries, on output (Lemma~\ref{oa:lem:rationing_flexible}). But as long as every posted price stays at or below its market-clearing value, sellers ration in proportion, and each buyer receives exactly what it would receive under flexible prices. On the balanced-growth path posted prices always stay at or below their market-clearing values, for every posted price is a market-clearing price from an earlier date and prices only rise. Along the transition posted prices do so as long as every price keeps rising.\footnote{Under the nominal-demand projection every price keeps rising when the transitory wave changes the price growth of every firm by less than inflation itself (Lemma~\ref{oa:lem:uniform_positivity}). This condition bounds the product of the magnitude of the smallest firms' excess degree, the injection misalignment, and network persistence $\lambda_2/(1-\lambda_2)$. It is therefore easiest to meet near degree neutrality and in quickly mixing networks, and it becomes harder to meet as the spectral gap closes or the degree tail thickens.} Sticky prices then change only the distribution of posted prices, and leave the allocation as it is under flexible prices. On the balanced-growth path, a posted price is nothing but today's market-clearing price multiplied by the vintage factor $(1+\pi)^{-u_i^{(t)}}$, which removes the inflation accumulated since its vintage (Corollary~\ref{oa:cor:sticky_bgp_vintage}).

When firm $i$ decides at date $t$ whether to reset, its posted price was set $u_i^{(t-1)}+1$ periods earlier, and both inflation and the transitory wave have moved its market-clearing price since then. Inflation has raised every price by $\log(1+\pi)$ per period, and we call the rise that inflation has accumulated over the price spell, $(u_i^{(t-1)}+1)\log(1+\pi)$, the `drift'. The transitory wave has also moved the firm's own price relative to the money stock, by an amount proportional to its Cantillon exposure $\mathcal C_i$, and we call this movement the `network correction'. Under Assumption~\ref{assu:small_returns} the log gap between the firm's market-clearing price and its posted price is the sum of the drift and the network correction (Lemma~\ref{lemma:hazard_sufficiency}),
\begin{equation}
\label{eq:gap_leading}
s_i^{(t)}:=\log\frac{p_i^{(t)}}{\bar p_i^{(t-1)}}
=\bigl(u_i^{(t-1)}+1\bigr)\log(1+\pi)
+\pi\,\mathcal C_i\bigl(\mathcal X_t-\mathcal X_{t-u_i^{(t-1)}-1}\bigr)
+o(\pi).
\end{equation}
We call $s_i^{(t)}$ the `pre-reset gap'.\footnote{The pre-reset gap is never negative on the balanced-growth path, nor along the transition as long as every price keeps rising, and we therefore define the price reset rule on nonnegative gaps.} We characterize local state-dependent pricing with Assumption~\ref{assu:hazard}, under which the probability that a firm resets its price depends on its pre-reset gap alone.

\begin{assumption}[Gap-triggered price hazard]\label{assu:hazard}
Below the maximal price age $\bar u$, a fixed positive integer, firm $i$ resets at date $t$ with probability $\eta_i^{(t)}=\eta\bigl(s_i^{(t)}\bigr)$, a function of the pre-reset gap alone. The hazard function $\eta:[0,\infty)\to[0,1]$ is continuous and nondecreasing, with $\eta(s)\sim c\,s^{\beta}$ as $s\to0$ for some $c>0$ and some state-dependence exponent $\beta\ge0$, and with $\limsup_{s\downarrow0}\eta(s)<1$ when $\beta=0$. Conditional on the pre-reset firm states, price reset draws are independent across firms. A firm whose price has reached the age $\bar u$ resets with probability one.
\end{assumption}

\noindent The state-dependence exponent $\beta$ measures how fast the price reset probability rises while the pre-reset gap is still small, which is what we mean by the degree of state dependence. The two workhorse rules of the literature are the two extremes of state dependence. At $\beta=0$ the hazard function is flat, and a firm resets at a constant rate whatever its pre-reset gap, as in the Calvo model. As $\beta\to\infty$ a firm almost never resets until its pre-reset gap is large, as in a menu-cost model, and we call this case the `menu-cost limit'. Whatever its exponent, the hazard function has a single argument, and therefore the network position of a firm can affect how long it keeps its price only through its pre-reset gap. Local state-dependent pricing differs from Calvo pricing and its multisector extensions precisely here \citep{pasten2020propagation}. In them, price reset rates are primitives chosen independently of network position. In our production network economy, the network itself shapes how long each firm keeps its price.

In the pre-reset gap~\eqref{eq:gap_leading}, the drift and the network correction behave differently. The drift $(u+1)\log(1+\pi)$ is the same for every firm and grows with the length of the price spell. But the network correction $\pi\mathcal C_i(\mathcal X_t-\mathcal X_{t-u-1})$ differs across firms and, while the propagation kernel is building, has the sign of the firm's Cantillon exposure. Two distances need to be kept apart here, for only one of them triggers a reset. The first is how far a firm's price stands from its equilibrium relative price, the level distortion $\ell_i^{(t)}=\pi\mathcal C_i\mathcal X_t$, which is what our measures of relative-price distortion record. The second is the pre-reset gap, how far the posted price of a firm stands from its market-clearing price. The level distortion enters the pre-reset gap only through its change since the last reset, $\ell_i^{(t)}-\ell_i^{(t-u_i^{(t-1)}-1)}$, which is the network correction. This is because a posted price is the firm's own market-clearing price from the date at which it was last set, and it therefore already carries the level distortion of that date. A level distortion that has stopped growing gives the firm no reason to reset its price, whereas a level distortion that is still growing widens its pre-reset gap and brings the next reset forward. Which means that a firm resets its price because it is falling behind its market, and not because its relative price stands far from equilibrium.

\subsection{The size of price change and the frequency of resets}
\label{subsec:distortion_sticky}

There are two ways in which firms can absorb inflation, by changing their prices more often or by changing them by more, and we call the first the `frequency margin' and the second the `size margin'. The margin that firms use decides what data on price changes can tell us. In the flexible-price regime there is no choice to be made, since every price changes every period. But in the sticky-price regime firms use both margins, and the shape of the hazard function decides how much they use each.

\begin{definition}[Average size of price change]
\label{def:size_price_change}
For a weight exponent $\zeta\ge0$, give firm $i$ the degree weight $\wgt{i}{\zeta}:=d_i^{\zeta}/\sum_j d_j^{\zeta}$. At a date at which at least one firm resets its price, the average size of price change conditional on adjustment is
\[
\phi_t\;:=\;
\frac{\sum_{i=1}^n \wgt{i}{\zeta}\,I_i^{(t)}\,\bigl|\Delta\log\bar p_i^{(t)}\bigr|}
{\sum_{i=1}^n \wgt{i}{\zeta}\,I_i^{(t)}}.
\]
\end{definition}

\noindent The average size of price change is what data on price changes can measure, for as long as every price keeps rising, a firm changes its posted price exactly when it resets. Each value of the weight exponent $\zeta$ gives a different average size of price change. At $\zeta=0$ the firms that reset are weighted equally, while at $\zeta=1$ they are weighted by size, since a firm's stationary balance is proportional to its degree.\footnote{In the flexible-price regime, $\phi_t$ equals $\log(1+\pi)$ in the steady state under every weighting. As long as every price keeps rising (Lemma~\ref{oa:lem:uniform_positivity}), the transient departure of $\phi_t$ from $\log(1+\pi)$ also vanishes to first order in $\pi$ at the size weighting $\zeta=1$. This is because size weights are stationary balances, and under stationary balances the Cantillon exposure averages to zero (Lemma~\ref{oa:lem:balance_centering}).} We suppress the weight exponent $\zeta$ whenever we do not compare weightings.

The hazard function decides how much firms use each margin through the length of their price spells. A firm that keeps its price for longer lets more drift accumulate before it resets, and so it changes its price less often but by more. And how long a firm keeps its price depends in turn on how fast its pre-reset gap widens and on how strongly its price reset probability responds to that gap. While the drift is small, the pre-reset gap of a price of age $u$ is of order $\pi u$, and under a hazard function with exponent $\beta$ the price reset probability at that age is of order $(\pi u)^{\beta}$. The cumulative price reset probability is the sum of these probabilities over ages, and it reaches order one at the age
\[
u_\pi^\star:=\pi^{-\beta/(\beta+1)}.
\]
We call $u_\pi^\star$ the `natural age scale', for it is the age at which the hazard function ends a typical price spell.

We consider inflation rates at which the drift over a typical price spell stays small and a typical price spell ends well before the maximal price age $\bar u$ of Assumption~\ref{assu:hazard}. Since $\bar u$ caps the age of a price, we call these inflation rates the `cap-nonbinding range' $\mathcal P_{\bar u}$ (Definition~\ref{oa:def:cap_range}).\footnote{The cap $\bar u$ is only a safeguard, for at these inflation rates the hazard function ends a typical price spell long before the cap would. The lower end of the range is strictly positive whenever the cap is finite. On the range, $X=\Theta(\pi^{a})$ therefore means that constants $0<c\le C<\infty$, set by the economy and not by $\pi$, bound $X$ between $c\,\pi^{a}$ and $C\,\pi^{a}$ at every $\pi$ in the range. For a given cap the range is empty once $\beta$ is large enough, but the deterministic inaction band of the computational experiments implements the menu-cost limit $\beta\to\infty$ directly.}

When pricing is state dependent, $\beta>0$, the natural age scale falls as inflation rises, for a pre-reset gap that widens faster reaches any given size sooner. There are therefore two ways in which an increase in inflation affects the size of price change. It widens the pre-reset gap faster, which raises the size of price change at a reset. But it also shortens the price spell over which the gap widens, which lowers the size of price change. The two effects partly offset each other.

The age of a firm's price at any date is random, with a probability law that depends on the firm only through its Cantillon exposure $\mathcal C_i$, which is fixed by the firm's position in the network. We write $u_i$ for the stationary post-decision price age of a firm with Cantillon exposure $\mathcal C_i$.

\begin{proposition}[Inflation is divided between the frequency and size margins]
\label{prop:frequency}
Let Assumptions~\ref{assu:primitivity}--\ref{assu:hazard} hold and let $\pi\in\mathcal P_{\bar u}$. Within the two-eigenmode degree representation, the stationary post-decision price age and the frequency of resets then satisfy
\[
\E[u_i]
=\Theta\!\left(\pi^{-\beta/(\beta+1)}\right),
\qquad
\Pr\!\left(u_i^{(t)}=0\right)
=\Theta\!\left(\pi^{\beta/(\beta+1)}\right),
\]
and, conditional on a price reset at $t$, the mean length of the completed price spell satisfies
\[
\E\!\left[u_i^{(t-1)}+1\mid I_i^{(t)}=1\right]
=\Theta\!\left(\pi^{-\beta/(\beta+1)}\right),
\]
with exponents that do not depend on the firm.
\end{proposition}

\noindent\textit{Proof in Appendix~\ref{app:proof_frequency}.}

\medskip
Note that the frequency of resets has elasticity $\beta/(\beta+1)$ with respect to inflation, and the size of price change has elasticity $1/(\beta+1)$. The two elasticities sum to one. This is because on the balanced-growth path every posted price rises on average by $\log(1+\pi)$ per period, and it can do so only through the frequency of resets and the size of price change. In the Calvo model, $\beta=0$, the frequency of resets does not respond at all, and the whole response goes into the size of price change, which is of order $\pi$. In the menu-cost limit, $\beta\to\infty$, the frequency of resets carries nearly the whole response, and the size of price change barely moves. So the more state dependent the pricing rule, the more of an increase in inflation the frequency margin absorbs, and the micro-price evidence in the Introduction emphasizes exactly this pattern.

\subsection{Price vintages and relative-price distortion under flexible and sticky prices}
\label{subsec:sticky_vs_flexible}

In the flexible-price regime every posted price is set at the current market-clearing price, so all posted prices share one vintage. But in the sticky-price regime resets are staggered, and the economy holds a mixture of vintages at every date. The posted price of firm $i$ is the market-clearing price it set $u_i$ periods earlier, and under Assumption~\ref{assu:small_returns} it carries the propagation kernel $\mathcal X_{t-u_i}$ of that date and not that of the current date. Staggered resets therefore add a dispersion of price vintages to the standing network distortion of Section~\ref{sec:flexible}. How much distortion they add depends on how dispersed price ages are and on which firms hold the old prices.

Note that on the balanced-growth path every price grows at exactly $\log(1+\pi)$ (Proposition~\ref{prop:balanced_growth_convergence}), so the network correction vanishes (Lemma~\ref{lemma:hazard_sufficiency}) and the pre-reset gap of every firm widens at the same rate. The age of every firm's price then follows the same law. But along the transition the price of firm $i$ grows at $\pi+\pi(\mathcal X_T-\mathcal X_{T-1})\mathcal C_i$ (Proposition~\ref{prop:size_price_change}), so pre-reset gaps widen at different rates across firms. While the propagation kernel is building, the market-clearing price of a firm above the threshold degree grows faster than inflation, so its pre-reset gap widens faster and its price spell tends to end sooner. And the market-clearing price of a firm below the threshold degree grows more slowly than inflation, so its price spell tends to end later. All of which is to say that the timing of price resets comes to depend on network position, although the price reset rule itself treats every firm alike.

\begin{lemma}[Price spells are ordered by Cantillon exposure along the transition]\label{lemma:network_vintages}
Let Assumptions~\ref{assu:primitivity}--\ref{assu:hazard} hold, and let the propagation kernel be building, $\mathcal X_t>\mathcal X_{t-1}$, at every date $t$ at which the price spells below can be in progress. Within the two-eigenmode degree representation, consider firms $i$ and $j$ that reset at a common date $t_0$, and write $L_i$ and $L_j$ for the lengths of the price spells they begin. Each length depends on the firm only through its Cantillon exposure, and $\mathcal C_i>\mathcal C_j$ implies
\[
\Pr(L_i>u)\;\le\;\Pr(L_j>u)
\]
at every $u$, strictly at every $u$ at which $\Pr(L_j>u)>0$ and the price reset probability of firm $i$ has exceeded that of firm $j$ at some age up to $u$. The stationary law of price age that the date-$t$ hazard profile would generate if it persisted is ordered by the monotone likelihood ratio in $\mathcal C_i$, so that under it the firm of larger exposure holds the stochastically younger price. On the balanced-growth path the law of the length of a price spell and the law of price age are the same for every firm.
\end{lemma}

\noindent\textit{Proof in Appendix~\ref{app:proof_network_vintages}.}

\medskip
The economic content of Lemma~\ref{lemma:network_vintages} is that along the transition the old prices sit with the small firms. This is because in a disassortative network the new money that lands on small firms flows on to their suppliers, which are large. The Cantillon exposure is therefore positive for firms above the threshold degree $d^{*}$ and negative for firms below it.\footnote{Among the very largest firms the ordering by size reverses. They sell to a representative cross-section of the economy, so their Cantillon exposure declines beyond the degree $\hat d=d^{*}(1-\nu^{2})^{-1/\nu^{2}}$ (Corollary~\ref{oa:cor:departure_peak}), and they keep their prices longer than firms at $\hat d$.} The timing of price resets depends on network position most strongly where the level distortion that a firm can reach, $\ell_i^{\max}=\pi\lvert\mathcal C_i\rvert\mathcal X_\infty$, is large compared with the drift over a typical price spell. And since $\mathcal X_\infty$ grows as the spectral gap narrows, the timing of price resets depends most on network position in slowly converging economies, which are also the economies with the largest standing network distortion.

We now have two versions of each measure of relative-price distortion: one for the flexible-price regime and the other for the sticky-price regime. We write $\omega_t^{\mathrm{flex}}$ and $\psi_t^{\mathrm{flex}}$ for Definitions~\ref{def:relative_price_gap} and~\ref{def:relative_price_entropy} computed at market-clearing prices, so that $(\omega_t^{\mathrm{flex}},\psi_t^{\mathrm{flex}})=(\omega_t,\psi_t)$. We write $\omega_t^{\mathrm{stick}}$ and $\psi_t^{\mathrm{stick}}$ for the same measures computed at posted prices. Without the time subscript, the same symbols denote long-run averages.

In the sticky-price regime each measure contains the standing network distortion and the effects of staggered resets. We fix a stationary horizon $T\ge\bar u$. And we write $\mathfrak x_i:=\log(1+\pi\mathcal C_i\mathcal X_{T-u_i})$ for the network displacement that the posted price of firm $i$ carries from its price reset date, and $\mathfrak x_i^{\mathrm{flex}}:=\log(1+\pi\mathcal C_i\mathcal X_T)$ for the network displacement it would carry had it been reset at $T$. Relative-price entropy at posted prices then splits into four terms, built from variances and covariances across firms under equilibrium price shares (Lemma~\ref{lemma:sticky_distortion}),
\begin{equation}
\label{eq:sticky_decomposition_main}
\begin{aligned}
\psi_T^{\mathrm{stick}}
&\approx\tfrac12\bigl\{
\underbrace{\Var_{\widehat r^*}(\mathfrak x^{\mathrm{flex}})}_{\text{flexible}}
+\underbrace{\Var_{\widehat r^*}(\mathfrak x)-\Var_{\widehat r^*}(\mathfrak x^{\mathrm{flex}})}_{\text{realized vintage}}\\[4pt]
&\qquad
+\underbrace{\log^2(1+\pi)\Var_{\widehat r^*}(u)}_{\text{vintage dispersion}}
-\underbrace{2\log(1+\pi)\Cov_{\widehat r^*}(u,\mathfrak x)}_{\text{interaction}}
\bigr\}.
\end{aligned}
\end{equation}

The flexible term is nothing but the standing network distortion of Theorem~\ref{theorem:price_deviation}, and no price reset rule removes it. The realized-vintage term is nonzero only along the transition, for posted prices carry the propagation kernel of their price reset dates, and once the propagation kernel has converged, every price reset date carries the same propagation kernel. The vintage-dispersion term is what staggered resets add by themselves. It depends only on price ages and on inflation, so it would be present even if every firm had zero Cantillon exposure. And the interaction term records whether the older prices sit at the more displaced firms, and along the transition they sit at the small firms (Lemma~\ref{lemma:network_vintages}). In the steady state the interaction term is the only one of the four terms that can take either sign.

The relative price gap has the same four terms under balance weights (Lemma~\ref{lemma:sticky_distortion}).\footnote{Neither measure of relative-price distortion depends on when the numeraire firm last reset its price, for the numeraire firm enters at its market-clearing price. Nor does either measure depend on the lag that all posted prices share, which relative-price entropy removes by normalizing prices into shares and the relative price gap removes by minimizing over a common rescaling.} In the steady state the realized-vintage term is zero, and since the age of every firm's price then follows the same law, the interaction term has no systematic part. What remains is the flexible term plus the vintage-dispersion term, which is never negative. Sticky prices therefore never undo the distortion that the network creates in the flexible-price regime. They can only add to it (Corollary~\ref{oa:cor:regime_ranking}).

As to how much more distorted relative prices are in the sticky-price regime than in the flexible-price regime, this depends on how long price spells last. Lower inflation slows the drift, but it also lengthens price spells, for a pre-reset gap that widens more slowly takes longer to trigger a reset. The mean age of a price rises as $\Theta(\pi^{-\beta/(\beta+1)})$ when inflation falls (Proposition~\ref{prop:frequency}), so the longer price spells partly offset the slower drift, and the drift over a typical price spell falls only as $\Theta(\pi^{1/(\beta+1)})$. The relative price gap in the sticky-price regime falls at the same rate, and relative-price entropy falls as its square (Lemma~\ref{lem:sticky_orders}). The relative price gap in the flexible-price regime, by contrast, falls one for one with inflation.

In the Calvo model, $\beta=0$, the measures of relative-price distortion are of the same order in inflation in the two regimes. But when $\beta>0$ they vanish more slowly in the sticky-price regime than in the flexible-price regime as inflation falls, and therefore how much more distorted the sticky-price regime is than the flexible-price regime depends on the rate of inflation.\footnote{The ranking can also change with the horizon. Early in the transition, posted prices still reflect conditions before the latest displacement, whereas flexible prices pass the displacement through at once, so the realized-vintage term can make the flexible-price regime the more distorted of the two. The sticky-price regime loses this advantage once the propagation kernel has converged. And it has the advantage only when the displacement that the old posted prices have not yet absorbed exceeds what the dispersion of price vintages adds (Corollary~\ref{oa:cor:regime_ranking}).}

\begin{proposition}[The ratio of sticky-price to flexible-price distortion rises as inflation falls]
\label{prop:sticky_flexible_ranking}
Let Assumptions~\ref{assu:primitivity}--\ref{assu:hazard} hold with $\beta>0$, and let $\pi\in\mathcal P_{\bar u}$. Within the two-eigenmode degree representation, the ratios of the steady-state sticky to flexible measures then satisfy
\[
\frac{\omega^{\mathrm{stick}}}{\omega^{\mathrm{flex}}}=\Theta\bigl(\pi^{-\beta/(\beta+1)}\bigr),
\qquad
\frac{\psi^{\mathrm{stick}}}{\psi^{\mathrm{flex}}}=\Theta\bigl(\pi^{-2\beta/(\beta+1)}\bigr).
\]
\end{proposition}

\noindent\textit{Proof in Appendix~\ref{app:proof_sticky_flexible_ranking}.}

\medskip
The broad lesson of Proposition~\ref{prop:sticky_flexible_ranking} is that a disinflation does less for relative prices when prices are sticky than when they are flexible. This is because firms answer lower inflation by keeping their prices longer, so the dispersion of price vintages falls more slowly than inflation, whereas the standing network distortion falls one for one with it. And the more state dependent the pricing rule, the more slowly the dispersion of price vintages falls. Which means that the lower the rate of inflation, the larger the share of relative-price distortion that staggered resets add to the standing network distortion.\footnote{This holds until the cap binds. From then on the age of a price cannot rise any further, the relative price gap in the sticky-price regime returns to order $\pi$, and the two ratios of Proposition~\ref{prop:sticky_flexible_ranking} level off at $\Theta(\bar u)$ and $\Theta(\bar u^{2})$. Both ratios stay at or above one throughout, for the vintage-dispersion term is never negative.}

\subsection{The size of price change as a measure of relative-price distortion}
\label{subsec:separation}

We now ask what the size of price change tells us about relative-price distortion.\footnote{The observation behind this question is seemingly simple but rather significant. A price change measures the distance between two prices posted by the same firm at different dates, whereas relative-price distortion measures the distance between the price a firm posts today and its general-equilibrium value. The first is a change and the second a level, so a large and lasting displacement of relative prices can coexist with small price changes.} Under local state-dependent pricing, firms `keep up with' inflation by resetting their prices more often. And the more state dependent the pricing rule, the more they rely on it, and the less the size of price change moves with inflation (Proposition~\ref{prop:frequency}). But keeping up with inflation in this way does not bring relative prices back to their equilibrium values. However often firms reset, relative prices stay at least as distorted as in the flexible-price regime. And as resets become frequent, this distortion keeps differing from one economy to the next, whereas the size of price change becomes the same in every economy. Which is perhaps the most significant insight of this paper, formalized in the following theorem.

\begin{theorem}[The size of price change does not identify relative-price distortion]
\label{thm:size_not_measure}
Let Assumptions~\ref{assu:primitivity}--\ref{assu:assortativity} hold, fix $\pi>0$ and $\theta\in(0,1)$, and suppose the Cantillon exposure is not constant across firms. Within the two-eigenmode degree representation, consider any sequence of price reset rules satisfying Assumption~\ref{assu:hazard} with a common finite cap $\bar u$ along which, in the steady state and for every firm,
\[
\Pr\!\left(u_i^{(t)}=0\right)\longrightarrow1,
\qquad
\E\!\left[u_i^{(t-1)}+1\mid I_i^{(t)}=1\right]\longrightarrow1.
\]
Then
\[
\phi_T\longrightarrow\log(1+\pi),
\qquad
\omega^{\mathrm{stick}}\longrightarrow\omega^{\mathrm{flex}}>0,
\qquad
\psi^{\mathrm{stick}}\longrightarrow\psi^{\mathrm{flex}}>0.
\]
At a fixed inflation rate the first limit is the same for every injection and network. The last two limits vary across injections and networks.
\end{theorem}

\noindent\textit{Proof in Appendix~\ref{app:proof_size_not_measure}.}

\medskip
The simple reading of the theorem is that frequent price adjustment removes the dispersion of price vintages but leaves the standing network distortion in place. As firms reset more and more often, each price change comes to cover a single period of inflation. At a given inflation rate the average size of price change therefore converges to $\log(1+\pi)$ in every economy, whatever its injection and its network. But relative-price distortion converges to its level in the flexible-price regime, which varies with the stationary misalignment, network persistence, and the dispersion of the Cantillon exposure. And this is precisely the sense in which our model differs from Calvo and menu-cost models \citep{calvo1983staggered,sheshinski1977inflation}. In them, the frequency of resets governs both the size of price change and relative-price entropy. Relative-price entropy is therefore strictly increasing in the size of price change (Lemma~\ref{oa:lem:workhorse_benchmark}). The production network economy severs this fortuitous alliance between relative-price entropy and the size of price change.

\section{Computational Experiments}
\label{sec:abm}

The relations between inflation and relative-price distortion in Sections~\ref{sec:framework}--\ref{sec:sticky} rest on approximations. We now ask whether these relations survive once the approximations are removed.\footnote{There are five such approximations, the first of which is the nominal-demand projection of Assumption~\ref{assu:small_returns}, under which quantities stay fixed and prices follow nominal demand. The second and the third are the two-eigenmode approximation and the degree representation, which retain only the first two eigenmodes of the propagation matrix and represent these eigenmodes by functions of degree. The fourth keeps only the leading order of each relation in the rate of inflation. And the fifth removes storage, so that a firm discards whatever output it cannot sell.} The computational economy dispenses with all of these approximations at once.\footnote{Dispensing with the nominal-demand projection is no formality. The projection is accurate when the response of output to new money is small beside the response of balances, that is, when the projection ratio $\varkappa$ of equation~\eqref{eq:projection_ratio} is small. But on the reconstructed United States production networks of Section~\ref{subsec:sim_us}, at the returns to scale $\varsigma=0.8$ of the computational economy, this ratio lies between $0.9$ and $1.9$. The computational experiments therefore run where Assumption~\ref{assu:small_returns} is far from accurate.} In it, production, inventories, and rationing move together with prices, on the exact realized network and at finite rates of inflation. Each firm follows the rules of the model, and we let the whole economy run forward in time and measure what it generates, as is usual in agent-based computational experiments \citep{axtell2000why,arthur2006out}.

These computational experiments serve a second, very different purpose, which is to estimate the quantitative significance of the relative-price distortion that small rates of inflation cause in a large economy. Note that our analytical results tell us how relative-price distortion scales with the rate of inflation and with the structure of the network. But how large the distortion is in a given economy is a question that depends on the specifics of the network. We estimate the magnitude of the distortion generated in the US economy by calibrating the model to a reconstructed production network with the universe of firms in the United States.

Section~\ref{subsec:sim_environment} describes the computational economy and the specific price reset rule followed by the firms that populate it. Section~\ref{subsec:sim_inflation} presents results from experiments on synthetic production networks that relax our analytical assumptions. Section~\ref{subsec:sim_us} presents estimates of relative-price distortion in the US economy.

\subsection{The computational economy}
\label{subsec:sim_environment}
\label{subsec:sim_calibration}

The computational economy is much the same as the analytical economy, except in two respects, the first of which is the price reset rule. The analytical economy says what firms do, but not the exact price reset rule they follow. A firm there resets its price with a probability that rises with its pre-reset gap, and the analytical results hold for a whole family of hazard functions. Here we assume that firms follow an $(S,s)$ rule, which is the menu-cost limit of that family.\footnote{The classical description of price setting under inflation is an $(S,s)$ rule \citep{sheshinski1977inflation,CaplinSpulber1987QJE,DotseyKingWolman1999,golosov2007menu}.} The second respect is storage. A firm in the analytical economy discards the output it cannot sell, whereas a firm in the computational economy carries that output forward as inventory.\footnote{In the steady state no firm posts above its market-clearing price, so no firm is left with unsold output, and the absence of storage in the analytical economy matters only along the transition. Along the transition some firms do post above their market-clearing prices, roughly half of all firms at some early dates on the full United States reconstruction, and inventories peak at about $0.2\%$ of output before they clear.}

Each firm $i$ trades at its posted price $\bar p_i^{(t)}$, the price it set at its most recent price reset. Since firms now hold inventories, the price that would clear a firm's market divides its nominal demand by its available supply, $p_i^{(t)}=\mathcal D_i^{(t)}/\widetilde q_i^{(t)}$. The available supply $\widetilde q_i^{(t)}=q_i^{(t)}+\mathcal J_i^{(t)}$ is the firm's current production plus the inventory $\mathcal J_i^{(t)}$ it has inherited. The firm resets its price when the magnitude of its pre-reset gap, $\lvert s_i^{(t)}\rvert=\lvert\log(p_i^{(t)}/\bar p_i^{(t-1)})\rvert$, exceeds the half-width $b_t$ of an inaction band, and it otherwise keeps the price it last posted,
\begin{equation}
  \bar p_i^{(t)}
  =
  \begin{cases}
    p_i^{(t)},
    &
    \left|
      \log\left(
        p_i^{(t)}/
        \bar p_i^{(t-1)}
      \right)
    \right|
    >
    b_t,
    \\[3pt]
    \bar p_i^{(t-1)},
    &
    \text{otherwise}
  \end{cases}.
  \label{eq:ssrule}
\end{equation}
The half-width of the inaction band responds to inflation,
\begin{equation}
  b_t
  =
  b_0
  \left[
    (1-b_w)
    +
    \frac{b_w}{1+b_\pi|\pi|}
  \right],
  \label{eq:band}
\end{equation}
where $b_0>0$ is the baseline half-width, $b_w\in[0,1]$ controls how far the band responds to inflation, and $b_\pi\ge0$, which we call the `band compression', controls how strongly it does so. When $b_w=0$ the band is fixed at $b_0$, and the rule is the standard $(S,s)$ rule. But when $b_w=1$ the band narrows as $1/(1+b_\pi|\pi|)$, so that higher inflation brings price resets forward through a narrower band as well as through a faster drift. And as the band compression $b_\pi$ grows without bound, the band closes and the rule approaches flexible pricing.

A period of the computational economy is one round of ordering, production, and settlement, and we read it as a quarter.\footnote{The length of a period is a property of the technology, and it cannot be shortened by dividing the calendar more finely. This is because the second eigenvalue $\lambda_2$ measures how much of the second eigenmode survives each round of trade. A quarter is the order of magnitude of days payable outstanding together with one production round, and we treat a month as the shorter end of the admissible range.} We calibrate the baseline half-width of the inaction band to the observed duration of price spells \citep{nakamura2008five}.\footnote{The calibration holds under either reading of a period, for the length of a price spell in years is, to first order in inflation, the baseline half-width divided by the annual rate of inflation.} Table~\ref{oa:tab:calibration} collects the calibration.\footnote{Except in the experiments on the size of the economy (Appendix~\ref{subsec:sim_size}), the synthetic production networks have a scale-free degree distribution, calibrated to the range documented for firm-level production networks. Appendix~\ref{oa:app:parameter_mapping} relates the firm-level evidence to the injection exponent $\theta$ and the assortativity exponent $\nu$.} Computational experiments on this calibrated economy are run after it has settled into the neighborhood of equilibrium. More specifically, we first let the economy settle into its zero-injection equilibrium, to numerical precision (Table~\ref{oa:tab:calibration}), so that its relative prices are the benchmark against which we measure relative-price distortion. We then inject new money at a constant rate and, once the economy has settled again, average both measures of relative-price distortion over the steady state.\footnote{The constant rate of injection is also the rate of price-index inflation, for a price index that weights firms by their sales grows at exactly this rate in the computational economy. And the averages do not depend on the horizon. On the reconstructed United States economies, averaging over the $200$ periods before date $600$ instead of the last $200$ of $1{,}200$ moves every average by less than $1\%$.} The synthetic economies on which we relax the approximations have $10^5$ firms on directed scale-free networks, with tail exponent $1.6$ and mean degree $25$, and their injection exponent is $\theta=0.5$. Their generator targets a Newman buyer--seller assortativity coefficient of $-0.2$, a statistic that shares the sign of the assortativity exponent $\nu$ but not its scale (Appendix~\ref{oa:app:nu_calibration}). Every curve we report for them is the mean over independent draws of the network, with a band at one standard deviation.

\subsection{Inflation, price adjustment, and relative-price distortion}
\label{subsec:sim_inflation}

\begin{figure}[H]
\centering
\includegraphics[width=0.72\textwidth]{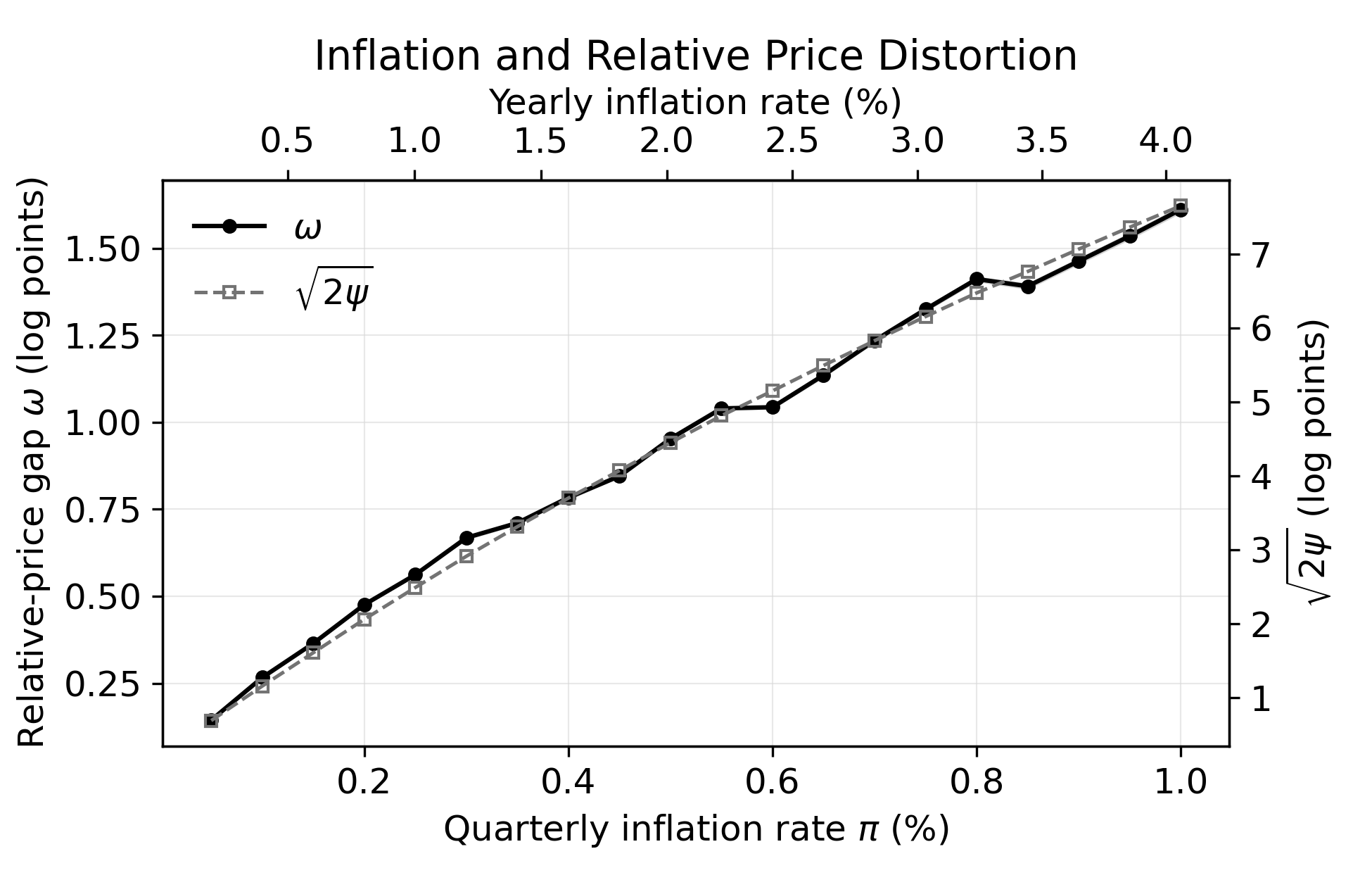}
\caption[Inflation and relative-price distortion]{\textbf{Inflation and relative-price distortion.}
Relative price gap $\omega$ (solid, left axis) and entropy-equivalent dispersion $\sqrt{2\psi}$ (dashed, right axis), in log points, against inflation, for twenty rates from $0.05\%$ to $1.0\%$ per period. The secondary axis annualizes at one period per quarter. Each curve is the mean over $20$ network draws, with shaded bands at $\pm1$ standard deviation.}
\label{fig:inflation_distortion}
\end{figure}

We begin by asking whether relative-price distortion grows with inflation in the computational economy as it does in Theorem~\ref{theorem:price_deviation}, the relative price gap in proportion to the rate of inflation and relative-price entropy in proportion to its square. We run the computational economy on twenty synthetic production networks of $10^5$ firms, at twenty rates of inflation from $0.05\%$ to $1.0\%$ per period, holding each network and the injection exponent fixed. At the quarterly reading these rates span $0.2\%$ to $4.1\%$ a year, which brackets the postwar experience of the United States. Figure~\ref{fig:inflation_distortion} shows that inflation distorts relative prices substantially, with the relative price gap reaching $1.6$ log points at $\pi=1\%$ per period.\footnote{The visible sawtooth is the lattice of price reset dates under the deterministic inaction band. Its dips sit exactly at the rates of inflation at which a price spell shortens by one period and the frequency of resets jumps.} And the two measures of relative-price distortion take the shapes that Theorem~\ref{theorem:price_deviation} predicts. The relative price gap is close to linear in inflation, and so is the entropy-equivalent dispersion $\sqrt{2\psi}$.\footnote{We compute both measures in log points and report relative-price entropy through the entropy-equivalent dispersion $\sqrt{2\psi}$, for small displacements the standard deviation of the log displacements under equilibrium price shares, which shares the scale of the relative price gap.} Which means that relative-price entropy itself rises with the square of inflation.

 For each of the twenty networks, we correlate the rate of inflation across the twenty rates with the frequency of resets and with the average size of price change. Inflation is almost perfectly correlated with the frequency of resets, at $0.98$, and essentially uncorrelated with the average size of price change, at $-0.08$.\footnote{These correlations compare economies across the grid of rates of inflation, and they are the same on every network, to two decimal places. This is because on the balanced-growth path the frequency of resets and the size of price change depend on a firm only through the age of its price, whose law is the same for every firm (Lemma~\ref{lemma:network_vintages}). Both are therefore properties of the inaction band alone. The correlation with the size of price change runs from $+0.79$ under a fixed band to $-0.62$ at a band compression of $b_\pi=500$, while the correlation with the frequency of resets stays above $+0.93$.} This is because under the inflation-responsive band firms answer higher inflation by resetting their prices sooner, and not by changing them by more. An economist who observed only the size of price change would therefore infer that inflation does little to disturb the price system. But both measures of relative-price distortion rise strongly with inflation.

\begin{figure}[H]
\centering
\includegraphics[width=0.72\textwidth]{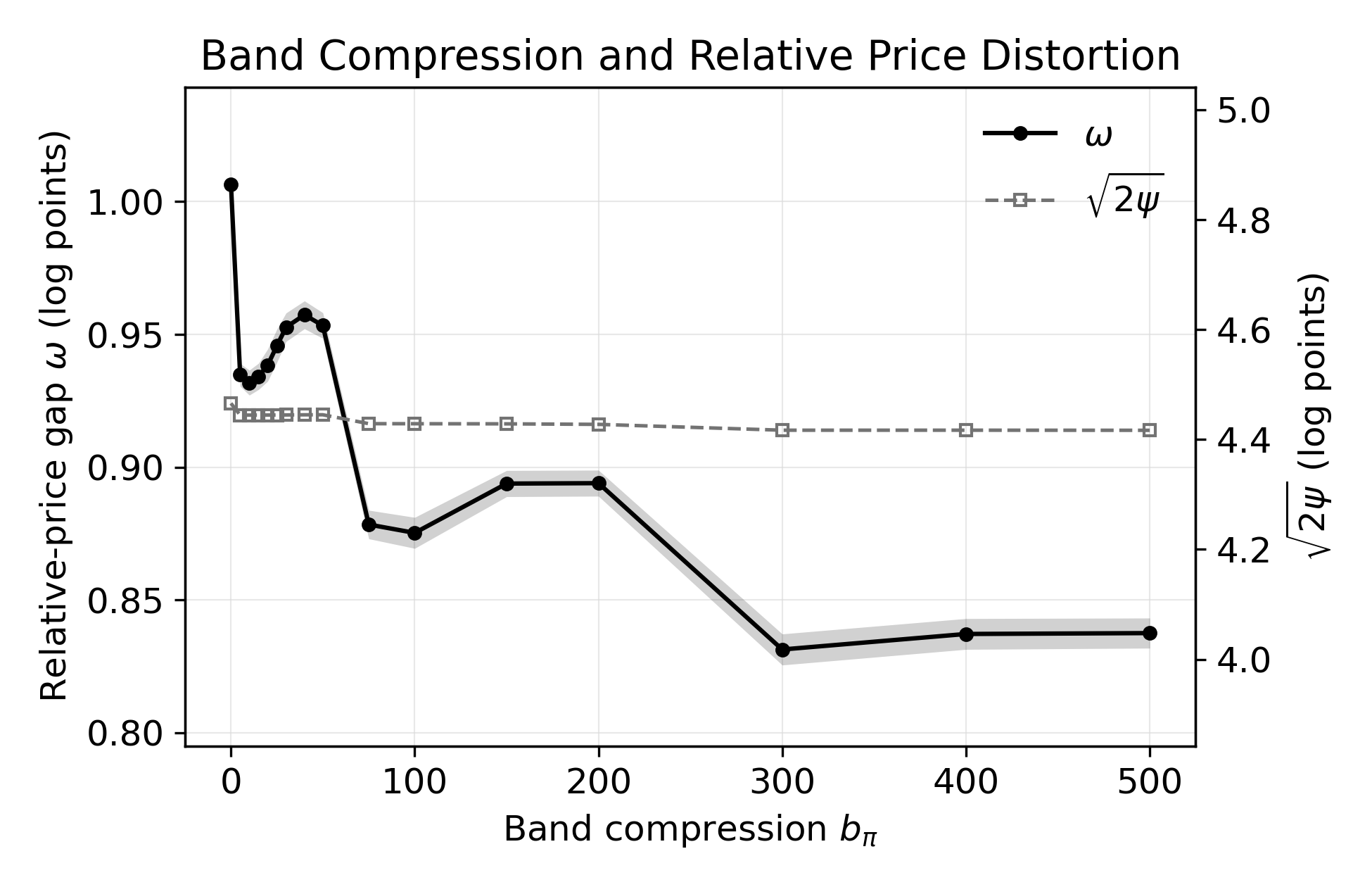}
\caption[Band compression and relative-price distortion]{\textbf{Band compression and relative-price distortion.}
Relative price gap $\omega$ (solid, left axis) and entropy-equivalent dispersion $\sqrt{2\psi}$ (dashed, right axis), in log points, against the band compression $b_\pi$, from $0$ to $500$, at $\pi=0.5\%$ per period. The two axes span the same proportion of their means. Each point is the mean over $20$ network draws, with a band at $\pm1$ standard deviation.}
\label{fig:band_distortion}
\end{figure}

Theorem~\ref{thm:size_not_measure} established that frequent price adjustment removes the dispersion of price vintages but leaves the standing network distortion in place. The band compression $b_\pi$ of the inaction band~\eqref{eq:band} lets us examine this computationally. Under the inflation-responsive band the half-width is $b_0/(1+b_\pi|\pi|)$, so at a given rate of inflation a larger band compression means a narrower band, a smaller pre-reset gap at which firms reset, and therefore shorter price spells. Which means that firms reset their prices more often the larger the band compression, and as $b_\pi$ grows without bound the band closes and firms reset every period. Raising $b_\pi$ walks the economy along the sequence of price reset rules of Theorem~\ref{thm:size_not_measure}, whose conclusion we can now test. Along that sequence the average size of price change falls toward a single period's inflation, and both measures of relative-price distortion fall toward their flexible-price values, which the network sets, and no further. We hold the network fixed, inflation at $\pi=0.5\%$ per period, and the injection exponent at $\theta=0.5$, and raise the band compression from $0$ to $500$, which carries the $(S,s)$ rule from sticky toward nearly flexible pricing. The average size of price change falls from $2.5\%$ to $1.0\%$, two periods' worth of inflation. But the relative price gap falls only from $1.01$ to $0.84$ log points, and relative-price entropy barely moves (Figure~\ref{fig:band_distortion}). Which is to say that an increase in the frequency of price change has but a modest impact on relative-price distortion.

To understand why relative-price distortion changes so little, note that under the fixed band the relative price gap has two parts. The first is the dispersion of price vintages, for firms that reset on different dates post prices that inflation has since carried apart. The second is the standing network distortion, which emerges from the fact that the market-clearing relative prices are not the general equilibrium relative prices in the absence of injection. As the band narrows, price spells shorten from five periods to two and the first part shrinks, in three steps, each at a band compression at which the price spell loses a period. That is the fall from $1.01$ to $0.84$ log points. But the second part stays however often prices are reset, for a reset brings a price to its market-clearing value and not to its general equilibrium value. Note that the two parts combine as variances do (Lemma~\ref{lemma:sticky_distortion}), i.e., by the sum of their squares. The relative price gap is the square root of the sum. Which means that removing the dispersion of price vintages, the smaller part, lowers the relative price gap by less than that dispersion itself. And relative-price entropy barely moves because under its weights, the equilibrium price shares, the standing network distortion is by far the larger of the two parts. All of which is to say that price flexibility takes away the part of relative-price distortion that staggered resets add, and leaves the part that the network creates.\footnote{The synthetic economies also bear out the comparative statics of Proposition~\ref{prop:comparative_statics} (Appendix~\ref{oa:sim_determinants}). Relative-price distortion vanishes as the injection approaches proportionality, falls roughly thirteen-fold as the network is rewired from disassortative to assortative mixing, and nearly triples as the tail of the degree distribution becomes heavier. Nor does it average away as the number of firms grows.}

\subsection{Relative-price distortion on the United States production network}
\label{subsec:sim_us}

We now turn to the second purpose of the computational experiments, which is to estimate the size of the relative-price distortion in a large economy.\footnote{This exercise is much in the spirit of structural estimation, though at a far finer grain and without an econometric objective \citep{reisswolak2007,lowmeghir2017}.} We calibrate the model on a reconstructed production network of the United States with $6.46\times10^6$ firms and about $3.4\times10^8$ directed buyer--seller links \citep{bhattathiripadveetil2026reconstruct}. The reconstruction preserves the empirical distribution of firm sizes, and its links between firms are calibrated so that, summed to sectors, they reproduce the observed input--output flows.\footnote{Like observed production networks, it is mildly disassortative \citep{FujiwaraAoyama2010,BMS19,bachilieri2023topology}.} We run the computational economy on the full network and on smaller reconstructions of it, with $10^4$, $10^5$, and $10^6$ firms.\footnote{The spectral gap $1-\lambda_2$ of their propagation matrices is $0.22$ at $10^4$ firms, $0.46$ at $10^5$, $0.37$ at $10^6$, and $0.34$ on the full network. The half-life of the propagation kernel therefore lies between one and three periods at every size.} Each firm's share of new money is proportional to its balance raised to the injection exponent $\theta$. We set $\theta=0.8$, inside the range $\theta<1$ that the evidence on how access to short-term and bank credit varies with firm size supports (Appendix~\ref{oa:theta_calibration}). With the network and the injection rule in place, we inject new money at each of twenty rates of inflation, from $0.05\%$ to $1.0\%$ per period, for $1{,}200$ periods. We average the two measures of relative-price distortion over the final $200$ periods.

\begin{figure}[H]
\centering
\includegraphics[width=0.72\textwidth]{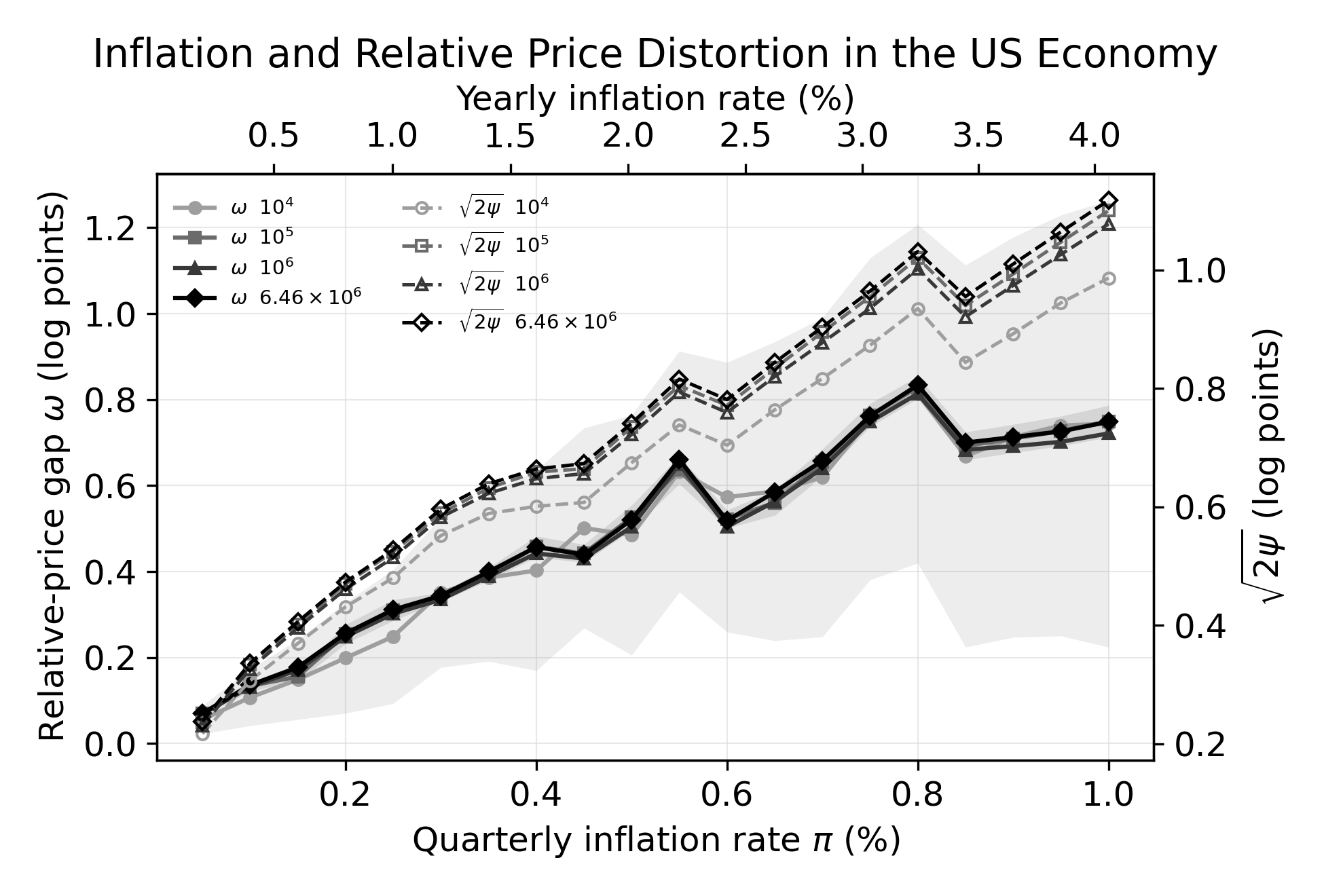}
\caption[Relative-price distortion in the United States]{\textbf{Relative-price distortion in the United States.}
Relative price gap $\omega$ (solid, left axis) and entropy-equivalent dispersion $\sqrt{2\psi}$ (dashed, right axis), in log points, against the rate of inflation per period, on reconstructed United States economies of $10^4$, $10^5$, and $10^6$ firms and on the full network of $6.46\times10^6$ firms. The secondary axis annualizes at one period per quarter. The $10^4$ curves are means over $100$ reconstruction draws and the $10^5$ curves means over $50$, with a shaded band at $\pm1$ standard deviation at $10^4$, while the two largest economies are single reconstructed networks.}
\label{fig:recon_us_inflation}
\end{figure}

The primary result of the experiment is that relative prices are roughly a quarter's worth of inflation away from their equilibrium values (Figure~\ref{fig:recon_us_inflation}). At the policy-relevant rate of $2\%$ a year, or $\pi=0.5\%$ per period at the quarterly reading, the relative price gap on the full network is $0.52$ log points. At $\pi=1\%$ per period, or $4.1\%$ a year, the relative price gap reaches $0.75$ log points, and the entropy-equivalent dispersion $\sqrt{2\psi}$ reaches $1.1$ log points, about one period's worth of inflation. Both measures rise almost linearly with the rate of inflation, as Theorem~\ref{theorem:price_deviation} predicts for the relative price gap and for the square root of relative-price entropy.\footnote{They rise with inflation under each of seven price reset rules. Four of them are the baseline band, the fixed band, the compressed band with $b_\pi=500$, and the exact flexible-price limit. The other three, calibrated to the frequency of resets of the baseline band, are inaction thresholds dispersed across firms and the hazard function of Assumption~\ref{assu:hazard} at $\beta=1$ and at $\beta=3$. At $\pi=1\%$ per period the relative price gap differs across the seven rules by a factor of about two. It is lowest where every firm resets every period, and highest under the hazard at $\beta=1$, for random resets leave the widest dispersion of price vintages.} Which means that the distortion we attribute to an annual rate of inflation is proportional to the length of a period. At the monthly reading, the shorter end of the admissible range, this distortion falls to a third, and the top of the range of inflation rises to $12.7\%$ a year.\footnote{The magnitude depends on how unevenly new money arrives. Over the band $\theta\in[0.70,0.86]$ around the calibrated value, the relative price gap at $\pi=1\%$ per period runs from $1.21$ to $0.51$ log points, a steeper response than the stationary misalignment $\xi_{\mathrm{ub}}$ alone predicts.}

Note that the curves of the relative price gap for the four reconstructed economies, from $10^4$ firms to the full network, nearly coincide. This is because each reconstructed economy is the same economy drawn at a different resolution, with the same distribution of firm sizes and the same pattern of who trades with whom. What matters for the macro question of relative-price distortion in the economy as a whole is the large structural features of the network, retained in each reconstruction, not the specifics of who trades with whom.\footnote{The two largest reconstructions are single draws, but their relative price gaps do not hinge on the particular draw. Rewiring $1\%$, $5\%$, or $10\%$ of their links, with every firm's numbers of suppliers and of customers fixed, lowers the relative price gap by about $0.5\%$ per percent of links rewired, for rewiring widens the spectral gap. Twenty independent rewirings at each fraction differ by less than $0.1\%$.}

\begin{figure}[H]
\centering
\includegraphics[width=0.95\textwidth]{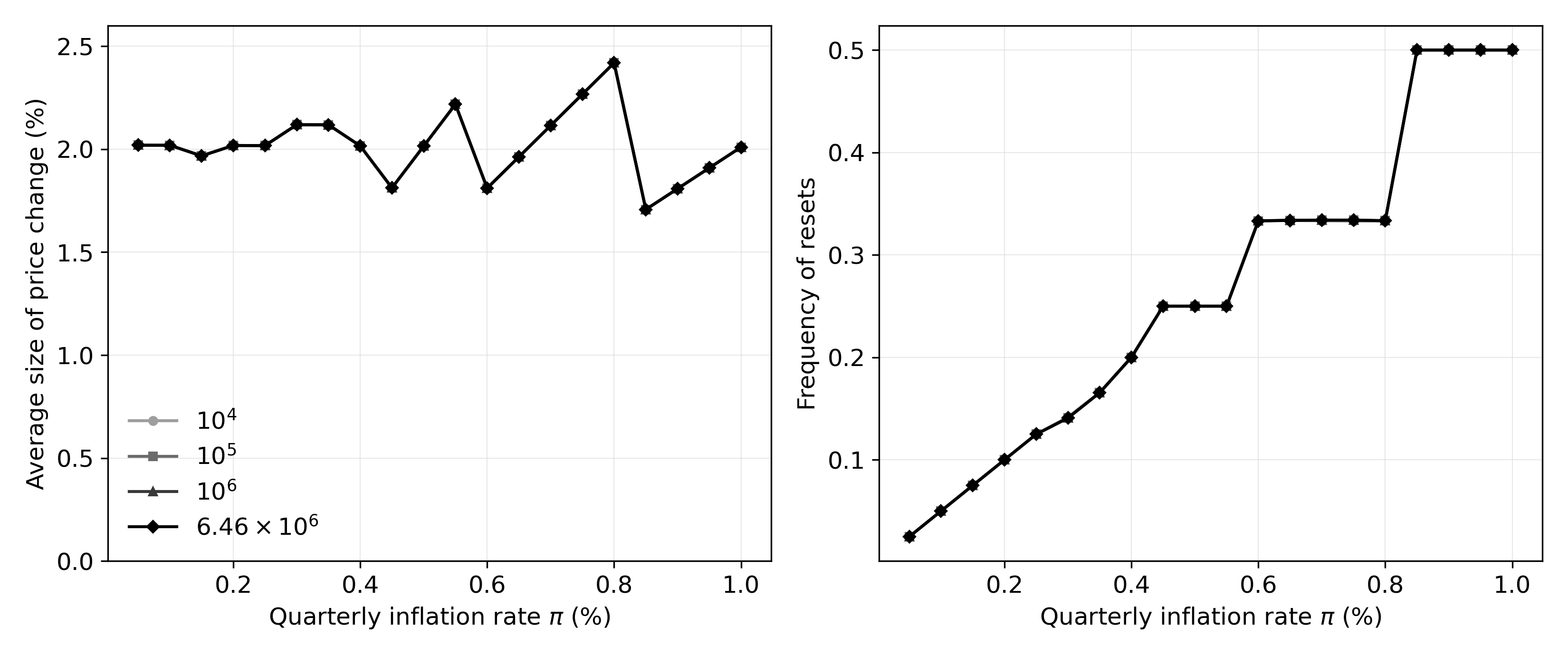}
\caption[Size of price change and frequency of resets on the United States network]{\textbf{Size of price change and frequency of resets on the United States network.}
The left panel reports the average size of price change among the firms that reset their prices, and the right panel the frequency of resets, the fraction of firms that reset their prices each period. Both are plotted against the rate of inflation per period, on reconstructed United States economies of $10^4$, $10^5$, and $10^6$ firms and on the full network of $6.46\times10^6$ firms. The $10^4$ and $10^5$ curves are means over $100$ and $50$ reconstruction draws, while the two largest economies are single reconstructed networks. The four curves coincide.}
\label{fig:recon_us_size_freq}
\end{figure}

It is also worth noting that between $10^4$ firms and the full network, a more than $600$-fold increase in the number of firms, the relative price gap at $\pi=1\%$ per period changes by less than $1\%$.\footnote{At $\pi=1\%$ per period the relative price gap is $0.74$ log points at $10^4$ firms, $0.75$ at $10^5$, $0.72$ at $10^6$, and $0.75$ on the full network. The entropy-equivalent dispersion $\sqrt{2\psi}$ rises from $0.97$ to $1.1$ log points between $10^4$ and $10^5$ firms, and then flattens.} The central-limit mechanism, on which so much of modern macroeconomics rests, makes itself felt only through its absence. Appendix~\ref{subsec:sim_size} reports computational experiments that measure the `scaling' of relative-price distortion with the number of firms, on networks in which the environment of each firm stays fixed as the economy grows. It finds that the distortion does not decay as the economy grows. The pattern observed at the different scales of the reconstructed United States network therefore has little to do with the peculiarities of the United States economy. The basic mechanism of monetary transmission within production networks is such that there are no forces that average one another with a growth in the number of firms.

Figure~\ref{fig:recon_us_size_freq} presents the change in the size of price change and in the frequency of price change with an increase in the rate of inflation. At every rate of inflation in the sweep, firms respond to inflation almost wholly with an increase in the frequency of price resets, while the average size of price change stays near the baseline half-width of the inaction band, about $2\%$, at every size of the economy.\footnote{Nor does the size of price change reveal how far a firm's posted price stands from its equilibrium value. In the steady state of the $10^5$-firm reconstructions, across the firms that reset at a date, the two are uncorrelated, within $\pm0.0004$, under the rules whose resets differ in size across firms. Along the transition they are correlated, and the correlation decays over $100$ to $150$ dates, an order of magnitude more slowly than relative-price distortion settles.} The workhorse models of price setting offer a yardstick with which we can compare these computational results. Appendix~\ref{oa:sim_magnitude} computes the size of price change and the dispersion of relative prices that the Calvo and menu-cost models generate at the same rate of inflation and the same frequency of resets. It does so from the laws of price age that the two rules imply (Lemma~\ref{oa:lem:workhorse_benchmark}). Our back-of-the-envelope calculations suggest that the network model generates relative-price distortions of the Calvo kind but a size of price change that, as in a menu-cost world, barely moves with inflation.

\section{Conclusion}
\label{sec:conclusion}
There was a certain tendency among an older generation of monetary theorists, those of the pre--World War II era, to view inflation as a process. It began with an unequal influx of money into certain parts of the economy, or equivalently into the hands of certain people or firms. It then unfolded over time through transactions between buyers and sellers of various goods in a multitude of markets. Money, as it were, took time to find its way through myriad interlinked markets. The price effects of monetary shocks, with all their peculiarities and specificities, were to be found along the transient routes that money took to dissolve itself into the economic system. The first proponent of this position was perhaps \citeauthor{Cantillon1755}. More than a quarter millennium ago, in his remarkable essay \textit{Essai sur la Nature du Commerce en G\'en\'eral}, he tells us that the increase of money circulating in a state is not always distributed equally among its inhabitants. In Cantillon's account, those who receive the new money first enjoy an advantage, for they can buy at the old prices. As their expenditure increases, prices begin to rise, and it is in this way that money spreads through the state and sustains circulation.\footnote{\citet[pp.~396, 409]{mises1949} notes that ``The quantity of money available in the whole market system cannot increase or decrease otherwise than by first increasing or decreasing the cash holdings of certain individual members''. It is the ``interplay of the reactions of these actors'', some of whom have gotten more money than others, ``that results in alterations of the price structure''.} On such a view of the inflationary process, prices change in a differential and successive manner in response to the disequilibrium flow of money through different markets. This implies that the relative-price distortions generated by monetary shocks depend upon the pattern of trading links through which money flows in the economic system. This view, however, does not imply anything about whether the size of price change in response to monetary shocks carries information about the relative-price distortions that the shocks generate. The position that the size of price change is a good measure of relative price gaps is founded on a wholly different view of the monetary pass-through, a view which originates from Cantillon's contemporary David Hume.

\citet{hume1752money}, in his essay ``Of Interest'', poses a thought experiment in which ``... by miracle, every man in Great Britain shou'd have five pounds slipt into his pocket in one night''. He goes on to claim that the monetary shock would generate a proportionate increase in the prices of all goods, leaving real economic activity unaffected. Unlike in Cantillon, economic actors in Hume's system receive an equal (more likely equiproportional) injection of money, thereby causing no changes in relative prices. It is this idea that \citet{ricardo1817principles} and \citet{mill1848principles}, in later years, develop into the analytical instrument of `neutral money'. Mill puts matters somewhat more accurately when he says that ``though the greater or less quantity of money makes in itself no difference in the rate of interest, a change from a less quantity to a greater, or from a greater to a less, may and does make a difference in it''. One does not find in Hume, Ricardo, and Mill a detailed discussion of the mechanisms by which money enters the economy and percolates through it. This, far from being a shortcoming, is perhaps a strength in pursuing the question they sought to answer: whether the quantity of money influences the wealth of a nation in the long run. All three, and economists since, answer in the negative \citep{lucas1996nobel}.

In the post-war years, monetary theorists introduced sticky prices within Hume's original long-run model to study the short-run effects of inflation, efforts that ultimately led to the literature on microfoundations of price stickiness \citep{ball1994sticky}. Price stickiness was of pivotal importance, for Hume's mental model (by construction) does not generate real effects if prices are fully flexible. The economy, as it were, jumps to a new equilibrium with a different price level but the same relative prices. Nearly all models of monetary non-neutrality, including contemporary workhorse New Keynesian models, ultimately build on a Humean approach to monetary dynamics augmented with pricing frictions of various sorts. Observe that a tight positive relation between the size of price change and relative-price distortions is a natural consequence of price stickiness in a Humean world. The longer it takes a price to change, the more it has deviated from optimal due to accumulated inflation, and the larger the size of the change will be when it ultimately changes. In practice, relative-price distortions are difficult to measure directly, since we never know enough about the real primitives to compute equilibrium prices. The size of price change can therefore serve as a proxy for the relative-price distortions that monetary shocks generate. The conjecture that the size of price change can gauge relative-price distortion is thus theory-dependent. A Humean world does give rise to such a conjecture, but a Cantillonian world does not. In this paper, we have developed a primitive model of a Cantillonian world with sticky prices, wherein inflationary dynamics can generate significant relative-price effects. Yet the size of price change does not identify the degree of relative-price distortions. 

Throughout the paper, we have made an attempt to retain analytical tractability. Perhaps the most significant assumption toward this end was that the production network is fixed, primitive, and diagonalizable. This allowed us to use the Perron--Frobenius theorem to separate monetary shocks into permanent and transient components whose dynamics decouple. We approximated the left and right eigenvectors with moments of the production network, ignored all eigenmodes beyond the second, and let prices follow nominal demand. These reductions allowed us to study inflationary dynamics as the interaction between transient waves of monetary changes that originate at successive dates. With them, we derive two theorems pertaining to the price effects of monetary shocks. The first states that inflation distorts relative prices even when prices are fully flexible: because new money enters unevenly and travels through the network at finite speed, it displaces relative prices while every market clears. The second concerns sticky prices. Stronger state dependence weakens the tie between the size of price change and the rate of inflation by shifting an increasing share of the response from size to frequency, yet inflation remains distortionary. In the steady state, staggered resetting layers a dispersion of price vintages upon the standing network distortion, so sticky prices are at least as distorted as flexible prices at every rate of inflation.

The natural question is whether the analytical separation between the size of price change and relative-price distortion survives once the approximations behind it are relaxed. We examine this question through agent-based computational experiments. Our analytical claims prove robust to relaxing some of the critical assumptions upon which they rest. On a reconstructed production network spanning the universe of United States firms, the computational economy produces a significant relative-price distortion conjointly with a size of price change that tracks inflation very weakly.

Nearly fifty years ago, \citet{Leijonhufvud1977} said ``we know very little about how inflations work their way through the economy''. What he had in mind was not the myriad of `channels' of inflation pass-through, which have since occupied such a prominent place in monetary theory and policy \citep{bernanke1995}. But the sequential and out-of-equilibrium process through which money percolates across real market transactions, one in which local interactions determine prices without a Walrasian auctioneer. In this paper, we hope to have taken some early steps in formally characterizing such an inflationary process. 

\newpage
\small
\singlespacing
\bibliographystyle{aea}
\bibliography{inflation}
\normalsize
\setstretch{1.4}

\newpage

\appendix
\section{Mathematical Appendix}\label{appendix:mathematics}

We now prove the two theorems of the paper. The proofs of the propositions and lemmas, and the statements of the lemmas that the text summarizes without displaying, are in Appendix~B, Sections~\ref{app:block_equilibrium}--\ref{app:block_sticky}, and Table~\ref{oa:tab:results_map} maps each result to its proof and supporting material. Both proofs maintain the standing assumptions of Sections~\ref{model} and~\ref{sec:framework}, including the nominal-demand projection of Assumption~\ref{assu:small_returns}. Except for the first part of the proof of Theorem~\ref{theorem:price_deviation}, they also maintain the two-eigenmode degree representation of Section~\ref{subsec:degree_closure}. Appendix~\ref{oa:sec:perron_degree_bound} bounds or records the spectral and degree approximations of this representation. The Cantillon exposure $\mathcal C_i$ is that of~\eqref{eq:cantillon_exposure}.

\subsection{Proof of Theorem~\ref{theorem:price_deviation}}\label{app:proof_thm2}

The proof starts on the exact network, where each log displacement is, to first order, the rate of inflation times the balance response $\mu_i$. The relative price gap, a standard deviation of these displacements, is then linear in inflation, and relative-price entropy, a divergence, is quadratic. Which leaves the two coefficients to be written in terms of the network. Within the two-eigenmode degree representation each log displacement is the product of the rate of inflation, the Cantillon exposure of the firm, and the propagation kernel. And the propagation kernel converges to the stationary misalignment times network persistence.

\begin{proof}
We first prove the two expansions of Theorem~\ref{theorem:price_deviation} on the exact network. By Proposition~\ref{prop:balanced_growth_convergence} normalized balances converge to $\widehat{\mathbf m}_{\mathrm{ub}}$, and under Assumption~\ref{assu:small_returns} detrended prices converge to $\widehat p_{i,\mathrm{ub}}=\widehat m_{i,\mathrm{ub}}/q_i^{*}$. The log displacements therefore converge to $g_i^{\mathrm{ub}}:=\log\widehat p_{i,\mathrm{ub}}-\log p_i^{*}=\log(\widehat m_{i,\mathrm{ub}}/m_i^{*})$, and the long-run averages $\omega$ and $\psi$ of Definitions~\ref{def:relative_price_gap} and~\ref{def:relative_price_entropy} equal the relative price gap and relative-price entropy of these limits. At $\pi=0$ the balanced-growth limit of normalized balances is $\mathbf m^{*}$, and it is differentiable in $\pi$ there by the implicit-function argument in the proof of Lemma~\ref{lem:first_order_price}. Hence
\[
\mathbf g^{\mathrm{ub}}=\pi\,\boldsymbol\mu+O(\pi^{2}).
\]
The relative price gap is the balance-weighted standard deviation of $\mathbf g^{\mathrm{ub}}$, so
\[
\omega=\pi\bigl(\Var_{\mathbf m^{*}}(\boldsymbol\mu)\bigr)^{1/2}+O(\pi^{2}).
\]
The current price shares are exponential tilts of the equilibrium price shares, $\widehat r_i=\widehat r_i^{*}e^{g_i}/\sum_j\widehat r_j^{*}e^{g_j}$. Writing $K(\mathbf g):=\log\sum_j\widehat r_j^{*}e^{g_j}$, the divergence is $\psi=\sum_i\widehat r_i\,g_i-K(\mathbf g)$, and expanding $K$ to third order gives
\[
\psi=\tfrac12\Var_{\widehat{\mathbf r}^{*}}\bigl(\mathbf g^{\mathrm{ub}}\bigr)+O\bigl(\lVert\mathbf g^{\mathrm{ub}}\rVert^{3}\bigr)
=\frac{\pi^{2}}{2}\Var_{\widehat{\mathbf r}^{*}}(\boldsymbol\mu)+O(\pi^{3}).
\]
Both coefficients, $\digamma_\omega$ and $\digamma_\psi$, are strictly positive. This is because the weights $\mathbf m^{*}$ and $\widehat{\mathbf r}^{*}$ are strictly positive, and the balance response $\boldsymbol\mu$ is not constant across firms when $\mathbf A(\boldsymbol\gamma_0-\mathbf m^{*})\ne\mathbf 0$ (Lemma~\ref{lem:first_order_price}).

We next derive the degree representation of the two coefficients $\digamma_\omega$ and $\digamma_\psi$. Equilibrium prices are $p_i^{*}=m_i^{*}/q_i^{*}=c_md_i/q_i^{*}$ under Assumption~\ref{assu:degree_structure}, with the common scale set by $\mathbf 1^\top\mathbf m^{*}=1$, and the equilibrium price shares are $\widehat r_i^{*}=p_i^{*}/\sum_jp_j^{*}$. Both measures of relative-price distortion are invariant to that common scale, the relative price gap because Definition~\ref{def:relative_price_gap} minimizes over it and relative-price entropy because its normalization removes it, so neither depends on the choice of numeraire. The date-$T$ prices carry the common balanced-growth level $(1+\pi)^{T}$, which is likewise removed. The output--degree map of Assumption~\ref{oa:ass:output_function}, $q_i^{*}\approx d_i^{\vartheta}$, gives $\widehat r_i^{*}\propto d_i^{1-\vartheta}$. It enters only the degree-moment form of the coefficient $\digamma_\psi$ in Appendix~\ref{app:proof_comparative_statics}, and nowhere in the relative price gap.

Under the nominal-demand projection the transient nominal dynamics give the market-clearing price to leading order, and under flexible prices the market-clearing price is also the posted price, $\bar p_i^{(T)}=p_i^{(T)}$:
\[
p_i^{(T)}
\approx
\frac{(1+\pi)^{T}}{q_i^{*}}
\Bigl(
m_i^{*}+\kappa\pi\,\delta_i\mathcal X_T
\Bigr).
\]
Here $\mathcal X_T$ is the propagation kernel in~\eqref{eq:master_kernel}. The right-hand side is $\mathcal D_i^{(T)}/q_i^{*}$, for Assumption~\ref{assu:small_returns} holds output at $q_i^{*}$. The numerator is the sum of the permanent trend and the transitory wave of Lemma~\ref{lemma:exposure}. Dividing by the common balanced-growth level $(1+\pi)^{T}$ and subtracting the equilibrium price $p_i^{*}=m_i^{*}/q_i^{*}$ gives the displacement of the detrended price,
\[
\frac{p_i^{(T)}}{(1+\pi)^{T}}-p_i^{*}
=
\kappa\pi\,\frac{\delta_i}{q_i^{*}}\,
\mathcal X_T.
\]

We take the relative price gap first. By Definition~\ref{def:relative_price_gap}, $\omega_T$ is the balance-weighted standard deviation of the log displacements $g_i^{(T)}$, since the minimization over a common rescaling removes the nominal scale. To leading order $g_i^{(T)}=\ell_i^{(T)}=\pi\mathcal C_i\mathcal X_T$, whose balance-weighted mean is zero by Lemma~\ref{oa:lem:balance_centering}. The weighted variance therefore reduces to the weighted second moment, and
\[
\omega_T
=
\Bigl(\sum_{i=1}^{n}m_i^{*}\bigl(\ell_i^{(T)}\bigr)^{2}\Bigr)^{1/2}+O(\pi^{2})
=
\pi\,\frac{\kappa}{c_m}\,\bigl|\mathcal X_T\bigr|\,
\Bigl(\sum_{i=1}^{n}m_i^{*}\,\widehat\delta_i^{\,2}\Bigr)^{1/2}+O(\pi^{2}).
\]
The last equality substitutes $\mathcal C_i=(\kappa/c_m)\widehat\delta_i$. The balance-weighted mean of the normalized excess degree is zero, so the last factor is its balance-weighted standard deviation, and $\sum_im_i^{*}\widehat\delta_i^{\,2}=\sum_i(\delta_i^{2}/d_i)/\sum_id_i$. Moreover, the propagation kernel $\mathcal X_T$ is strictly positive for $T\ge1$, because the injection misalignments $\xi_t$ are strictly positive (Lemma~\ref{lem:Ct_positive}), and so are the weights $\lambda_2^{\,T-t}(1+\pi)^{-(T-t)}$. The relative price gap $\omega_T$ is increasing in $\mathcal X_T$, for its leading term is proportional to $\mathcal X_T$.

As the economy converges to its steady state, $\boldsymbol{\gamma}_t \to \boldsymbol{\gamma}_{\mathrm{ub}}$ and $\xi_t\to \xi_{\mathrm{ub}}>0$. With the geometric weights of the propagation kernel, these limits give $\mathcal X_T\to\xi_{\mathrm{ub}}\,\lambda_2(1+\pi-\lambda_2)^{-1}$. In the small-$\pi$ regime this limit is $\xi_{\mathrm{ub}}\,\lambda_{2}(1-\lambda_{2})^{-1}>0$ up to an $O(\pi)$ remainder, which the $o(\pi)$ term of the expansion for $\omega$ absorbs. Substituting $\mathcal X_T\to \xi_{\mathrm{ub}}\,\lambda_{2}(1-\lambda_{2})^{-1}+O(\pi)$ gives
\[
\omega
=
\pi\,\digamma_\omega+o(\pi),
\qquad
\digamma_\omega
=
\frac{\kappa}{c_m}\,
\xi_{\mathrm{ub}}\,\frac{\lambda_{2}}{1-\lambda_{2}}\,
\Bigl(\Var_{\mathbf m^{*}}\bigl(\widehat{\boldsymbol\delta}\bigr)\Bigr)^{1/2}.
\]
The coefficient $\digamma_\omega$ is strictly positive whenever the stationary misalignment $\xi_{\mathrm{ub}}$ is strictly positive (Lemma~\ref{lem:Ct_positive}) and the excess degree is not identically zero, which holds for $\nu\ne0$ on any degree sequence taking at least two values.

The degree law of Assumption~\ref{assu:degree_distribution} turns $\digamma_\omega$ into a function of the tail exponent $\alpha$ and the assortativity exponent $\nu$. The law of large numbers, applied to the numerator and the denominator of $\Var_{\mathbf m^{*}}(\widehat{\boldsymbol\delta})=\sum_i(\delta_i^{2}/d_i)/\sum_id_i$, gives in the large-cutoff Pareto approximation
\[
\frac{\E\bigl[d^{-1}\bigl(d^{\nu^{2}}-\E[d^{\nu^{2}}]\bigr)^{2}\bigr]}{\E[d]}
=\;Q(\alpha,\nu)\,(\alpha-1),
\]
where the dispersion functional is
\[
Q(\alpha,\nu)
:=
\frac{1}{\alpha+1-2\nu^{2}}
-\frac{2\alpha}{(\alpha-\nu^{2})(\alpha+1-\nu^{2})}
+\frac{\alpha^{2}}{(\alpha-\nu^{2})^{2}(\alpha+1)},
\]
so that $\alpha\,Q(\alpha,\nu)=\E\bigl[d^{-1}\bigl(d^{\nu^{2}}-\E[d^{\nu^{2}}]\bigr)^{2}\bigr]$ under the large-cutoff Pareto moments $\E[d^{k}]=\alpha/(\alpha-k)$. At a finite cutoff the same display holds with the corresponding truncated moments. The required moments exist under balance weights whenever $\alpha>1$. In the large-cutoff Pareto approximation, therefore,
\[
\digamma_\omega
=
\frac{\kappa}{c_m}\,
\xi_{\mathrm{ub}}\,\frac{\lambda_{2}}{1-\lambda_{2}}\,
\Bigl(\,Q(\alpha,\nu)\,(\alpha-1)\,\Bigr)^{1/2}.
\]

We now turn to relative-price entropy. Since $\mathcal C_i=\kappa\delta_i/m_i^{*}$ and $p_i^{*}=m_i^{*}/q_i^{*}$, the detrended price can be written as
\[
\frac{p_i^{(T)}}{(1+\pi)^{T}}=p_i^{*}\bigl(1+\pi\,\mathcal X_T\,\mathcal C_i\bigr).
\]
Once the propagation kernel has converged, $\mathcal X_T\to \xi_{\mathrm{ub}}\,\frac{\lambda_{2}}{1-\lambda_{2}}+O(\pi)$, and we drop the time index. The steady-state price shares are then
\[
\widehat{r}_i^{(\infty)}
=
\frac{\widehat{r}_i^{\,*}\bigl(1+\pi\,\xi_{\mathrm{ub}}\frac{\lambda_{2}}{1-\lambda_{2}}\,\mathcal C_i\bigr)}{\sum_{j=1}^n \widehat{r}_j^{\,*}\bigl(1+\pi\,\xi_{\mathrm{ub}}\frac{\lambda_{2}}{1-\lambda_{2}}\,\mathcal C_j\bigr)}.
\]
The limit of the propagation kernel, $\xi_{\mathrm{ub}}\,\lambda_2/(1-\lambda_2)$, is the same for every firm, so it passes unchanged through every centering in the entropy expansion.

Since the price shares sum to one, their first-order expansion is
\[
\widehat{r}_{i}^{(\infty)}
=
\widehat{r}_{i}^{\,*}\Bigl\{1+\pi\,\xi_{\mathrm{ub}}\frac{\lambda_{2}}{1-\lambda_{2}}\Bigl(\mathcal C_{i}-\sum_{j=1}^{n}\widehat{r}_{j}^{\,*}\mathcal C_{j}\Bigr)\Bigr\}+O(\pi^{2}).
\]
The first-order perturbation $\mathcal C_{i}-\sum_{j}\widehat{r}_{j}^{\,*}\mathcal C_{j}$ has zero mean under $\widehat{r}^{\,*}$, so the linear term cancels in relative-price entropy, which starts at second order:
\[
\psi
=
\frac{\pi^{2}}{2}\Bigl(\xi_{\mathrm{ub}}\,\frac{\lambda_{2}}{1-\lambda_{2}}\Bigr)^{\!2}\sum_{i=1}^{n}\widehat{r}_{i}^{\,*}\Bigl(\mathcal C_{i}-\sum_{j=1}^{n}\widehat{r}_{j}^{\,*}\mathcal C_{j}\Bigr)^{\!2}+o(\pi^{2}).
\]
The coefficient of $\pi^{2}$ is therefore
\[
\psi
=
\pi^{2}\,\digamma_\psi+o(\pi^{2}),
\qquad
\digamma_\psi
=
\frac{\bigl(\xi_{\mathrm{ub}}\,\frac{\lambda_{2}}{1-\lambda_{2}}\bigr)^{2}}{2}\sum_{i=1}^n \widehat{r}_i^{\,*}
\Biggl(\mathcal C_i-\sum_{j=1}^n \widehat{r}_j^{\,*}\mathcal C_j\Biggr)^{\!2}
=\frac12\Bigl(\xi_{\mathrm{ub}}\,\frac{\lambda_{2}}{1-\lambda_{2}}\Bigr)^{2}\Var_{\widehat{\mathbf r}^{*}}\bigl(\boldsymbol{\mathcal C}\bigr).
\]
Moreover, $\digamma_\psi>0$ whenever the Cantillon exposure varies across firms. And since the date-$T$ relative price gap and relative-price entropy converge, their long-run averages $\omega$ and $\psi$ (Definitions~\ref{def:relative_price_gap} and~\ref{def:relative_price_entropy}) coincide with these limits.

\end{proof}

\subsection{Proof of Theorem~\ref{thm:size_not_measure}}\label{app:proof_size_not_measure}

The proof lets firms reset more and more often once the economy has reached its steady state. Each price change then covers a single period of inflation, so the average size of price change converges to $\log(1+\pi)$ in every economy. The dispersion of price vintages vanishes at the same time, and both measures of relative-price distortion converge to their levels in the flexible-price regime. These levels are strictly positive (Theorem~\ref{theorem:price_deviation}) and differ across economies (Proposition~\ref{prop:comparative_statics}).

\begin{proof}
We fix $\pi>0$ and $\theta\in(0,1)$, so that the stationary misalignment $\xi_{\mathrm{ub}}$ is strictly positive (Lemma~\ref{lem:Ct_positive}), and suppose the Cantillon exposure is not constant across firms. We send the horizon to infinity first, so that every statement is a steady-state one, and then take any sequence of price reset rules of the theorem. Along that sequence,
\begin{equation}\label{eq:continuous_adjustment_limit}
\begin{aligned}
\Pr\bigl(u_i^{(t)}=0\bigr)&\longrightarrow1,\\
\E\!\left[u_i^{(t-1)}+1\mid I_i^{(t)}=1\right]&\longrightarrow1
\qquad\text{for every firm}.
\end{aligned}
\end{equation}
The order of limits matters, for the distortion is the standing object, and the price reset rule is what we vary around it.

We first prove the limit of the average size of price change $\phi_T$. Because the horizon limit is taken first, Corollary~\ref{oa:cor:sticky_bgp_vintage} gives the exact price change of a firm that resets its price on the balanced-growth path,
\[
\Delta\log\bar p_i^{(T)}
=\bigl(u_i^{(T-1)}+1\bigr)\log(1+\pi)>0,
\qquad I_i^{(T)}=1.
\]
The stationary firm-specific displacement cancels exactly between the two price reset dates. The absolute value in Definition~\ref{def:size_price_change} may therefore be dropped without a separate small-injection sign condition. Along the sequence of price reset rules, the limits~\eqref{eq:continuous_adjustment_limit} make the weighted mean length of completed price spells, conditional on adjustment, converge to one for every weight exponent $\zeta\ge0$, and hence
\[
\phi_T\longrightarrow\log(1+\pi).
\]
The average size of price change therefore carries no network or injection primitive in the limit.

We next prove the limits of the two measures of relative-price distortion. We apply the same limit to the decomposition of Lemma~\ref{lemma:sticky_distortion}. The monotone hazard strengthens the first limit in~\eqref{eq:continuous_adjustment_limit} to the required second-moment convergence. For each firm, stationarity gives
\[
0\le\mathcal S_i(1)
\le\Pr(u_i=0)^{-1}-1
\longrightarrow0,
\]
where $\mathcal S_i(u)$ is the probability that firm $i$'s price survives beyond $u$ periods, the survival function of Appendix~\ref{app:proof_frequency}. Because the hazard function is nondecreasing,
$\mathcal S_i(u+1)/\mathcal S_i(u)\le\mathcal S_i(1)$ and hence
$\mathcal S_i(u)\le\mathcal S_i(1)^u$. Therefore
\[
\E[u_i^2]
=\Pr(u_i=0)\sum_{u\ge1}u^2\mathcal S_i(u)
\le
\sum_{u\ge1}u^2\mathcal S_i(1)^u
=\frac{\mathcal S_i(1)\{1+\mathcal S_i(1)\}}
{\{1-\mathcal S_i(1)\}^3}
\longrightarrow0.
\]
It follows that $\E[\Var_{\widehat r^*}(u)]\to0$, and likewise under balance weights, so the vintage-dispersion terms vanish. Since $\mathfrak x$ is bounded on the truncated support, the Cauchy--Schwarz inequality also gives $\E[|\Cov_{\widehat r^*}(u,\mathfrak x)|]\to0$, and likewise under balance weights, so the interaction terms vanish too. And since the horizon limit is taken first, the propagation kernel is constant, $\mathcal X_{T-u_i}=\mathcal X_T$, so the realized-vintage term is zero. Both measures in the sticky-price regime therefore converge to their levels in the flexible-price regime,
\[
\omega^{\mathrm{stick}}\longrightarrow\omega^{\mathrm{flex}}
=\pi\digamma_\omega+o(\pi),
\qquad
\psi^{\mathrm{stick}}\longrightarrow\psi^{\mathrm{flex}}
=\pi^{2}\digamma_\psi+o(\pi^{2}),
\]
where the expansions are those of Theorem~\ref{theorem:price_deviation}. Within the two-eigenmode degree representation, $\omega^{\mathrm{flex}}$ is the balance-weighted standard deviation of the displacements $\log(1+\pi\mathcal C_i\mathcal X_\infty)$, with $\mathcal X_\infty=\xi_{\mathrm{ub}}\lambda_2(1+\pi-\lambda_2)^{-1}>0$, so it is strictly positive at the fixed $\pi$, for the Cantillon exposure is not constant across firms. And $\psi^{\mathrm{flex}}$ is strictly positive by Gibbs' inequality, for the same displacements move the price shares away from the equilibrium price shares. Hence $\omega^{\mathrm{flex}}>0$ and $\psi^{\mathrm{flex}}>0$ in the very limit in which $\phi_T$ becomes uninformative.

We finally prove that $\phi_T$ identifies neither measure of relative-price distortion. In that limit $\phi_T=\log(1+\pi)$ is a function of $\pi$ alone. By Theorem~\ref{theorem:price_deviation}, by contrast, $\omega^{\mathrm{flex}}$ depends on the injection and the network through $\xi_{\mathrm{ub}}$, $\lambda_2$, $\alpha$, and $\nu$. And $\psi^{\mathrm{flex}}$ depends on them through the same limit of the propagation kernel and through the dispersion of the Cantillon exposure under equilibrium price shares. Proposition~\ref{prop:comparative_statics} shows that the two coefficients $\digamma_\omega$ and $\digamma_\psi$ vary across the economies that the assumptions admit. At a fixed $\pi$, these economies therefore share $\phi_T=\log(1+\pi)$ but can carry different $\omega^{\mathrm{flex}}$ and $\psi^{\mathrm{flex}}$. Which means that no function of $\phi_T$ can return either measure.
\end{proof}

\onlineappendix{B}{Analytical Results and Proofs}

\begin{table}[H]
\centering
\caption{Where each main-text result is proved, and what it rests on}
\label{oa:tab:results_map}
\footnotesize
\setlength{\tabcolsep}{4pt}
\renewcommand{\arraystretch}{1.15}
\begin{tabularx}{\textwidth}{@{}>{\raggedright\arraybackslash}p{0.42\textwidth}>{\raggedright\arraybackslash}p{0.17\textwidth}>{\raggedright\arraybackslash}X@{}}
\toprule
Main-text result & Proof & Rests on\\
\midrule
Proposition~\ref{prop:equilibrium_existence} (existence and uniqueness of equilibrium) & Section~\ref{app:proof_equilibrium_existence} & --\\
Proposition~\ref{prop:balanced_growth_convergence} (balanced-growth convergence) & Section~\ref{oa:proof_balanced_growth} & convergence of the injection profile \citep{bhattathiripad2026}\\
Lemma~\ref{lemma:exposure} (permanent trend and transitory wave) & Section~\ref{app:money_injection_and_transient} & Lemma~\ref{oa:lem:eigenvector_proxies}\\
Proposition~\ref{prop:size_price_change} (transient price growth by firm size) & Section~\ref{app:proof_thm1} & Lemmas~\ref{lemma:exposure} and~\ref{lem:Ct_positive}\\
Theorem~\ref{theorem:price_deviation} (relative-price distortion under flexible prices) & Appendix~\ref{app:proof_thm2} & Lemmas~\ref{lem:first_order_price}, \ref{lem:Ct_positive}, and~\ref{oa:lem:balance_centering}, Assumption~\ref{oa:ass:output_function}\\
Proposition~\ref{prop:comparative_statics} (network comparative statics) & Section~\ref{app:proof_comparative_statics} & Lemmas~\ref{lem:Ct_positive} and~\ref{oa:lem:dispersion_growth}\\
Proposition~\ref{prop:frequency} (frequency and size margins) & Section~\ref{app:proof_frequency} & Lemmas~\ref{oa:lem:level_amplitude}, \ref{oa:lem:uniform_positivity}, and~\ref{lemma:hazard_sufficiency}\\
Lemma~\ref{lemma:network_vintages} (price spells ordered by Cantillon exposure) & Section~\ref{app:proof_network_vintages} & Lemma~\ref{lemma:hazard_sufficiency}\\
Proposition~\ref{prop:sticky_flexible_ranking} (ratio of sticky-price to flexible-price distortion) & Section~\ref{app:proof_sticky_flexible_ranking} & Theorem~\ref{theorem:price_deviation}, Lemma~\ref{lem:sticky_orders}\\
Theorem~\ref{thm:size_not_measure} (the size of price change does not identify relative-price distortion) & Appendix~\ref{app:proof_size_not_measure} & Corollary~\ref{oa:cor:sticky_bgp_vintage}, Lemma~\ref{lemma:sticky_distortion}\\
\bottomrule
\end{tabularx}
\end{table}

\newpage

\noindent Throughout this appendix the propagation matrix $\mathbf A$ is primitive and diagonalizable, and its second eigenvalue is real, simple, and separated from the rest of the spectrum (Assumptions~\ref{assu:primitivity}, \ref{assu:spectral_gap}, and~\ref{assu:diagonalisable}). We call the replacement of the Perron vector by normalized degree, $m_i^{*}=c_md_i$, the `Perron--degree representation'. And we write $\operatorname{cond}(\mathbf V):=\lVert\mathbf V\rVert\,\lVert\mathbf V^{-1}\rVert$ for the condition number, in the spectral norm, of the matrix $\mathbf V$ whose columns are the right eigenvectors of $\mathbf A$.

\section{Equilibrium and Convergence}\label{app:block_equilibrium}

\subsection{Existence and uniqueness of equilibrium}\label{app:proof_equilibrium_existence}

\begin{proof}[Proof of Proposition~\ref{prop:equilibrium_existence}]
By the Perron--Frobenius theorem, the column-stochastic and primitive matrix $\mathbf A$ (Assumption~\ref{assu:primitivity}) has a unique positive right eigenvector $\mathbf m^{*}$ at eigenvalue $1$, fixed by the normalization $\mathbf 1^\top\mathbf m^{*}=1$. For the output block, we take logs of the feasibility and market-clearing conditions. With $p_i^{*}=m_i^{*}/q_i^{*}$,
\[
\begin{aligned}
\log q_j^{*}
&=\varsigma\sum_i a_{ij}\bigl(\log a_{ij}+\log m_j^{*}-\log p_i^{*}\bigr)\\
&=\varsigma\sum_i a_{ij}\log a_{ij}+\varsigma\log m_j^{*}-\varsigma\bigl(\mathbf A^\top\log\mathbf m^{*}\bigr)_j+\varsigma\bigl(\mathbf A^\top\log\mathbf q^{*}\bigr)_j,
\end{aligned}
\]
which rearranges to
\[
(\mathbf I-\varsigma\mathbf A^\top)\log\mathbf q^{*}
=\varsigma\!\left[\mathbf b_A+(\mathbf I-\mathbf A^\top)\log\mathbf m^{*}\right].
\]
Since $\mathbf A^\top$ is row-stochastic, the spectral radius of $\varsigma\mathbf A^\top$ is $\rho(\varsigma\mathbf A^\top)=\varsigma<1$, so $\mathbf I-\varsigma\mathbf A^\top$ is a nonsingular $M$-matrix with convergent Neumann inverse $\sum_{k\ge0}\varsigma^{k}(\mathbf A^\top)^{k}\ge0$. Hence $\log\mathbf q^{*}$ exists and is unique, and $\mathbf q^{*}=\exp(\log\mathbf q^{*})>0$. Prices follow from $p_i^{*}=m_i^{*}/q_i^{*}>0$, and they are unique because $\mathbf I-\varsigma\mathbf A^\top$ is invertible and $\mathbf m^{*}$ is unique.
\end{proof}

\subsection{Convergence without injection}\label{app:proof_monotone_convergence}

Without injection, balances, output, and prices return to the stationary equilibrium of Proposition~\ref{prop:equilibrium_existence}. The balance deviation ($\mathbf m_t-\mathbf m^{*}$) aligns asymptotically with the second eigenvector of the propagation matrix ($\mathbf v_2$) whenever it loads on that eigenvector. The main text summarizes this result after the real-block recursion~\eqref{eq:zero_injection_real_dynamics}.

\begin{oalemma}[Convergence and second-eigenmode alignment]\label{lemma:monotone_convergence}
Let the rate of inflation of Section~\ref{subsec:monetary_process} be zero, $\pi=0$, let prices be flexible, and let $\varsigma\in(0,1)$. From every positive initial configuration of balances, orders in transit, and output with $\mathbf 1^\top\mathbf m_0=1$, the economy converges to the stationary equilibrium of Proposition~\ref{prop:equilibrium_existence},
\[
\mathbf m_t\longrightarrow\mathbf m^{*},
\qquad
\mathbf q_t\longrightarrow\mathbf q^{*},
\qquad
\mathbf p_t\longrightarrow\mathbf p^{*}.
\]
The balance deviation satisfies $\lVert\mathbf m_t-\mathbf m^{*}\rVert=O(\lambda_2^{t})$. Whenever its projection on the eigenspace of $\lambda_2$ is nonzero, it becomes asymptotically aligned with that eigendirection, so its second-eigenmode component declines without oscillation. Output and prices converge at the geometric factor $\max\{\lambda_2,\varsigma\}$, and their deviations are $O\bigl(\max\{\lambda_2,\varsigma\}^{t}\bigr)$ when $\lambda_2\ne\varsigma$ and $O(t\lambda_2^{t})$ at equality.
\end{oalemma}

\begin{proof}
With zero injection the balances evolve by $\mathbf m_{t+1}=\mathbf A\mathbf m_t$, and $\mathbf 1^\top\mathbf A=\mathbf 1^\top$ conserves the money stock, $\mathbf 1^\top\mathbf m_t=\mathbf 1^\top\mathbf m_0$. Since $\mathbf A$ is column-stochastic and primitive (Assumption~\ref{assu:primitivity}), Perron--Frobenius gives a simple eigenvalue $\lambda_1=1$ with right eigenvector $\mathbf m^{*}$ (normalized $\mathbf 1^\top\mathbf m^{*}=1$) and constant left eigenvector $\mathbf u_1=\mathbf 1$. Every other eigenvalue satisfies $|\lambda_j|\le\lambda_2<1$. We fix an initial balance vector with $\mathbf 1^\top\mathbf m_0=1$, so that its Perron component is exactly $\mathbf m^{*}$, and expand the remainder in the transitory eigenbasis, $\mathbf m_0-\mathbf m^{*}=\sum_{j\ge2}c_j\mathbf v_j$. Applying $\mathbf A^{t}$,
\[
\mathbf m_t-\mathbf m^{*}=\sum_{j\ge2}c_j\lambda_j^{t}\,\mathbf v_j,
\qquad
\bigl\lVert\mathbf m_t-\mathbf m^{*}\bigr\rVert=O(\lambda_2^{t}),
\]
so the deviation decays geometrically at rate no slower than $\lambda_2$, with relaxation time of order $[-\log\lambda_2]^{-1}=\Theta\bigl((1-\lambda_2)^{-1}\bigr)$ as $\lambda_2\to1$. If $c_2\ne0$, then, because $\lambda_2$ is real, simple, and strictly dominant in modulus among the transitory eigenvalues (Assumption~\ref{assu:spectral_gap}),
\[
\mathbf m_t-\mathbf m^{*}=c_2\lambda_2^{t}\mathbf v_2+o(\lambda_2^{t}),
\]
so the deviation becomes asymptotically aligned with $\mathbf v_2$. Since $\lambda_2\in(0,1)$, the magnitude of this second-eigenmode component declines monotonically and carries no oscillation. If $c_2=0$, the leading eigenmode among those with a nonzero coefficient determines the asymptotic path, and its orientation may oscillate. In all cases the $O(\lambda_2^t)$ bound holds.

It remains to establish convergence of output and prices. Define $\boldsymbol\mu_t:=\log\mathbf m_t-\log\mathbf m^{*}$ and $\mathbf y_t:=\log\mathbf q_t-\log\mathbf q^{*}$. Since $\mathbf m_t\to\mathbf m^{*}$ and $\min_i m_i^{*}>0$, the coordinatewise logarithm is Lipschitz on a neighborhood of $\mathbf m^{*}$, so $\lVert\boldsymbol\mu_t\rVert\le C\lVert\mathbf m_t-\mathbf m^{*}\rVert=O(\lambda_2^{t})$. Subtracting the stationary counterpart of the real recursion~\eqref{eq:zero_injection_real_dynamics} gives the forced deviation
\[
\mathbf y_{t+1}=\varsigma\mathbf A^\top\mathbf y_t+\varsigma\bigl(\boldsymbol\mu_{t-1}-\mathbf A^\top\boldsymbol\mu_t\bigr),
\]
with $\boldsymbol\mu_{-1}$ fixed by the initial orders in transit. Row-stochasticity of $\mathbf A^\top$ bounds the forcing by a constant multiple of $\lVert\boldsymbol\mu_{t-1}\rVert_\infty+\lVert\boldsymbol\mu_t\rVert_\infty=O(\lambda_2^{t})$, and the variation-of-constants solution is
\[
\mathbf y_t=(\varsigma\mathbf A^\top)^{t}\,\mathbf y_0+\varsigma\sum_{s=0}^{t-1}(\varsigma\mathbf A^\top)^{t-1-s}\bigl(\boldsymbol\mu_{s-1}-\mathbf A^\top\boldsymbol\mu_s\bigr).
\]
Since $\mathbf A^\top$ is nonnegative with unit row sums, $\lVert(\varsigma\mathbf A^\top)^{k}\rVert_\infty=\varsigma^{k}$ for every $k\ge0$, so
\[
\lVert\mathbf y_t\rVert_\infty
=O\left(\varsigma^{t}+\sum_{s=0}^{t-1}\varsigma^{t-1-s}\lambda_2^{s}\right),
\qquad
\sum_{s=0}^{t-1}\varsigma^{t-1-s}\lambda_2^{s}
=\begin{cases}
\dfrac{\varsigma^{t}-\lambda_2^{t}}{\varsigma-\lambda_2}, & \varsigma\ne\lambda_2,\\[9pt]
t\,\varsigma^{t-1}, & \varsigma=\lambda_2.
\end{cases}.
\]
Hence $\lVert\mathbf y_t\rVert=O(\max\{\varsigma,\lambda_2\}^{t})$, with the polynomial correction at equality, and $\mathbf q_t\to\mathbf q^{*}$. Finally, flexible market clearing gives $\log\mathbf p_t-\log\mathbf p^{*}=\boldsymbol\mu_t-\mathbf y_t$, so prices converge at the same joint rate.
\end{proof}

\subsection{The spectrum of the adjustment dynamics}
\label{oa:stability_full}

In Lemma~\ref{lemma:monotone_convergence} balances converge at the geometric factor $\lambda_2$, and output at the larger of $\lambda_2$ and the returns to scale ($\varsigma$). Both $\lambda_2$ and $\varsigma$ are eigenvalues of the joint nominal--real dynamics, linearized under zero injection.

\begin{oalemma}[Spectrum of the adjustment dynamics]\label{oa:lem:full_stability}
Suppose $\varsigma\in(0,1)$, prices are flexible, and injection is zero. Linearize at the zero-injection equilibrium normalized by $\mathbf 1^\top\mathbf m^*=1$, using the unrestricted augmented state $(\boldsymbol\mu_t,\boldsymbol\mu_{t-1},\mathbf y_t)$. The linearized propagator of the joint nominal--real economy is block triangular, with spectrum $\{\lambda_j\}_{j=1}^{n}\cup\{\varsigma\lambda_j\}_{j=1}^{n}$ together with $n$ zero eigenvalues contributed by the settlement-lag register. The nominal eigenvalue $\lambda_1=1$ corresponds to a common rescaling of money balances and is excluded from perturbations that preserve the money stock.
\end{oalemma}

\begin{proof}
The zero-injection log-balance Jacobian
$\bigl(\operatorname{diag}(\mathbf m^*)\bigr)^{-1}\mathbf A\operatorname{diag}(\mathbf m^*)$
is similar to $\mathbf A$. The settlement lag makes the real block read the balance deviation at two dates, so the propagator acts on the state $(\boldsymbol\mu_t,\boldsymbol\mu_{t-1},\mathbf y_t)$ and equals
\[
\begin{pmatrix}
\bigl(\operatorname{diag}(\mathbf m^*)\bigr)^{-1}\mathbf A\operatorname{diag}(\mathbf m^*) & \mathbf 0 & \mathbf 0\\[2pt]
\mathbf I & \mathbf 0 & \mathbf 0\\[2pt]
-\varsigma\mathbf A^\top & \varsigma\mathbf I & \varsigma\mathbf A^\top
\end{pmatrix}.
\]
This propagator is block triangular, so its spectrum is $\{\lambda_j\}\cup\{\varsigma\lambda_j\}$ together with the $n$ zeros of the settlement-lag register. The nominal block contributes $\{\lambda_j\}$, whose leading transient is $\lambda_2$, and decreasing returns contribute $\{\varsigma\lambda_j\}$, whose leading modulus is $\varsigma$.
\end{proof}

\noindent Cobb--Douglas shares are independent of $\varsigma$, so $\mathbf A$ and $\lambda_2$ do not depend on it. The returns to scale $\varsigma$ re-enter prices only through output, in $p_i^{(t)}=\mathcal D_i^{(t)}/q_i^{(t)}$. The nominal-demand projection of Assumption~\ref{assu:small_returns} lets prices follow nominal demand and so projects the price map onto the nominal block. Every analytical relation for prices and for the distortion of relative prices in Sections~\ref{sec:flexible}--\ref{sec:sticky} then depends on $\mathbf A$ and $\lambda_2$. But the real spectrum $\{\varsigma\lambda_j\}$ still governs the dynamics of output, and the computational economy retains it. Under sticky prices full settlement preserves the same one-way nominal-to-real ordering, although the real block becomes piecewise nonlinear through rationing.

\subsection{Balanced growth under sustained injection}
\label{oa:proof_balanced_growth}

The proof of Proposition~\ref{prop:balanced_growth_convergence} takes the convergence of the injection profile, $\boldsymbol\gamma_t\to\boldsymbol\gamma_{\mathrm{ub}}$, as given \citep[Proposition~5 and Supplementary Material, Corollary~S1.3]{bhattathiripad2026}.

\begin{proof}[Proof of Proposition~\ref{prop:balanced_growth_convergence}]
We write $M_t=(1+\pi)^{t}$ and $\widehat{\mathbf m}_t=\mathbf m_t/M_t$, so that the normalized recursion~\eqref{eq:normalised_balance_dynamics} governs the distribution of money across firms.

We begin with the fixed point. Rearranging~\eqref{eq:balanced_growth_money_fixed_point} gives $\bigl((1+\pi)\mathbf I-\mathbf A\bigr)\widehat{\mathbf m}_{\mathrm{ub}}=\pi\mathbf A\boldsymbol\gamma_{\mathrm{ub}}$. Since $\rho(\mathbf A)=1<1+\pi$, the inverse exists and has the Neumann representation
\[
\bigl((1+\pi)\mathbf I-\mathbf A\bigr)^{-1}
=\frac{1}{1+\pi}\sum_{k=0}^{\infty}\Bigl(\frac{\mathbf A}{1+\pi}\Bigr)^{k},
\]
nonnegative term by term and, by primitivity, entrywise positive once the powers of $\mathbf A$ are positive. Hence $\widehat{\mathbf m}_{\mathrm{ub}}=\pi\bigl((1+\pi)\mathbf I-\mathbf A\bigr)^{-1}\mathbf A\boldsymbol\gamma_{\mathrm{ub}}$ exists, is unique, and is strictly positive. Left-multiplying the fixed-point equation by $\mathbf 1^\top$ and using column-stochasticity gives $\mathbf 1^\top\widehat{\mathbf m}_{\mathrm{ub}}=\frac{1}{1+\pi}\bigl(\mathbf 1^\top\widehat{\mathbf m}_{\mathrm{ub}}+\pi\bigr)$, so $\mathbf 1^\top\widehat{\mathbf m}_{\mathrm{ub}}=1$, and the limit lies in the interior of the simplex.

The same fixed point gives the limit of normalized balances $\widehat{\mathbf m}_{\mathrm{ub}}$ at low inflation. In the biorthonormal eigensystem of $\mathbf A$, with $\mathbf v_1=\mathbf m^*$ and $\mathbf u_1=\mathbf1$,
\[
\widehat{\mathbf m}_{\mathrm{ub}}
=
\mathbf m^*
+\pi\sum_{j=2}^{n}
\frac{\lambda_j}{1+\pi-\lambda_j}\,
\mathbf v_j\mathbf u_j^\top\boldsymbol\gamma_{\mathrm{ub}}.
\]
On a fixed network the second term is $O(\pi)$, with a constant set by the spectral gap $1-\lambda_2$ and the eigenbasis condition number $\operatorname{cond}(\mathbf V)$. Smoothness of the power map on the interior of the simplex therefore gives
\[
\gamma_{i,\mathrm{ub}}
=
\frac{(m_i^*)^\theta}{\sum_j(m_j^*)^\theta}
+O(\pi)
\;\approx\;
\frac{d_i^\theta}{\sum_jd_j^\theta}
+O(\pi),
\]
under the Perron--degree representation, again on a fixed network \citep[Supplementary Material, Proposition~S2.1]{bhattathiripad2026}. This leading profile, $d_i^\theta/\sum_jd_j^\theta$, is the stationary injection profile of the low-inflation coefficients of the main text. But the constant in this step also depends on the smallest stationary balance $\min_i m_i^*$, so a spectral gap bounded away from zero does not by itself make the approximation uniform along a sequence of growing sparse networks.

For convergence, let $\mathbf e_t:=\widehat{\mathbf m}_t-\widehat{\mathbf m}_{\mathrm{ub}}$ be the deviation of normalized balances from their limit. Subtracting the fixed-point equation from the recursion gives
\begin{equation}
\label{oa:eq:normalised_money_error}
\mathbf e_{t+1}
=\frac{\mathbf A}{1+\pi}\,\mathbf e_t
+\frac{\pi\,\mathbf A}{1+\pi}\bigl(\boldsymbol\gamma_t-\boldsymbol\gamma_{\mathrm{ub}}\bigr),
\end{equation}
whose iteration is
\[
\mathbf e_t
=\Bigl(\frac{\mathbf A}{1+\pi}\Bigr)^{t}\mathbf e_0
+\frac{\pi}{1+\pi}\sum_{s=0}^{t-1}\Bigl(\frac{\mathbf A}{1+\pi}\Bigr)^{t-1-s}\mathbf A\bigl(\boldsymbol\gamma_s-\boldsymbol\gamma_{\mathrm{ub}}\bigr).
\]
Both $\mathbf e_t$ and $\boldsymbol\gamma_t-\boldsymbol\gamma_{\mathrm{ub}}$ have zero coordinate sum, and column-stochasticity preserves the zero-sum subspace, on which the homogeneous operator has asymptotic factor $\lambda_2/(1+\pi)<1$. Since $\boldsymbol\gamma_t-\boldsymbol\gamma_{\mathrm{ub}}\to\mathbf 0$, the stable convolution vanishes and $\widehat{\mathbf m}_t\to\widehat{\mathbf m}_{\mathrm{ub}}$. If the injection profile converges at a geometric rate, $\lVert\boldsymbol\gamma_t-\boldsymbol\gamma_{\mathrm{ub}}\rVert=O(\rho_\gamma^{t})$ with $\rho_\gamma<1$, the convolution bound gives the factor $\rho_m=\max\{\lambda_2/(1+\pi),\rho_\gamma\}$, with the usual polynomial correction when the two coincide.

For the real block, the post-injection balances $\widetilde{\mathbf m}_{t-1}$ finance the orders of date $t-1$, which settle at date $t$, so the recursion~\eqref{eq:zero_injection_real_dynamics} reads
\[
\log\mathbf q_{t+1}
=\varsigma\mathbf A^\top\log\mathbf q_t
+\varsigma\,\mathbf b_A
+\varsigma\bigl(\log\widetilde{\mathbf m}_{t-1}-\mathbf A^\top\log\mathbf m_t\bigr).
\]
We write $\widehat{\mathbf n}_t:=\widetilde{\mathbf m}_t/M_{t+1}=(\widehat{\mathbf m}_t+\pi\boldsymbol\gamma_t)/(1+\pi)$, so that $\mathbf A\widehat{\mathbf n}_t=\widehat{\mathbf m}_{t+1}$, $\log\widetilde{\mathbf m}_{t-1}=\log M_t\,\mathbf 1+\log\widehat{\mathbf n}_{t-1}$, and $\log\mathbf m_t=\log M_t\,\mathbf 1+\log\widehat{\mathbf m}_t$. Because $\mathbf A^\top\mathbf 1=\mathbf 1$, the common trend enters the settled orders and the settling prices at the same date and cancels,
\[
\log\mathbf q_{t+1}
=\varsigma\mathbf A^\top\log\mathbf q_t
+\varsigma\,\mathbf b_A
+\varsigma\bigl(\log\widehat{\mathbf n}_{t-1}-\mathbf A^\top\log\widehat{\mathbf m}_t\bigr).
\]
The real block is therefore driven only by the distribution of money across firms. Since $\widehat{\mathbf m}_t\to\widehat{\mathbf m}_{\mathrm{ub}}$ and $\boldsymbol\gamma_t\to\boldsymbol\gamma_{\mathrm{ub}}$, the forcing converges to $\varsigma(\log\widehat{\mathbf n}_{\mathrm{ub}}-\mathbf A^\top\log\widehat{\mathbf m}_{\mathrm{ub}})$. The fixed point of the output recursion is then the $\log\mathbf q_{\mathrm{ub}}$ displayed in Proposition~\ref{prop:balanced_growth_convergence}, unique because $\mathbf I-\varsigma\mathbf A^\top$ is invertible. Subtracting the fixed-point equation gives
\[
\log\mathbf q_{t+1}-\log\mathbf q_{\mathrm{ub}}
=\varsigma\mathbf A^\top\bigl(\log\mathbf q_t-\log\mathbf q_{\mathrm{ub}}\bigr)
+\varsigma\bigl[(\log\widehat{\mathbf n}_{t-1}-\log\widehat{\mathbf n}_{\mathrm{ub}})-\mathbf A^\top(\log\widehat{\mathbf m}_t-\log\widehat{\mathbf m}_{\mathrm{ub}})\bigr].
\]
The limits lie in the interior of the simplex, so the logarithms are Lipschitz on a neighborhood of them and the forcing converges at the rate of $(\widehat{\mathbf m}_t,\boldsymbol\gamma_t)$. Variation of constants, as in the proof of Lemma~\ref{lemma:monotone_convergence}, gives the joint factor $\max\{\varsigma,\rho_m\}$ and $\mathbf q_t\to\mathbf q_{\mathrm{ub}}$.

Finally,
\[
\widehat p_i^{(t)}
=\frac{p_i^{(t)}}{M_t}
=\frac{\widehat m_i^{(t)}}{q_i^{(t)}}
\longrightarrow
\frac{\widehat m_{i,\mathrm{ub}}}{q_{i,\mathrm{ub}}},
\]
so detrended prices converge, and
\[
\Delta\log p_i^{(t)}
=\log(1+\pi)
+\Delta\log\widehat m_i^{(t)}
-\Delta\log q_i^{(t)}
\longrightarrow
\log(1+\pi).
\]
\end{proof}

\section{Monetary Transmission through the Network}\label{app:block_transmission}

\subsection{The first-order price response of the full model}\label{app:proof_first_order_price}

In the full model, output also responds to new money, so the first-order response of prices to inflation is the balance response ($\boldsymbol\mu$) of the main text net of an output response. The main text summarizes this price response in a footnote to Assumption~\ref{assu:small_returns}. Alongside $\boldsymbol\mu$, we define
\[
\mathbf y:=\partial_\pi\log\mathbf q_{\mathrm{ub}}\big|_{\pi=0},
\qquad
\iota_i:=\frac{\gamma_{0,i}}{m_i^{*}}\ \ (i\in N),
\]
and call them the `output response' and the `incidence ratio'. The incidence ratio records how much of a new injection a firm receives per unit of its stationary size. The departure of the incidence ratio from proportionality ($\boldsymbol\iota-\mathbf 1$) is positive at small firms and negative at large ones whenever $\theta<1$, and it vanishes identically at $\theta=1$. Because detrended prices satisfy $\log\widehat p_{i,\mathrm{ub}}=\log\widehat m_{i,\mathrm{ub}}-\log q_{i,\mathrm{ub}}$, the first-order price response of the full model is $\boldsymbol\mu-\mathbf y$, and the nominal-demand projection reports $\boldsymbol\mu$ in its place.

\begin{oalemma}[First-order price response of the full model]
\label{lem:first_order_price}
Let $\varsigma\in(0,1)$ and $\theta\in(0,1]$, and let prices be flexible. The balance response is the unique solution of
\[
(\mathbf I-\mathbf A)\bigl(\boldsymbol\mu\circ\mathbf m^{*}\bigr)=\mathbf A\boldsymbol\gamma_0-\mathbf m^{*},
\qquad
\mathbf 1^\top\bigl(\boldsymbol\mu\circ\mathbf m^{*}\bigr)=0,
\]
and it differs across firms if and only if $\mathbf A(\boldsymbol\gamma_0-\mathbf m^{*})\neq\mathbf 0$. The output and price responses are
\begin{equation}
\label{eq:output_response}
\mathbf y=\varsigma(\mathbf I-\varsigma\mathbf A^\top)^{-1}\bigl[(\mathbf I-\mathbf A^\top)\boldsymbol\mu+(\boldsymbol\iota-\mathbf 1)\bigr],
\end{equation}
\begin{equation}
\label{eq:price_response}
\boldsymbol\mu-\mathbf y=(\mathbf I-\varsigma\mathbf A^\top)^{-1}\bigl[(1-\varsigma)\boldsymbol\mu-\varsigma(\boldsymbol\iota-\mathbf 1)\bigr].
\end{equation}
The output response satisfies
\begin{equation}
\label{eq:projection_bound}
\lVert\mathbf y\rVert_\infty
\;\le\;
\frac{\varsigma}{1-\varsigma}\,
\bigl\lVert(\mathbf I-\mathbf A^\top)\boldsymbol\mu+(\boldsymbol\iota-\mathbf 1)\bigr\rVert_\infty .
\end{equation}
At $\theta=1$ all three responses vanish identically. For $\theta\in(0,1)$ the price response is real analytic in $\varsigma$ on $[0,1)$ and equals $\boldsymbol\mu$ at $\varsigma=0$. When stationary balances are not all equal, it does not vanish identically on that interval.
\end{oalemma}

The price response~\eqref{eq:price_response} has three features, the first of which is that output modifies the balance response through~\eqref{eq:output_response}, by an amount that carries the returns to scale ($\varsigma$) as a prefactor. When $\theta<1$ and stationary balances are not all equal, the modified price response still differs across firms for all but at most $n-1$ values of $\varsigma$ (Lemma~\ref{lem:non_degeneracy}). A firm that receives new money spends it on inputs, produces more, and returns part of its nominal gain as a lower price. The second feature is that injection incidence enters the price response directly, through $\boldsymbol\iota-\mathbf 1$. Relative prices are therefore displaced by the differential output that unevenly received money finances, as well as by the percolation of balances. And the third is that the whole price response vanishes with the departure of the incidence ratio from proportionality ($\boldsymbol\iota-\mathbf 1$), so that at $\theta=1$ no relative price moves. The proportional-injection benchmark of Section~\ref{subsec:monetary_process} of the main text then holds exactly in the full model.

\begin{proof}
We fix $\varsigma\in(0,1)$ and $\theta\in(0,1]$ and differentiate the balanced-growth fixed points of Proposition~\ref{prop:balanced_growth_convergence} at $\pi=0$, where $\widehat{\mathbf m}_{\mathrm{ub}}=\widehat{\mathbf n}_{\mathrm{ub}}=\mathbf m^{*}$, $\boldsymbol\gamma_{\mathrm{ub}}=\boldsymbol\gamma_0$, and $\mathbf q_{\mathrm{ub}}=\mathbf q^{*}$. Our argument invokes the implicit function theorem, whose two hypotheses hold here. The fixed-point map~\eqref{eq:balanced_growth_money_fixed_point} is smooth in $(\pi,\widehat{\mathbf m})$ on the interior of the simplex. And its Jacobian in $\widehat{\mathbf m}$ at $(0,\mathbf m^{*})$ is $\mathbf I-\mathbf A$ restricted to the zero-sum subspace, which is invertible, since primitivity makes the eigenvalue one of $\mathbf A$ simple.

Consider first the balance response. Writing the fixed point~\eqref{eq:balanced_growth_money_fixed_point} as $(1+\pi)\widehat{\mathbf m}_{\mathrm{ub}}=\mathbf A(\widehat{\mathbf m}_{\mathrm{ub}}+\pi\boldsymbol\gamma_{\mathrm{ub}})$ and differentiating at $\pi=0$ gives
\[
\mathbf m^{*}+\partial_\pi\widehat{\mathbf m}_{\mathrm{ub}}
=\mathbf A\,\partial_\pi\widehat{\mathbf m}_{\mathrm{ub}}+\mathbf A\boldsymbol\gamma_0,
\qquad\text{that is}\qquad
(\mathbf I-\mathbf A)\,\partial_\pi\widehat{\mathbf m}_{\mathrm{ub}}=\mathbf A\boldsymbol\gamma_0-\mathbf m^{*}.
\]
The derivative of the injection profile enters this display multiplied by $\pi$ and therefore drops out at $\pi=0$. The right side has zero coordinate sum, since $\mathbf 1^\top\mathbf A=\mathbf 1^\top$ and both $\boldsymbol\gamma_0$ and $\mathbf m^{*}$ lie in the unit simplex. The normalization $\mathbf 1^\top\widehat{\mathbf m}_{\mathrm{ub}}=1$ holds at every $\pi$, so $\mathbf 1^\top\partial_\pi\widehat{\mathbf m}_{\mathrm{ub}}=0$. And since $\mathbf I-\mathbf A$ is invertible on the zero-sum subspace, the displayed equation and this zero-sum condition determine $\partial_\pi\widehat{\mathbf m}_{\mathrm{ub}}$ uniquely. Dividing coordinatewise by $\mathbf m^{*}$, which is strictly positive, gives $\boldsymbol\mu$ and the first display of the lemma. Since $\mathbf 1^\top(\boldsymbol\mu\circ\mathbf m^{*})=0$ and $\mathbf 1^\top\mathbf m^{*}=1$, the balance response is constant across firms only if $\boldsymbol\mu=\mathbf 0$. And by the first display $\boldsymbol\mu=\mathbf 0$ if and only if $\mathbf A\boldsymbol\gamma_0-\mathbf m^{*}=\mathbf A(\boldsymbol\gamma_0-\mathbf m^{*})=\mathbf 0$, where the first equality uses $\mathbf A\mathbf m^{*}=\mathbf m^{*}$.

The response of normalized post-injection balances ($\widehat{\mathbf n}_{\mathrm{ub}}$) follows. Differentiating $\widehat{\mathbf n}_{\mathrm{ub}}=(\widehat{\mathbf m}_{\mathrm{ub}}+\pi\boldsymbol\gamma_{\mathrm{ub}})/(1+\pi)$ at $\pi=0$ gives
\[
\partial_\pi\widehat{\mathbf n}_{\mathrm{ub}}
=\partial_\pi\widehat{\mathbf m}_{\mathrm{ub}}+\boldsymbol\gamma_0-\mathbf m^{*},
\]
and dividing coordinatewise by $\mathbf m^{*}$ yields $\partial_\pi\log\widehat{\mathbf n}_{\mathrm{ub}}=\boldsymbol\mu+\boldsymbol\iota-\mathbf 1$.

For the output response, the technological constant $\mathbf b_A$ does not depend on $\pi$, so differentiating the output fixed point of Proposition~\ref{prop:balanced_growth_convergence} in the form $(\mathbf I-\varsigma\mathbf A^\top)\log\mathbf q_{\mathrm{ub}}=\varsigma\mathbf b_A+\varsigma(\log\widehat{\mathbf n}_{\mathrm{ub}}-\mathbf A^\top\log\widehat{\mathbf m}_{\mathrm{ub}})$ gives
\[
(\mathbf I-\varsigma\mathbf A^\top)\,\mathbf y
=\varsigma\bigl[(\boldsymbol\mu+\boldsymbol\iota-\mathbf 1)-\mathbf A^\top\boldsymbol\mu\bigr]
=\varsigma\bigl[(\mathbf I-\mathbf A^\top)\boldsymbol\mu+(\boldsymbol\iota-\mathbf 1)\bigr].
\]
Since $\mathbf A^\top$ is row-stochastic, the spectral radius of $\varsigma\mathbf A^\top$ is $\rho(\varsigma\mathbf A^\top)=\varsigma<1$. So $\mathbf I-\varsigma\mathbf A^\top$ is invertible, and~\eqref{eq:output_response} follows.

The price response follows by subtraction. Detrended prices satisfy $\log\widehat p_{i,\mathrm{ub}}=\log\widehat m_{i,\mathrm{ub}}-\log q_{i,\mathrm{ub}}$ (Proposition~\ref{prop:balanced_growth_convergence}), so the first-order price response is $\boldsymbol\mu-\mathbf y$. Substituting the operator identity
\[
\varsigma(\mathbf I-\mathbf A^\top)=(\mathbf I-\varsigma\mathbf A^\top)-(1-\varsigma)\mathbf I
\]
into~\eqref{eq:output_response} and expanding,
\[
\mathbf y=\boldsymbol\mu-(1-\varsigma)(\mathbf I-\varsigma\mathbf A^\top)^{-1}\boldsymbol\mu
+\varsigma(\mathbf I-\varsigma\mathbf A^\top)^{-1}(\boldsymbol\iota-\mathbf 1),
\]
and subtracting this from $\boldsymbol\mu$ gives~\eqref{eq:price_response}.

The bound on the output response follows from row stochasticity. Since $\mathbf A^\top\ge0$ and $\mathbf A^\top\mathbf 1=\mathbf 1$, the induced norm satisfies $\lVert\mathbf A^\top\rVert_\infty=1$, so the Neumann series gives
\[
\bigl\lVert(\mathbf I-\varsigma\mathbf A^\top)^{-1}\bigr\rVert_\infty\le\sum_{k\ge0}\varsigma^{k}=\frac{1}{1-\varsigma}.
\]
Applied to~\eqref{eq:output_response}, this bound yields~\eqref{eq:projection_bound}. The remainder of the nominal-demand projection therefore carries $\varsigma$ as a prefactor, for this remainder is nothing but the output response $\mathbf y$. And the bound~\eqref{eq:projection_bound} requires no separation between $\varsigma$ and $\lambda_2$.

It remains to treat the two endpoints, $\theta=1$ and $\varsigma=0$. At $\theta=1$ the injection profile is $\boldsymbol\gamma_0=\mathbf m^{*}$, so $\boldsymbol\iota=\mathbf 1$ and $\mathbf A\boldsymbol\gamma_0-\mathbf m^{*}=\mathbf A\mathbf m^{*}-\mathbf m^{*}=\mathbf 0$. Hence $\boldsymbol\mu=\mathbf 0$, and~\eqref{eq:output_response} and~\eqref{eq:price_response} give $\mathbf y=\mathbf 0$ and $\boldsymbol\mu-\mathbf y=\mathbf 0$. For $\theta\in(0,1)$ the entries of $(\mathbf I-\varsigma\mathbf A^\top)^{-1}$ are rational functions of $\varsigma$ whose common denominator $\det(\mathbf I-\varsigma\mathbf A^\top)$ has no zero on $[0,1)$, so the right side of~\eqref{eq:price_response} is real analytic in $\varsigma$ there. At $\varsigma=0$ it equals $\boldsymbol\mu$. We write $\boldsymbol\tau:=\boldsymbol\gamma_0-\mathbf m^{*}$ for the injection tilt. For $\theta\in(0,1)$ the injection tilt is nonzero when stationary balances are not all equal. This is because the concave injection rule then moves the injection profile off $\mathbf m^{*}$. If $\mathbf A\boldsymbol\tau\ne\mathbf 0$, then $\boldsymbol\mu\ne\mathbf 0$, so the right side of~\eqref{eq:price_response} is nonzero at $\varsigma=0$. A real analytic function that is nonzero at a point of an interval does not vanish identically on that interval. If instead $\mathbf A\boldsymbol\tau=\mathbf 0$, then $\boldsymbol\mu=\mathbf 0$ and~\eqref{eq:price_response} reduces to $-\varsigma(\mathbf I-\varsigma\mathbf A^\top)^{-1}(\boldsymbol\iota-\mathbf 1)$, which is nonzero at every $\varsigma\in(0,1)$ because the inverse is injective and $\boldsymbol\iota-\mathbf 1=\boldsymbol\tau/\mathbf m^{*}\ne\mathbf 0$. This proves the final claim.
\end{proof}

\subsection{Non-degeneracy of the price response in the full model}\label{app:proof_non_degeneracy}

A price response that is the same for every firm would move the price level and leave relative prices where they are. Such a uniform response arises only at isolated values of the returns to scale.

\begin{oalemma}[Non-degeneracy of the price response]\label{lem:non_degeneracy}
Let $\varsigma\in(0,1)$ and $\theta\in(0,1)$, and write $\boldsymbol\tau:=\boldsymbol\gamma_0-\mathbf m^{*}$ for the injection tilt. The first-order price response $\boldsymbol\mu-\mathbf y$ is constant across firms if and only if
\[
\mathbf A\boldsymbol\tau=\varsigma\boldsymbol\tau .
\]
When stationary balances are not all equal, it therefore has strictly positive cross-sectional dispersion whenever $\varsigma$ is not an eigenvalue of $\mathbf A$, and hence for all but at most $n-1$ values of $\varsigma$ in $(0,1)$.
\end{oalemma}

\begin{proof}
We fix $\varsigma\in(0,1)$ and $\theta\in(0,1)$, and write $\mathbf x:=\boldsymbol\mu\circ\mathbf m^{*}$.

Suppose first that the price response is constant across firms, $\boldsymbol\mu-\mathbf y=c\mathbf 1$ for a scalar $c$. Since $\mathbf A^\top\mathbf 1=\mathbf 1$ gives $(\mathbf I-\varsigma\mathbf A^\top)c\mathbf 1=(1-\varsigma)c\mathbf 1$, applying $\mathbf I-\varsigma\mathbf A^\top$ to~\eqref{eq:price_response} turns the supposition into $(1-\varsigma)\boldsymbol\mu-\varsigma(\boldsymbol\iota-\mathbf 1)=(1-\varsigma)c\mathbf 1$, that is
\[
\boldsymbol\mu-c\mathbf 1=\frac{\varsigma}{1-\varsigma}\,(\boldsymbol\iota-\mathbf 1).
\]
Multiplying coordinatewise by $\mathbf m^{*}$ and using $\boldsymbol\iota-\mathbf 1=\boldsymbol\tau/\mathbf m^{*}$ gives $\mathbf x-c\,\mathbf m^{*}=\varsigma(1-\varsigma)^{-1}\boldsymbol\tau$. Taking $\mathbf 1^\top$ and using $\mathbf 1^\top\mathbf x=0$, $\mathbf 1^\top\boldsymbol\tau=0$, and $\mathbf 1^\top\mathbf m^{*}=1$ forces $c=0$, so the supposition is equivalent to $\mathbf x=\varsigma(1-\varsigma)^{-1}\boldsymbol\tau$.

This equation converts into an eigenvalue condition on the injection tilt ($\boldsymbol\tau$). Lemma~\ref{lem:first_order_price} gives $(\mathbf I-\mathbf A)\mathbf x=\mathbf A\boldsymbol\gamma_0-\mathbf m^{*}=\mathbf A\boldsymbol\tau$, the second equality using $\mathbf A\mathbf m^{*}=\mathbf m^{*}$. Substituting $\mathbf x=\varsigma(1-\varsigma)^{-1}\boldsymbol\tau$,
\[
\frac{\varsigma}{1-\varsigma}(\boldsymbol\tau-\mathbf A\boldsymbol\tau)=\mathbf A\boldsymbol\tau,
\qquad\text{so}\qquad
\frac{\varsigma}{1-\varsigma}\,\boldsymbol\tau=\Bigl(1+\frac{\varsigma}{1-\varsigma}\Bigr)\mathbf A\boldsymbol\tau=\frac{1}{1-\varsigma}\,\mathbf A\boldsymbol\tau,
\]
which is $\mathbf A\boldsymbol\tau=\varsigma\boldsymbol\tau$. Every step reverses, so the two conditions are equivalent.

The final claim follows. If $\varsigma$ is not an eigenvalue of $\mathbf A$, then $\mathbf A\boldsymbol\tau=\varsigma\boldsymbol\tau$ forces $\boldsymbol\tau=\mathbf 0$. But $\boldsymbol\tau\ne\mathbf 0$ for $\theta\in(0,1)$ when stationary balances are not all equal, since the concave injection rule then moves the injection profile off $\mathbf m^{*}$. Hence the price response is not constant across firms. Since $\mathbf A$ has at most $n$ eigenvalues and $\lambda_1=1$ lies outside $(0,1)$, at most $n-1$ values of $\varsigma$ in $(0,1)$ are excluded.
\end{proof}

\subsection{Relative-price distortion in the full model}\label{oa:app:full_model_distortion}

When output also responds to new money, as it does in the full model, the relative price gap ($\omega$) remains of the order of $\pi$ and relative-price entropy ($\psi$) of the order of $\pi^{2}$.

\begin{oacorollary}[Relative-price distortion in the full model]\label{coro:full_model_distortion}
Let prices be flexible, let Assumptions~\ref{assu:primitivity} and~\ref{assu:spectral_gap} hold, let $\theta,\varsigma\in(0,1)$, and let the injection tilt $\boldsymbol\tau=\boldsymbol\gamma_0-\mathbf m^{*}$ satisfy $\mathbf A\boldsymbol\tau\ne\varsigma\boldsymbol\tau$, the non-degeneracy condition of Lemma~\ref{lem:non_degeneracy}. Without Assumption~\ref{assu:small_returns}, the steady-state relative price gap and relative-price entropy are
\[
\omega=\pi\Bigl(\Var_{\mathbf m^{*}}\bigl(\boldsymbol\mu-\mathbf y\bigr)\Bigr)^{1/2}+o(\pi),
\qquad
\psi=\frac{\pi^{2}}{2}\Var_{\widehat{\mathbf r}^{*}}\bigl(\boldsymbol\mu-\mathbf y\bigr)+o(\pi^{2}),
\]
and both leading coefficients are strictly positive.
\end{oacorollary}

\begin{proof}
The first part of the proof of Theorem~\ref{theorem:price_deviation} (Appendix~\ref{app:proof_thm2}) carries over with a single change. Without Assumption~\ref{assu:small_returns}, output enters the limits of detrended prices, $\widehat p_{i,\mathrm{ub}}=\widehat m_{i,\mathrm{ub}}/q_{i,\mathrm{ub}}$ (Proposition~\ref{prop:balanced_growth_convergence}), and with them the limits $g_i^{\mathrm{ub}}:=\log\widehat p_{i,\mathrm{ub}}-\log p_i^{*}$ of the log displacements. The limits of normalized balances and output equal $\mathbf m^{*}$ and $\mathbf q^{*}$ at $\pi=0$ and are differentiable in $\pi$ there, as the proof of Lemma~\ref{lem:first_order_price} shows. Hence
\[
\mathbf g^{\mathrm{ub}}=\pi\,(\boldsymbol\mu-\mathbf y)+O(\pi^{2}),
\]
and the two expansions of the proof of Theorem~\ref{theorem:price_deviation} hold with the price response $\boldsymbol\mu-\mathbf y$ in place of the balance response $\boldsymbol\mu$. Both leading coefficients stay strictly positive, for under the non-degeneracy condition the price response is not constant across firms (Lemma~\ref{lem:non_degeneracy}), and every firm carries a positive weight.
\end{proof}

\subsection{The projection ratio}
\label{oa:sec:projection_ratio}

The bound~\eqref{eq:projection_bound} of Lemma~\ref{lem:first_order_price} on the output response ($\mathbf y$) carries $\varsigma$ as a prefactor and vanishes at $\varsigma=0$. Measured against the balance response ($\boldsymbol\mu$), the output response also depends on the spectral gap ($1-\lambda_2$). The returns to scale ($\varsigma$) and the spectral gap enter through a single ratio, the projection ratio of Section~\ref{subsec:nominal_projection} of the main text,
\begin{equation}
\label{eq:projection_ratio}
\varkappa\;=\;\frac{\varsigma}{1-\varsigma}\,(1-\lambda_2).
\end{equation}
In relative-balance coordinates the propagation matrix becomes $\widetilde{\mathbf A}:=\operatorname{diag}(\mathbf m^{*})^{-1}\mathbf A\operatorname{diag}(\mathbf m^{*})$, which is row-stochastic and has the eigenvalues of $\mathbf A$. Row $i$ of $\widetilde{\mathbf A}$ averages over the customers of firm $i$, while row $i$ of $\mathbf A^\top$ averages over its suppliers. Dividing the balance equation of Lemma~\ref{lem:first_order_price} coordinatewise by $\mathbf m^{*}$ gives $(\mathbf I-\widetilde{\mathbf A})\boldsymbol\mu=\widetilde{\mathbf A}(\boldsymbol\iota-\mathbf 1)$, so that $(\mathbf I-\mathbf A^\top)\boldsymbol\mu=\widetilde{\mathbf A}(\boldsymbol\iota-\mathbf 1)+(\widetilde{\mathbf A}-\mathbf A^\top)\boldsymbol\mu$. Substituting into~\eqref{eq:projection_bound} gives, for $\boldsymbol\mu\ne\mathbf 0$,
\[
\frac{\lVert\mathbf y\rVert_\infty}{\lVert\boldsymbol\mu\rVert_\infty}
\;\le\;
\frac{\varsigma}{1-\varsigma}
\left(
2\,\frac{\lVert\boldsymbol\iota-\mathbf 1\rVert_\infty}{\lVert\boldsymbol\mu\rVert_\infty}
+\frac{\bigl\lVert(\widetilde{\mathbf A}-\mathbf A^\top)\boldsymbol\mu\bigr\rVert_\infty}{\lVert\boldsymbol\mu\rVert_\infty}
\right).
\]
The first ratio carries the spectral gap. This is because $(\mathbf I-\widetilde{\mathbf A})^{-1}\widetilde{\mathbf A}$ multiplies the second eigenmode by $\lambda_2/(1-\lambda_2)$, so a departure of the incidence ratio from proportionality ($\boldsymbol\iota-\mathbf 1$) whose projection on the second eigenmode is bounded away from zero gives $\lVert\boldsymbol\iota-\mathbf 1\rVert_\infty=O\bigl((1-\lambda_2)\lVert\boldsymbol\mu\rVert_\infty\bigr)$. The second ratio measures how differently the customers and the suppliers of a firm respond to new money. It vanishes on reversible networks, where $a_{ij}m_j^{*}=a_{ji}m_i^{*}$ for every pair of firms and hence $\widetilde{\mathbf A}=\mathbf A^\top$. It also vanishes within the degree representation, whose mean-field closure averages over customers and over suppliers alike (Lemma~\ref{oa:lem:eigenvector_proxies}). There the ratio $\lVert\mathbf y\rVert_\infty/\lVert\boldsymbol\mu\rVert_\infty$ is at most of the order of the projection ratio $\varkappa$.

\subsection{The permanent trend and the transitory wave}
\label{app:money_injection_and_transient}

The proof of Lemma~\ref{lemma:exposure} follows each injection vintage through two eigenmodes of the propagation matrix ($\mathbf A$), the Perron eigenmode, which carries the permanent trend, and the second eigenmode, which carries the transitory wave.

\begin{proof}[Proof of Lemma~\ref{lemma:exposure}]
We fix $T\ge1$. Under the pre-propagation timing of Section~\ref{subsec:monetary_process}, the period-$t$ injection $\boldsymbol\epsilon_t:=\pi(1+\pi)^{t-1}\boldsymbol\gamma_{t-1}$ is added to $\mathbf m_{t-1}$ before propagation. It therefore contributes $\mathbf A^{\,T-t+1}\boldsymbol\epsilon_t$ to $\mathbf m_T$, one round of trade on entry and $T-t$ rounds thereafter. Since the economy starts from the stationary balances, $\mathbf m_0=\mathbf m^{*}$, balances at horizon $T$ are
\[
\mathbf m_T=\mathbf m^{*}+\sum_{t=1}^{T}\mathbf A^{\,T-t+1}\boldsymbol\epsilon_t .
\]

Diagonalizability and the biorthogonal normalization of Assumption~\ref{assu:diagonalisable} give the exact spectral representation
\[
\mathbf A^{\,s}=\mathbf v_1\mathbf u_1^\top+\sum_{j=2}^{n}\lambda_j^{\,s}\,\mathbf v_j\mathbf u_j^\top,
\]
so that $\mathbf A^{\,T-t+1}\boldsymbol\epsilon_t=\mathbf v_1(\mathbf u_1^\top\boldsymbol\epsilon_t)+\sum_{j\ge2}\lambda_j^{\,T-t+1}\mathbf v_j(\mathbf u_j^\top\boldsymbol\epsilon_t)$ is the full modal representation of the disturbance.

The Perron components give the permanent trend. Since $\mathbf v_1=\mathbf m^{*}$ and $\mathbf u_1=\mathbf 1$, the Perron components of the initial balances and of the injections add up to
\[
\mathbf m^{*}\Bigl(1+\sum_{t=1}^{T}\mathbf 1^\top\boldsymbol\epsilon_t\Bigr)=\mathbf m^{*}\,\mathbf 1^\top\mathbf m_T=\mathbf m^{*}(1+\pi)^{T},
\]
which is the permanent trend $P_i^{(T)}=m_i^{*}(1+\pi)^{T}$ of the lemma.

Retaining the second eigenmode among the transitory components gives
\[
\mathbf A^{\,T-t+1}\boldsymbol\epsilon_t
=\mathbf v_1(\mathbf u_1^\top\boldsymbol\epsilon_t)
+\lambda_2^{\,T-t+1}\,\mathbf v_2(\mathbf u_2^\top\boldsymbol\epsilon_t)
+\mathbf R_{T-t+1,t},
\]
where $\mathbf R_{h,t}=\sum_{j\ge3}\lambda_j^{\,h}\mathbf v_j\mathbf u_j^\top\boldsymbol\epsilon_t$, so that $\lVert\mathbf R_{h,t}\rVert\le\operatorname{cond}(\mathbf V)\,|\lambda_3|^{\,h}\,\lVert\boldsymbol\epsilon_t\rVert$ after $h$ rounds of trade. Summing these bounds over injection dates controls the higher eigenmodes at every finite horizon. We write $S_{i\leftarrow t}^{(T)}:=\lambda_2^{\,T-t+1}(\mathbf v_2)_i(\mathbf u_2^\top\boldsymbol\epsilon_t)$ for the contribution of the period-$t$ injection to firm $i$'s transitory wave at horizon $T$.

Finally, Lemma~\ref{oa:lem:eigenvector_proxies} represents the second-eigenmode response as a whole,
\[
\begin{aligned}
\bigl(\mathbf v_2\mathbf u_2^\top\boldsymbol\epsilon_t\bigr)_i
&\approx
\kappa\,\delta_i\,\pi(1+\pi)^{t-1}
\left(
\E_{\boldsymbol\gamma_{t-1}}[d^{\nu}]
-\frac{\E_n[d^{1+\nu}]}{\E_n[d]}
\right)\\
&=
\kappa\,\delta_i\,\pi(1+\pi)^{t-1}\xi_{t-1}.
\end{aligned}
\]
The term $\E_n[d^{1+\nu}]/\E_n[d]$ centers the injection-side degree function on the realized degree sequence. The amplitude $\kappa$ scales the combined response of the injection side and the receiving side. It follows that $S_{i\leftarrow t}^{(T)}\approx\kappa\,\delta_i\,\lambda_2^{\,T-t+1}\pi(1+\pi)^{t-1}\xi_{t-1}$. Summing over $t=1,\dots,T$,
\[
S_i^{(T)}=\sum_{t=1}^{T}S_{i\leftarrow t}^{(T)}
\approx\kappa\,\delta_i\sum_{t=1}^{T}\lambda_2^{\,T-t+1}\pi(1+\pi)^{t-1}\xi_{t-1}
=\kappa\pi(1+\pi)^{T}\,\delta_i\,\mathcal X_T,
\]
with the propagation kernel $\mathcal X_T$ of~\eqref{eq:master_kernel}. This is the transitory wave of the lemma.
\end{proof}

\noindent The error of the two-eigenmode approximation of the transitory wave ($S_i^{(T)}$) has two sources, the first of which is the higher-eigenmode remainder, of relative order $(|\lambda_3|/\lambda_2)^{T-t+1}$ beside the second eigenmode. Corollary~\ref{oa:cor:second_mode_remainder} bounds this remainder, and the ratio ($r$) of~\eqref{oa:eq:truncation_ratio} tells when the remainder is small beside the second eigenmode. The second source is the error of the degree representation, which in turn separates into two parts. Lemma~\ref{oa:lem:perron_degree_distortion} bounds the Perron--degree part. But the other part comes from the mean-field closure itself, by which Lemma~\ref{oa:lem:eigenvector_proxies} represents the second eigenmode through degree.

Under the large-network scaling of Lemma~\ref{oa:lem:eigenvector_proxies}, $\kappa=\Theta(1/n)$, each vintage contribution $S_{i\leftarrow t}^{(T)}$ is $O(1/n)$, and for fixed degree classes $P_i^{(T)}=m_i^{*}(1+\pi)^{T}=\Theta(1/n)$. The level distortion of firm $i$'s nominal demand~\eqref{eq:level_distortion}, the ratio of its transitory wave to its permanent trend ($\ell_i^{(T)}=S_i^{(T)}/P_i^{(T)}$), is therefore intensive. It does not vanish as the network grows when the Cantillon exposure ($\mathcal C_i$) varies across firms. Lemma~\ref{lemma:newest_dominates} shows that the amplitude of the second-eigenmode response ($\kappa$) is positive.

\subsection{The injection misalignment}\label{app:Ct_sign}

When new money lands disproportionately on small firms ($\theta<1$) and mixing is disassortative ($\nu<0$), the injection misalignment ($\xi_t$) is positive at every date at low inflation. Every injection vintage then moves a firm's transitory wave in the same direction, and the propagation kernel ($\mathcal X_T$) builds up toward its limit.

\begin{oalemma}[Sign and build-up of the injection misalignment]\label{lem:Ct_positive}
Let Assumptions~\ref{assu:primitivity}, \ref{assu:spectral_gap}, and~\ref{assu:diagonalisable} hold, let $\theta\in(0,1)$, and let the economy start from $\mathbf m_0=\mathbf m^{*}$. We write $\tilde\lambda:=\lambda_2/(1+\pi)$ for the decay factor of the propagation kernel, $\xi(\boldsymbol\gamma):=\E_{\boldsymbol\gamma}[d^{\nu}]-\E_n[d^{1+\nu}]\,\E_n[d]^{-1}$ for the misalignment of a profile $\boldsymbol\gamma$, and $\xi_0:=\xi(\boldsymbol\gamma_0)$ for the misalignment of the injection profile evaluated at the zero-injection equilibrium $\boldsymbol\gamma_0\propto(\mathbf m^{*})^{\circ\theta}$. Under the Perron--degree representation of Section~\ref{subsec:degree_closure} and Assumption~\ref{assu:assortativity}, $\xi_0>0$. For every $\pi\ge0$ and every $t\ge0$,
\[
\bigl\lVert\widehat{\mathbf m}_t-\mathbf m^{*}\bigr\rVert
\;\le\;
\frac{2\operatorname{cond}(\mathbf V)\,\pi}{1+\pi-\lambda_2}.
\]
For all sufficiently small $\pi$, $\xi_t=\xi_0+O(\pi)$ uniformly in $t$, so that $\xi_t>0$ at every date and $\xi_{\mathrm{ub}}=\xi_0+O(\pi)$. For these values of $\pi$, the propagation kernel satisfies, uniformly in $T$,
\[
\mathcal X_T=\xi_0\sum_{h=1}^{T}\tilde\lambda^{h}+O(\pi),
\qquad
\mathcal X_T-\mathcal X_{T-1}=\xi_0\tilde\lambda^{T}+O(\pi).
\]
There is consequently a constant $K$, independent of $T$ and $\pi$, such that the propagation kernel is strictly increasing at every horizon $T$ with $\tilde\lambda^{T}>K\pi$. At a fixed positive spectral gap, under the Perron--degree representation,
\[
\xi_{\mathrm{ub}}
=
\frac{\E_n[d^{\theta+\nu}]}{\E_n[d^\theta]}-\frac{\E_n[d^{1+\nu}]}{\E_n[d]}
+O(\pi).
\]
\end{oalemma}

\noindent We call the set of horizons $\{T:\tilde\lambda^{T}>K\pi\}$ the early-transient window. Because $\tilde\lambda^{T}$ decays geometrically, this window is an initial segment of horizons whose length grows like $\log(1/\pi)/\log(1/\tilde\lambda)$. On this window the propagation kernel builds monotonically to first order in $\pi$. Beyond the early-transient window the increments of the propagation kernel are of order $\pi$. The propagation kernel enters price growth multiplied by $\pi$, so the contribution of these increments to price growth is of second order.

\begin{proof}
The sign of $\xi_0$ comes from a stochastic-dominance comparison. Under the Perron--degree representation the proportional injection profile is $\gamma_i^{\mathrm{prop}}=d_i/\sum_jd_j$, so the benchmark term of $\xi(\boldsymbol\gamma)$, $\E_n[d^{1+\nu}]/\E_n[d]$, equals $\E_{\boldsymbol\gamma^{\mathrm{prop}}}[d^{\nu}]$ and $\xi(\boldsymbol\gamma)=\E_{\boldsymbol\gamma}[d^{\nu}]-\E_{\boldsymbol\gamma^{\mathrm{prop}}}[d^{\nu}]$ compares the same function under two profiles. The likelihood ratio of $\boldsymbol\gamma_0\propto d^{\theta}$ against the benchmark is proportional to $d^{\theta-1}$, which is decreasing in $d$ for $\theta<1$. Hence $\boldsymbol\gamma_0$ is dominated by $\boldsymbol\gamma^{\mathrm{prop}}$ in the monotone likelihood-ratio order and therefore in first-order stochastic dominance, strictly whenever the degree sequence takes at least two values. On the maintained domain $\nu\in(-1,0)$ (Assumption~\ref{assu:assortativity}) the function $x\mapsto x^{\nu}$ is strictly decreasing, so the profile that places more mass on smaller degrees carries the larger value of $\E[d^{\nu}]$, and $\xi_0>0$.

For the bound on normalized balances, let $\mathbf e_t:=\widehat{\mathbf m}_t-\mathbf m^{*}$ be their deviation from stationary balances. Subtracting $\mathbf m^{*}=(\mathbf A\mathbf m^{*}+\pi\mathbf m^{*})/(1+\pi)$ from the normalized recursion~\eqref{eq:normalised_balance_dynamics} gives
\[
\mathbf e_{t+1}=\frac{1}{1+\pi}\bigl[\mathbf A\mathbf e_t+\pi(\mathbf A\boldsymbol\gamma_t-\mathbf m^{*})\bigr],
\qquad
\mathbf e_0=\mathbf 0,
\]
and hence
\[
\mathbf e_t=\frac{\pi}{1+\pi}\sum_{s=0}^{t-1}\Bigl(\frac{\mathbf A}{1+\pi}\Bigr)^{t-1-s}(\mathbf A\boldsymbol\gamma_s-\mathbf m^{*}).
\]
Every forcing term has zero coordinate sum, because $\mathbf A\boldsymbol\gamma_s$ and $\mathbf m^{*}$ both lie in the unit simplex. The Euclidean norm of every forcing term is at most its $\ell_1$ norm, which is at most two. A zero-sum vector has no Perron component, so diagonalizability (Assumption~\ref{assu:diagonalisable}) gives $\lVert\mathbf A^{k}\mathbf z\rVert\le\operatorname{cond}(\mathbf V)\lambda_2^{k}\lVert\mathbf z\rVert$ for every zero-sum $\mathbf z$ and every $k\ge0$, since $|\lambda_j|\le\lambda_2$ for $j\ge2$ (Assumption~\ref{assu:spectral_gap}). Summing the geometric series,
\[
\lVert\mathbf e_t\rVert
\le\frac{2\operatorname{cond}(\mathbf V)\,\pi}{1+\pi}\sum_{k\ge0}\Bigl(\frac{\lambda_2}{1+\pi}\Bigr)^{k}
=\frac{2\operatorname{cond}(\mathbf V)\,\pi}{1+\pi-\lambda_2},
\]
which is the bound of the lemma. The bound holds at every date and uses no property of the injection path beyond its lying in the simplex.

Consider next the injection misalignment ($\xi_t$). The power map $\widehat{\mathbf m}\mapsto\boldsymbol\gamma(\widehat{\mathbf m})=\widehat{\mathbf m}^{\circ\theta}/\mathbf 1^\top\widehat{\mathbf m}^{\circ\theta}$ is continuously differentiable on the interior of the simplex, and $\mathbf m^{*}$ is interior, so the map is Lipschitz on a neighborhood of $\mathbf m^{*}$ with a constant that depends on $\theta$ and on $\min_im_i^{*}$. For $\pi$ small enough that the ball of radius $2\operatorname{cond}(\mathbf V)\pi/(1+\pi-\lambda_2)$ around $\mathbf m^{*}$ lies inside that neighborhood, $\lVert\boldsymbol\gamma_t-\boldsymbol\gamma_0\rVert=O(\pi)$ uniformly in $t$. Since $0<d^{\nu}\le1$ on the degree support, the injection misalignment then satisfies $|\xi_t-\xi_0|\le\sum_i|\gamma_i^{(t)}-\gamma_{0,i}|=O(\pi)$ uniformly in $t$, with a constant that depends on the network but not on $t$ or $\pi$. Reducing $\pi$ further if necessary, we obtain $\xi_t>0$ at every date, since $\xi_0>0$. The injection profile converges, $\boldsymbol\gamma_t\to\boldsymbol\gamma_{\mathrm{ub}}$, in the maintained mixing--forcing regime \citep[Proposition~5]{bhattathiripad2026}, so the limit $\xi_{\mathrm{ub}}=\xi(\boldsymbol\gamma_{\mathrm{ub}})$ also satisfies $\xi_{\mathrm{ub}}=\xi_0+O(\pi)$.

The propagation kernel follows. By definition~\eqref{eq:master_kernel}, $\mathcal X_T=\sum_{t=0}^{T-1}\xi_t\tilde\lambda^{T-t}$, so
\[
\mathcal X_T-\xi_0\sum_{h=1}^{T}\tilde\lambda^{h}
=\sum_{t=0}^{T-1}(\xi_t-\xi_0)\tilde\lambda^{T-t},
\]
whose absolute value is at most $\sup_t|\xi_t-\xi_0|\,\tilde\lambda/(1-\tilde\lambda)=O(\pi)$ uniformly in $T$. Differencing,
\[
\mathcal X_T-\mathcal X_{T-1}-\xi_0\tilde\lambda^{T}
=(\xi_{T-1}-\xi_0)\tilde\lambda+\sum_{t=0}^{T-2}(\xi_t-\xi_0)\bigl(\tilde\lambda^{T-t}-\tilde\lambda^{T-1-t}\bigr),
\]
and since $\sum_{t=0}^{T-2}|\tilde\lambda^{T-t}-\tilde\lambda^{T-1-t}|=(1-\tilde\lambda)\sum_{h=1}^{T-1}\tilde\lambda^{h}\le\tilde\lambda$, the right side is at most $2\tilde\lambda\sup_t|\xi_t-\xi_0|$ in absolute value, which is $O(\pi)$ uniformly in $T$. We write $C_\xi\pi$ for the uniform bound on $|\xi_t-\xi_0|$. Since $\tilde\lambda\le\lambda_2$, the increment is positive whenever $\xi_0\tilde\lambda^{T}>2\lambda_2C_\xi\pi$, which is the early-transient window with $K:=2\lambda_2C_\xi/\xi_0$.

The moment representation of $\xi_{\mathrm{ub}}$ follows from substituting the stationary injection profile of Section~\ref{subsec:master_equation}, $\boldsymbol\gamma_{\mathrm{ub}}\propto d^{\theta}+O(\pi)$, into $\xi(\boldsymbol\gamma)$. The benchmark term of $\xi(\boldsymbol\gamma)$ is $\E_n[d^{1+\nu}]\,\E_n[d]^{-1}$. Population moments replace these empirical moments only in the large-$n$ relations between the distortion of relative prices and the degree law. Wherever $\xi_{\mathrm{ub}}$ enters the coefficient of an $O(\pi)$ or $O(\pi^2)$ distortion in the main text, it denotes the leading term of~\eqref{eq:Cub_empirical}, and the $O(\pi)$ profile correction is absorbed by the corresponding remainder.
\end{proof}

By Lemma~\ref{lem:Ct_positive} the propagation kernel ($\mathcal X_T$) is strictly increasing on the early-transient window. The relative price gap ($\omega_T$) is proportional to $\mathcal X_T$ (proof of Theorem~\ref{theorem:price_deviation}), so it too is strictly increasing in $T$ on that window. Since every $\xi_t$ is positive and $\sup_t\xi_t=\xi_0+O(\pi)$, the bound $0<\mathcal X_s\le\sup_t\xi_t\,\lambda_2(1+\pi-\lambda_2)^{-1}=\xi_{\mathrm{ub}}\lambda_2(1+\pi-\lambda_2)^{-1}+O(\pi)$ holds at every date $s$.

\section{Inflation under Flexible Prices}\label{app:block_flexible}

\subsection{Transient price growth and firm size}\label{app:proof_thm1}

Proposition~\ref{prop:size_price_change} of the main text signs the departure of a firm's price growth from inflation ($\Delta\log p_i^{(T)}-\pi$) by the firm's degree. While the propagation kernel ($\mathcal X_T$) builds, this departure rises with degree only up to a peak, and its ordering by size reverses once the propagation kernel declines.

\begin{oacorollary}[Peak of the transient departure from inflation]\label{oa:cor:departure_peak}
Under the hypotheses of Proposition~\ref{prop:size_price_change}, and within the two-eigenmode degree representation, the Cantillon exposure rises with degree up to $\hat d:=d^{*}(1-\nu^{2})^{-1/\nu^{2}}$ and declines beyond it. If the propagation kernel is building, $\mathcal X_T>\mathcal X_{T-1}$, the departure of price growth from inflation therefore rises with degree up to $\hat d$ and declines beyond it. But if the propagation kernel declines, the ordering of Proposition~\ref{prop:size_price_change} reverses.
\end{oacorollary}

\begin{proof}[Proof of Proposition~\ref{prop:size_price_change} and Corollary~\ref{oa:cor:departure_peak}]

Let $P_i^{(T)}$ and $S_i^{(T)}$ denote firm $i$'s permanent trend and transitory wave. Then
\[
P_i^{(T)} = m_i^{*}\,(1+\pi)^T.
\]
By Lemma~\ref{lemma:exposure}, the two-eigenmode approximation separates the transitory wave into a firm-specific, time-invariant factor and a common, time-varying one:
\[
S_i^{(T)}\approx\kappa\pi(1+\pi)^{T}\delta_i\,\mathcal X_T,
\]
where $\mathcal X_T$ is the propagation kernel~\eqref{eq:master_kernel}. The prefactor $\pi(1+\pi)^{T}\mathcal X_T$, common to all firms, depends on $\lambda_2$ through the per-period factor $\lambda_2(1+\pi)^{-1}$.

Under the nominal-demand projection of Assumption~\ref{assu:small_returns} output is fixed, so price growth equals the growth of nominal demand:
\[
\Delta\log p_i^{(T)}=\log \Bigl(\frac{\mathcal D_i^{(T)}}{\mathcal D_i^{(T-1)}}\Bigr).
\]
Since $\mathcal D_i^{(T)}\approx P_i^{(T)}+S_i^{(T)}$, the level distortion of firm $i$'s nominal demand~\eqref{eq:level_distortion} is $\ell_i^{(T)}=S_i^{(T)}/P_i^{(T)}\approx\pi\mathcal C_i\mathcal X_T$. The propagation kernel is bounded at every date (Section~\ref{app:Ct_sign}) and the Cantillon exposure is bounded on the truncated degree support, so $\sup_{i,T}|\ell_i^{(T)}|=O(\pi)$ and $1+\ell_i^{(T)}\ge1/2$ for all sufficiently small $\pi$. Expanding the two logarithms gives
\[
\Delta\log p_i^{(T)}
\approx
\log(1+\pi)
+\bigl(\ell_i^{(T)}-\ell_i^{(T-1)}\bigr)+O(\pi^2)
\]
and substituting the level distortions $\ell_i^{(T)}$ and $\ell_i^{(T-1)}$ yields
\[
\Delta\log p_i^{(T)}
\approx
\pi+\pi(\mathcal X_T-\mathcal X_{T-1})\,\mathcal C_i+O(\pi^2).
\]
Here $\pi$ times the increment of the propagation kernel, $\pi\,(\mathcal X_T-\mathcal X_{T-1})$, is a transient factor common to all firms, and $\mathcal C_i=\kappa\delta_i/m_i^{*}$ is firm $i$'s Cantillon exposure.

The Cantillon exposure ($\mathcal C_i$) therefore governs the cross-sectional profile of the departure of price growth from inflation, and the claims of Proposition~\ref{prop:size_price_change} and Corollary~\ref{oa:cor:departure_peak} follow from its shape as a function of degree. By~\eqref{eq:cantillon_exposure}, $\mathcal C(d)=(\kappa/c_m)\widehat\delta(d)$, where $\widehat\delta(d):=\bigl(d^{\nu^{2}}-\E_n[d^{\nu^{2}}]\bigr)/d$ is the normalized excess degree of a firm of degree $d$. The factor $\kappa/c_m$ does not affect signs or thresholds, for it is positive. Since $d\mapsto d^{\nu^{2}}$ is strictly increasing, $\mathcal C<0$ on $[1,d^{*})$ and $\mathcal C>0$ on $(d^{*},\infty)$ with threshold degree $d^{*}=\bigl(\E_n[d^{\nu^{2}}]\bigr)^{1/\nu^{2}}$. As $\nu\to0$, $\log d^{*}=\nu^{-2}\log\E_n[e^{\nu^{2}\log d}]\to\E_n[\log d]$, so the threshold tends to the empirical geometric-mean degree. Under the large-$n$, large-cutoff Pareto approximation of Assumption~\ref{assu:degree_distribution}, $\E_n[d^{\nu^{2}}]$ tends to $\alpha/(\alpha-\nu^{2})$, so $d^{*}$ tends to $\bigl(\alpha/(\alpha-\nu^{2})\bigr)^{1/\nu^{2}}$, which approaches $e^{1/\alpha}$ as $\nu\to0$. The derivative of the Cantillon exposure with respect to degree,
\[
\mathcal C'(d)\;=\;\frac{\kappa}{c_m}\cdot\frac{\E_n[d^{\nu^{2}}]-(1-\nu^{2})\,d^{\nu^{2}}}{d^{2}},
\]
is positive precisely when $d<\hat d$, where $\hat d=\bigl(\E_n[d^{\nu^{2}}]/(1-\nu^{2})\bigr)^{1/\nu^{2}}=d^{*}(1-\nu^{2})^{-1/\nu^{2}}\to e\,d^{*}$ as $\nu\to0$. Hence, on $[1,d_{\max}]$, $\mathcal C$ increases up to $\min\{\hat d,d_{\max}\}$ and, if $d_{\max}>\hat d$, decreases thereafter. The maximizer of $\mathcal C$ on the maintained support is therefore $\min\{\hat d,d_{\max}\}$. The extension of $\mathcal C$ to unbounded support has a single interior maximum at $\hat d$ and satisfies $\mathcal C(d)\sim d^{\nu^{2}-1}\to0$ as $d\to\infty$. On the negative side the same derivative is positive throughout, because $[1,d^{*})\subset[1,\hat d)$. Hence $|\mathcal C|$ decreases and the most negative value sits at the minimum degree, the worst case that binds in Lemma~\ref{oa:lem:uniform_positivity}. These properties of the Cantillon exposure are the threshold claim of Proposition~\ref{prop:size_price_change} and the first claim of Corollary~\ref{oa:cor:departure_peak}. The excess degree $\delta_i$ is monotone in degree. But the normalized excess degree, and with it the Cantillon exposure, need not be. The other two claims of the corollary follow from the display for $\Delta\log p_i^{(T)}$. This is because the departure of firm $i$'s price growth from inflation, $\pi(\mathcal X_T-\mathcal X_{T-1})\mathcal C_i$, is the Cantillon exposure scaled by a factor common to all firms, which is positive while the propagation kernel builds and negative when it declines.
\end{proof}

\subsection{Amplitude of the level distortion and positive price changes}
\label{oa:uniform_positivity}

The level distortion of a firm's nominal demand, the ratio of its transitory wave to its permanent trend ($\ell_i^{(s)}=S_i^{(s)}/P_i^{(s)}$), is bounded at every date by a multiple of the magnitude of the firm's normalized excess degree ($|\widehat\delta_i|$). A sufficient condition on the injection and the spectrum keeps every firm's flexible-price change positive to first order in $\pi$. Under this condition the average size of price change ($\phi_T$) equals its signed counterpart.

\begin{oalemma}[Amplitude of the level distortion]\label{oa:lem:level_amplitude}
Under the hypotheses of Lemma~\ref{lem:Ct_positive}, the level distortion of firm $i$'s nominal demand ($\ell_i^{(s)}$) satisfies at every date $s$
\[
\bigl|\ell_i^{(s)}\bigr|
=|\mathcal C_i|\,\pi\,\mathcal X_s
\;\le\;
|\widehat\delta_i|\cdot\frac{\pi\,\kappa\,\xi_{\mathrm{ub}}\,\lambda_{2}}{c_m(1+\pi-\lambda_{2})}\bigl(1+O(\pi)\bigr).
\]
\end{oalemma}

\begin{proof}
The propagation kernel satisfies $0<\mathcal X_s\le\sup_t\xi_t\,\lambda_2(1+\pi-\lambda_2)^{-1}$ at every date $s$, with $\sup_t\xi_t=\xi_{\mathrm{ub}}+O(\pi)$ (Section~\ref{app:Ct_sign}). The Perron--degree representation $m_i^{*}=c_md_i$ gives $|\mathcal C_i|=(\kappa/c_m)|\widehat\delta_i|$, and the bound follows.
\end{proof}

\begin{oalemma}[A sufficient condition for positive flexible-price changes]\label{oa:lem:uniform_positivity}
Let $c_m=1/\sum_{j}d_j$ and let $\kappa>0$ be the amplitude of the second-eigenmode response, both as in Section~\ref{subsec:degree_closure}. Suppose
\begin{equation}
\label{oa:eq:uniform_positivity_condition}
\bigl(\E_n[d^{\nu^{2}}] - 1\bigr)
\frac{\kappa\,\xi_{\mathrm{ub}}\,\lambda_{2}}
{c_m(1-\lambda_{2})}<1.
\end{equation}
Then for all sufficiently small $\pi>0$ every firm has a strictly positive flexible-price change,
\[
\Delta\log p_i^{(T)} > 0 \qquad \text{for all } i \in N \text{ and } T \ge 1,
\]
so that the average size of price change ($\phi_T$) equals its signed degree-weighted counterpart to first order in $\pi$,
\[
\phi_T = \sum_{i\in N} \wgt{i}{\zeta}\,\Delta\log p_i^{(T)}.
\]
\end{oalemma}

\begin{proof}
By Proposition~\ref{prop:size_price_change} and Lemma~\ref{lemma:exposure}, the flexible-price change of firm $i$ admits the linear-in-$\pi$ representation
\[
\Delta\log p_i^{(T)} \;\approx\; \pi + \mathcal C_i\,\pi(\mathcal X_T-\mathcal X_{T-1}),
\]
where $\mathcal C_i$ is the firm's Cantillon exposure.
To leading order, the transient scale $\pi(\mathcal X_T-\mathcal X_{T-1})$ is bounded by $\pi$ times the limit of the propagation kernel. By Lemma~\ref{lem:Ct_positive}, $\mathcal X_T$ is positive, and on the early-transient window it increases with increment $\xi_0\tilde\lambda^{T}+O(\pi)$ toward a limit within $O(\pi)$ of $\xi_{\mathrm{ub}}\,\frac{\lambda_{2}}{1-\lambda_{2}}$. Each one-step increment is therefore at most the total increase of the propagation kernel up to an $O(\pi)$ term, $\pi\,|\mathcal X_T-\mathcal X_{T-1}|\le\pi\,\xi_{\mathrm{ub}}\,\frac{\lambda_{2}}{1-\lambda_{2}}+O(\pi^{2})$. With the Perron--degree representation $m_i^{*}=c_m d_i$, the departure of firm $i$'s price growth from inflation is bounded by
\[
|\mathcal C_i\,\pi(\mathcal X_T-\mathcal X_{T-1})| \;\le\; |\widehat\delta_i|\cdot \frac{\pi\,\kappa\,\xi_{\mathrm{ub}}\,\lambda_{2}}{c_m(1-\lambda_{2})},
\]
which, under the large-network scaling of Lemma~\ref{oa:lem:eigenvector_proxies}, does not grow with the network size $n$. The upper bound on $\kappa/c_m$ offsets the extensive factor $1/c_m=\sum_j d_j=\Theta(n)$ carried by $\frac{\delta_i}{m_i^{*}}$. The lower bound is needed only for the separate claim, in Section~\ref{app:money_injection_and_transient}, that the level distortion of a firm's nominal demand does not vanish as the network grows.

The binding firm has the smallest degree. On the early-transient window the increments of the propagation kernel are positive, so the most negative departure from inflation comes from the firms of negative Cantillon exposure. The magnitude of this departure is largest at $d_i=1$, where $\delta_i=1-\E_n[d^{\nu^{2}}]<0$. This is because the magnitude $-\widehat\delta(d)=(\E_n[d^{\nu^2}]-d^{\nu^2})/d$ of the normalized excess degree decreases throughout the negative-exposure interval. At leading order in $\pi$, the magnitude of the most negative departure satisfies
\[
|\mathcal C_i\pi(\mathcal X_T-\mathcal X_{T-1})|_{\max}
\;\le\;
\bigl(\E_n[d^{\nu^{2}}] - 1\bigr)
\frac{\kappa\,\xi_{\mathrm{ub}}\,\lambda_{2}}
{c_m(1-\lambda_{2})}\,\pi.
\]
The coefficient multiplying $\pi$ is strictly below one by~\eqref{oa:eq:uniform_positivity_condition}. The omitted remainder is uniform on the truncated support, so $\Delta\log p_i^{(T)}>0$ uniformly across $i$ for all sufficiently small $\pi$. This covers the early-transient window. Beyond it, $\mathcal X_T-\mathcal X_{T-1}=O(\pi)$ uniformly in $T$ (Lemma~\ref{lem:Ct_positive}), so $\Delta\log p_i^{(T)}=\pi+O(\pi^{2})>0$ uniformly across firms and horizons for all sufficiently small $\pi$.

\medskip
Uniform positivity $\Delta\log p_i^{(T)} > 0$ implies $|\Delta\log p_i^{(T)}| = \Delta\log p_i^{(T)}$ for every $i$, hence
\[
\phi_T = \sum_{i\in N} \wgt{i}{\zeta}\,|\Delta\log p_i^{(T)}| = \sum_{i\in N} \wgt{i}{\zeta}\,\Delta\log p_i^{(T)},
\]
identically across the cross-section.
\end{proof}

The positivity condition~\eqref{oa:eq:uniform_positivity_condition} takes an explicit form under an untruncated Pareto approximation with lower cutoff one. Then $\E[d^{\nu^{2}}]=\alpha/(\alpha-\nu^{2})$ for $\alpha>\nu^{2}$ and $\E[d^{\nu^{2}}]-1=\nu^{2}/(\alpha-\nu^{2})$. Combining this expression with the leading form $\xi_{\mathrm{ub}}\approx-\nu(1-\theta)/[(\alpha-1)(\alpha-\theta)]$ of Section~\ref{subsec:comparative_statics} and holding $\kappa/c_m$ fixed gives, as $\nu\to0$ at a fixed spectral gap,
\[
\bigl(\E_n[d^{\nu^{2}}] - 1\bigr)
\frac{\kappa\,\xi_{\mathrm{ub}}\,\lambda_{2}}
{c_m(1-\lambda_{2})}
\;\lesssim\;
\frac{|\nu|^{3}(1-\theta)}
{(\alpha-\nu^{2})(\alpha-1)(\alpha-\theta)}
\frac{\kappa\,\lambda_2}{c_m(1-\lambda_2)}
=O(|\nu|^{3}).
\]
The positivity condition therefore holds for sufficiently small $|\nu|$, with the intensive response scale $\kappa/c_m$ and the spectral gap fixed. At any finite calibration we can evaluate it directly from the realized degree moments and spectrum. The computational experiments do without it, for they run on the exact network. We check the positivity condition economy by economy, and not uniformly in $n$. This is because on network families whose spectral gap closes with size, the factor $\lambda_2/(1-\lambda_2)$ grows. The condition also degrades as $\alpha\to1$, since the factor $(\alpha-1)^{-1}$ diverges.

\subsection{Stationary balances and the measurement of relative-price distortion}
\label{oa:balance_centering}

Weighting firms by their stationary balances centers the log displacements of Section~\ref{subsec:distortion_flex} of the main text, the log deviations of detrended prices from their stationary values ($g_i$). The centering holds to first order in $\pi$ on the balanced-growth path of any network, and at every horizon within the degree representation.

\begin{oalemma}[Stationary balances center the log displacements]\label{oa:lem:balance_centering}
Let prices be flexible, let Assumptions~\ref{assu:primitivity}--\ref{assu:small_returns} hold, and let $\theta\in(0,1)$. The balance response then has zero mean under stationary balances, $\sum_{i=1}^{n}m_i^{*}\mu_i=0$, so the balance-weighted mean of the log displacements on the balanced-growth path is $O(\pi^{2})$. If Assumptions~\ref{assu:diagonalisable}, \ref{assu:degree_structure}, and~\ref{assu:assortativity} also hold, the Cantillon exposure has zero mean under stationary balances on every realized degree sequence,
\begin{equation}
\label{eq:share_weighted_exposure}
\sum_{i=1}^{n}m_i^{*}\,\mathcal C_i=\sum_{i=1}^{n}c_md_i\cdot\frac{\kappa}{c_m}\,\widehat\delta_i=\kappa\sum_{i=1}^{n}\delta_i=0,
\end{equation}
so within the two-eigenmode degree representation the balance-weighted mean of the log displacements is $O(\pi^{2})$ at every horizon.
\end{oalemma}

\begin{proof}
Normalized balances sum to one at every rate of inflation, $\mathbf 1^\top\widehat{\mathbf m}_{\mathrm{ub}}=1$, and equal the stationary balances $\mathbf m^{*}$ at $\pi=0$. Since $\widehat{\mathbf m}_{\mathrm{ub}}$ is differentiable in $\pi$ at $\pi=0$ (the implicit-function argument in the proof of Lemma~\ref{lem:first_order_price}), differentiating the normalization there gives $\sum_im_i^{*}\mu_i=\mathbf 1^\top\partial_\pi\widehat{\mathbf m}_{\mathrm{ub}}\big|_{\pi=0}=0$. Under Assumption~\ref{assu:small_returns} detrended prices on the balanced-growth path are $\widehat m_{i,\mathrm{ub}}/q_i^{*}$, so the log displacements there are $g_i=\log(\widehat m_{i,\mathrm{ub}}/m_i^{*})=\pi\mu_i+O(\pi^{2})$, and their balance-weighted mean is $O(\pi^{2})$.

Within the degree representation $m_i^{*}=c_md_i$ and $\mathcal C_i=(\kappa/c_m)\widehat\delta_i$ with $\widehat\delta_i=\delta_i/d_i$, so the $i$th term of~\eqref{eq:share_weighted_exposure} is $\kappa\delta_i$. Excess degrees sum to zero on every realized degree sequence, since $\sum_i\delta_i=\sum_id_i^{\nu^{2}}-n\,\E_n[d^{\nu^{2}}]=0$. Under Assumption~\ref{assu:small_returns} the master equation~\eqref{eq:level_distortion} gives $g_i^{(T)}=\pi\,\mathcal C_i\,\mathcal X_T+O(\pi^{2})$ at every horizon $T$, and the factor $\pi\mathcal X_T$ is common to all firms, so the balance-weighted mean of the log displacements is $O(\pi^{2})$.
\end{proof}

\subsection{Elasticities of the measures of relative-price distortion}\label{oa:proof_cor2a}

To leading order the relative price gap ($\omega$) is linear in inflation and relative-price entropy ($\psi$) is quadratic (Theorem~\ref{theorem:price_deviation}). Their elasticities with respect to inflation, which the main text summarizes after that theorem, are one and two up to terms of order $\pi$.

\begin{oacorollary}[Relative-price distortion scales with inflation]
\label{coro:inflation_elasticity_distortion}
Let prices be flexible, let Assumptions~\ref{assu:primitivity}--\ref{assu:assortativity} hold, let $\theta\in(0,1)$, and let $\mathbf A(\boldsymbol\gamma_0-\mathbf m^{*})\neq\mathbf 0$. The elasticities of the relative price gap and relative-price entropy with respect to inflation are then
\[
\frac{\partial\log\omega}{\partial\log\pi}=1+O(\pi),
\qquad
\frac{\partial\log\psi}{\partial\log\pi}=2+O(\pi).
\]
\end{oacorollary}

\begin{proof}
Theorem~\ref{theorem:price_deviation} yields the steady-state expansions
\[
\omega
=
\pi\,\digamma_\omega \;+\; o(\pi),
\qquad
\psi
=
\pi^{2}\,\digamma_\psi \;+\; o(\pi^{2}),
\]
where the coefficients $\digamma_\omega$ and $\digamma_\psi$ depend on the network primitives and on the stationary injection profile, but not on $\pi$.
Moreover, $\digamma_\omega>0$ and $\digamma_\psi>0$ on the admissible domain. In particular, the limit of the propagation kernel, $\xi_{\mathrm{ub}}\,\frac{\lambda_{2}}{1-\lambda_{2}}$, is strictly positive. This is because the injection misalignments $\xi_t$ are positive and bounded away from zero (Lemma~\ref{lem:Ct_positive}), and $\xi_t\to \xi_{\mathrm{ub}}$.

To differentiate, we need more than the leading-order statement of Theorem~\ref{theorem:price_deviation}. But at a fixed positive spectral gap, the fixed-point expansion implies that $\xi_{\mathrm{ub}}$ differs from its zero-inflation limit by $O(\pi)$. And with the intensive response scale $\kappa/c_m$ held fixed, network persistence and the logarithms in relative-price entropy are real analytic for $\pi$ near zero. Hence
\[
\omega(\pi)=\pi\digamma_\omega+O(\pi^2),
\qquad
\psi(\pi)=\pi^2\digamma_\psi+O(\pi^3),
\]
with one-sided derivatives at $\pi=0$. These expansions differentiate termwise,
\[
\frac{\partial \omega}{\partial \pi}
=
\digamma_\omega \;+\; O(\pi),
\qquad
\frac{\partial \psi}{\partial \pi}
=
2\pi\,\digamma_\psi \;+\; O(\pi^{2}),
\]
and therefore
\[
\frac{\pi}{\omega}\,\frac{\partial \omega}{\partial \pi}
=
1+O(\pi),
\qquad
\frac{\pi}{\psi}\,\frac{\partial \psi}{\partial \pi}
=
2+O(\pi).
\]
\end{proof}

\subsection{Comparative statics}\label{app:proof_comparative_statics}

\begin{proof}[Proof of Proposition~\ref{prop:comparative_statics}]
We fix $\alpha>1$, $\vartheta>0$, $\theta\in(0,1)$, $\lambda_2\in(0,1)$, and $\nu\in(-1,0)$. By Theorem~\ref{theorem:price_deviation}, evaluated at the large-cutoff Pareto moments of Assumption~\ref{assu:degree_distribution} (Appendix~\ref{app:proof_thm2}), the leading coefficients of the relative price gap and of relative-price entropy are
\[
\digamma_\omega=\frac{\kappa}{c_m}\,\xi_{\mathrm{ub}}\,\frac{\lambda_2}{1-\lambda_2}\bigl(Q(\alpha,\nu)(\alpha-1)\bigr)^{1/2},
\qquad
\digamma_\psi=\frac12\Bigl(\frac{\kappa}{c_m}\Bigr)^{\!2}\Bigl(\xi_{\mathrm{ub}}\,\frac{\lambda_2}{1-\lambda_2}\Bigr)^{\!2}\mathcal V(\alpha,\nu;\vartheta).
\]
Here $\xi_{\mathrm{ub}}>0$ is the stationary misalignment (Lemma~\ref{lem:Ct_positive}) and $\lambda_2/(1-\lambda_2)$ is network persistence. The dispersion functional $Q(\alpha,\nu)$ is defined in the proof of Theorem~\ref{theorem:price_deviation}, and $Q(\alpha,\nu)(\alpha-1)$ is the variance of the normalized excess degree under balance weights. The counterpart of $Q$ for relative-price entropy, $\mathcal V(\alpha,\nu;\vartheta)$, is the variance of the normalized excess degree under equilibrium price shares. The functional $\mathcal V$ is the population form, under the law of large numbers, of the variance displayed in that theorem,
\[
\mathcal V(\alpha,\nu;\vartheta)
:=\Var_{\widehat r^{*}}\!\left[\frac{d^{\nu^{2}}-\E[d^{\nu^{2}}]}{d}\right],
\qquad
\widehat r^{*}\propto d^{1-\vartheta}.
\]
The variance of the Cantillon exposure in that theorem is then $\Var_{\widehat{\mathbf r}^{*}}(\boldsymbol{\mathcal C})=(\kappa/c_m)^{2}\mathcal V$. The two measures of relative-price distortion use different weights, stationary balances for $\digamma_\omega$ and equilibrium price shares for $\digamma_\psi$. But what they share is the limit of the propagation kernel and a dispersion of the Cantillon exposure that depends on neither $\theta$ nor $\lambda_2$. We hold $\kappa/c_m$, $\alpha$, and $\vartheta$ fixed throughout. Every quantity in $\digamma_\omega$ and $\digamma_\psi$ is exact in $\nu$ on the maintained range, so our argument uses no expansion in $\nu$.

Consider first the injection exponent ($\theta$). Only the stationary misalignment ($\xi_{\mathrm{ub}}$) carries $\theta$, and under the Pareto substitution its leading fixed-gap value is
\[
\xi_{\mathrm{ub}}
=
\frac{\E[d^{\theta+\nu}]}{\E[d^\theta]}
-
\frac{\E[d^{1+\nu}]}{\E[d]}
=
\frac{-\nu(1-\theta)}{(\alpha-\theta-\nu)(\alpha-1-\nu)}.
\]
Since the factor $\alpha-1-\nu$ does not depend on $\theta$, differentiating gives
\[
\frac{\partial \xi_{\mathrm{ub}}}{\partial\theta}=\frac{\nu}{(\alpha-\theta-\nu)^{2}}<0
\]
at every $\nu\in(-1,0)$, so $\partial_\theta\omega<0$ and $\partial_\theta\psi<0$. The magnitude of $\nu$ plays no part in this step.

Consider next the second eigenvalue ($\lambda_2$). In the leading coefficients the stationary misalignment ($\xi_{\mathrm{ub}}$) is its leading term, the degree-moment expression of Lemma~\ref{lem:Ct_positive}, which does not depend on $\lambda_2$. Holding these moments and the intensive response scale fixed,
\[
\frac{\partial}{\partial\lambda_2}\Bigl(\xi_{\mathrm{ub}}\,\frac{\lambda_2}{1-\lambda_2}\Bigr)
=\frac{\xi_{\mathrm{ub}}}{(1-\lambda_2)^{2}}>0.
\]
Hence $\partial_{\lambda_2}\omega>0$ and $\partial_{\lambda_2}\psi>0$. But the order of the two limits matters as $\lambda_2\to1$. At a fixed rate of inflation the limit $\xi_{\mathrm{ub}}\lambda_2/(1+\pi-\lambda_2)$ of the propagation kernel stays below $\xi_{\mathrm{ub}}/\pi$, so the low-inflation form, with network persistence $\lambda_2/(1-\lambda_2)$, applies as $\lambda_2\to1$ only when $\pi/(1-\lambda_2)\to0$.

The assortativity exponent ($\nu$) reaches $\digamma_\omega$ and $\digamma_\psi$ through two channels, the stationary misalignment and the dispersion of the Cantillon exposure. We write $x:=-\nu\in(0,1)$ and $y:=\nu^{2}=x^{2}$, and abbreviate $a:=\alpha-\theta>0$ and $b:=\alpha-1>0$. Differentiating the exact stationary misalignment gives
\begin{equation}
\label{eq:xi_nu_derivative}
\frac{\partial \xi_{\mathrm{ub}}}{\partial\nu}
=
\frac{(\theta-1)\bigl[(\alpha-\theta)(\alpha-1)-\nu^{2}\bigr]}
{(\alpha-1-\nu)^{2}(\alpha-\theta-\nu)^{2}}.
\end{equation}
Because $\theta<1$, the sign of~\eqref{eq:xi_nu_derivative} is the sign of $\nu^{2}-(\alpha-\theta)(\alpha-1)$. The stationary misalignment therefore rises in $|\nu|$ up to the turning point
\begin{equation}
\label{eq:nu_turning_point}
\nu^{\dagger}:=\sqrt{(\alpha-\theta)(\alpha-1)}
\end{equation}
and declines beyond it. The stationary misalignment peaks when the partner-degree gradient is steep enough to seat every small firm beside a large one. Read in $x$, the stationary misalignment is $\xi_{\mathrm{ub}}=x(1-\theta)/[(a+x)(b+x)]$, single-peaked with maximum at $x=\sqrt{ab}=\nu^{\dagger}$. The misalignment channel therefore saturates within the maintained range exactly when $(\alpha-\theta)(\alpha-1)<1$, which requires $\alpha<2$. Whenever $\alpha\ge2$, the misalignment channel saturates outside the maintained range, since $\nu^{\dagger}>1$ in that case. At $\nu=0$ the cross derivative of the stationary misalignment is
\[
\frac{\partial^{2}\xi_{\mathrm{ub}}}{\partial\nu\,\partial\theta}
\;=\;
(\alpha-\theta)^{-2}>0,
\]
so the marginal effect of disassortative mixing declines in magnitude as $\theta\to1$ and vanishes at proportional injection.

The dispersion channel supplies what the misalignment channel loses, for the dispersion of the Cantillon exposure rises with $|\nu|$ on the whole range. By Lemma~\ref{oa:lem:dispersion_growth}, which requires only $\alpha>1$ and $\vartheta>0$, each of the two dispersion functionals $D\in\{Q(\alpha,\nu),\mathcal V(\alpha,\nu;\vartheta)\}$, read as a function of $y$, satisfies the elasticity bound
\begin{equation}
\label{eq:dispersion_elasticity}
\varepsilon_D(y):=\frac{y\,D'(y)}{D(y)}\;\ge\;2,
\qquad
y\in(0,1),
\end{equation}
or equivalently, $y\mapsto D(y)/y^{2}$ is nondecreasing. Both functionals vanish at $y=0$ at the exact rate $y^{2}$, which recovers the orders $\digamma_\omega=O(|\nu|^{3})$ and $\digamma_\psi=O(\nu^{6})$ near degree neutrality. In words, neither dispersion functional grows more slowly than $y^{2}$ anywhere on the maintained range.

Combining the two channels signs the derivatives of $\digamma_\omega$ and $\digamma_\psi$ in $\nu$. Since $a>0$ and $b>0$, the elementary bound
\[
\frac{1}{a+x}+\frac{1}{b+x}<\frac{2}{x}
\]
holds at every $x>0$. The logarithm of $\digamma_\omega$ differs from $\log\xi_{\mathrm{ub}}+\tfrac12\log Q$ by terms free of $\nu$, so~\eqref{eq:dispersion_elasticity} gives
\[
\frac{\partial\log\digamma_\omega}{\partial x}
=\frac{1+\varepsilon_Q(x^{2})}{x}-\frac{1}{a+x}-\frac{1}{b+x}
\;\ge\;\frac{3}{x}-\frac{1}{a+x}-\frac{1}{b+x}
\;>\;\frac{3}{x}-\frac{2}{x}
=\frac{1}{x}>0.
\]
In $\digamma_\psi$ the stationary misalignment enters squared and $\log\mathcal V(x^{2})$ replaces $\tfrac12\log Q(x^{2})$, so the same two inequalities give
\[
\frac{\partial\log\digamma_\psi}{\partial x}
=\frac{2+2\varepsilon_{\mathcal V}(x^{2})}{x}-\frac{2}{a+x}-\frac{2}{b+x}
\;\ge\;\frac{6}{x}-\frac{2}{a+x}-\frac{2}{b+x}
\;>\;\frac{6}{x}-\frac{4}{x}
=\frac{2}{x}>0.
\]
Both leading coefficients are therefore strictly increasing in $x$, and $x=-\nu$ gives $\partial_\nu\omega<0$ and $\partial_\nu\psi<0$ at every $\nu\in(-1,0)$. Neither inequality uses the magnitude of $\nu$, and neither restricts $\alpha$, $\theta$, or $\vartheta$ beyond the maintained conditions. The margin in each display also records how far the dispersion channel outruns the misalignment channel, which is why the turning point in $\xi_{\mathrm{ub}}$ leaves no trace in either measure of relative-price distortion.

Proposition~\ref{prop:comparative_statics} holds the tail exponent ($\alpha$) fixed. Since
\[
\xi_{\mathrm{ub}}=\frac{-\nu(1-\theta)}{(\alpha-\theta-\nu)(\alpha-1-\nu)},
\]
the stationary misalignment falls with $\alpha$. But the dispersion functional and the equilibrium-price normalization also move. For the relative price gap,
\[
\frac{\partial\log\omega}{\partial\alpha}
=\frac{\partial_\alpha\xi_{\mathrm{ub}}}{\xi_{\mathrm{ub}}}
+\frac12\frac{\partial_\alpha Q(\alpha,\nu)}{Q(\alpha,\nu)}
+\frac{1}{2(\alpha-1)}.
\]
The three terms have no ordering that holds at every calibration, which is why Proposition~\ref{prop:comparative_statics} holds $\alpha$ fixed.
\end{proof}

\section{Inflation under Sticky Prices}\label{app:block_sticky}

\subsection{Rationing and the sticky balanced-growth allocation}
\label{oa:sec:sticky_allocation}

Under sticky prices the goods market need not clear, so we verify that the flexible balanced-growth allocation remains feasible. Nominal commitments settle in full, so the nominal block remains the autonomous $\mathbf A$-recursion and drives the real block without receiving feedback from it.

We write $p_i^{(t)}=\mathcal D_i^{(t)}/q_i^{(t)}$ for the price that clears current output. Under the timing of Section~\ref{subsec:sticky_prices}, the posted price at which firm $i$ trades at date $t$ is $\bar p_i^{(t)}=p_i^{(t-u_i^{(t)})}$. Buyer $j$'s nominal commitment to firm $i$ is $a_{ij}\widetilde m_j^{(t-1)}$. Under the settlement convention of Section~\ref{subsec:sticky_prices}, in which each buyer pays what it committed whatever quantity it receives, rationing changes the physical input delivered but not this nominal payment:
\begin{equation}
\label{oa:eq:delivery_max}
x_{ij}^{(t)}=\frac{a_{ij}\,\widetilde m_j^{(t-1)}}{\max\{\bar p_i^{(t)},\,p_i^{(t)}\}} .
\end{equation}

\begin{oalemma}[Rationing preserves the flexible allocation]
\label{oa:lem:rationing_flexible}
In the sticky-price economy of Section~\ref{sec:sticky}, under the settlement convention~\eqref{oa:eq:delivery_max}, the following hold.
\begin{enumerate}[label=(\roman*)]
\item On the rationing margin, where firm $i$ posts at or below its market-clearing price, $\bar p_i^{(t)}\le p_i^{(t)}$, every buyer $j$ receives $x_{ij}^{(t)}=a_{ij}\widetilde m_j^{(t-1)}/p_i^{(t)}$. The entire output is allocated, $\sum_jx_{ij}^{(t)}=q_i^{(t)}$, in the same proportions as in the flexible-price economy at the same balances and output.
\item On the disposal margin, where firm $i$ posts strictly above its market-clearing price, $\bar p_i^{(t)}>p_i^{(t)}$, buyer $j$ receives $a_{ij}\widetilde m_j^{(t-1)}/\bar p_i^{(t)}$ and the unsold amount $q_i^{(t)}-\mathcal D_i^{(t)}/\bar p_i^{(t)}>0$ is discarded.
\item The flexible balanced-growth path of Proposition~\ref{prop:balanced_growth_convergence}, together with any admissible configuration of bounded price ages, is also a sticky-price balanced-growth allocation. Every posted price lies at or below its market-clearing price by the accumulated drift, part~(i) gives exactly the flexible allocation, and no output is discarded.
\end{enumerate}
\end{oalemma}

\begin{proof}
(i) If $\bar p_i^{(t)}\le p_i^{(t)}$, the maximum in~\eqref{oa:eq:delivery_max} is $p_i^{(t)}$. Summing deliveries over buyers gives $\mathcal D_i^{(t)}/p_i^{(t)}=q_i^{(t)}$, and buyer $j$'s share is $a_{ij}\widetilde m_j^{(t-1)}/\mathcal D_i^{(t)}$, exactly as under flexible clearing. (ii) If $\bar p_i^{(t)}>p_i^{(t)}$, the maximum is $\bar p_i^{(t)}$, so total delivery is $\mathcal D_i^{(t)}/\bar p_i^{(t)}<q_i^{(t)}$ and the difference is discarded. (iii) On the flexible balanced-growth path, $p_i^{(t)}=M_t\widehat p_{i,\mathrm{ub}}$ and $q_i^{(t)}=q_{i,\mathrm{ub}}$. The date-$t$ posted price is therefore
\[
\bar p_i^{(t)}
=p_i^{(t-u_i^{(t)})}
=M_{t-u_i^{(t)}}\widehat p_{i,\mathrm{ub}},
\qquad
\frac{\bar p_i^{(t)}}{p_i^{(t)}}=(1+\pi)^{-u_i^{(t)}}\le1.
\]
Part~(i) then reproduces the flexible inputs and hence the same output $\mathbf q_{\mathrm{ub}}$. The allocation is therefore self-consistent under sticky prices, and no output is discarded.
\end{proof}

\begin{oacorollary}[Standing posted prices on the sticky balanced-growth path]
\label{oa:cor:sticky_bgp_vintage}
On the sticky balanced-growth path,
\[
\log\frac{\bar p_i^{(t)}}{M_t p_i^*}
=\log\frac{\widehat p_{i,\mathrm{ub}}}{p_i^*}
-u_i^{(t)}\log(1+\pi),
\]
and therefore
\[
\log\frac{\bar r_i^{(t)}}{r_i^*}
=\log\left(
\frac{\widehat p_{i,\mathrm{ub}}/\widehat p_{k,\mathrm{ub}}}
{p_i^*/p_k^*}
\right)
-(u_i^{(t)}-u_k^{(t)})\log(1+\pi).
\]
Here $\bar r_i^{(t)}:=\bar p_i^{(t)}/\bar p_k^{(t)}$ and $r_i^{*}:=p_i^{*}/p_k^{*}$ for an arbitrary firm $k$, which cancels from both measures of relative-price distortion of the main text. The first term is the exact displacement of firm $i$'s relative price on the flexible balanced-growth path, and the second is the exact vintage drift, the inflation accumulated over the difference in price ages ($u_i^{(t)}-u_k^{(t)}$).
\end{oacorollary}

\begin{proof}
By Lemma~\ref{oa:lem:rationing_flexible}(iii), $p_i^{(t)}=M_t\widehat p_{i,\mathrm{ub}}$ and $\bar p_i^{(t)}=M_{t-u_i^{(t)}}\widehat p_{i,\mathrm{ub}}$. Dividing the latter by $M_t p_i^*$ and using $M_{t-u_i^{(t)}}/M_t=(1+\pi)^{-u_i^{(t)}}$ gives the first identity, and subtracting the corresponding identity for firm $k$ yields the second.
\end{proof}

\subsection{Attraction to sticky balanced growth under sustained injection}
\label{oa:sticky_bgp_attraction}

Full settlement keeps the nominal block independent of posted prices. Under sustained injection, balances therefore converge as they do under flexible prices, and posted prices converge to the age-indexed balanced-growth configuration of Corollary~\ref{oa:cor:sticky_bgp_vintage}.

\begin{oalemma}[Sticky-price attraction under sustained injection]
\label{oa:lem:sticky_bgp_attraction}
Let $\pi>0$ and suppose that the injection profile converges, $\boldsymbol\gamma_t\to\boldsymbol\gamma_{\mathrm{ub}}$, as in Proposition~\ref{prop:balanced_growth_convergence}. In the analytical sticky-price economy, suppose nominal commitments settle in full, unsold output is discarded, a reset posts the current market-clearing price before trade, and price age is bounded by the fixed cap $\bar u<\infty$. Then every positive initial configuration satisfies
\[
\widehat{\mathbf m}_t\longrightarrow\widehat{\mathbf m}_{\mathrm{ub}},
\qquad
\mathbf q_t\longrightarrow\mathbf q_{\mathrm{ub}},
\qquad
\frac{\mathbf p_t}{M_t}\longrightarrow\widehat{\mathbf p}_{\mathrm{ub}}.
\]
Posted prices approach the balanced-growth vintage carrying their realized ages:
\begin{equation}
\label{oa:eq:sticky_bgp_attraction}
\max_i
\left|
\log\frac{\bar p_i^{(t)}/M_t}
{(1+\pi)^{-u_i^{(t)}}\widehat p_{i,\mathrm{ub}}}
\right|
\longrightarrow0.
\end{equation}
Under Assumption~\ref{assu:hazard}, the joint age process converges to its product stationary law. Hence the detrended sticky-price economy converges in distribution to the age-indexed balanced-growth configuration of Corollary~\ref{oa:cor:sticky_bgp_vintage}.

If $\|\boldsymbol\gamma_t-\boldsymbol\gamma_{\mathrm{ub}}\|=O(\rho_\gamma^t)$ for some $\rho_\gamma<1$, then with
\[
\rho_m:=\max\left\{\frac{\lambda_2}{1+\pi},\rho_\gamma\right\}
\]
the balance--output--clearing-price block has geometric factor
\[
\max\left\{\varsigma^{1/(\bar u+1)},\rho_m\right\}<1,
\]
up to the usual polynomial factors when component rates coincide.
\end{oalemma}

\begin{proof}
Full settlement leaves the normalized nominal recursion independent of posted prices. Proposition~\ref{prop:balanced_growth_convergence} therefore gives $\widehat{\mathbf m}_t\to\widehat{\mathbf m}_{\mathrm{ub}}$ and $\widehat{\mathbf n}_t\to\widehat{\mathbf n}_{\mathrm{ub}}$, where $\widehat{\mathbf n}_t:=\widetilde{\mathbf m}_t/M_{t+1}$. Define
\[
\boldsymbol\mu_t
:=\log\widehat{\mathbf m}_t-\log\widehat{\mathbf m}_{\mathrm{ub}},
\qquad
\mathbf y_t
:=\log\mathbf q_t-\log\mathbf q_{\mathrm{ub}},
\qquad
\mathbf f_t:=\boldsymbol\mu_t-\mathbf y_t.
\]
Because $\boldsymbol{\mathcal D}_t=\mathbf m_t$, $\mathbf f_t=\log(\mathbf p_t/M_t)-\log\widehat{\mathbf p}_{\mathrm{ub}}$. Once every firm has reset, which the cap guarantees within a bounded number of dates, let the detrended posted-price deviation be
\[
\bar{\mathbf f}_t
:=\log(\bar{\mathbf p}_t/M_t)-\log\widehat{\mathbf p}_{\mathrm{ub}}.
\]
Its components satisfy
\[
\bar f_{i,t}
=f_{i,t-u_i^{(t)}}-u_i^{(t)}\log(1+\pi).
\]
The balanced-growth value of $\bar f_{i,t}$ at the same age is $-u_i^{(t)}\log(1+\pi)$. Since the coordinatewise maximum is nonexpansive and $\max\{-u_i^{(t)}\log(1+\pi),0\}=0$,
\[
\left\|\max\{\bar{\mathbf f}_t,\mathbf f_t\}\right\|_\infty
\le
\max_{0\le u\le\bar u}\|\mathbf f_{t-u}\|_\infty.
\]

Under sustained injection the exact production identity is
\[
\log\mathbf q_{t+1}
=
\varsigma\left[
\mathbf b_A+\log\widetilde{\mathbf m}_{t-1}
-\mathbf A^\top\max\{\log\bar{\mathbf p}_t,\log\mathbf p_t\}
\right].
\]
Using $\widetilde{\mathbf m}_{t-1}=M_t\widehat{\mathbf n}_{t-1}$, detrending, and subtracting the balanced-growth identity of Lemma~\ref{oa:lem:rationing_flexible}(iii) gives
\[
\mathbf y_{t+1}
=
\varsigma\left[
\log\widehat{\mathbf n}_{t-1}
-\log\widehat{\mathbf n}_{\mathrm{ub}}
-\mathbf A^\top\max\{\bar{\mathbf f}_t,\mathbf f_t\}
\right].
\]
Because $\mathbf A^\top$ is row stochastic, $z_t:=\|\mathbf y_t\|_\infty$ consequently obeys
\[
z_{t+1}
\le
\varsigma\max_{0\le u\le\bar u}z_{t-u}
+e_t,
\]
where
\[
e_t
:=
\varsigma\left[
\left\|\log\widehat{\mathbf n}_{t-1}
-\log\widehat{\mathbf n}_{\mathrm{ub}}\right\|_\infty
+
\max_{0\le u\le\bar u}\|\boldsymbol\mu_{t-u}\|_\infty
\right]
\longrightarrow0.
\]
Iterating over blocks of $\bar u+1$ dates, we obtain $z_t\to0$, since $\varsigma<1$. When the forcing ($e_t$) decays at the geometric factor $\rho_m$, the same block iteration yields the factor $\max\{\varsigma^{1/(\bar u+1)},\rho_m\}$ of the lemma. Hence $\mathbf q_t\to\mathbf q_{\mathrm{ub}}$, and $\mathbf f_t\to0$ implies $\mathbf p_t/M_t\to\widehat{\mathbf p}_{\mathrm{ub}}$. The identity for $\bar f_{i,t}$ then delivers~\eqref{oa:eq:sticky_bgp_attraction}.

Finally, the joint age process converges to its product stationary law. Clearing prices converge and the cap bounds price age, so the pre-reset gap of a firm at age $u$ converges to $(u+1)\log(1+\pi)$, uniformly in $u\le\bar u$ and across firms. By continuity of $\eta$, the conditional price reset probabilities converge uniformly to those of the limiting chain, in which firms reset independently, with probability $\eta\bigl((u+1)\log(1+\pi)\bigr)$ at age $u<\bar u$ and with probability one at age $\bar u$. In the limiting chain the positive price reset probability at age zero gives a self-loop, and the forced reset at $\bar u$ makes age zero accessible from every age within $\bar u+1$ dates. The power $\mathbf H^{\bar u+1}$ of its transition matrix $\mathbf H$ is therefore a strict contraction in total variation, with some factor $\mathsf q<1$, toward its stationary law $\mathscr L_*$, the product of the stationary laws of price age of the firms. Write $\mathscr L_t$ for the joint law of ages at date $t$. Uniform convergence of the price reset probabilities gives $\lVert\mathscr L_{t+1}-\mathscr L_t\mathbf H\rVert_{\mathrm{TV}}\to0$, and
\[
\lVert\mathscr L_{t+\bar u+1}-\mathscr L_*\rVert_{\mathrm{TV}}
\le\mathsf q\,\lVert\mathscr L_t-\mathscr L_*\rVert_{\mathrm{TV}}
+\sum_{h=0}^{\bar u}\lVert\mathscr L_{t+h+1}-\mathscr L_{t+h}\mathbf H\rVert_{\mathrm{TV}}.
\]
Iterating gives $\mathscr L_t\to\mathscr L_*$. Combining this convergence with~\eqref{oa:eq:sticky_bgp_attraction} proves the distributional conclusion.
\end{proof}

\subsection{Convergence after injection stops}
\label{oa:proof_sticky_convergence}

Once injection stops, $\pi=0$, and the economy starts from the balances, output, and posted prices that the preceding inflation has left. We index that date by zero and write $M:=\mathbf 1^\top\mathbf m_0$ for the money stock it inherits. The zero-injection equilibrium at this nominal scale, $(M\mathbf m^*,\mathbf q^*,M\mathbf p^*)$, has the real allocation and relative prices of the equilibrium $(\mathbf m^*,\mathbf q^*,\mathbf p^*)$ of Proposition~\ref{prop:equilibrium_existence} and differs from it only in its nominal scale.

Our argument invokes two properties of the price reset rule of Section~\ref{sec:sticky}, the first of which is that a firm that resets its price does so before trade and posts its current market-clearing price, $\bar p_i^{(t)}=p_i^{(t)}=\mathcal D_i^{(t)}/q_i^{(t)}$. And the second is that no posted price is older than $\bar u$ (Assumption~\ref{assu:hazard}), so every firm resets at least once in every $\bar u+1$ consecutive dates. Full settlement makes the nominal block autonomous, $\mathbf m_{t+1}=\mathbf A\mathbf m_t$, while nominal orders and posted prices determine delivered inputs and hence next-period output. The system is therefore block triangular, as under flexible prices, with a bounded delay introduced by price ages.

Define the equilibrium-normalized log deviations
\[
\begin{aligned}
\boldsymbol\mu_t &:= \log\mathbf m_t-\log(M\mathbf m^*),
&\qquad
\mathbf y_t &:= \log\mathbf q_t-\log\mathbf q^*,\\
\mathbf f_t &:= \log\mathbf p_t-\log(M\mathbf p^*),
&\qquad
\bar{\mathbf f}_t &:= \log\bar{\mathbf p}_t-\log(M\mathbf p^*),
\end{aligned}
\]
and write $z_t:=\|\mathbf y_t\|_\infty$. Under zero injection the demand indexing gives $\boldsymbol{\mathcal D}_t=\mathbf m_t$, so
\begin{equation}
\label{oa:eq:sticky_price_deviations}
\mathbf f_t=\boldsymbol\mu_t-\mathbf y_t,
\qquad
\|\bar{\mathbf f}_t\|_\infty
\le
\max_{0\le u\le\bar u}\|\mathbf f_{t-u}\|_\infty,
\end{equation}
after every firm has reset at least once. The second relation follows because each coordinate of $\bar{\mathbf f}_t$ is the corresponding coordinate of $\mathbf f$ at that firm's most recent price reset date.

Under the settlement convention of the main text, buyer $j$'s realized input of good $i$ is
\[
x_{ij}^{(t)}
=\frac{a_{ij}m_j^{(t-1)}}{\max\{\bar p_i^{(t)},p_i^{(t)}\}}.
\]
Cobb--Douglas production therefore gives the exact identity
\begin{equation}
\label{oa:eq:production_identity_global}
\log\mathbf q_{t+1} = \varsigma\left[\mathbf b_A+\log\mathbf m_{t-1}-\mathbf A^\top\max\{\log\bar{\mathbf p}_t,\log\mathbf p_t\}\right].
\end{equation}
Subtracting the equilibrium identity and using that $\mathbf A^\top$ is row stochastic and the coordinatewise maximum is nonexpansive yields
\begin{equation}
\label{oa:eq:one_step_output_bound}
\|\mathbf y_{t+1}\|_\infty \le \varsigma\left[\|\boldsymbol\mu_{t-1}\|_\infty+\max\{\|\bar{\mathbf f}_t\|_\infty,\|\mathbf f_t\|_\infty\}\right].
\end{equation}
Combining~\eqref{oa:eq:sticky_price_deviations} and~\eqref{oa:eq:one_step_output_bound} gives, for all sufficiently large $t$,
\begin{equation}
\label{oa:eq:sticky_delay_contraction}
z_{t+1}
\le
\varsigma\max_{0\le u\le\bar u}z_{t-u}
+e_t,
\qquad
e_t
:=
\varsigma\left[
\|\boldsymbol\mu_{t-1}\|_\infty
+\max_{0\le u\le\bar u}\|\boldsymbol\mu_{t-u}\|_\infty
\right].
\end{equation}
Money therefore affects output as a decaying forcing term, while the homogeneous real block under sticky prices is a bounded-delay contraction.

\begin{oalemma}[Bounded-delay contraction]
\label{oa:lem:bounded_delay_contraction}
Let a nonnegative sequence $(z_t)_{t\ge0}$ satisfy
\[
z_{t+1}\le a\max_{0\le u\le\bar u}z_{t-u}+e_t,
\qquad
0<a<1,
\qquad
e_t=O(b^t),
\]
for some $b\in(0,1)$. Then
\[
z_t
=
O\!\left((1+t)\rho^t\right),
\qquad
\rho:=\max\{a^{1/(\bar u+1)},b\}<1.
\]
The factor $1+t$ is needed only when $a=b^{\bar u+1}$.
\end{oalemma}

\begin{proof}
Let $L:=\bar u+1$ and $Z_k:=\max_{kL\le t<(k+1)L}z_t$. Iterating the one-step inequality within the next block gives
\[
Z_{k+1}
\le
aZ_k+\frac{E_k}{1-a},
\qquad
E_k:=\max_{(k+1)L-1\le t<(k+2)L-1}e_t
=O(b^{kL}).
\]
This is because, if all preceding observations in the new block are bounded by $aZ_k+E_k/(1-a)$, the next observation is bounded by
$a[Z_k+E_k/(1-a)]+E_k=aZ_k+E_k/(1-a)$. The first observation starts the induction. Iterating the block recursion yields
\[
Z_k
=
O\!\left((1+k)\max\{a,b^L\}^{k}\right),
\]
with the polynomial factor only at equality. Translating from blocks to dates gives the result.
\end{proof}

\begin{oalemma}[Convergence with sticky prices]
\label{lemma:sticky_convergence}
Let $\pi=0$ and $\varsigma\in(0,1)$, and let $M:=\mathbf 1^\top\mathbf m_0$. In the sticky-price economy of Section~\ref{sec:sticky}, every positive initial configuration of balances, output, and posted prices converges to the equilibrium $(M\mathbf m^*,\mathbf q^*,M\mathbf p^*)$ at the inherited monetary scale:
\[
\begin{aligned}
&\|\log\mathbf m_t-\log(M\mathbf m^*)\|_\infty
+\|\log\mathbf q_t-\log\mathbf q^*\|_\infty
+\|\log\bar{\mathbf p}_t-\log(M\mathbf p^*)\|_\infty\\
&\qquad=O\!\left((1+t)
\left(\max\bigl\{\varsigma^{1/(\bar u+1)},\lambda_2\bigr\}\right)^{\!t}\right),
\end{aligned}
\]
where the polynomial factor is needed only when $\varsigma=\lambda_2^{\bar u+1}$ and may otherwise be omitted.
\end{oalemma}

\begin{proof}
Full settlement gives $\mathbf m_{t+1}=\mathbf A\mathbf m_t$ from every configuration. The money stock is conserved at $M$, and primitivity and the spectral assumptions give $\|\boldsymbol\mu_t\|_\infty=O(\lambda_2^t)$. Consequently the forcing term in~\eqref{oa:eq:sticky_delay_contraction} is $e_t=O(\lambda_2^t)$. Lemma~\ref{oa:lem:bounded_delay_contraction}, with $a=\varsigma$ and $b=\lambda_2$, gives $\mathbf q_t\to\mathbf q^*$ at the displayed rate. The market-clearing prices then converge, $\mathbf p_t\to M\mathbf p^*$, through $\mathbf p_t=\boldsymbol{\mathcal D}_t/\mathbf q_t$, and bounded reset ages carry the posted prices with them, $\bar{\mathbf p}_t\to M\mathbf p^*$, through~\eqref{oa:eq:sticky_price_deviations}. The only limit at the inherited monetary scale is the equilibrium. This is because $\mathbf m=\mathbf A\mathbf m$ and $\mathbf1^\top\mathbf m=M$ force $\mathbf m=M\mathbf m^*$, and the zero-injection production identity then gives the unique $\mathbf q^*$ and $M\mathbf p^*$ of Proposition~\ref{prop:equilibrium_existence}.
\end{proof}

Discarding unsold output, as the analytical economy does, removes inventory as a state, but not the effect of money on production. In the computational economy, by contrast, unsold output stays with the firm as inventory. The convergence argument of Lemma~\ref{lemma:sticky_convergence} extends to such carryover under two conditions, the first of which is that resets price the entire available stock to clear within the finite cap $\bar u$. A reset then clears the firm's inventory, so every inventory returns to zero at least once in every $\bar u+1$ dates, and inventory is a state only over that window. The second condition is that the real block over such a window be contractive, and the bounded-delay contraction of Lemma~\ref{oa:lem:bounded_delay_contraction} then applies as before. Sufficiently strong inventory depreciation provides an alternative route. And every computational economy we report relaxes to its zero-injection equilibrium before injection begins (Table~\ref{oa:tab:calibration}).

\subsection{The pre-reset gap}\label{app:proof_hazard_sufficiency}

\begin{oalemma}[Conditional separation of the leading pre-reset gap]
\label{lemma:hazard_sufficiency}
At date $t$, the price inherited immediately before firm $i$'s price reset decision is $\bar p_i^{(t-1)}=p_i^{(t-u_i^{(t-1)}-1)}$. Under the nominal-demand projection of Assumption~\ref{assu:small_returns}, the leading pre-reset log gap between firm $i$'s market-clearing price and its inherited posted price separates into a common accumulated drift and a firm-specific network correction,
\[
\log\frac{p_i^{(t)}}{\bar p_i^{(t-1)}}
\;=\;
\underbrace{\bigl(u_i^{(t-1)}+1\bigr)\log(1+\pi)}_{\text{drift}}
\;+\;
\underbrace{\log(1+\ell_i^{(t)})-\log(1+\ell_i^{(t-u_i^{(t-1)}-1)})}_{\text{network correction}},
\]
where $\ell_i^{(s)}\approx\pi\mathcal C_i\mathcal X_s$ is the level distortion of firm $i$'s nominal demand~\eqref{eq:level_distortion}, and the network correction is $O(\pi|\mathcal C_i|)$. At a fixed date and conditional on the common propagation path $(\mathcal X_s)_s$, cross-firm variation in this leading gap is indexed by the drift accumulated over the price spell that a reset would complete and by the firm's Cantillon exposure, $\bigl(\pi[u_i^{(t-1)}+1],\mathcal C_i\bigr)$. In the balanced-growth steady state the network correction vanishes exactly.
\end{oalemma}

The nominal-demand projection sets aside an output remainder, the change in firm $i$'s log output over the price spell, since $p_i^{(s)}=\mathcal D_i^{(s)}/q_i^{(s)}$ at every date. This remainder is exactly zero on the balanced-growth path, where output is stationary (Lemma~\ref{oa:lem:rationing_flexible}), and off it the remainder is a real-block term over a bounded window (Lemma~\ref{oa:lem:full_stability}). But during the transition the two firm-level indices of the lemma do not make a complete aggregate state. This is because the network correction also varies with the common calendar state.

\begin{proof}
Immediately before the date-$t$ price reset decision, the inherited posted price is
$\bar p_i^{(t-1)}=p_i^{(t-u_i^{(t-1)}-1)}$. Since $p_i^{(s)}=\mathcal D_i^{(s)}/q_i^{(s)}$ at every date, the exact pre-reset gap is the nominal-demand ratio minus the corresponding log output ratio. We split firm $i$'s nominal demand at each date into its permanent trend and its level distortion (Lemma~\ref{lemma:exposure}), $\mathcal D_i^{(s)}=P_i^{(s)}(1+\ell_i^{(s)})$, where $P_i^{(s)}=m_i^{*}(1+\pi)^{s}$ and $\ell_i^{(s)}\approx\pi\mathcal C_i\mathcal X_s$. Taking logs,
\[
\log\frac{p_i^{(t)}}{\bar p_i^{(t-1)}}
=\underbrace{\log\!\left(\frac{P_i^{(t)}}{P_i^{(t-u_i^{(t-1)}-1)}}\right)}_{=\,[u_i^{(t-1)}+1]\log(1+\pi)}
\;+\;\log(1+\ell_i^{(t)})-\log(1+\ell_i^{(t-u_i^{(t-1)}-1)})
\;-\;\log\!\left(\frac{q_i^{(t)}}{q_i^{(t-u_i^{(t-1)}-1)}}\right),
\]
which holds exactly. The nominal-demand projection retains the common drift and the network correction and sets aside the output remainder, which leaves the two-term gap of the lemma. Let $\ell_i^{\max}:=\sup_s|\ell_i^{(s)}|$. The amplitude bound of Lemma~\ref{oa:lem:level_amplitude} gives
\[
\ell_i^{\max}
\le
|\widehat\delta_i|\,
\frac{\pi\kappa \xi_{\mathrm{ub}}\lambda_2}{c_m(1+\pi-\lambda_2)}\bigl(1+O(\pi)\bigr)
=O(\pi|\mathcal C_i|).
\]
For sufficiently small $\pi$, $\ell_i^{\max}<1$, and the mean-value theorem for $\log(1+x)$ yields
\[
|\log(1+\ell_i^{(t)})-\log(1+\ell_i^{(t-u_i^{(t-1)}-1)})|
\le
\frac{|\ell_i^{(t)}-\ell_i^{(t-u_i^{(t-1)}-1)}|}{1-\ell_i^{\max}}
\le
\frac{2\ell_i^{\max}}{1-\ell_i^{\max}}
=O(\pi|\mathcal C_i|).
\]
At a fixed date and conditional on the common propagation path, the drift enters the nominal leading term through $u_i^{(t-1)}+1$ alone, while the network correction carries firm identity only through $\mathcal C_i$. But the level of the network correction depends on the common calendar state through $\mathcal X_t$ and $\mathcal X_{t-u_i^{(t-1)}-1}$. The pair $\bigl(u_i^{(t-1)}+1,\mathcal C_i\bigr)$ is therefore a cross-sectional reduction, and not a complete state of the transition. In the balanced-growth steady state the propagation kernel is constant and output is stationary, so the retained network correction vanishes.
\end{proof}

\subsection{The frequency and size margins}\label{app:proof_frequency}

Propositions~\ref{prop:frequency} and~\ref{prop:sticky_flexible_ranking} of the main text hold at rates of inflation at which the drift over a typical price spell stays small and a typical price spell ends well before the maximal price age ($\bar u$).

\begin{oadefinition}[Cap-nonbinding range and cap-nonbinding sequences]\label{oa:def:cap_range}
Fix small constants $s_0,\chi\in(0,1)$, chosen where the proofs use them. The cap-nonbinding range is
\begin{equation}
\label{eq:cap_nonbinding_range}
\mathcal P_{\bar u}
:=\bigl\{\pi>0:\pi u_\pi^\star\le s_0,\ u_\pi^\star\le\chi\bar u\bigr\}.
\end{equation}
A sequence $(\pi_k,\bar u_k)$ is cap-nonbinding if $\pi_k\downarrow0$, $\bar u_k\to\infty$, and $u_{\pi_k}^\star/\bar u_k\to0$.
\end{oadefinition}

\noindent The first inequality keeps the drift accumulated over a typical price spell at the small gaps where the hazard function is close to $c\,s^{\beta}$ (Assumption~\ref{assu:hazard}), and the second keeps the typical price spell well inside the cap. Since the natural age scale is $u_\pi^\star=\pi^{-\beta/(\beta+1)}$, the cap-nonbinding range is the interval
\begin{equation}
\label{eq:cap_range_interval}
\mathcal P_{\bar u}=\bigl[(\chi\bar u)^{-(\beta+1)/\beta},\ s_0^{\beta+1}\bigr],
\end{equation}
which is bounded away from zero at a finite cap.\footnote{The interval~\eqref{eq:cap_range_interval} is nonempty if and only if $\bar u\ge s_0^{-\beta}/\chi$, so the cap needed to accommodate a state-dependence exponent $\beta$ grows exponentially in $\beta$. At a fixed rate of inflation the first inequality in~\eqref{eq:cap_nonbinding_range} gives $\beta\le\log\pi/\log s_0-1$. The menu-cost limit $\beta\to\infty$ therefore lies outside the cap-nonbinding range, both at a fixed cap and at a fixed rate of inflation.} On this interval an order $\Theta(\pi^{a})$ has constants that depend on the economy but not on $\pi$. Along a cap-nonbinding sequence, $\pi_k\in\mathcal P_{\bar u_k}$ for all large $k$, since $\pi_k\le s_0^{\beta+1}$ and $u_{\pi_k}^\star\le\chi\bar u_k$ eventually. The same constants therefore apply at every term of the sequence, and the orders in inflation become ordinary asymptotics. Letting the cap grow is also the natural limit economically. This is because a cap held fixed while inflation vanishes ends every price spell for a mechanical reason rather than a pricing one.

\begin{proof}[Proof of Proposition~\ref{prop:frequency}]
The size of a price change is the change in the firm's market-clearing price accumulated over the completed price spell. A firm that resets at $T$ replaces a price last set at $T-u_i^{(T-1)}-1$, and its price change is
\[
\Delta\log\bar p_i^{(T)}
=\sum_{t=T-[u_i^{(T-1)}+1]+1}^{T}\Delta\log p_i^{(t)}.
\]
Under the nominal-demand projection of Assumption~\ref{assu:small_returns}, nominal demand is $\mathcal D_i^{(t)}=P_i^{(t)}(1+\ell_i^{(t)})$ with $P_i^{(t)}=m_i^{*}(1+\pi)^{t}$ (Lemma~\ref{lemma:exposure}), where $\ell_i^{(t)}$ is the level distortion of firm $i$'s nominal demand (Section~\ref{subsec:master_equation} of the main text). Each market-clearing price change is therefore $\Delta\log p_i^{(t)}=\log(1+\pi)+\log(1+\ell_i^{(t)})-\log(1+\ell_i^{(t-1)})$, and the sum telescopes exactly to
\[
\Delta\log\bar p_i^{(T)}
=
\bigl(u_i^{(T-1)}+1\bigr)\log(1+\pi)
+
\log\bigl(1+\ell_i^{(T)}\bigr)-\log\bigl(1+\ell_i^{(T-u_i^{(T-1)}-1)}\bigr).
\]
The first term is the drift accumulated over the completed price spell, and the second is $O(\pi)$. The mean-value bound in the proof of Lemma~\ref{lemma:hazard_sufficiency} controls the second term by the amplitude bound on the level distortion (Lemma~\ref{oa:lem:level_amplitude}), which does not depend on the length of the price spell. Hence no per-period remainder accumulates over the price spell. Under the positivity condition of Lemma~\ref{oa:lem:uniform_positivity}, the per-period changes are positive, so the absolute and signed sizes of price change coincide. The output remainder that the nominal-demand projection sets aside is exactly zero on the balanced-growth path (Lemma~\ref{oa:lem:rationing_flexible}).

We fix a firm with Cantillon exposure $\mathcal C_i$. For $u\ge0$, define
\[
\mathcal S_i(u):=
\begin{cases}
\displaystyle\prod_{j=1}^{u}\bigl(1-\eta(s_i(j))\bigr),
&0\le u\le\bar u,\\[2mm]
0,&u>\bar u,
\end{cases}
\qquad
\mathcal S_i(0)=1,
\]
the probability that a price spell survives beyond $u$ periods, including the certain reset after age $\bar u$. Renewal accounting gives
\[
\Pr\bigl(u_i^{(t)}=0\bigr)
=\left(\sum_{v=0}^{\bar u}\mathcal S_i(v)\right)^{-1},
\qquad
\Pr(u_i=u)
=\frac{\mathcal S_i(u)}{\sum_{v=0}^{\bar u}\mathcal S_i(v)},
\quad 0\le u\le\bar u,
\]
and
\[
\E\!\left[u_i^{(t-1)}+1\mid I_i^{(t)}=1\right]
=\sum_{u=0}^{\bar u}\mathcal S_i(u).
\]
The same survival function therefore determines the frequency of resets, the mean length of a completed price spell, and the stationary law of price age, without a second age state. But along the transition the pre-reset gap $s_i(j)$ carries the increments of the propagation kernel at the date $t$ of observation. With those increments held fixed, these formulas describe the stationary law of the age process, the law of Lemma~\ref{lemma:network_vintages}. On the balanced-growth path the increments vanish, so the formulas describe the actual stationary law.

Take first $\beta>0$ and $\pi\in\mathcal P_{\bar u}$, and write
\[
s_i(j):=j\log(1+\pi)+\log\bigl(1+\pi\mathcal C_i\mathcal X_t\bigr)-\log\bigl(1+\pi\mathcal C_i\mathcal X_{t-j}\bigr)
\]
for firm $i$'s pre-reset gap at age $j$ (Lemma~\ref{lemma:hazard_sufficiency}). The level distortions $\ell_i^{(t)}$ and $\ell_i^{(t-j)}$ share the sign of $\mathcal C_i$ and are at most $\ell_i^{\max}=O(\pi|\mathcal C_i|)$ in magnitude (Lemma~\ref{oa:lem:level_amplitude}). The mean-value theorem therefore gives $|s_i(j)-j\log(1+\pi)|\le\ell_i^{\max}/(1-\ell_i^{\max})\le2\ell_i^{\max}$ once $\ell_i^{\max}\le1/2$, uniformly over the truncated support. Since $\eta$ is nondecreasing, $\eta(j\log(1+\pi)-2\ell_i^{\max})\le \eta(s_i(j))\le \eta(j\log(1+\pi)+2\ell_i^{\max})$. For $u=\lfloor u_\pi^\star y\rfloor$ and $y$ in a fixed bounded interval,
\[
-\log\mathcal S_i(u)
=-\sum_{j=1}^{u}\log\bigl(1-\eta(s_i(j))\bigr)
\asymp
\varrho_i\,y^{\beta+1},
\]
uniformly over the stated range. This is because $\pi u_\pi^\star\le s_0$ keeps these hazards at the small gaps where the hazard function is close to $c\,s^{\beta}$, and the sum of their squares is $O(s_0^\beta)$ on this scale. Choosing $s_0$ small gives uniform upper and lower Weibull bounds. The condition $u_\pi^\star\le\chi\bar u$ leaves a fixed multiple of the natural scale inside the support. The monotonicity of $\eta$ controls the ages beyond the local window. At every age $u$ with $u\log(1+\pi)>s_0$ the hazard is at least $\eta(s_0-2\ell_i^{\max})>0$, so $\mathcal S_i(u)$ decays geometrically from the edge of the local window and the ages outside it contribute $O(1)$ to $\sum_{u}\mathcal S_i(u)$, with a constant independent of $\pi$. The exponential-tail bound on the scale $u_\pi^\star$ in the proof of Lemma~\ref{lem:sticky_orders} controls the higher age moments in the same way. The bounds hold with constants depending on the economy but not on $\pi$, which is the uniform sense fixed at~\eqref{eq:cap_range_interval}. It follows that
\[
\sum_{u=0}^{\bar u}\mathcal S_i(u)
=\Theta(u_\pi^\star)
=\Theta\!\left(\pi^{-\beta/(\beta+1)}\right),
\]
uniformly on $\mathcal P_{\bar u}$. The order $\Theta(u_\pi^\star)$ also holds as an asymptotic along any cap-nonbinding sequence, since every term of such a sequence lies in the cap-nonbinding range of its own cap and the constants do not vary along the sequence.
The stationary-mass identity also gives, for every fixed $q>0$,
\[
\E[u_i^q]
=\frac{\sum_{u=0}^{\bar u}u^q\mathcal S_i(u)}
{\sum_{u=0}^{\bar u}\mathcal S_i(u)}
=\Theta\bigl((u_\pi^\star)^q\bigr)
=\Theta\!\left(\pi^{-q\beta/(\beta+1)}\right).
\]
In particular the stationary post-decision price age and the mean length of a completed price spell have the same inflation exponent, although their constants differ.

At the endpoint $\beta=0$, the hazard function $\eta$ stays bounded above and below by positive constants at small arguments. The fixed cap therefore produces a mean length of completed price spells and a mean post-decision price age of order one. This is exactly what the exponent $\pi^{-\beta/(\beta+1)}$ gives at $\beta=0$. If, in the Calvo special case, $\eta\equiv c$, the truncated laws converge to geometric laws as $\bar u$ grows, with mean $c^{-1}$ for the length of a completed price spell and mean $(1-c)/c$ for the post-decision price age. The gap then enters only through a hazard that has become flat. For every $\beta\ge0$ the exponent therefore does not depend on the firm. For $\beta>0$ the constants of the order $\Theta(u_\pi^\star)$ can depend on the firm through its network correction. The uniform bound $\ell_i^{\max}=O(\pi|\mathcal C_i|)$ on the network correction, together with the bounded Cantillon exposure on the truncated degree support, keeps them finite and positive uniformly across firms. As $\pi\to0$ along a cap-nonbinding sequence, the network correction, of order $\pi$, becomes negligible beside the drift at ages of order $u_\pi^\star$, which is of order $\pi^{1/(\beta+1)}$. Hence $-\log\mathcal S_i(\lfloor u_\pi^\star y\rfloor)\to c\,y^{\beta+1}/(\beta+1)$ for every fixed $y>0$ and every firm, and the leading coefficient is common to all firms.

The orders in inflation of the frequency of resets and of the average size of price change follow directly. The frequency of resets satisfies $\Pr(u_i^{(t)}=0)=\Theta\bigl(\pi^{\beta/(\beta+1)}\bigr)$, and the average size of price change satisfies
\[
\pi\,\E\!\left[u_i^{(t-1)}+1\mid I_i^{(t)}=1\right]
=\Theta\!\left(\pi^{1/(\beta+1)}\right),
\]
so the frequency of resets and the average size of price change scale with exponents $\beta/(\beta+1)$ and $1/(\beta+1)$, respectively. The two exponents sum to one. In the Calvo case, $\beta=0$, the leading frequency exponent is zero and inflation passes one for one into the size of price change. As $\beta\to\infty$ the leading size exponent vanishes and inflation is absorbed entirely on the frequency margin. These are leading-order statements within $\mathcal P_{\bar u}$. The network enters through the bounded endpoint correction $\ell_i^{(T)}-\ell_i^{(T-u_i^{(T-1)}-1)}$, not by amplifying the drift over the completed price spell. If instead $\pi\to0$ with $\bar u$ fixed and $\beta>0$, then $\mathcal S_i(u)\to1$ for $0\le u\le\bar u$, so $\Pr(u_i^{(t)}=0)\to(\bar u+1)^{-1}$, $\E[u_i]\to\bar u/2$, and the mean length of a completed price spell tends to $\bar u+1$, so inflation falls entirely on the size margin.
\end{proof}

Along the transition the network correction in the pre-reset gap moves with the calendar. The renewal formulas of this proof then describe the stationary law of the age process with the increments of the propagation kernel held at their date-$t$ values. Under this law the constants of the bounds in Proposition~\ref{prop:frequency} can depend on the firm's Cantillon exposure at a given rate of inflation. But on the balanced-growth path the law is exact and common to all firms, since the network correction vanishes there.

\subsection{Price spells ordered by Cantillon exposure}\label{app:proof_network_vintages}

Along the transition the reset probability that a price spell faces depends on the date at which it reaches each age, so a firm's price age at a given date has no stationary law of its own. Our argument therefore compares price spells begun at a common date. And for laws of price age it holds the hazard profile at its date-$t$ values, which gives the law that the proof of Proposition~\ref{prop:frequency} describes along the transition.

\begin{proof}[Proof of Lemma~\ref{lemma:network_vintages}]
We fix the common price reset date $t_0$. A price set at date $t_0$ has age $j-1$ before the price reset decision of date $t_0+j$. Within the nominal-demand projection and the two-eigenmode degree representation, the level distortion of firm $i$'s nominal demand is $\ell_i^{(s)}=\pi\mathcal C_i\mathcal X_s$. By Lemma~\ref{lemma:hazard_sufficiency} with $t=t_0+j$ and $u_i^{(t-1)}+1=j$, the pre-reset gap that the price faces at that decision is therefore exactly
\[
s_i(t_0+j)=j\log(1+\pi)+\log\bigl(1+\pi\mathcal C_i\mathcal X_{t_0+j}\bigr)-\log\bigl(1+\pi\mathcal C_i\mathcal X_{t_0}\bigr),
\qquad j\ge1 .
\]
Under Assumption~\ref{assu:hazard} the price reset decision depends on $(i,t_0+j)$ only through this gap, so the law of the length of the price spell depends on $i$ only through $\mathcal C_i$.

First, the pre-reset gaps are ordered by Cantillon exposure. Read as a function of the exposure $C$, the gap at age $j$ has the derivative
\[
\frac{\partial s}{\partial C}
=\frac{\pi\bigl(\mathcal X_{t_0+j}-\mathcal X_{t_0}\bigr)}{\bigl(1+\pi C\mathcal X_{t_0+j}\bigr)\bigl(1+\pi C\mathcal X_{t_0}\bigr)},
\]
whose denominator is positive for all sufficiently small $\pi$ (Lemma~\ref{oa:lem:level_amplitude}). Because the propagation kernel is building at every date at which the price spell can be in progress, $\mathcal X_{t_0+j}-\mathcal X_{t_0}>0$ at every age $j\ge1$. A firm of larger Cantillon exposure therefore faces a strictly larger pre-reset gap at every age of a price spell begun at $t_0$. And since $\eta$ is nondecreasing, it resets with a weakly larger probability at every age.

The order of the price reset probabilities passes to the lengths of the price spells. The probability that the price spell begun at $t_0$ survives beyond $u$ periods is
\[
\Pr(L_i>u)=\prod_{j=1}^{u}\bigl(1-\eta(s_i(t_0+j))\bigr),
\qquad 0\le u\le\bar u,
\]
and zero above the cap. For $\mathcal C_i>\mathcal C_k$ every factor of $\Pr(L_i>u)$ is at most the corresponding factor of $\Pr(L_k>u)$, strictly below it at each age at which $\eta(s_i(t_0+j))>\eta(s_k(t_0+j))$. Hence $\Pr(L_i>u)\le\Pr(L_k>u)$ at every $u$, strictly at every $u$ at which $\Pr(L_k>u)>0$ and such an age has occurred. This is the hazard-rate order of the two lengths. It holds for every pair of price spells begun at $t_0$ and needs no stationarity.

The law of price age follows from the same factors. We hold the increments of the propagation kernel at their date-$t$ values, so that a price of age $j$ faces the pre-reset gap $s_i(j)$ of the proof of Proposition~\ref{prop:frequency} at every date. Let $\mathcal S_i(u):=\prod_{j=1}^{u}(1-\eta(s_i(j)))$ be the survival function of that proof. The stationary law of price age generated by this hazard profile is $\Pr(u_i=u)\propto\mathcal S_i(u)$, and the likelihood ratio of two such laws is
\[
\frac{\Pr(u_i=u)}{\Pr(u_k=u)}\;\propto\;\frac{\mathcal S_i(u)}{\mathcal S_k(u)}=\prod_{j=1}^{u}\frac{1-\eta(s_i(j))}{1-\eta(s_k(j))}.
\]
For $\mathcal C_i>\mathcal C_k$ and a building propagation kernel, $\mathcal X_t-\mathcal X_{t-j}>0$ at every $j\ge1$, so $s_i(j)>s_k(j)$ by the derivative above. At every age at which $\mathcal S_k(u)>0$, every factor of the ratio therefore lies in $[0,1]$, and the ratio is nonincreasing in $u$. At the remaining ages both laws put no mass. This is the monotone likelihood-ratio order in $\mathcal C_i$. Because both laws live on the common finite state space $\{0,\ldots,\bar u\}$, this order implies first-order stochastic dominance of the ages. But the realized distribution of ages at date $t$ is not this law. This is because it also carries the price reset probabilities of earlier dates, which differ across firms, and the age process is not stochastically monotone in the hazard. The ordering therefore holds for price spells begun together and for the stationary law of the date-$t$ hazard profile, and it can fail for realized ages at a particular date.

Finally, on the balanced-growth path the propagation kernel is constant, so $\mathcal X_{t_0+j}-\mathcal X_{t_0}=0$ and $\mathcal X_t-\mathcal X_{t-j}=0$, and every pre-reset gap reduces to $j\log(1+\pi)$ whatever the firm, so the survival functions built from these gaps coincide across firms. The ordering is a property of the transition.
\end{proof}

\subsection{The sticky-price decomposition and the orders of the sticky-price measures}\label{app:proof_sticky_distortion}

Under sticky prices each measure of relative-price distortion splits into four components, the flexible-price distortion, a realized-vintage term, the dispersion of price vintages, and an interaction between price ages and network displacements. Section~\ref{subsec:sticky_vs_flexible} of the main text summarizes this second-order decomposition of the nominal-demand leading term. Every equality involving posted prices or measures of relative-price distortion in this subsection holds within the nominal-demand projection. Replacing the network displacements ($\mathfrak x_i$) by the market-clearing displacements gives the same decomposition for the full price. Lemma~\ref{lem:sticky_orders} is the resulting low-inflation reading.

\begin{oalemma}[Nominal-leading decomposition of the relative-price distortion under sticky prices]
\label{lemma:sticky_distortion}
Let $T$ be a horizon, and let $u_i:=u_i^{(T)}$ be firm $i$'s realized price age at that horizon. Recall the network displacements of Section~\ref{subsec:sticky_vs_flexible},
\[
\mathfrak x_i=\log\!\left(1+\pi\mathcal C_i\mathcal X_{T-u_i}\right),
\qquad
\mathfrak x_i^{\mathrm{flex}}=\log\!\left(1+\pi\mathcal C_i\mathcal X_T\right).
\]
For weights $\varpi$, let
\[
\Var_\varpi(x):=\sum_i \varpi_i\left(x_i-\sum_j \varpi_jx_j\right)^2,
\qquad
\Cov_\varpi(x,y):=\sum_i \varpi_i
\left(x_i-\sum_j \varpi_jx_j\right)
\left(y_i-\sum_j \varpi_jy_j\right)
\]
be the weighted cross-sectional variance and covariance, and let $\varpi^{(1)}:=\mathbf m^{*}$ be the balance weights. Firm $i$'s log displacement at its posted price is $\bar g_i=-u_i\log(1+\pi)+\mathfrak x_i$, where $\bar g_i$ abbreviates $\bar g_i^{(T)}$. The realized relative price gap then satisfies, exactly within the nominal-demand projection,
\begin{equation}\label{eq:sticky_interaction}
\begin{aligned}
\bigl(\omega_T^{\mathrm{stick}}\bigr)^2=\Var_{\varpi^{(1)}}\bigl(\bar g\bigr)
&=\underbrace{\Var_{\varpi^{(1)}}(\mathfrak x^{\mathrm{flex}})}_{\text{flexible}}+\underbrace{\Var_{\varpi^{(1)}}(\mathfrak x)-\Var_{\varpi^{(1)}}(\mathfrak x^{\mathrm{flex}})}_{\text{realized vintage}}\\[4pt]
&\qquad+\underbrace{\log^2(1+\pi)\Var_{\varpi^{(1)}}(u)}_{\text{vintage dispersion}}-\underbrace{2\log(1+\pi)\Cov_{\varpi^{(1)}}(u,\mathfrak x)}_{\text{interaction}}.
\end{aligned}
\end{equation}
Relative-price entropy has the same four components under its own weights. Its normalization removes the common nominal scale before any approximation, and its nominal-leading second-order decomposition is
\begin{equation}\label{eq:sticky_entropy_interaction}
\begin{aligned}
\psi_T^{\mathrm{stick}}
&=\tfrac12\bigl\{\underbrace{\Var_{\widehat r^*}(\mathfrak x^{\mathrm{flex}})}_{\text{flexible}}+\underbrace{\Var_{\widehat r^*}(\mathfrak x)-\Var_{\widehat r^*}(\mathfrak x^{\mathrm{flex}})}_{\text{realized vintage}}\\[4pt]
&\qquad+\underbrace{\log^2(1+\pi)\Var_{\widehat r^*}(u)}_{\text{vintage dispersion}}-\underbrace{2\log(1+\pi)\Cov_{\widehat r^*}(u,\mathfrak x)}_{\text{interaction}}\bigr\},
\end{aligned}
\end{equation}
with remainder $o\!\left(\|-u\log(1+\pi)+\mathfrak x\|^2\right)$.
\end{oalemma}

\begin{proof}
We fix a realized horizon $T$ and abbreviate $u_i:=u_i^{(T)}$. The vintage formula used in Lemma~\ref{lemma:network_vintages}, together with $p_i^{*}=m_i^{*}/q_i^{*}$ and $\mathcal C_i=\kappa\delta_i/m_i^{*}$, gives the nominal-demand leading term of the realized posted price
\[
\bar p_i^{(T)}
=p_i^{(T-u_i)}
=(1+\pi)^{T-u_i}p_i^*
\left(1+\pi\mathcal C_i\mathcal X_{T-u_i}\right).
\]
With $\mathfrak x_i$ and $\mathfrak x_i^{\mathrm{flex}}$ as in the lemma, detrending by $M_T=(1+\pi)^T$ removes the common factor, so the realized log displacement of firm $i$'s posted price is exactly
\begin{equation}\label{eq:realized_relative_log_gap}
\bar g_i^{(T)}=\log\frac{\bar p_i^{(T)}/M_T}{p_i^{*}}
=-u_i\log(1+\pi)+\mathfrak x_i.
\end{equation}
The identity holds for the realized ages. This is because we take no expectation and use no firm's vintage as a reference.

For the relative price gap, Definition~\ref{def:relative_price_gap} minimizes over a common nominal rescaling. The square of the realized relative price gap is therefore the $\varpi^{(1)}$-weighted variance of the realized log displacements $\bar g_i=-u_i\log(1+\pi)+\mathfrak x_i$, and the common factor $(1+\pi)^{T}$ drops out before any approximation. Hence $(\omega_T^{\mathrm{stick}})^2=\Var_{\varpi^{(1)}}(\bar g)$ exactly, since no expansion is needed for a measure defined on log displacements. Bilinearity of the variance, with $\Var_{\varpi^{(1)}}(\mathfrak x^{\mathrm{flex}})$ added and subtracted, gives the four components of~\eqref{eq:sticky_interaction}. The realized age of any single firm therefore enters only through the cross-section, and no firm's vintage is privileged.

For relative-price entropy, the normalization removes the common scale too, since the price shares are
\[
\widehat{\bar r}_i^{(T)}
=\frac{p_i^*\exp\{\bar g_i\}}
{\sum_jp_j^*\exp\{\bar g_j\}}
=\frac{\widehat r_i^*\exp\!\left\{-u_i\log(1+\pi)+\mathfrak x_i\right\}}
{\sum_j\widehat r_j^*\exp\!\left\{-u_j\log(1+\pi)+\mathfrak x_j\right\}},
\]
so the common vintage factor is absent before any approximation. The standard second-order expansion of this exponential tilt gives
\[
\psi_T^{\mathrm{stick}}
=\frac12\Var_{\widehat r^*}\!\left(-u\log(1+\pi)+\mathfrak x\right)
+o\!\left(\|-u\log(1+\pi)+\mathfrak x\|^2\right).
\]
Variance bilinearity and the addition and subtraction of $\Var_{\widehat r^*}(\mathfrak x^{\mathrm{flex}})$ yield~\eqref{eq:sticky_entropy_interaction}. The two measures of relative-price distortion therefore differ only in their weights, stationary balances for the relative price gap and equilibrium price shares for relative-price entropy.

\end{proof}

The decomposition gives the orders in inflation of the two measures of relative-price distortion in the sticky-price regime, which Section~\ref{subsec:sticky_vs_flexible} of the main text summarizes.

\begin{oalemma}[Sticky-price distortion in the cap-nonbinding range]
\label{lem:sticky_orders}
Under the hypotheses of Proposition~\ref{prop:sticky_flexible_ranking}, and within the two-eigenmode degree representation, the steady-state sticky-price measures satisfy
\[
\omega^{\mathrm{stick}}=\Theta\bigl(\pi^{1/(\beta+1)}\bigr),
\qquad
\psi^{\mathrm{stick}}=\Theta\bigl(\pi^{2/(\beta+1)}\bigr).
\]
\end{oalemma}

\begin{proof}
Let $\beta>0$ and $\pi\in\mathcal P_{\bar u}$, and let $u_\pi^\star=\pi^{-\beta/(\beta+1)}$ be the scale of the length of price spells (Proposition~\ref{prop:frequency}). Since the decompositions of Lemma~\ref{lemma:sticky_distortion} involve second moments of realized ages, we work directly from the survival function $\mathcal S_i(u)$ of the proof of Proposition~\ref{prop:frequency}. For $u=\lfloor u_\pi^\star y\rfloor$ inside the fixed support, the local form of $\eta$ and a Riemann-sum approximation give
\[
-\log \mathcal S_i(\lfloor u_\pi^\star y\rfloor)
\asymp \varrho_i\,y^{\beta+1}.
\]
As in the proof of Proposition~\ref{prop:frequency}, the two inequalities of Definition~\ref{oa:def:cap_range} give uniform logarithmic bounds and keep a fixed multiple of the natural scale inside the support. At an arbitrary stationary date the age mass is proportional to $\mathcal S_i(u)$. Ages beyond the local scale $u_\pi^\star$ need a separate bound. For $u_\pi^\star\ge2$, the local lower bound on $\eta$ and $(u_\pi^\star)^{\beta+1}=\pi^{-\beta}$ give a constant $C_0>0$, uniform across firms, such that
\[
\eta\bigl(s_i(\lfloor u_\pi^\star\rfloor)\bigr)
\ge\frac{C_0}{u_\pi^\star}.
\]
Monotonicity of $\eta$ then controls every age beyond the local scale,
\[
\mathcal S_i(u)
\le
\exp\!\left[-C_0\,
\frac{u-\lfloor u_\pi^\star\rfloor}{u_\pi^\star}\right],
\qquad u\ge\lfloor u_\pi^\star\rfloor.
\]
Beyond the cap the inequality is immediate, because $\mathcal S_i(u)=0$ there. Together with the local Riemann-sum bound, this exponential tail implies, for every fixed $q>0$ and some $a>0$,
\[
\E\!\left[u_i^q e^{a u_i/u_\pi^\star}\right]
=O\!\left((u_\pi^\star)^q\right),
\]
uniformly over the truncated degree support and the cap-nonbinding range. We also use the exponential tail to control the upper moments of price age that the remainder of the entropy expansion involves. Consequently, for every fixed $q>0$,
\begin{equation}\label{eq:stationary_age_moments}
\E[u_i^q]=\Theta\bigl((u_\pi^\star)^q\bigr),
\qquad
\Var(u_i)=\Theta\bigl((u_\pi^\star)^2\bigr),
\end{equation}
uniformly over the truncated degree support. The scaled law of price age is non-degenerate uniformly over the truncated degree support and the cap-nonbinding range. At $\beta=0$ all age moments are $\Theta(1)$. In the Calvo special case $\eta(s)\to c$, the limiting law is truncated geometric.

The conditional independence in Assumption~\ref{assu:hazard} also guarantees that expected cross-sectional age dispersion has the order $(u_\pi^\star)^2$ of the stationary age variance. For any positive weights $\varpi_i$,
\begin{equation}\label{eq:expected_weighted_age_variance}
\E[\Var_\varpi(u)]
=\sum_i \varpi_i(1-\varpi_i)\Var(u_i)
+\Var_\varpi\bigl((\E u_i)_{i=1}^n\bigr),
\end{equation}
which is strictly positive and $\Theta\bigl((u_\pi^\star)^2\bigr)$ when at least two firms have positive weight. The same identity gives the weighted age dispersion in the vintage-dispersion term of~\eqref{eq:sticky_interaction} an expectation of order $(u_\pi^\star)^2$.

Now $\log(1+\pi)=\pi+O(\pi^2)$, while $\mathfrak x_i$ and $\mathfrak x_i^{\mathrm{flex}}$ are uniformly $O(\pi)$ under the truncated support and uniform positivity. Hence for $\beta>0$ the vintage-dispersion term in either measure of relative-price distortion is
\[
\log^2(1+\pi)\,(u_\pi^\star)^2=\Theta\!\left(\pi^{2/(\beta+1)}\right),
\]
the flexible and realized-vintage terms are $O(\pi^2)$, and the interaction is
\[
O\!\left(\pi u_\pi^\star\log(1+\pi)\right)
=O\!\left(\pi^{(\beta+2)/(\beta+1)}\right).
\]
Both of these exponents exceed $2/(\beta+1)$ when $\beta>0$. The vintage factor that carries the leading term is exact by Corollary~\ref{oa:cor:sticky_bgp_vintage}. The nominal vintage-dispersion term therefore dominates the standing network distortion. At $\beta=0$ the vintage-dispersion term returns to order $\pi^{2}$, the same order as the flexible benchmark. The identity~\eqref{eq:sticky_interaction} for the relative price gap is exact, so only the entropy expansion carries a Taylor remainder, and the exponential tail above, rather than the deterministic maximum age, controls it. Uniform boundedness of the network displacements ($\mathfrak x_i$) gives
\[
\left|\bar g_i\right|
\le \log(1+\pi)\,u_i+C\pi.
\]
The exponential-tilt remainder in~\eqref{eq:sticky_entropy_interaction} is bounded by a constant multiple of $\sum_i\widehat r_i^{*}\,|\bar g_i|^{3}\exp(2|\bar g_i|)$. After reducing $s_0$ if necessary, the exponential-moment bound gives an expected third-order remainder $O((u_\pi^\star\log(1+\pi))^3)$. Since the leading squared terms are $\Theta((u_\pi^\star\log(1+\pi))^2)$, the relative remainder is $O(u_\pi^\star\log(1+\pi))=O(s_0)$ uniformly and vanishes along cap-nonbinding low-inflation sequences. The finite age cap makes the joint age process positive recurrent, and conditional independence makes its stationary law the product of the exposure-specific laws of price age. The ergodic theorem therefore converts the time averages in Definitions~\ref{def:relative_price_gap} and~\ref{def:relative_price_entropy} into stationary expectations. For the relative price gap, division by $u_\pi^\star\log(1+\pi)$ leaves the leading random norm
\[
\left\{\sum_i \varpi_i^{(1)}
\left(\frac{u_i-\E_{\varpi^{(1)}}[u]}{u_\pi^\star}\right)^2\right\}^{1/2}.
\]
The scaled ages $u_i/u_\pi^\star$ have non-degenerate laws over the stated range, so this norm has a finite, strictly positive mean. The moment bounds in~\eqref{eq:stationary_age_moments} give uniform integrability. Together with~\eqref{eq:expected_weighted_age_variance}, the positive mean of this norm and its uniform integrability supply positive leading constants for both measures of relative-price distortion, so
\[
\omega^{\mathrm{stick}}
=\Theta\!\left(\pi^{1/(\beta+1)}\right),
\qquad
\psi^{\mathrm{stick}}
=\Theta\!\left(\pi^{2/(\beta+1)}\right),
\]
with constants uniform in $\pi$ on $\mathcal P_{\bar u}$. These orders also hold as asymptotics along any cap-nonbinding sequence.
At $\beta=0$, $u_i=O_p(1)$ and every displayed component has the flexible order, $\omega^{\mathrm{stick}}=\Theta(\pi)$ and $\psi^{\mathrm{stick}}=\Theta(\pi^2)$.

Finally, we compare relative-price distortion in the sticky-price regime and in the flexible-price regime across horizons through the realized-vintage term. On the early-transient window, $\mathcal X_{T-u_i}\le\mathcal X_T$, so $\mathfrak x_i^2\le\bigl(\mathfrak x_i^{\mathrm{flex}}\bigr)^2$ on the uniform-positivity region. Early in the transition, when the realized-vintage term dominates the vintage dispersion and the interaction, relative prices can therefore be more distorted in the flexible-price regime than in the sticky-price regime. But in the steady state $\mathcal X_{T-u_i}$ and $\mathcal X_T$ converge to the same limit, so $\mathfrak x_i-\mathfrak x_i^{\mathrm{flex}}\to0$. The attenuation of $\mathfrak x_i$ relative to $\mathfrak x_i^{\mathrm{flex}}$ then vanishes.
\end{proof}

The decomposition ranks the sticky-price regime against the flexible-price regime, in the steady state and along the transition. Section~\ref{subsec:sticky_vs_flexible} of the main text draws on this ranking.

\begin{oacorollary}[Ranking of the two regimes]
\label{oa:cor:regime_ranking}
Let the hypotheses of Proposition~\ref{prop:sticky_flexible_ranking} hold, and read both measures of relative-price distortion within the nominal-demand projection and the two-eigenmode degree representation.

(i) In the steady state the price ages $u_1,\dots,u_n$ have a common law, with variance $\Var(u)>0$, and
\begin{gather*}
\omega^{\mathrm{stick}}\;\ge\;\omega^{\mathrm{flex}},\\[2pt]
\E\bigl[(\omega_T^{\mathrm{stick}})^{2}\bigr]-(\omega^{\mathrm{flex}})^{2}
=\log^{2}(1+\pi)\,\Var(u)\Bigl(1-\sum_{i}(m_i^{*})^{2}\Bigr),\\[2pt]
\psi^{\mathrm{stick}}-\psi^{\mathrm{flex}}
=\tfrac12\log^{2}(1+\pi)\,\Var(u)\Bigl(1-\sum_{i}(\widehat r_i^{*})^{2}\Bigr),
\end{gather*}
where $\E$ is the expectation under the stationary law of price ages and the last equality holds to the second order of Lemma~\ref{lemma:sticky_distortion}.

(ii) At any horizon $T$, relative prices are more distorted in the flexible-price regime than in the sticky-price regime, $\omega_T^{\mathrm{flex}}>\omega_T^{\mathrm{stick}}$, if and only if
\begin{equation}\label{eq:reversal_condition}
\Var_{\varpi^{(1)}}(\mathfrak x^{\mathrm{flex}})-\Var_{\varpi^{(1)}}(\mathfrak x)
\;>\;
\log^{2}(1+\pi)\Var_{\varpi^{(1)}}(u)-2\log(1+\pi)\Cov_{\varpi^{(1)}}(u,\mathfrak x),
\end{equation}
and the same holds for relative-price entropy with $\widehat r^{*}$ in place of $\varpi^{(1)}$, to the second order of Lemma~\ref{lemma:sticky_distortion}. On the early-transient window of Section~\ref{app:Ct_sign} the left side of~\eqref{eq:reversal_condition} is nonnegative, and to leading order in $\pi$
\[
\Var_{\varpi^{(1)}}(\mathfrak x^{\mathrm{flex}})-\Var_{\varpi^{(1)}}(\mathfrak x)
=\pi^{2}\Bigl[\sum_{i}m_i^{*}\mathcal C_i^{2}\bigl(\mathcal X_T^{2}-\mathcal X_{T-u_i}^{2}\bigr)
+\Bigl(\sum_{i}m_i^{*}\mathcal C_i\mathcal X_{T-u_i}\Bigr)^{2}\Bigr]+O(\pi^{3}).
\]
At every horizon $T$ whose reset dates $T-\bar u,\dots,T$ all lie beyond the early-transient window, the left side of~\eqref{eq:reversal_condition} is $O(\pi^{3})$, while the vintage-dispersion term on the right is $\pi^{2}\Var_{\varpi^{(1)}}(u)\,(1+O(\pi))$.
\end{oacorollary}

\begin{proof}
(i) On the balanced-growth path the propagation kernel is constant, so $\mathfrak x_i=\mathfrak x_i^{\mathrm{flex}}$ for every firm and the realized-vintage term of Lemma~\ref{lemma:sticky_distortion} vanishes. By Lemma~\ref{lemma:hazard_sufficiency} the pre-reset gap at age $j$ is $j\log(1+\pi)$ for every firm, so the survival functions in the proof of Lemma~\ref{lemma:network_vintages} coincide across firms and the price ages have a common law. The variance of this common law is positive by the non-degeneracy of the law of price age established in the proof of Lemma~\ref{lem:sticky_orders}. Under Assumption~\ref{assu:hazard} the ages are independent across firms, and they are independent of the network displacements ($\mathfrak x^{\mathrm{flex}}$), which are fixed.

We write $\mathsf P_\varpi z:=z-(\sum_j\varpi_jz_j)\mathbf 1$ for centering under the weights $\varpi$ and $\lVert z\rVert_\varpi:=(\sum_i\varpi_iz_i^{2})^{1/2}$ for the weighted norm, so that $\Var_\varpi(z)=\lVert\mathsf P_\varpi z\rVert_\varpi^{2}$. With $\varpi=\varpi^{(1)}$, $\omega_T^{\mathrm{stick}}=\lVert\mathsf P_\varpi(\mathfrak x^{\mathrm{flex}}-\log(1+\pi)\,u)\rVert_\varpi$ and $\omega^{\mathrm{flex}}=\lVert\mathsf P_\varpi\mathfrak x^{\mathrm{flex}}\rVert_\varpi$. Since every firm has the same mean age, $\E[\mathsf P_\varpi u]=\mathsf P_\varpi(\E[u_1]\mathbf 1)=\mathbf 0$. A norm is convex, so Jensen's inequality gives
\[
\E\,\omega_T^{\mathrm{stick}}
\;\ge\;
\bigl\lVert\mathsf P_\varpi\mathfrak x^{\mathrm{flex}}-\log(1+\pi)\,\E[\mathsf P_\varpi u]\bigr\rVert_\varpi
=\omega^{\mathrm{flex}},
\]
and the ergodic theorem, as in the proof of Lemma~\ref{lem:sticky_orders}, identifies $\omega^{\mathrm{stick}}$ with the expectation on the left. For the mean square we take expectations in~\eqref{eq:sticky_interaction}. The realized-vintage term is zero. The interaction term has zero mean, since for any fixed weights
\[
\E\,\Cov_\varpi(u,\mathfrak x^{\mathrm{flex}})
=\sum_{i}\varpi_i\Bigl(\E u_i-\sum_{j}\varpi_j\E u_j\Bigr)\Bigl(\mathfrak x_i^{\mathrm{flex}}-\sum_{j}\varpi_j\mathfrak x_j^{\mathrm{flex}}\Bigr)=0
\]
when the means $\E u_i$ coincide. And~\eqref{eq:expected_weighted_age_variance} with a common law gives $\E\Var_\varpi(u)=\Var(u)\,(1-\sum_i\varpi_i^{2})$. Together these give the identity for the relative price gap. The same three steps in~\eqref{eq:sticky_entropy_interaction}, with $\varpi=\widehat r^{*}$, give the difference of the entropies, up to the remainder that the proof of Lemma~\ref{lem:sticky_orders} controls.

(ii) The equivalence is the identity~\eqref{eq:sticky_interaction} read at horizon $T$, since $(\omega_T^{\mathrm{flex}})^{2}=\Var_{\varpi^{(1)}}(\mathfrak x^{\mathrm{flex}})$. The entropy version follows from~\eqref{eq:sticky_entropy_interaction} in the same way. For the left side of~\eqref{eq:reversal_condition} we expand $\mathfrak x_i=\pi\mathcal C_i\mathcal X_{T-u_i}+O(\pi^{2})$ and $\mathfrak x_i^{\mathrm{flex}}=\pi\mathcal C_i\mathcal X_T+O(\pi^{2})$, uniformly in $i$ on the truncated degree support, with $\mathcal X_s:=0$ for $s\le0$. Lemma~\ref{oa:lem:balance_centering} gives $\sum_im_i^{*}\mathcal C_i=0$, so
\begin{gather*}
\Var_{\varpi^{(1)}}(\mathfrak x^{\mathrm{flex}})=\pi^{2}\mathcal X_T^{2}\sum_im_i^{*}\mathcal C_i^{2}+O(\pi^{3}),\\[2pt]
\Var_{\varpi^{(1)}}(\mathfrak x)=\pi^{2}\Bigl[\sum_im_i^{*}\mathcal C_i^{2}\mathcal X_{T-u_i}^{2}-\Bigl(\sum_im_i^{*}\mathcal C_i\mathcal X_{T-u_i}\Bigr)^{2}\Bigr]+O(\pi^{3}),
\end{gather*}
and subtracting gives the displayed expansion. On the early-transient window the propagation kernel is increasing (Lemma~\ref{lem:Ct_positive}), so $0\le\mathcal X_{T-u_i}\le\mathcal X_T$ for every firm and both bracketed terms are nonnegative. When every reset date $T-\bar u,\dots,T$ lies beyond the early-transient window, $\tilde\lambda^{T-u_i}\le K\pi$, and Lemma~\ref{lem:Ct_positive} gives
\[
\bigl|\mathcal X_T-\mathcal X_{T-u_i}\bigr|
\le\xi_0\sum_{h=T-u_i+1}^{T}\tilde\lambda^{h}+O(\pi)
\le\xi_0\,\frac{\tilde\lambda^{T-u_i+1}}{1-\tilde\lambda}+O(\pi)
=O(\pi),
\]
with a constant that does not depend on the cap. Hence $\mathfrak x_i^{\mathrm{flex}}-\mathfrak x_i=O(\pi^{2})$ uniformly. Keeping $\varpi=\varpi^{(1)}$, the reverse triangle inequality then gives
\[
\bigl|\Var_{\varpi}(\mathfrak x^{\mathrm{flex}})-\Var_{\varpi}(\mathfrak x)\bigr|
\le\lVert\mathsf P_{\varpi}(\mathfrak x^{\mathrm{flex}}-\mathfrak x)\rVert_{\varpi}
\bigl(\lVert\mathsf P_{\varpi}\mathfrak x^{\mathrm{flex}}\rVert_{\varpi}+\lVert\mathsf P_{\varpi}\mathfrak x\rVert_{\varpi}\bigr)
=O(\pi^{3}),
\]
while $\log^{2}(1+\pi)=\pi^{2}(1+O(\pi))$ gives the order of the vintage-dispersion term.
\end{proof}

Corollary~\ref{oa:cor:regime_ranking} is the formal content of the claim in Section~\ref{subsec:sticky_vs_flexible} of the main text that, in the steady state, staggered resets add to the standing network distortion. Part (i) says so at every rate of inflation. Under the stationary law of price age, the expected square of the relative price gap exceeds its flexible-price counterpart by the vintage-dispersion term, and the time-averaged relative price gap is at least as large as its flexible-price counterpart. Relative-price entropy has the corresponding excess to second order. At a realized horizon the ranking is the condition~\eqref{eq:reversal_condition} of part (ii), interaction term included. Beyond the early-transient window its left side, by which the dispersion of the network displacements of posted prices falls short of its flexible-price counterpart, is of order $\pi^{3}$, while the vintage-dispersion term is of order $\pi^{2}$. To leading order the rate of inflation cancels from both sides of~\eqref{eq:reversal_condition}, so the condition compares the standing network distortion accumulated so far, in units of $\pi$, with the dispersion of price ages in periods.

\subsection{Sticky-price against flexible-price distortion}\label{app:proof_sticky_flexible_ranking}

\begin{proof}[Proof of Proposition~\ref{prop:sticky_flexible_ranking}]
Both measures of relative-price distortion are read in the steady state. Theorem~\ref{theorem:price_deviation} gives the orders in the flexible-price regime, $\omega^{\mathrm{flex}}=\Theta(\pi)$ and $\psi^{\mathrm{flex}}=\Theta(\pi^{2})$, with positive nominal-demand leading coefficients. For $\pi\in\mathcal P_{\bar u}$, Lemma~\ref{lem:sticky_orders} gives the orders in the sticky-price regime, $\Theta(\pi^{1/(\beta+1)})$ and $\Theta(\pi^{2/(\beta+1)})$. Division therefore yields
\[
\frac{\omega^{\mathrm{stick}}}{\omega^{\mathrm{flex}}}
=\Theta\!\left(\pi^{-\beta/(\beta+1)}\right),
\qquad
\frac{\psi^{\mathrm{stick}}}{\psi^{\mathrm{flex}}}
=\Theta\!\left(\pi^{-2\beta/(\beta+1)}\right).
\]

It remains to identify what happens below the cap-nonbinding range. We hold $\bar u$ fixed and let $\pi\to0$ with $\beta>0$. Then $\eta(s_i(j))\to0$ for every $1\le j\le\bar u$, so every price spell converges to length $\bar u+1$ and stationary age converges to the uniform law on $\{0,\ldots,\bar u\}$. The standard deviation of this uniform law is $\Theta(\bar u)$. The vintage-dispersion term therefore gives
\[
\omega^{\mathrm{stick}}=\Theta(\pi\bar u),
\qquad
\psi^{\mathrm{stick}}=\Theta(\pi^2\bar u^2).
\]
Relative to the orders in the flexible-price regime, the two ratios level off at $\Theta(\bar u)$ and $\Theta(\bar u^2)$ rather than diverging. Finally, the growth exponent $\beta/(\beta+1)$ rises strictly from zero to one, since its derivative is $(\beta+1)^{-2}>0$. This growth exponent holds inside $\mathcal P_{\bar u}$, where the cap does not bind.
\end{proof}

\subsection{Calvo and menu-cost benchmarks}
\label{oa:proof_workhorse}

The Calvo model \citep{calvo1983staggered} and the menu-cost model \citep{sheshinski1977inflation} share a coupling between the size of price change and relative-price distortion. The coupling is exact in the drift-only benchmark, in which inflation supplies a common drift and there are no idiosyncratic shocks, so adjustment reduces to renewal accounting. The only random object is the length of the price spell, and both the size of price change and relative-price entropy are functionals of the stationary law of price age.\footnote{Richer menu-cost economies add endogenous selection. Because a firm adjusts precisely when the gap between its price and its target has grown large, the firms observed resetting are those furthest from their targets \citep{CaplinSpulber1987QJE,golosov2007menu}. Selection changes the formulas and generally introduces state variables beyond the frequency of resets. But Lemma~\ref{oa:lem:workhorse_benchmark} isolates the drift-only benchmark, in which the trigger fixes the length of the price spell and the frequency of resets remains the sole margin of adjustment.}

\begin{oalemma}[Workhorse benchmark: the size of price change proxies relative-price distortion]\label{oa:lem:workhorse_benchmark}
Let the drift-only Calvo economy have price reset probability $c\in(0,1)$, and let the drift-only menu-cost economy have completed price spells of deterministic length $\mathsf T_K\in\mathbb N$, both under a constant drift $\pi$. In both economies let the cross-sectional distribution of price ages equal the stationary law of price age, with the price reset dates of the menu-cost economy uniformly staggered across firms. At fixed reset frequencies, as $\pi\to0$,
\[
\phi^{\mathrm{C}}=\frac{\log(1+\pi)}{c},
\qquad
\sqrt{2\psi^{\mathrm{C}}}
=\pi\,\frac{\sqrt{1-c}}{c}+O(\pi^{2}),
\]
\[
\phi^{\mathrm{MC}}=\mathsf T_K\log(1+\pi),
\qquad
\sqrt{2\psi^{\mathrm{MC}}}
=\frac{\pi}{\sqrt{12}}\sqrt{\mathsf T_K^{2}-1}+O(\pi^{2}).
\]
At every fixed $\pi>0$, $\phi$ and $\psi$ are strictly decreasing in the frequency of resets and fall to $\log(1+\pi)$ and $0$ as price adjustment becomes continuous. Eliminating that frequency therefore leaves $\psi$ strictly increasing in $\phi$.
\end{oalemma}

\begin{proof}
In the drift-only benchmark the frictionless log price is $p_t^{\circ}=p_0^{\circ}+t\log(1+\pi)$ and a firm last resetting at $t-u$ posts $p_{t-u}^{\circ}$, so its log gap is
\begin{equation}\label{eq:workhorse_gap}
p_t^{\circ}-\bar p_t=u\log(1+\pi).
\end{equation}
Firms are ex ante identical, so the degree weights $\wgt{i}{\zeta}$ of Definition~\ref{def:size_price_change} sum out and the stationary cross-section is represented by the law of price age alone.

Both the size of price change ($\phi$) and relative-price entropy ($\psi$) reduce to renewal accounting. Let the price reset dates form a stationary renewal process with i.i.d.\ lengths of price spells $\mathsf T\ge1$, $\E[\mathsf T]<\infty$, and let $u$ be the stationary age. By~\eqref{eq:workhorse_gap} a price reset closes the log gap accumulated over a completed price spell, so the price change at a reset is $\mathsf T\log(1+\pi)>0$ and
\begin{equation}\label{eq:workhorse_phi}
\phi=\E[\mathsf T]\log(1+\pi).
\end{equation}
A firm of age $u$ carries the price factor $(1+\pi)^{-u}$ relative to the common frictionless benchmark. The normalized price-share density of this firm, with respect to the equilibrium cross-section, is therefore
\[
s_\pi(u)
=\frac{(1+\pi)^{-u}}{\E[(1+\pi)^{-u}]}.
\]
Definition~\ref{def:relative_price_entropy} gives
\[
\psi=\E\!\left[s_\pi(u)\log s_\pi(u)\right].
\]
If $\E[u^3]<\infty$, expansion around $\pi=0$ yields
\begin{equation}\label{eq:workhorse_entropy}
\psi
=\frac12\log^2(1+\pi)\Var(u)+O(\pi^3)
=\frac12\pi^2\Var(u)+O(\pi^3).
\end{equation}
The entropy-equivalent dispersion is therefore $\sqrt{2\psi}=\pi\sqrt{\Var(u)}+O(\pi^2)$ whenever the stationary law of price age is non-degenerate. And this is the relative-price entropy of the network economy itself, rather than the root-mean-square gap of each price from its own target.

In the Calvo economy, price reset draws are i.i.d.\ Bernoulli$(c)$, so $\Pr(\mathsf T=t)=c(1-c)^{t-1}$ for $t\ge1$ and $\E[\mathsf T]=1/c$. By~\eqref{eq:workhorse_phi}, $\phi^{\mathrm{C}}=\log(1+\pi)/c$. The stationary age is geometric from zero, $\Pr(u=t)=c(1-c)^{t}$ for $t\ge0$, with
\[
\E[u]=\frac{1-c}{c},
\qquad
\Var(u)=\frac{1-c}{c^{2}}.
\]
Equation~\eqref{eq:workhorse_entropy} therefore gives
\[
\sqrt{2\psi^{\mathrm{C}}}
=\pi\frac{\sqrt{1-c}}{c}+O(\pi^2).
\]
The exact entropy is
\[
\psi^{\mathrm{C}}
=\log\frac{\pi+c}{c(1+\pi)}
-\frac{1-c}{\pi+c}\log(1+\pi).
\]
It is finite for every $\pi>0$ and $c\in(0,1)$.

The menu-cost rule resets the price when the log gap reaches a trigger $\bar s(K)$, which increases in the menu cost $K$ and vanishes as $K\to0$. By~\eqref{eq:workhorse_gap} the log gap climbs deterministically from zero by $\log(1+\pi)$ each period, so the length of the price spell is $\mathsf T_K=\lceil\bar s(K)/\log(1+\pi)\rceil$ and the stationary age is uniform on $\{0,\dots,\mathsf T_K-1\}$. The frequency of resets is then $1/\mathsf T_K$, and~\eqref{eq:workhorse_phi} gives $\phi^{\mathrm{MC}}=\mathsf T_K\log(1+\pi)$ and
\[
\Var(u)
=\frac{\mathsf T_K^{2}-1}{12}.
\]
At fixed $\mathsf T_K$, Equation~\eqref{eq:workhorse_entropy} gives
\[
\sqrt{2\psi^{\mathrm{MC}}}
=\frac{\pi}{\sqrt{12}}
\sqrt{\mathsf T_K^{2}-1}+O(\pi^2).
\]
The corresponding exact entropy is
\[
\psi^{\mathrm{MC}}
=\log\frac{\mathsf T_K\pi}
{(1+\pi)\left[1-(1+\pi)^{-\mathsf T_K}\right]}
-\log(1+\pi)
\left(
\frac{1}{\pi}
-\frac{\mathsf T_K(1+\pi)^{-\mathsf T_K}}
{1-(1+\pi)^{-\mathsf T_K}}
\right).
\]
The expansion above is conditional on the induced length of the price spell. If $K$ is instead held fixed as $\pi\to0$, then $\mathsf T_K$ varies with $\pi$, and the exact expression, rather than a fixed-$\mathsf T_K$ remainder, governs the limit.

Finally, the size of price change and relative-price entropy move together. By~\eqref{eq:workhorse_phi}, $\phi$ is strictly increasing in the length of the completed price spell, and so is relative-price entropy. In the Calvo economy, differentiation of the exact formula with respect to the price reset probability gives
\[
-\frac{\partial\psi^{\mathrm{C}}}{\partial c}
=\frac{\pi}{c(\pi+c)}
-\frac{(1+\pi)\log(1+\pi)}{(\pi+c)^2}>0.
\]
The inequality follows from $e^z>1+z$ and $e^{-z}>1-z$ at $z=\log(1+\pi)>0$. In the menu-cost economy, differentiating the exact expression while treating $\mathsf T_K$ as continuous gives
\[
\frac{\partial\psi^{\mathrm{MC}}}{\partial\mathsf T_K}
=\frac{1}{\mathsf T_K}
-\frac{\mathsf T_K\log^2(1+\pi)(1+\pi)^{\mathsf T_K}}
{\left[(1+\pi)^{\mathsf T_K}-1\right]^2}>0,
\]
because $2\sinh\!\left(\mathsf T_K\log(1+\pi)/2\right)>\mathsf T_K\log(1+\pi)$. Relative-price entropy is therefore strictly decreasing in the frequency of resets in both economies. At the continuous-adjustment limits $c\uparrow1$ and $\mathsf T_K\downarrow1$, the laws of price age concentrate at zero, so $\psi\to0$ and $\phi\to\log(1+\pi)$. Eliminating the common frequency consequently makes $\psi$ strictly increasing in $\phi$ within each benchmark family.
\end{proof}

Within each drift-only benchmark family, relative-price distortion is therefore a strictly increasing function of the size of price change, so that the observable size of price change measures the unobservable relative-price distortion.

\section{The Degree Representation}
\label{oa:sec:perron_degree_bound}

\subsection{Perron--degree alignment and the omitted eigenmodes}

The degree representation of the main text replaces the Perron vector by normalized degree and the second eigenvectors by centered degree functions, and the two replacements are distinct approximations. Lemma~\ref{oa:lem:perron_degree_distortion} treats the first replacement without imposing the second, and Corollary~\ref{oa:cor:second_mode_remainder} bounds separately the error from discarding the eigenmodes $j\ge3$. The operator bounds of this section are exact under diagonalizability. Their degree-based corollaries also use the mean-field approximation of the main text. For a general column-stochastic matrix, degree need not equal the Perron vector. Equality $\mathbf A\mathbf d=\mathbf d$ obtains under even expenditure shares and exact firm-level agreement of supplier and customer counts, which Assumption~\ref{assu:degree_structure} maintains. But realized networks meet the firm-level agreement of supplier and customer counts only approximately. Lemma~\ref{oa:lem:perron_degree_distortion} bounds the effect of departures from the identity $\mathbf A\mathbf d=\mathbf d$. The complementary bound on the angle between degree and the leading eigenvector (Section~\ref{subsec:degree_closure}) bears on this Perron--degree step alone. Size means the stationary balance $\mathbf m^{*}$ throughout, and degree stands in for it under this representation. We parametrize the output--degree relation separately, by the output--degree exponent $\vartheta$ (Section~\ref{oa:app:output_degree}).

Throughout this section $\lVert\cdot\rVert$ denotes the Euclidean norm on vectors and the induced spectral norm on matrices. We write $\mathbf d^{\mathrm{sh}}:=\mathbf d/\mathbf 1^\top\mathbf d$ for the normalized degree vector, $\mathbf m^{*}$ for the right Perron eigenvector of the column-stochastic matrix $\mathbf A$, normalized so that $\mathbf 1^\top\mathbf m^{*}=1$, and $\boldsymbol{\Pi}_1:=\mathbf m^{*}\mathbf 1^\top$ for the Perron projector. The operator $\mathbf B:=\mathbf A-\boldsymbol{\Pi}_1$ then contains all transitory eigenmodes of the network.

\begin{oalemma}[Perron--degree alignment and the accumulated non-Perron response]
\label{oa:lem:perron_degree_distortion}
Let $\mathbf A\in\R^{n\times n}$ be column-stochastic, irreducible, aperiodic, and diagonalizable, with eigenvalues $\lambda_1=1$ and $\rho_\perp:=\max_{j\ge2}|\lambda_j|<1$, and let $\mathbf V$ be an eigenvector matrix of $\mathbf A$. For a stationary injection profile $\boldsymbol\gamma\in\R_+^n$ with $\mathbf 1^\top\boldsymbol\gamma=1$, define the accumulated non-Perron response to the injection by
\[
\mathbf h_\perp(\boldsymbol\gamma)
:=\sum_{h=1}^{\infty}\left(\frac{\mathbf A-\boldsymbol{\Pi}_1}{1+\pi}\right)^{\!h}\bigl(\boldsymbol\gamma-\mathbf m^{*}\bigr),
\qquad \pi\ge0.
\]
Then
\[
\bigl\lVert\mathbf h_\perp(\boldsymbol\gamma)\bigr\rVert
\le\operatorname{cond}(\mathbf V)\,\frac{\rho_\perp}{1+\pi-\rho_\perp}\,\bigl\lVert\boldsymbol\gamma-\mathbf m^{*}\bigr\rVert.
\]
If, in addition, the Perron vector is close to the normalized degree vector in the sense that $\lVert\mathbf m^{*}-\mathbf d^{\mathrm{sh}}\rVert\le\varepsilon_d$, then
\[
\bigl\lVert\mathbf h_\perp(\boldsymbol\gamma)\bigr\rVert
\le\operatorname{cond}(\mathbf V)\,\frac{\rho_\perp}{1+\pi-\rho_\perp}\Bigl(\bigl\lVert\boldsymbol\gamma-\mathbf d^{\mathrm{sh}}\bigr\rVert+\varepsilon_d\Bigr).
\]
\end{oalemma}

\begin{proof}
In the biorthonormal eigensystem of the diagonalizable $\mathbf A$ ($\mathbf v_1=\mathbf m^{*}$, $\mathbf u_1=\mathbf 1$, $\mathbf u_j^\top\mathbf v_k=\mathbf 1\{j=k\}$), the Perron projector is $\boldsymbol{\Pi}_1=\mathbf v_1\mathbf u_1^\top$, so
\[
\mathbf B=\mathbf A-\boldsymbol{\Pi}_1=\sum_{j=2}^{n}\lambda_j\,\mathbf v_j\mathbf u_j^\top
\]
has eigenvalues $0,\lambda_2,\ldots,\lambda_n$ and spectral radius $\rho_\perp<1$. Since $\rho(\mathbf B/(1+\pi))=\rho_\perp/(1+\pi)<1$, the series defining $\mathbf h_\perp$ converges to $\tfrac{\mathbf B}{1+\pi}\bigl(\mathbf I-\tfrac{\mathbf B}{1+\pi}\bigr)^{-1}(\boldsymbol\gamma-\mathbf m^{*})$. In the eigenbasis this operator is $\mathbf V\operatorname{diag}\bigl(0,\tfrac{\lambda_2}{1+\pi-\lambda_2},\ldots,\tfrac{\lambda_n}{1+\pi-\lambda_n}\bigr)\mathbf V^{-1}$, so its spectral norm is at most $\operatorname{cond}(\mathbf V)\max_{j\ge2}|\lambda_j/(1+\pi-\lambda_j)|$. For each $j\ge2$, $|1+\pi-\lambda_j|\ge 1+\pi-|\lambda_j|\ge 1+\pi-\rho_\perp>0$, and hence $|\lambda_j/(1+\pi-\lambda_j)|\le\rho_\perp/(1+\pi-\rho_\perp)$. This gives the first bound, and substituting the triangle inequality $\|\boldsymbol\gamma-\mathbf m^{*}\|\le\|\boldsymbol\gamma-\mathbf d^{\mathrm{sh}}\|+\varepsilon_d$ into it gives the second.
\end{proof}

Lemma~\ref{oa:lem:perron_degree_distortion} holds for second eigenvectors of any shape, degree-aligned or not. If the normalized degree vector is close to the Perron vector, degree-centering approximates exact Perron-centering. The Perron--degree error, multiplied by the persistence factor of the transitory eigenmodes, $\operatorname{cond}(\mathbf V)\,\rho_\perp/(1+\pi-\rho_\perp)$, then bounds the resulting error in the accumulated non-Perron response. Corollary~\ref{oa:cor:second_mode_remainder} separates this Perron--degree representation from the two-eigenmode approximation behind the analytical relations of Sections~\ref{sec:framework}--\ref{sec:sticky}.

\begin{oacorollary}[Two-eigenmode approximation and higher-eigenmode remainder]
\label{oa:cor:second_mode_remainder}
Under the assumptions of Lemma~\ref{oa:lem:perron_degree_distortion}, suppose in addition that $\lambda_2\in(0,1)$ is real and simple and that $\lambda_2>|\lambda_3|\ge\cdots\ge|\lambda_n|$. Let $\mathbf v_j$ and $\mathbf u_j$ denote right and left eigenvectors normalized so that $\mathbf u_j^\top\mathbf v_k=\mathbf 1\{j=k\}$. Then
\[
\mathbf h_\perp(\boldsymbol\gamma)
=\frac{\lambda_2}{1+\pi-\lambda_2}\,\mathbf v_2\mathbf u_2^\top\bigl(\boldsymbol\gamma-\mathbf m^{*}\bigr)+\mathbf R_3(\boldsymbol\gamma),
\qquad
\bigl\lVert\mathbf R_3(\boldsymbol\gamma)\bigr\rVert
\le\operatorname{cond}(\mathbf V)\,\frac{|\lambda_3|}{1+\pi-|\lambda_3|}\,\bigl\lVert\boldsymbol\gamma-\mathbf m^{*}\bigr\rVert.
\]
If, in addition, $\lVert\mathbf m^{*}-\mathbf d^{\mathrm{sh}}\rVert\le\varepsilon_d$, then
\[
\bigl\lVert\mathbf R_3(\boldsymbol\gamma)\bigr\rVert
\le\operatorname{cond}(\mathbf V)\,\frac{|\lambda_3|}{1+\pi-|\lambda_3|}\Bigl(\bigl\lVert\boldsymbol\gamma-\mathbf d^{\mathrm{sh}}\bigr\rVert+\varepsilon_d\Bigr).
\]
\end{oacorollary}

\begin{proof}
The operator $\mathbf B$ has the spectral decomposition
\[
\mathbf A-\boldsymbol{\Pi}_1
=
\sum_{j=2}^n
\lambda_j\mathbf v_j\mathbf u_j^\top.
\]
Applying the resolvent term by term gives
\[
\mathbf h_\perp(\boldsymbol\gamma)
=
\sum_{j=2}^n
\frac{\lambda_j}
     {1+\pi-\lambda_j}\,
\mathbf v_j\mathbf u_j^\top
\bigl(
\boldsymbol\gamma-\mathbf m^{*}
\bigr).
\]
Separating the $j=2$ term yields the stated decomposition, with
\[
\mathbf R_3(\boldsymbol\gamma)
:=
\sum_{j=3}^n
\frac{\lambda_j}
     {1+\pi-\lambda_j}\,
\mathbf v_j\mathbf u_j^\top
\bigl(
\boldsymbol\gamma-\mathbf m^{*}
\bigr).
\]
The same diagonalization argument used in the proof of
Lemma~\ref{oa:lem:perron_degree_distortion}, restricted to the eigenmodes $j\ge3$, gives
\[
\bigl\lVert
\mathbf R_3(\boldsymbol\gamma)
\bigr\rVert
\le
\operatorname{cond}(\mathbf V)\,
\max_{j\ge3}
\left|
\frac{\lambda_j}{1+\pi-\lambda_j}
\right|
\bigl\lVert
\boldsymbol\gamma-\mathbf m^{*}
\bigr\rVert
\le
\operatorname{cond}(\mathbf V)\,
\frac{|\lambda_3|}
     {1+\pi-|\lambda_3|}\,
\bigl\lVert
\boldsymbol\gamma-\mathbf m^{*}
\bigr\rVert.
\]
Substituting the triangle inequality
$\lVert\boldsymbol\gamma-\mathbf m^{*}\rVert\le\lVert\boldsymbol\gamma-\mathbf d^{\mathrm{sh}}\rVert+\varepsilon_d$ then gives the second bound of the corollary.
\end{proof}

The accumulated non-Perron response $\mathbf h_\perp$ is precisely what the transitory components of the monetary process of the main text accumulate over all rounds of trade. The transitory wave of the main text keeps only the slowest of these components, the second eigenmode. Under the pre-propagation timing of Section~\ref{subsec:monetary_process} and a stationary injection profile $\boldsymbol\gamma$, the normalized balances $\widehat{\mathbf m}_t=\mathbf m_t/(1+\pi)^{t}$ satisfy
\[
\widehat{\mathbf m}_{t+1}-\mathbf m^{*}
=
\frac{\mathbf B}{1+\pi}\bigl(\widehat{\mathbf m}_{t}-\mathbf m^{*}\bigr)
+
\frac{\pi}{1+\pi}\,\mathbf B\boldsymbol\gamma,
\]
whose fixed point is $\widehat{\mathbf m}_{\infty}-\mathbf m^{*}=\pi\,\mathbf h_\perp(\boldsymbol\gamma)$, since $\mathbf B\mathbf m^{*}=\mathbf 0$. The $j=2$ term of Corollary~\ref{oa:cor:second_mode_remainder} then carries the coefficient $\lambda_2(1+\pi-\lambda_2)^{-1}$, and under the degree representation of Lemma~\ref{oa:lem:eigenvector_proxies} the stationary misalignment $\xi_{\mathrm{ub}}$ summarizes the loading $\mathbf u_2^\top(\boldsymbol\gamma-\mathbf m^{*})$ of the stationary injection profile on the second eigenmode. This term therefore reproduces the limit of the propagation kernel, $\lim_{T\to\infty}\mathcal X_T=\xi_{\mathrm{ub}}\,\lambda_2(1+\pi-\lambda_2)^{-1}$, of Lemma~\ref{lemma:exposure}. Under the standing assumptions of the main text $\rho_\perp=\lambda_2$, so the persistence factor of Lemma~\ref{oa:lem:perron_degree_distortion} is the familiar $\lambda_2(1+\pi-\lambda_2)^{-1}$, inflated by the eigenbasis condition number.

Corollary~\ref{oa:cor:second_mode_remainder} bounds the omitted eigenmodes $j\ge3$ in absolute terms. But the analytical relations of Sections~\ref{sec:framework}--\ref{sec:sticky} need these eigenmodes small beside the second eigenmode. The second-eigenmode term has norm $\lambda_2(1+\pi-\lambda_2)^{-1}\,|\mathbf u_2^\top(\boldsymbol\gamma-\mathbf m^{*})|\,\lVert\mathbf v_2\rVert$, so the two-eigenmode approximation is accurate in the relative sense when the ratio
\begin{equation}
\label{oa:eq:truncation_ratio}
r
:=
\operatorname{cond}(\mathbf V)\,
\frac{|\lambda_3|\,(1+\pi-\lambda_2)}{\lambda_2\,(1+\pi-|\lambda_3|)}\cdot
\frac{\lVert\boldsymbol\gamma-\mathbf m^{*}\rVert}{|\mathbf u_2^\top(\boldsymbol\gamma-\mathbf m^{*})|\,\lVert\mathbf v_2\rVert}
\end{equation}
is small. Whether $r$ is small depends on where the bulk of the spectrum sits. On sparse random graphs the transitory eigenvalues concentrate at a modulus of order one over the square root of the mean degree. On the synthetic tail sweep of Appendix~\ref{subsec:sim_exponent}, the measured $\lambda_2$ lies close to that bulk, so the spectral separation that the two-eigenmode approximation requires is slight there.

\subsection{Degree representation of the second-eigenmode response}
\label{oa:subsec:eigenvector_proxies}

A mean-field closure gives the degree functions that stand in for the second eigenvectors in the analytical approximation of the main text. The mean-field closure sets partner-degree averages at the conditional mean that the assortativity relation of Assumption~\ref{assu:assortativity} implies, and it reads cross-sectional moments under the degree law, by the law of large numbers.

We take the degree sequence to be drawn independently from the truncated power law of Assumption~\ref{assu:degree_distribution}, conditional on graphicality and on the equality of aggregate in- and out-stubs. We assume that this conditioning leaves unchanged the first-order empirical-process limit of the finite collection of degree functionals the analysis uses. Empirical moments of continuous functions of degree then converge to their population values by the law of large numbers, at the rate $\sqrt n$ of the central limit theorem.

\begin{oalemma}[Degree representation of the second-eigenmode response]\label{oa:lem:eigenvector_proxies}
Let Assumptions~\ref{assu:primitivity}, \ref{assu:spectral_gap}, and~\ref{assu:degree_structure}--\ref{assu:assortativity} hold, and let partner-degree averages be closed at their conditional means. Then the degree dependence of the left and right factors of the second-eigenmode response is
\[
d_j^{\nu}-\frac{\E_n[d^{1+\nu}]}{\E_n[d]}
\qquad\text{and}\qquad
\delta_i=d_i^{\nu^2}-\E_n[d^{\nu^2}],
\]
the second of which is the excess degree. The displayed constants are the unique centerings of these two raw degree shapes that are biorthogonal to the Perron pair. The response to an injection profile $\boldsymbol\gamma$ is accordingly represented as
\[
\bigl(\mathbf v_2\mathbf u_2^\top\boldsymbol\gamma\bigr)_i
\;\approx\;
\kappa\,\delta_i
\sum_j
\left(
d_j^\nu-\frac{\E_n[d^{1+\nu}]}{\E_n[d]}
\right)\gamma_j,
\]
with amplitude $\kappa>0$ in this orientation. For the large-network claims, we maintain $\kappa/c_m$ bounded away from zero and infinity. Under the finite-mean degree law, this bound is equivalent to $\kappa=\Theta(n^{-1})$.
\end{oalemma}

\begin{proof}
The shapes of the two factors follow from iterating the assortativity relation. Under the even expenditure shares of Assumption~\ref{assu:degree_structure} a round of trade averages degree functions over trading partners, and under its degree symmetry the assortativity relation closes such averages on either side of a link. For a degree power $d^{k}$, we close the partner average at $\bar d^{\,k}\propto d^{\nu k}$ (Assumption~\ref{assu:assortativity}), so one mean-field round carries the exponent $k$ into $\nu k$. The substitution is exact at $k=1$ and a Jensen approximation otherwise. The curvature of the power signs the Jensen gap, as in the proof of Lemma~\ref{lemma:newest_dominates}. Applied twice to a degree tilt in balances, $k=1$, the map $k\mapsto\nu k$ therefore gives the exponents $\nu$ and $\nu^{2}$ of the reading shape and of the receiving shape. On the maintained domain $\nu\in(-1,0)$ the shape with iterated exponent $\nu^{s}$ flattens geometrically toward the constant function, so $(d^{\nu},d^{\nu^{2}})$ is the slowest nonconstant pair of shapes the mean-field map sustains. The two-eigenmode approximation retains this pair of shapes.

The centerings come from the Perron pair, which is known within the degree representation, $\mathbf v_1\propto\mathbf d$ and $\mathbf u_1=\mathbf 1$ (Assumption~\ref{assu:degree_structure}). Biorthogonality to this pair determines the centerings uniquely for the two prescribed raw shapes. The centered reading shape satisfies
\[
\sum_j
\left(
d_j^\nu-\frac{\E_n[d^{1+\nu}]}{\E_n[d]}
\right)d_j=0,
\]
while $\sum_i\delta_i=0$. These empirical centerings hold exactly on every realized degree sequence. We appeal to the law of large numbers to replace these empirical centerings by population moments, and we do so only in the large-$n$ analytical relations of the main text.

The orientation of the second-eigenmode response is the sign of $\kappa$. The rank-one projector $\mathbf v_2\mathbf u_2^\top$ is invariant to a joint sign change of its factors, so this sign is a property of the response, not of either eigenvector separately. The assortativity relation alone leaves the sign of $\kappa$ open. Lemma~\ref{lemma:newest_dominates} computes the first round of trade directly. Under a concave injection profile, nominal demand per unit of stationary size lies above its proportional path at high-degree firms and below it at low-degree firms. The response therefore loads positively on $\delta_i$ when the injection loads positively on the centered reading shape, which is the orientation $\kappa>0$.

It remains to set the magnitude of $\kappa$. For a fixed network, the preceding representation requires only $\kappa>0$. Mass conservation leaves this magnitude open. This is because $\sum_i\delta_i=0$ makes the represented second-eigenmode response zero-sum for every $\kappa$. The bound on $\kappa/c_m$ in the lemma is therefore the weakest nondegenerate intensive condition, one that makes the second-eigenmode response commensurate with the permanent trend and keeps the Cantillon exposure intensive without forcing it to vanish.
\end{proof}

Lemma~\ref{oa:lem:eigenvector_proxies} is a mean-field closure and not a spectral identity, so we retain the exact spectral decomposition of the main text and replace only its second-eigenmode response with the represented one. How faithfully the centered degree functions stand in for the exact eigenvectors depends on features that the assortativity exponent leaves out, such as communities, core--periphery patterns, and cycles. But the computational experiments of Section~\ref{sec:abm} run on the exact eigenvectors of the realized network, so the quantitative magnitudes reported there depend on neither $\kappa$ nor the degree representation.

At the boundary $\nu=0$ of the maintained domain both centered degree functions vanish identically. More precisely,
\[
d_i^\nu-\frac{\E_n[d^{1+\nu}]}{\E_n[d]}
=\nu\left(\log d_i-\frac{\E_n[d\log d]}{\E_n[d]}\right)+O(\nu^2),
\qquad
\delta_i
=\nu^2\bigl(\log d_i-\E_n[\log d]\bigr)+O(\nu^4),
\]
so the represented response $\mathbf v_2\mathbf u_2^\top\boldsymbol\gamma$ is $O(\nu^3)$ and equals zero at neutrality. But the exact response does not inherit this zero. With $\mathbf m^{*}$ and a stationary injection profile $\boldsymbol\gamma$ normalized to unit mass, the accumulated non-Perron response of Lemma~\ref{oa:lem:perron_degree_distortion} has the modal form
\[
\mathbf h_\perp(\boldsymbol\gamma)
=
\sum_{j=2}^{n}
\frac{\lambda_j}{1+\pi-\lambda_j}\,
\mathbf v_j\mathbf u_j^\top
\bigl(\boldsymbol\gamma-\mathbf m^*\bigr)
\]
(Corollary~\ref{oa:cor:second_mode_remainder}). A flat conditional mean degree pins down neither $\lambda_2$ nor $\mathbf u_2^\top(\boldsymbol\gamma-\mathbf m^*)$. When $\theta<1$, the departure $\boldsymbol\gamma-\mathbf m^*$ of the injection profile from the stationary balances is generically nonzero and can project onto the second eigenmode of the realized network. This projection alone suffices for a residual response, which eigenmodes $j\ge3$ can reinforce. An exact second-eigenmode response can therefore remain after its degree representation has collapsed. The analytical zero at $\nu=0$ is a degeneracy of the degree representation, and not a property of the exact spectrum.

\subsection{Approximation of equilibrium output with degree}
\label{oa:app:output_degree}

\begin{oaassumption}[Output--degree map]
\label{oa:ass:output_function}
In the Cobb--Douglas network economy within the degree representation (Lemma~\ref{oa:lem:eigenvector_proxies}), stationary balances satisfy $m_i^{*}\approx c_m d_i$, and firm identity enters the stationary configuration only through degree. The analytical approximation represents equilibrium output as
\[
q_i^{*} \;\approx\; d_i^{\vartheta},
\qquad \vartheta>0,
\]
where we normalize the output scale to one because it cancels in every relative-price object.
\end{oaassumption}

Perron--Frobenius gives the stationary balance vector exactly, and Lemma~\ref{oa:lem:perron_degree_distortion} bounds the effect of the separate approximation $m_i\approx c_m d_i$. Conditional on the equilibrium price vector, Cobb--Douglas production gives
\[
q_i^{*}=m_i^{\varsigma}\prod_j\Bigl(\frac{a_{ji}}{p_j}\Bigr)^{\varsigma a_{ji}}.
\]
The first factor rises with stationary balances, and the second factor is the input-price aggregate of a firm's suppliers. Within the degree representation this aggregate resolves into a power law in degree, and Lemma~\ref{oa:lem:output_degree} computes its exponent from the returns to scale and the assortativity exponent. All analytical formulas carry $\vartheta$ explicitly. The computational experiments of Section~\ref{sec:abm} instead use equilibrium output on the realized network.

\begin{oalemma}[Degree representation of equilibrium output]
\label{oa:lem:output_degree}
Let Assumptions~\ref{assu:primitivity}, \ref{assu:degree_structure}, and~\ref{assu:assortativity} hold, let $\varsigma\in(0,1)$, and let the Perron--degree representation $m_i^{*}=c_md_i$ hold. Let the mean-field map $\mathbf A^\top d^{k}\propto d^{\nu k}$ of Lemma~\ref{oa:lem:eigenvector_proxies} act through its logarithmic member $\mathbf A^\top\log\mathbf d=\nu\log\mathbf d+\mathrm{const}$.\footnote{The exponent $\nu$ is the conditional-mean assortativity exponent of Assumption~\ref{assu:assortativity}, which enters through the operator action $\mathbf A^\top\log\mathbf d$. On heavy-tailed degree sequences it differs from the Newman coefficient $\rho_{\mathrm{BS}}$ (Appendix~\ref{oa:app:nu_calibration}).} The zero-injection equilibrium output of Proposition~\ref{prop:equilibrium_existence} then satisfies $\log q_i^{*}=\vartheta_{\star}\log d_i+\mathrm{const}$, and the derived output--degree exponent is
\[
\vartheta_{\star}=\frac{-\varsigma\nu}{1-\varsigma\nu}=\frac{\varsigma|\nu|}{1+\varsigma|\nu|}\in(0,1).
\]
The exponent $\vartheta_{\star}$ increases in the returns to scale $\varsigma$ and in the strength $|\nu|$ of disassortativity, stays below one, and vanishes at degree-neutral mixing.
\end{oalemma}

\begin{proof}
Expenditure shares are even, $a_{ij}=1/d_j$, by Assumption~\ref{assu:degree_structure}.

The forcing term of the output block depends on degree alone. The technology constant is exact under even expenditure shares, $(b_A)_j=\sum_i a_{ij}\log a_{ij}=-\log d_j$. The Perron--degree representation gives $\log m_i^{*}=\log c_m+\log d_i$, so the sum $(b_A)_j+\log m_j^{*}=\log c_m$ is firm-independent. The output block of Proposition~\ref{prop:equilibrium_existence} reduces, through $\mathbf A^\top\mathbf 1=\mathbf 1$, to
\[
(\mathbf I-\varsigma\mathbf A^\top)\log\mathbf q^{*}
=\varsigma\bigl[\mathbf b_A+(\mathbf I-\mathbf A^\top)\log\mathbf m^{*}\bigr]
=-\varsigma\,\mathbf A^\top\log\mathbf d.
\]

The solution lies in the plane spanned by $\mathbf 1$ and $\log\mathbf d$. The mean-field map sends $\log\mathbf d$ to $\nu\log\mathbf d+c_0\mathbf 1$ with a firm-independent constant $c_0$, so $\mathbf A^\top$ leaves this plane invariant. The affine candidate $\log q_i^{*}=c_q+b_q\log d_i$ gives
\[
(\mathbf I-\varsigma\mathbf A^\top)\log\mathbf q^{*}
=c_q(1-\varsigma)-\varsigma b_q c_0+b_q(1-\varsigma\nu)\log\mathbf d.
\]
Matching the coefficient of $\log\mathbf d$ against the forcing $-\varsigma\nu\log\mathbf d$ gives $b_q(1-\varsigma\nu)=-\varsigma\nu$, hence $b_q=\vartheta_{\star}$. The constant $c_q$ absorbs the output-scale normalization.

Finally, on the maintained domain $\nu\in(-1,0)$ and $\varsigma\in(0,1)$ the numerator $-\varsigma\nu$ is positive and below the denominator $1-\varsigma\nu$, so $\vartheta_{\star}\in(0,1)$. The derivatives $\partial\vartheta_{\star}/\partial\varsigma$ and $\partial\vartheta_{\star}/\partial|\nu|$ are positive, and $\vartheta_{\star}=0$ at $\nu=0$.
\end{proof}

The derived exponent $\vartheta_{\star}$ is small because even expenditure shares make each input order $x_{ij}^{*}=a_{ij}m_j^{*}/p_i^{*}=c_m/p_i^{*}$ independent of the buyer's degree. Physical output therefore responds to degree only through the prices a firm's suppliers post. Revenue, by contrast, is exactly linear in degree, for the identity $p_i^{*}q_i^{*}=m_i^{*}=c_md_i$ holds whatever the value of $\vartheta$. A firm-level fit of revenue on degree therefore does not identify $\vartheta$, which governs how revenue divides between price and quantity. We read Lemma~\ref{oa:lem:output_degree} as the closure-consistent form of Assumption~\ref{oa:ass:output_function}.

Since $|\nu|<1$, the derived exponent lies below $\varsigma/(1+\varsigma)$ and hence below one half. The analytical relations of Sections~\ref{sec:framework}--\ref{sec:sticky} carry $\vartheta$ as a parameter, and the technology of the main text sets it to $\vartheta_\star$.

The signs of the comparative statics are the same at a fixed exponent $\vartheta$ and at the derived exponent $\vartheta_\star$. With $\vartheta$ held fixed, Proposition~\ref{prop:comparative_statics} gives $\partial_\nu\psi<0$, $\partial_\theta\psi<0$, and $\partial_{\lambda_2}\psi>0$. At $\vartheta_\star$ the signs in $\theta$ and $\lambda_2$ are unchanged, because $\vartheta_\star$ depends on neither. But the exponent $\vartheta_\star$ moves with $\nu$, and along $\vartheta=\vartheta_\star(\varsigma,\nu)$ the derivative of $\log\digamma_\psi$ in $x:=-\nu$ acquires the additional term $(\partial_\vartheta\log\mathcal V)\,\partial\vartheta_\star/\partial x$. The substitution $\alpha=1+p$, $\vartheta=q$, and $y=t/(1+t)$ maps the domain $\alpha>1$, $\vartheta>0$, $y=\nu^{2}\in(0,1)$ onto the open orthant $p,q,t>0$. After the substitution, and after clearing the powers of $1+t$, every coefficient of the numerator of $\partial_\vartheta\mathcal V+\mathcal V$ is negative and its denominator is negative, so $\partial_\vartheta\mathcal V+\mathcal V>0$ on the whole domain, which gives $\partial_\vartheta\log\mathcal V>-1$. Since $0<\partial\vartheta_\star/\partial x=\varsigma/(1+\varsigma x)^{2}<1$, the additional term exceeds $-1$. The proof of Proposition~\ref{prop:comparative_statics} bounds the remaining terms of the derivative below by $2/x$, which exceeds $2$ on the maintained range. The derivative in $x$ is therefore positive, and $\partial_\nu\psi<0$ holds along $\vartheta_\star$ at every $\nu\in(-1,0)$.

\subsection{Orientation of the second-eigenmode response}\label{app:newest_dominates}

The exact rank-one spectral projector $\mathbf v_2\mathbf u_2^\top$ has a fixed sign, since reversing both eigenvectors leaves the projector unchanged. What remains is its alignment with the degree coordinate once the degree representation of Lemma~\ref{oa:lem:eigenvector_proxies} replaces the finite-network projector. Within that representation the alignment follows from the assortativity relation and the shape of the injection profile.

\begin{oalemma}[Large-firm orientation of the second-eigenmode response]
\label{lemma:newest_dominates}
Let $\nu\in(-1,0)$ and $\theta\in(0,1)$, and let the leading low-inflation stationary injection profile be $\gamma_j\propto d_j^{\,\theta}$. Within the degree representation, one round of trade delivers nominal demand per unit of stationary size
\[
\frac{(\mathbf A\boldsymbol\gamma)_i}{d_i}\;\propto\;d_i^{\,-\nu(1-\theta)},\qquad -\nu(1-\theta)>0,
\]
so nominal demand lies above its proportional path at high-degree firms and below it at low-degree firms. Equivalently, the amplitude $\kappa$ in Lemma~\ref{oa:lem:eigenvector_proxies} is positive in the orientation $\delta_i=d_i^{\nu^2}-\E[d^{\nu^2}]$. The exponent vanishes at $\nu=0$, where the second-eigenmode response carries no degree orientation.
\end{oalemma}

\begin{proof}
Each firm spreads its expenditure evenly across its $d_j$ suppliers (Assumption~\ref{assu:degree_structure}), so $a_{ij}=1/d_j$ for every supplier $i$ of $j$. One round of trade therefore gives
\[
(\mathbf A\boldsymbol\gamma)_i
=\sum_{j:\,a_{ij}>0}\frac{\gamma_j}{d_j}
\;\propto\;
\sum_{j:\,a_{ij}>0}d_j^{\,\theta-1}.
\]
Since $\theta<1$ on the maintained domain, the summand is strictly decreasing in the customer's degree. A smaller customer passes on more per link. This is because its receipt under the tilted injection profile is less than proportional to its degree and it spreads that receipt over fewer suppliers. Firm $i$ has $d_i$ customers whose mean degree is $\bar d_i\propto d_i^{\nu}$ (Assumption~\ref{assu:assortativity}), so replacing the average of $d_j^{\,\theta-1}$ across those customers by its degree mean-field value $\bar d_i^{\,\theta-1}$ gives
\[
(\mathbf A\boldsymbol\gamma)_i\;\propto\;d_i\,\bar d_i^{\,\theta-1}\;\propto\;d_i^{\,1-\nu(1-\theta)}.
\]
Dividing by $d_i$ yields the display of Lemma~\ref{lemma:newest_dominates}. On $\nu\in(-1,0)$ and $\theta\in(0,1)$, $-\nu(1-\theta)>0$, and the exponent vanishes at $\nu=0$. Because $x\mapsto x^{\,\theta-1}$ is convex for $\theta\in(0,1)$, Jensen's inequality gives the direction of the pointwise mean-field gap. But the size of that gap can vary across degree classes. The displayed power is therefore a mean-field representation of the cross-sectional orientation.
\end{proof}

\subsection{Growth of the exposure dispersion in the assortativity exponent}
\label{oa:app:dispersion_growth}

The two leading coefficients of Theorem~\ref{theorem:price_deviation}, $\digamma_\omega$ and $\digamma_\psi$, each multiply the limit of the propagation kernel by a dispersion of the Cantillon exposure. The relative price gap uses stationary balances as weights and the functional $Q(\alpha,\nu)$ defined in Appendix~\ref{app:proof_thm2}. Relative-price entropy uses equilibrium price shares and the variance of the normalized excess degree under those shares,
\[
\mathcal V(\alpha,\nu;\vartheta)
:=\Var_{\widehat r^{*}}\!\left[\frac{d^{\nu^{2}}-\E[d^{\nu^{2}}]}{d}\right],
\qquad
\widehat r^{*}\propto d^{1-\vartheta}.
\]
Both functionals $Q$ and $\mathcal V$ are exact functions of $\nu^{2}$ under the Pareto degree law, and both vanish at degree neutrality. Our argument for the comparative static in $\nu$ (Section~\ref{app:proof_comparative_statics}) uses the rate at which they grow away from degree neutrality.

\begin{oalemma}[Quadratic growth of the exposure dispersion]
\label{oa:lem:dispersion_growth}
Let $y:=\nu^{2}$, so that $Q(\alpha,\nu)$ and $\mathcal V(\alpha,\nu;\vartheta)$ are functions $D(y)$ of $y$ alone, and let $\alpha>1$ and $\vartheta>0$. Each functional is then strictly positive on $y\in(0,1)$ and satisfies
\[
\varepsilon_D(y):=\frac{y\,D'(y)}{D(y)}\;\ge\;2,
\qquad y\in(0,1),
\]
equivalently $y\mapsto D(y)/y^{2}$ is nondecreasing. The bound holds with equality only in the limit $y\to0$, at which each functional is exactly of order $y^{2}$.
\end{oalemma}

\begin{proof}
We write $\mathfrak a:=\alpha-y$ and $\mathfrak b:=\alpha+1-2y$, both strictly positive on $y\in(0,1)$ because $\alpha>1$.

Consider first the functional $Q$ of the relative price gap. Substituting the Pareto moments $\E[d^{k}]=\alpha/(\alpha-k)$ into $\alpha Q=\E[(d^{y}-\E[d^{y}])^{2}d^{-1}]$ and clearing denominators gives
\[
y\,Q'(y)-2\,Q(y)
=
\frac{-2y^{3}\Bigl[\mathfrak b\bigl(3\mathfrak a^{3}-\mathfrak a^{2}\mathfrak b-\mathfrak a\mathfrak b^{2}+\mathfrak b^{3}\bigr)
+\bigl(2\mathfrak a^{3}-\mathfrak a\mathfrak b^{2}+2\mathfrak b^{3}\bigr)y
+\mathfrak b^{2}y^{2}\Bigr]}
{-\mathfrak a^{3}\mathfrak b^{2}(\mathfrak b+y)^{2}(\mathfrak b+2y)}.
\]
The denominator is strictly negative. For the numerator, we set $t:=\mathfrak b/\mathfrak a>0$. The two cubics in the bracket are then $\mathfrak a^{3}(t^{3}-t^{2}-t+3)$ and $\mathfrak a^{3}(2t^{3}-t^{2}+2)$. The first has derivative $(t-1)(3t+1)$ and therefore attains its minimum over $t>0$ at $t=1$, where it equals $2$. The second has derivative $2t(3t-1)$ and attains its minimum at $t=1/3$, where it equals $53/27$. Both cubics are strictly positive, as is the remaining term $\mathfrak b^{2}y^{2}$, so the numerator is strictly negative. A negative numerator over a negative denominator gives $yQ'(y)-2Q(y)>0$, which proves the bound for the relative price gap.

The functional $\mathcal V$ of relative-price entropy weights firms by equilibrium price shares, $\widehat r^{*}\propto d^{1-\vartheta}$. Under these weights the Pareto moments keep their form, $\E_{\widehat r^{*}}[d^{k}]=\E[d^{k+1-\vartheta}]/\E[d^{1-\vartheta}]=\mathfrak c/(\mathfrak c-k)$ with $\mathfrak c:=\alpha-1+\vartheta>0$, while the centering $\E[d^{y}]=\alpha/(\alpha-y)$ is unweighted. Writing the normalized excess degree as $d^{y-1}-\E[d^{y}]\,d^{-1}$ and expanding the variance gives
\[
\mathcal V(y)
=\frac{\mathfrak c}{\mathfrak c+2-2y}
-\frac{2\alpha}{\alpha-y}\,\frac{\mathfrak c}{\mathfrak c+2-y}
+\Bigl(\frac{\alpha}{\alpha-y}\Bigr)^{2}\frac{\mathfrak c}{\mathfrak c+2}
-\Bigl(\frac{\mathfrak c}{\mathfrak c+1-y}-\frac{\alpha}{\alpha-y}\,\frac{\mathfrak c}{\mathfrak c+1}\Bigr)^{2},
\]
a rational function of $y$ whose coefficients are polynomials in $(\alpha,\mathfrak c)$. Clearing denominators yields
\[
y\,\mathcal V'(y)-2\,\mathcal V(y)
=\frac{N(\alpha,\mathfrak c,y)}
{(\alpha-y)^{3}(\mathfrak c+1)^{2}(\mathfrak c+2)(\mathfrak c+2-y)^{2}(\mathfrak c+1-y)^{3}(\mathfrak c+2-2y)^{2}},
\]
and every factor of the denominator is strictly positive on the domain, since $y<1<\alpha$ and $\mathfrak c>0$. The sign is therefore that of the numerator $N$, a polynomial of degree nine in $y$. We substitute $\alpha=1+p$, $\mathfrak c=p+q$, and $y=t/(1+t)$, which maps the domain $\alpha>1$, $\vartheta>0$, $y\in(0,1)$ onto the open orthant $p,q,t>0$, and multiply by $(1+t)^{9}$. The substitution turns the numerator into a polynomial in $(p,q,t)$ of total degree nineteen with $377$ monomials, each with a positive integer coefficient, the smallest being $2$. A polynomial with positive coefficients is strictly positive on the open orthant, so $y\mathcal V'(y)-2\mathcal V(y)>0$ on the whole domain. The same substitution applied to $\mathcal V$ itself gives a polynomial with $112$ positive monomials, so $\mathcal V>0$ there as well. The bound holds with equality in the limit $y\to0$. Expanding the excess degree as $d^{y}-\E[d^{y}]=y(\log d-\E[\log d])+O(y^{2})$ gives $\mathcal V(y)=y^{2}\Var_{\widehat r^{*}}[(\log d-\E[\log d])/d]+O(y^{3})$, so $\varepsilon_{\mathcal V}(0^{+})=2$. This proves the bound for relative-price entropy.
\end{proof}

\noindent Both functionals $Q$ and $\mathcal V$ are of order $y^{2}$ at degree neutrality, which is where the orders $\digamma_\omega=O(|\nu|^{3})$ and $\digamma_\psi=O(\nu^{6})$ of Section~\ref{app:proof_comparative_statics} come from.

\section{Asymptotic Normality across Network Draws}
\label{oa:gaussian_concentration}

We now characterize the distribution of the average size of price change ($\phi_T$), the relative price gap ($\omega_T$), and relative-price entropy ($\psi_T$) when the production network is stochastic. Money enters deterministically, as in the main text, so the only randomness in prices comes from the draw of the network, which influences both the initial impact of new money and its percolation. When degrees are drawn from the truncated power law of Assumption~\ref{assu:degree_distribution}, the leading terms of the three statistics in $\pi$ are asymptotically normal about deterministic limits across network draws. Though the degree distribution is fat-tailed, the variation it induces in these statistics is thin-tailed. When the degree cutoff grows like $n^{1/\alpha}$, the limits hold whenever the tail exponent of the degree distribution ($\alpha$) and the weight exponent of the average size of price change ($\zeta$) satisfy $\alpha>\max\{2,2\zeta\}$. The relations between the size of price change, relative-price distortion, and inflation are therefore robust to randomness in the production network.

We prove the two lemmas with a multivariate central limit theorem for a finite vector of empirical degree moments, followed by the delta method. Throughout, we hold the date fixed and vary the draw of the network, within the two-eigenmode degree representation, and the fluctuations we find are of order $n^{-1/2}$. Lemma~\ref{oa:lem:gaussian_phiT} treats the average size of price change of Proposition~\ref{prop:size_price_change}, and Lemma~\ref{oa:lem:gaussian_omega_psi} the two measures of relative-price distortion of Theorem~\ref{theorem:price_deviation} and Proposition~\ref{prop:comparative_statics}.

We adopt three conventions. First, the degree sequence is drawn as in Section~\ref{oa:subsec:eigenvector_proxies}, independently from the truncated power law and conditional on graphicality and on the equality of aggregate in- and out-stubs. The conditioning leaves the first-order empirical-process limit of the finite collection of degree functionals unchanged. Second, $\E_n[d^{k}]$ denotes an empirical moment and $\E[d^{k}]$ its population limit, and the excess degree keeps its empirical centering, $\delta_i=d_i^{\nu^{2}}-\E_n[d^{\nu^{2}}]$. Third, the degree cutoff $d_{\max}$ is either fixed or diverges with $n$. With a fixed cutoff every degree moment is finite and the central limit theorem needs no tail restriction. With a diverging cutoff the moment condition stated in each lemma keeps the relevant second moments finite, and the untruncated Pareto moments $\E[d^{k}]=\alpha/(\alpha-k)$ are the limits of the truncated ones.

The spectral statistics $(\lambda_2,\xi_{\mathrm{ub}},\kappa/c_m)$ and the empirical degree moments come from the same network draw. We state the results under an assumption with a conditional and a joint form.

\begin{oaassumption}[Joint spectral--moment limit]\label{oa:ass:joint_spectral_clt}
One of the following holds. (i) The spectral statistics $(\lambda_2,\xi_{\mathrm{ub}},\kappa/c_m)$ are held fixed as coefficients of the degree representation, and only the degrees vary across draws. (ii) The finite vector of the empirical degree moments used in Lemmas~\ref{oa:lem:gaussian_phiT} and~\ref{oa:lem:gaussian_omega_psi}, together with $\lambda_2$ and $\kappa/c_m$, has a joint Gaussian limit at rate $\sqrt n$ about its population value, with a finite covariance matrix. The limiting spectral gap is positive. Under (ii) the stationary misalignment $\xi_{\mathrm{ub}}$ is evaluated as its moment functional~\eqref{eq:Cub_empirical}.
\end{oaassumption}

By the law of large numbers and the central limit theorem, every continuous degree functional converges, and its fluctuations are of order $n^{-1/2}$. The $\pm1$-standard-deviation bands of Section~\ref{sec:abm} of the main text measure the total variability across draws, which form (ii) describes.\footnote{The literature gives form (ii) only limited support. In the random graphs with given expected degrees of \citet{chunglu2002}, the largest eigenvalue of the adjacency matrix and the bulk of the spectrum of the normalized Laplacian concentrate around deterministic values \citep{chungluvu2003}. These results give $\lambda_2$ a limit in that model and say nothing on the rate or the shape of its fluctuations. For symmetric random matrices with independent entries, the largest eigenvalue fluctuates at rate $n^{-1/2}$ with a Gaussian law when the entries have a positive mean \citep{furedi1981eigenvalues}, and the eigenvalues that a finite-rank perturbation separates from the bulk fluctuate at the same rate with a law that depends on the perturbation \citep{capitaine2009largest}. An eigenvalue at the edge of the bulk fluctuates at rate $n^{-2/3}$ with a Tracy--Widom law, as the second eigenvalue of a random regular graph does \citep{bauerschmidt2020edge}. Assumption~\ref{assu:spectral_gap} separates $\lambda_2$ from the bulk, but no result of either kind covers the column-stochastic propagation matrix of a network with a given degree sequence, or the eigenvector quantity $\kappa/c_m$. Our computational experiments present some empirical evidence to support the assumption that the spectral statistics vary across draws together with the degree moments, and that their joint variation falls with the number of firms. More specifically, the experiments on synthetic scale-free networks of $10^5$ firms and on reconstructions of the United States network reported in Section~\ref{sec:abm} of the main paper carry bands of one standard deviation, across twenty draws of the synthetic network from the same degree law (Figures~\ref{fig:inflation_distortion} and~\ref{fig:band_distortion}) and across one hundred reconstructions at $10^4$ firms (Figure~\ref{fig:recon_us_inflation}). And since each draw runs the computational economy on its own realized network, these bands reflect the variation of $\lambda_2$ and $\kappa/c_m$ together with that of the degrees.}

\begin{oalemma}[Asymptotic normality of the average size of price change]\label{oa:lem:gaussian_phiT}
Let prices be flexible, let Assumptions~\ref{assu:primitivity}--\ref{assu:assortativity} hold, let $\theta\in(0,1)$, and let the positivity condition~\eqref{oa:eq:uniform_positivity_condition} hold. Fix a horizon $2\le T<\infty$, hold the injection misalignment at $\xi_{\mathrm{ub}}$, and let Assumption~\ref{oa:ass:joint_spectral_clt} hold. Write $\widetilde\phi_T$ for the first-order approximation in $\pi$ of the average size of price change, within the two-eigenmode degree representation. Then there are a deterministic limit $\phi_T^{\infty}$ and a variance $\sigma_{\phi,T}^{2}<\infty$ such that
\[
\sqrt n\,\bigl(\widetilde\phi_T-\phi_T^{\infty}\bigr)\;\Rightarrow\;\mathcal N\bigl(0,\sigma_{\phi,T}^{2}\bigr).
\]
With a fixed degree cutoff no tail restriction is needed. If the cutoff grows like $n^{1/\alpha}$, $\alpha>\max\{2,2\zeta\}$ suffices. At the size weighting $\zeta=1$, $\sigma_{\phi,T}^{2}=0$ and $\widetilde\phi_T=\log(1+\pi)$ on every draw.
\end{oalemma}

\begin{proof}
Under the positivity condition every flexible-price change is positive (Lemma~\ref{oa:lem:uniform_positivity}), so $\phi_T=\sum_{i\in N}\wgt{i}{\zeta}\Delta\log p_i^{(T)}$. By the display for $\Delta\log p_i^{(T)}$ in the proof of Proposition~\ref{prop:size_price_change}, the first-order approximation of the average size of price change is
\[
\widetilde\phi_T=\log(1+\pi)+\pi\,(\mathcal X_T-\mathcal X_{T-1})\,Z(\zeta),
\qquad
Z(\zeta):=\sum_{i\in N}\wgt{i}{\zeta}\,\mathcal C_i,
\]
and substituting $\wgt{i}{\zeta}=d_i^{\zeta}/\sum_jd_j^{\zeta}$ and $\mathcal C_i=(\kappa/c_m)\delta_i/d_i$ gives
\[
Z(\zeta)=\frac{\kappa}{c_m}\cdot\frac{\E_n[d^{\zeta-1+\nu^{2}}]-\E_n[d^{\nu^{2}}]\,\E_n[d^{\zeta-1}]}{\E_n[d^{\zeta}]}.
\]
With the injection misalignment held at $\xi_{\mathrm{ub}}$, the increment of the propagation kernel is $\mathcal X_T-\mathcal X_{T-1}=\xi_{\mathrm{ub}}\tilde\lambda^{T}$, so
\[
\widetilde\phi_T=\log(1+\pi)+\pi\,\xi_{\mathrm{ub}}\,\tilde\lambda^{T}\,Z(\zeta).
\]
The loading $Z(\zeta)$ is a smooth function of the vector of empirical moments
\[
\mathbf M_n:=\bigl(\E_n[d^{\zeta}],\ \E_n[d^{\zeta-1+\nu^{2}}],\ \E_n[d^{\zeta-1}],\ \E_n[d^{\nu^{2}}]\bigr)
\]
on a neighborhood of its population limit $\mathbf M$, since $\E[d^{\zeta}]>0$. The four moments are computed on the same draw, so their fluctuations are correlated and we need their joint limit. With a fixed cutoff each entry of $\mathbf M_n$ is a bounded function of degree. Under the first convention of this section the multivariate central limit theorem therefore gives $\sqrt n(\mathbf M_n-\mathbf M)\Rightarrow\mathcal N(\mathbf 0,\Sigma)$ with a finite covariance matrix $\Sigma$, and the law of large numbers gives $\mathbf M_n\to\mathbf M$ in probability. Under form (i) of Assumption~\ref{oa:ass:joint_spectral_clt}, the delta method then gives
\[
\sqrt n\,\bigl(Z(\zeta)-Z^{\infty}(\zeta)\bigr)\Rightarrow\mathcal N(0,\sigma_Z^{2}),
\qquad
\phi_T^{\infty}=\log(1+\pi)+\pi\,\xi_{\mathrm{ub}}\,\tilde\lambda^{T}Z^{\infty}(\zeta),
\qquad
\sigma_{\phi,T}^{2}=\bigl(\pi\,\xi_{\mathrm{ub}}\,\tilde\lambda^{T}\bigr)^{2}\sigma_Z^{2},
\]
where $Z^{\infty}(\zeta)$ is $Z(\zeta)$ evaluated at the population moments and $\sigma_Z^{2}=\nabla Z(\mathbf M)^\top\Sigma\,\nabla Z(\mathbf M)$. Under form (ii), we augment $\mathbf M_n$ by $\lambda_2$, by $\kappa/c_m$, and by the moments $\E_n[d^{\theta+\nu}]$, $\E_n[d^{\theta}]$, $\E_n[d^{1+\nu}]$, and $\E_n[d]$ that enter the moment functional~\eqref{eq:Cub_empirical} of $\xi_{\mathrm{ub}}$. The display for $\widetilde\phi_T$ is smooth in the augmented vector while the limiting spectral gap is positive, and the same delta-method argument gives the Gaussian limit, with $\sigma_{\phi,T}^{2}$ now carrying the variances and covariances of the augmented vector.

If the cutoff grows like $n^{1/\alpha}$, the scale of the largest of $n$ Pareto draws, the entries of $\mathbf M_n$ form a triangular array. Their variances stay bounded when $\alpha>\max\{2\zeta,2\nu^{2}\}$, and $\alpha>\max\{2,2\zeta\}$ implies this because $\nu^{2}<1$. Under the same condition the Lyapunov ratio of the array vanishes, so we invoke the Lyapunov central limit theorem in place of the fixed-cutoff argument. The truncation moves a moment $\E[d^{k}]$ by a term of order $n^{(\max\{k,0\}-\alpha)/\alpha}$, which is $o(n^{-1/2})$ whenever $\alpha>2k$. The same condition therefore lets the untruncated Pareto moments serve as the centering. At $\zeta=1$, finally, $Z(1)=(\kappa/c_m)\bigl(\E_n[d^{\nu^{2}}]-\E_n[d^{\nu^{2}}]\bigr)/\E_n[d]=0$ on every degree sequence (Lemma~\ref{oa:lem:balance_centering}), so $\widetilde\phi_T=\log(1+\pi)$ on every draw and $\sigma_{\phi,T}^{2}=0$. The average size of price change itself departs from $\log(1+\pi)$ only at second order, by $-\tfrac{\pi^{2}}{2}\bigl(\mathcal X_T^{2}-\mathcal X_{T-1}^{2}\bigr)\sum_im_i^{*}\mathcal C_i^{2}+O(\pi^{3})$.
\end{proof}

\begin{oalemma}[Asymptotic normality of the two measures of relative-price distortion]\label{oa:lem:gaussian_omega_psi}
Let prices be flexible, let Assumptions~\ref{assu:primitivity}--\ref{assu:assortativity} and the output--degree map of Assumption~\ref{oa:ass:output_function} hold, let $\theta\in(0,1)$, and let Assumption~\ref{oa:ass:joint_spectral_clt} hold. Fix a horizon $T\ge1$ or the steady state, and hold the injection misalignment at $\xi_{\mathrm{ub}}$. Write $\widetilde\omega_T$ and $\widetilde\psi_T$ for the leading terms in $\pi$ of the relative price gap and of relative-price entropy, within the two-eigenmode degree representation. Then there are deterministic limits $\omega_T^{\infty}$ and $\psi_T^{\infty}$ and variances $\sigma_{\omega,T}^{2},\sigma_{\psi,T}^{2}<\infty$ such that
\[
\sqrt n\,\bigl(\widetilde\omega_T-\omega_T^{\infty}\bigr)\;\Rightarrow\;\mathcal N\bigl(0,\sigma_{\omega,T}^{2}\bigr),
\qquad
\sqrt n\,\bigl(\widetilde\psi_T-\psi_T^{\infty}\bigr)\;\Rightarrow\;\mathcal N\bigl(0,\sigma_{\psi,T}^{2}\bigr).
\]
With a fixed degree cutoff no tail restriction is needed. If the cutoff grows like $n^{1/\alpha}$, $\alpha>2$ suffices.
\end{oalemma}

\begin{proof}
By the proof of Theorem~\ref{theorem:price_deviation} in Appendix~\ref{app:proof_thm2} of the main text, within the two-eigenmode degree representation the leading terms of the two measures of relative-price distortion at horizon $T$ are
\[
\omega_T=\pi\,\frac{\kappa}{c_m}\,\mathcal X_T\Bigl(\frac{\E_n[\delta^{2}/d]}{\E_n[d]}\Bigr)^{1/2},
\qquad
\psi_T=\frac{\pi^{2}}{2}\,\mathcal X_T^{2}\Bigl(\frac{\kappa}{c_m}\Bigr)^{2}\Bigl[\frac{\E_n[d^{1-\vartheta}\widehat\delta^{2}]}{\E_n[d^{1-\vartheta}]}-\Bigl(\frac{\E_n[d^{1-\vartheta}\widehat\delta]}{\E_n[d^{1-\vartheta}]}\Bigr)^{2}\Bigr],
\]
where $\sum_im_i^{*}\widehat\delta_i^{2}=\E_n[\delta^{2}/d]/\E_n[d]$ under the Perron--degree representation, and the equilibrium price shares are $\widehat r_i^{*}\propto d_i^{1-\vartheta}$ under Assumption~\ref{oa:ass:output_function}. In the steady state $\mathcal X_T$ is replaced by its limit $\xi_{\mathrm{ub}}\lambda_2(1+\pi-\lambda_2)^{-1}$. Expanding $\widehat\delta=d^{\nu^{2}-1}-\E_n[d^{\nu^{2}}]\,d^{-1}$, every average in the two displays is a polynomial in the entries of the vector of empirical moments
\begin{align*}
\mathbf M_n:=\bigl(&\E_n[d],\ \E_n[d^{-1}],\ \E_n[d^{\nu^{2}}],\ \E_n[d^{\nu^{2}-1}],\ \E_n[d^{2\nu^{2}-1}],\ \E_n[d^{1-\vartheta}],\\
&\E_n[d^{-\vartheta}],\ \E_n[d^{\nu^{2}-\vartheta}],\ \E_n[d^{-1-\vartheta}],\ \E_n[d^{\nu^{2}-1-\vartheta}],\ \E_n[d^{2\nu^{2}-1-\vartheta}]\bigr).
\end{align*}
The two measures are therefore smooth functions of $\mathbf M_n$ on a neighborhood of its population limit $\mathbf M$. The denominators $\E[d]$ and $\E[d^{1-\vartheta}]$ are positive. And the square root in $\omega_T$ is smooth, for $\E[\delta^{2}/d]$ is a positively weighted average of $(d^{\nu^{2}}-\E[d^{\nu^{2}}])^{2}$, which is positive when $\nu\ne0$, since $d^{\nu^{2}}$ is then not constant on the support of the degree law.

With a fixed cutoff each entry of $\mathbf M_n$ is a bounded function of degree, so the multivariate central limit theorem gives $\sqrt n(\mathbf M_n-\mathbf M)\Rightarrow\mathcal N(\mathbf 0,\Sigma)$ with a finite covariance matrix, and the law of large numbers gives $\mathbf M_n\to\mathbf M$ in probability. Under form (i) of Assumption~\ref{oa:ass:joint_spectral_clt}, the delta method gives the two Gaussian limits, with
\[
\omega_T^{\infty}=\pi\,\frac{\kappa}{c_m}\,\mathcal X_T\Bigl(\frac{\E[H_\omega(d)]}{\E[d]}\Bigr)^{1/2},
\qquad
\psi_T^{\infty}=\frac{\pi^{2}}{2}\,\mathcal X_T^{2}\Bigl(\frac{\kappa}{c_m}\Bigr)^{2}\Bigl[\E[H_\psi(d)]-\Bigl(\frac{\E[d^{-\vartheta}(d^{\nu^{2}}-\E[d^{\nu^{2}}])]}{\E[d^{1-\vartheta}]}\Bigr)^{2}\Bigr],
\]
where
\[
H_\omega(d):=\frac{\bigl(d^{\nu^{2}}-\E[d^{\nu^{2}}]\bigr)^{2}}{d},
\qquad
H_\psi(d):=\frac{d^{1-\vartheta}}{\E[d^{1-\vartheta}]}\Bigl(\frac{d^{\nu^{2}}-\E[d^{\nu^{2}}]}{d}\Bigr)^{2},
\]
and each variance is $\nabla f(\mathbf M)^\top\Sigma\,\nabla f(\mathbf M)$ for the corresponding function $f$ of the moments. Under form (ii), we augment $\mathbf M_n$ by $\lambda_2$, by $\kappa/c_m$, and by the moments $\E_n[d^{\theta+\nu}]$, $\E_n[d^{\theta}]$, and $\E_n[d^{1+\nu}]$ of the moment functional~\eqref{eq:Cub_empirical}. The propagation kernel $\mathcal X_T=\xi_{\mathrm{ub}}\sum_{h=1}^{T}\tilde\lambda^{h}$ and its limit $\xi_{\mathrm{ub}}\lambda_2(1+\pi-\lambda_2)^{-1}$ are smooth in $(\lambda_2,\xi_{\mathrm{ub}})$ while the limiting spectral gap is positive, so the delta method applies to the augmented vector and gives the Gaussian limits with the augmented covariance. The limit of the propagation kernel responds to $\lambda_2$ at the rate $(1-\lambda_2)^{-2}$, which grows as the spectral gap closes, so the joint form is a fixed-gap result.

If the cutoff grows like $n^{1/\alpha}$, the entries of $\mathbf M_n$ form a triangular array. Every degree power in $\mathbf M_n$ has exponent at most one, since $\nu^{2}<1$ and $\vartheta>0$. Hence $\alpha>2$ keeps every variance bounded, makes the Lyapunov ratio of the array vanish, and makes the truncation bias of every moment $o(n^{-1/2})$. We then apply the Lyapunov central limit theorem in place of the fixed-cutoff argument, with the untruncated Pareto moments as the centering. The exponent one belongs to the mean degree $\E_n[d]$, which enters the relative price gap alone, so for relative-price entropy by itself $\alpha>2\max\{\nu^{2},1-\vartheta\}$ suffices.
\end{proof}

\onlineappendix{C}{Calibration, Computational Results, and Notation}

\section{Mapping Model Parameters to Empirical Objects}
\label{oa:app:parameter_mapping}

\begin{table}[p]
\centering
\caption{Parameters: analytical restriction, computational baseline, and evidence}
\label{oa:tab:calibration}
\footnotesize
\setlength{\tabcolsep}{4pt}
\renewcommand{\arraystretch}{1.15}
\begin{tabularx}{\textwidth}{@{}>{\raggedright\arraybackslash}p{0.16\textwidth}>{\raggedright\arraybackslash}p{0.22\textwidth}>{\raggedright\arraybackslash}p{0.22\textwidth}>{\raggedright\arraybackslash}X@{}}
\toprule
Parameter & Analytical restriction & Computational baseline & Robustness range / evidence\\
\midrule
$\theta$ (injection exponent: $\theta=1$ injection proportional to balances, $\theta<1$ shares less concentrated than balances) & $\theta\in(0,1)$ & $0.8$ (United States), $0.5$ (synthetic sweeps, high-distortion benchmark) & Credit-flow studies place $\theta$ below one without fixing its value (Section~\ref{oa:theta_calibration}). The United States exercise is reported over $[0.70,0.86]$, and the synthetic experiments sweep $[0.5,1]$\\
$\nu$ (assortativity exponent, $\bar d_i\propto d_i^{\nu}$) & $\nu\in(-1,0)$, on which the analytical relations of the main text hold & imposed per exercise & Newman $\rho_{\mathrm{BS}}$ from $-0.21$ to $-0.04$ on observed networks, and from $-0.12$ at $10^4$ firms to $-0.008$ on the full U.S.\ reconstruction, gives the sign of $\nu$ only. The direct nearest-neighbor slope gives $\nu\approx-0.7$ (Section~\ref{oa:app:nu_calibration})\\
$\vartheta$ (output--degree exponent, $q_i^*\approx d_i^{\vartheta}$) & identified within the model as $\vartheta_\star=-\varsigma\nu/(1-\varsigma\nu)<\tfrac12$ (Lemma~\ref{oa:lem:output_degree}), and enters $\digamma_\psi$ only, not $\digamma_\omega$ & Analytical formulas only, at the derived $\vartheta_\star$. The experiments use equilibrium output on the realized network & Not identified by firm-level revenue data, which the model makes linear in degree whatever the value of $\vartheta$ (Appendix~\ref{oa:app:output_degree})\\
$\alpha$ (tail exponent) & $\alpha>1$ & $1.6$ (synthetic), implied (United States) & Production networks $\alpha\in[1,3]$ \citep{bachilieri2023topology,BMS19}\\
$\varsigma$ (returns to scale) & $\varsigma\in(0,1)$ & $0.8$ & Decreasing returns\\
$(b_0,b_w,b_\pi)$ (inaction band: baseline half-width, responsiveness weight, band compression) & -- & $(0.02,\,1,\,30)$ & Band compression $b_\pi$ swept $0$--$500$ (Section~\ref{subsec:sim_inflation})\\
$\pi$ (rate of inflation) & small-$\pi$ regime & swept $0.05\%$--$1.0\%$ & --\\
Network & primitive, column-stochastic & $n=10^5$ and mean degree $25$ (synthetic), $6.46\times10^6$ firms (United States) & \citet{bhattathiripadveetil2026reconstruct}\\
$1-\lambda_2$ (spectral gap, measured) & separation of Assumption~\ref{assu:spectral_gap} & $0.22$ ($10^4$), $0.46$ ($10^5$), $0.37$ ($10^6$), $0.34$ ($6.46\times10^6$) & Measured on the propagation matrices of the reconstructions, with a half-life of the propagation kernel between one and three periods at $\pi=1\%$\\
$\varkappa$ (projection ratio, equation~\eqref{eq:projection_ratio}) & $\varkappa\ll1$ makes Assumption~\ref{assu:small_returns} accurate within the degree representation (Section~\ref{oa:sec:projection_ratio}) & $0.9$--$1.9$ at $\varsigma=0.8$ on the reconstructions & Order one at the computational calibration, so the experiments relax the analytical regime rather than instantiate it (Section~\ref{sec:abm})\\
Horizons & -- & relax to residual $10^{-12}$ (fp32 floor at $10^6$, $6.46\times10^6$) / inject $1{,}200$ / average final $200$ & Horizon audit: the averages over the $200$ periods before dates $600$ and $1{,}200$ differ by less than $1\%$ at every size and rate (Section~\ref{subsec:sim_environment})\\
Moments of price changes (measured) & -- & Frequency of resets $0.25$ at $\pi=0.5\%$ and $0.50$ at $1\%$, price spells of $12$ and $6$ months at the quarterly reading, average size of price change $2.0\%$, index inflation equal to $\pi$ & Against price spells of $8$--$11$ months and regular-price changes of $8.5\%$ median and $11\%$ mean \citep{nakamura2008five,KlenowMalin2010} (Section~\ref{subsec:sim_calibration})\\
\bottomrule
\end{tabularx}
\end{table}

\subsection{Mapping credit dependence into \texorpdfstring{$\theta$}{theta}}
\label{oa:theta_calibration}

A single structural primitive on the injection side, the injection exponent $\theta$ of Section~\ref{subsec:monetary_process}, feeds the relative-price distortion measured in Section~\ref{subsec:sim_us}. It scales both measures of relative-price distortion, through the stationary misalignment $\xi_{\mathrm{ub}}$ (Theorem~\ref{theorem:price_deviation}). But no prior study estimates $\theta$. This is because $\theta$ is structural to this injection rule. What the empirical literature delivers instead is the size gradient of firms' access to fresh credit and liquidity. We therefore set out the mapping from that gradient to $\theta$, and we read the evidence through it.

The mapping has two steps, the first of which identifies the model object for which the data provide a proxy. The model object on the injection side is the cross-sectional profile of fresh liquidity, which $\theta$ parametrizes. An aggregate expansion at rate $\pi$ gives firm $i$ the amount $\pi(\mathbf 1^\top\mathbf m)\gamma_i\propto z_i^\theta$, where $z_i$ is firm size, the empirical proxy for the stationary balance that degree represents. The injection exponent $\theta$ is therefore the cross-sectional elasticity of fresh-liquidity intake with respect to size. The impulse-response studies that record how firms' credit or borrowing reactions vary with size are the closest empirical proxies for this first-round incidence before it propagates. But they observe the intake of the model only indirectly. The relevant outcomes are flows of new short-term or working-capital credit after an easing, proxied by short-term debt, bank borrowing, or line drawdowns. Investment responses are one step further removed from fresh-liquidity intake. We therefore read each reported gradient as indirect information about the size profile of fresh-liquidity intake.

The second step converts the size gradient into the elasticity. Since fresh-liquidity intake is proportional to $z_i^\theta$, its amount per unit of firm size is proportional to $z_i^{\theta-1}$. Hence
\[
\theta-1
=
\frac{d\log(\text{fresh-liquidity intake per unit of size})}
{d\log z_i},
\]
the elasticity of size-normalized intake with respect to size. The same elasticity appears in new borrowing per unit of assets, in the change of the loan-to-assets ratio, and in the other size-deflated flows that the regressions report. A regression that interacts the shock with $\log$ size estimates the derivative of the response of size-normalized intake with respect to $\log$ size, and not its elasticity. Its interaction coefficient gives the elasticity only after division by the average response,
\[
\theta-1
=
\frac{\text{coefficient on the shock interacted with }\log z_i}
{\text{average response of size-normalized intake to the shock}}.
\]
Where a study instead reports only a contrast between a small-firm group (size $z_s$) and a large-firm group (size $z_\ell>z_s$), let $\mathfrak r>0$ denote the ratio of their per-size intakes, small relative to large. We then read the gradient as the finite difference
\begin{equation}
\theta\;\approx\;1-\frac{\log\mathfrak r}{\log(z_\ell/z_s)}. \tag{$\star$}
\end{equation}
A flat per-size intake ($\mathfrak r=1$) returns the proportional benchmark $\theta=1$. A steeper small-firm tilt (a larger $\mathfrak r$, or the same $\mathfrak r$ sustained over a narrower size range) drives $\theta$ below one. A ratio $\mathfrak r<1$, with larger firms taking in more per unit of size, returns $\theta>1$. Under the maintained injection rule $\boldsymbol\gamma\propto m^{\theta}$ the log--log slope is constant. Every pair of size bins then returns the same $\theta$ when the bins are narrow and firm size measures the stationary balance without error. But coarse bins and a noisy proxy for size move the reading of $(\star)$ even under the maintained rule.

The investment-margin studies call for a caution of another kind. They gauge responsiveness conditional on access to finance, and not the incidence of liquidity itself. The elasticity of investment per unit of size with respect to size is $\theta-1$ plus the elasticity, with respect to size, of the investment that each unit of new credit finances. This second elasticity can take either sign, so an investment gradient bounds $\theta$ in neither direction.

Note that the network leaves the mapping intact at low inflation. This is because at a fixed positive spectral gap the leading stationary injection profile has the same degree exponent $\theta$ as the first-round rule. Network structure, the assortativity exponent $\nu$ included, enters only the $O(\pi)$ fixed-point correction of the stationary injection profile.

Table~\ref{oa:tab:theta_estimates} collects the empirical literature on the firm-size gradient of monetary transmission. Its dotted rule separates the direct credit-flow studies from those that gauge responsiveness conditional on access to finance.

The direct credit-flow studies of Panel~A place $\theta$ below one, with the one exception of the credit surge of 2020 in \citet{chodorow2022bank}. But their gradients do not map into a single value of $\theta$. Two of them measure the gradient along another dimension, bank size in \citet{kashyap2000million} and firm age and dividend status in \citet{cloyne2023monetary}. And in \citet{gertler1994monetary} the short-term debt of small firms falls while that of large firms rises, so the ratio $\mathfrak r$ is negative and $(\star)$ does not apply. The evidence therefore places $\theta$ below one without fixing its value. For the reconstructed United States exercise of Section~\ref{subsec:sim_us} we set $\theta=0.8$, a moderate tilt of new money toward small firms, and we report the relative price gap over the band $\theta\in[0.70,0.86]$ around this value (Section~\ref{subsec:sim_us}). The computational experiments sweep $\theta\in[0.5,1]$, and they retain $\theta=0.5$ as a deliberate high-distortion benchmark.

\begin{table}[H]
\centering
\caption{The literature on the firm-size gradient of monetary transmission and its reading for the injection exponent $\theta$}
\label{oa:tab:theta_estimates}
\footnotesize
\setlength{\tabcolsep}{4pt}
\renewcommand{\arraystretch}{1.15}
\begin{tabularx}{\textwidth}{@{}>{\raggedright\arraybackslash}p{0.205\textwidth}>{\raggedright\arraybackslash}p{0.235\textwidth}>{\raggedright\arraybackslash}X>{\centering\arraybackslash}p{0.085\textwidth}@{}}
\toprule
\multicolumn{4}{@{}l}{\textbf{Panel A.\ Direct measures of fresh-liquidity incidence}}\\
\midrule
Study & Data / sample & Documented size gradient in access to fresh credit / liquidity & Reading of $\theta$\\
\midrule
\citet{gertler1994monetary} & QFR manufacturing, tight-money episodes & Small firms' short-term debt falls while large firms' rises, and small firms' sales and inventories fall far more & $\theta<1$\\
\citet{kashyap2000million} & $\sim$1M bank-quarter Call Report obs.\ & Lending by small, illiquid banks, the lenders serving bank-dependent small firms, is the most policy-sensitive & $\theta<1$\\
\citet{cloyne2023monetary} & US/UK firm panels, high-frequency MP shocks & Young, non-dividend-paying firms adjust borrowing and capex far more than mature payers & $\theta<1$\\
\citet{chodorow2022bank} & US supervisory loan-level data (Y-14), $60\%$ of corporate loans & SMEs reliant on short-maturity bank credit ($\theta<1$), while the 2020 credit surge came from large-firm line drawdowns ($\theta\gtrsim1$) & mixed\\
\bottomrule
\end{tabularx}

\smallskip
\noindent\makebox[\textwidth]{\dotfill}
\smallskip

\begin{tabularx}{\textwidth}{@{}>{\raggedright\arraybackslash}p{0.205\textwidth}>{\raggedright\arraybackslash}p{0.235\textwidth}>{\raggedright\arraybackslash}X>{\centering\arraybackslash}p{0.085\textwidth}@{}}
\toprule
\multicolumn{4}{@{}l}{\textbf{Panel B.\ Responsiveness conditional on access, and analogues}}\\
\midrule
\citet{crouzet2020small} & US Census firm panel, business cycle & Top-$1\%$ firms least cyclically sensitive, but the gradient is largely non-financial (industry scope) & none\\
\citet{ottonello2020financial} & Compustat, high-frequency MP shocks & Low-default-risk firms (low debt burdens, high distance to default) are the most investment-responsive & none\\
\citet{KaplanMollViolante2018,Auclert2019} & HANK, household surveys & Policy redistributes toward high-MPC, liquidity-constrained (smaller-balance) agents & $\theta<1$\\
\midrule
\multicolumn{4}{@{}l}{\emph{This paper}: calibration inside the range $\theta<1$ of Panel~A}\\
\textbf{This paper (calibration)} & Reconstructed US network \citep{bhattathiripadveetil2026reconstruct}, with external evidence on short-term liquidity by firm size & $\theta$ read as the first-round size-elasticity of fresh short-term or bank liquidity, set at $0.8$ and reported over $[0.70,0.86]$, with $0.5$ kept as a high-distortion benchmark & $0.8$\\
\bottomrule
\end{tabularx}
\\[4pt]
\parbox{\textwidth}{\footnotesize \emph{Notes}: No listed study estimates $\theta$. The last column reports the side of the proportional benchmark $\theta=1$ on which each documented gradient places $\theta$, `mixed' where the side changes across episodes, and `none' where the gradient does not bear on $\theta$. Studies differ in the shock (monetary vs cyclical), in the outcome (credit, borrowing, investment), and in the dimension along which the gradient runs (firm size, bank size, firm age). Section~\ref{oa:theta_calibration} explains the two panels and the dotted rule that separates them.}
\end{table}

\subsection{Mapping assortativity statistics into \texorpdfstring{$\nu$}{nu}}\label{oa:app:nu_calibration}

The assortativity exponent $\nu$ of Assumption~\ref{assu:assortativity} is the slope of the mean nearest-neighbor degree $\bar d_i\propto d_i^{\nu}$, and the empirical literature documents the disassortative case $\nu<0$. The analytical approximation uses this slope on both directed sides of a link, for the degree-symmetry restriction gives both sides the same slope. Where data permit, the preferred implementation therefore checks both $\E[d_{\mathrm{supplier}}\mid d_{\mathrm{buyer}}=d]$ and $\E[d_{\mathrm{customer}}\mid d_{\mathrm{seller}}=d]$. We read this production-network primitive from the data rather than estimate it from price behavior.

The empirical literature and this paper measure assortativity with different statistics. Empirical studies overwhelmingly report Newman's degree-assortativity coefficient $\rho_{\mathrm{BS}}$ (the Pearson correlation of buyer and seller degrees across an edge, \citealp{newman2002,newman2003}), a scalar in $[-1,1]$ with $\rho_{\mathrm{BS}}<0$ disassortative. The assortativity exponent $\nu$ is a different object, for it is the exponent of the mean nearest-neighbor degree, the elasticity also known as downstream and upstream degree assortativity \citep{BDMMM22}. Newman's $\rho_{\mathrm{BS}}$ and the assortativity exponent $\nu$ share a sign but not a scale. In large scale-free networks the diverging third degree moment in the normalization of $\rho_{\mathrm{BS}}$ can drive it toward zero even when the underlying assortativity is held fixed \citep{litvak2013uncovering}. A measured $\rho_{\mathrm{BS}}$ therefore cannot be inverted into $\nu$ without a model of the full mixing matrix. Under the power-law heuristic of the paper, the leading relation is $\rho_{\mathrm{BS}}\approx C\nu$, where the positive degree-moment factor $C$ is attenuated by the edge-end variance $\sigma_e^{2}=\langle d^{3}\rangle/\langle d\rangle-(\langle d^{2}\rangle/\langle d\rangle)^{2}$. We consequently read $\rho_{\mathrm{BS}}$ for its sign only. The direct slope of the mean nearest-neighbor degree $\bar d_i$ is the preferred measure.

\begin{table}[H]
\centering
\caption{Reported degree assortativity in firm-level production networks, and what it gives for $\nu$}
\label{oa:tab:nu_estimates}
\footnotesize
\setlength{\tabcolsep}{5pt}
\renewcommand{\arraystretch}{1.2}
\begin{tabularx}{\textwidth}{@{}>{\raggedright\arraybackslash}p{0.27\textwidth}>{\raggedright\arraybackslash}p{0.30\textwidth}>{\raggedright\arraybackslash}X@{}}
\toprule
Network / source & Reported statistic & What it gives for $\nu$ ($\bar d_i\propto d_i^{\nu}$)\\
\midrule
Ecuador, administrative records \citep{bachilieri2023topology} & Newman $\rho_{\mathrm{BS}}=-0.12$ & sign only, $\nu<0$\\
Hungary, administrative records \citep{bachilieri2023topology} & Newman $\rho_{\mathrm{BS}}=-0.04$ to $-0.08$ & sign only, $\nu<0$\\
Japan \citep{FujiwaraAoyama2010} & Newman $\rho_{\mathrm{BS}}=-0.075$ & sign only, $\nu<0$\\
Japan, listed firms, as compiled by \citet{bachilieri2023topology} & Newman $\rho_{\mathrm{BS}}=-0.21$ & sign only, $\nu<0$\\
Japan, inter-firm \citep{watanabe2014mean} & $\bar d_i\propto d_i^{-0.7}$ (direct slope) & $\nu\approx-0.7$\\
United States, reconstruction \citep{bhattathiripadveetil2026reconstruct}, this paper & Newman out--in $\rho_{\mathrm{BS}}=-0.12$ at $10^{4}$ firms, $-0.024$ at $10^{6}$, $-0.008$ at the full network & sign only, $\nu<0$\\
\bottomrule
\end{tabularx}
\\[4pt]
\parbox{\textwidth}{\footnotesize \emph{Notes}: Fat-tailed networks attenuate Newman's $\rho_{\mathrm{BS}}$ \citep{litvak2013uncovering}, so the $\rho_{\mathrm{BS}}$-based rows give the sign of $\nu$ but neither its value nor a bound on it. The one direct fit of the mean nearest-neighbor degree \citep{watanabe2014mean} places $\nu$ near $-0.7$.}
\end{table}

The one direct fit of Table~\ref{oa:tab:nu_estimates}, $\nu\approx-0.7$, lies inside the analytical domain. This is because the relation between the rate of inflation and the distortion of relative prices (Theorem~\ref{theorem:price_deviation}) and the signed comparative statics of Proposition~\ref{prop:comparative_statics} hold on the whole maintained range $\nu\in(-1,0)$. What does move with $|\nu|$ is the balance between two channels, the stationary misalignment $\xi_{\mathrm{ub}}$ and the dispersion of the Cantillon exposure. The stationary misalignment peaks at $|\nu|=\nu^{\dagger}=\sqrt{(\alpha-\theta)(\alpha-1)}$, which equals $0.69$ at $\alpha=1.6$ and $\theta=0.8$. Beyond that point the comparative static in $\nu$ comes from the dispersion of the Cantillon exposure (Lemma~\ref{oa:lem:dispersion_growth}). As $|\nu|$ grows, the uniform-positivity condition of Lemma~\ref{oa:lem:uniform_positivity} tightens, admitting values of $(\kappa/c_m)\lambda_2/(1-\lambda_2)$ up to about $5{,}000$ at $|\nu|=0.1$ but only about $32$ at $|\nu|=0.7$. And the small-$|\nu|$ expansions used for interpretation, such as $d^{*}\to\exp(\E_n[\log d])$, lose their accuracy.

\section{Injection Incidence, Network Structure, and the Size of the Economy}
\label{oa:sim_determinants}

Relative-price distortion needs two things, new money that enters the economy unevenly and a production network that carries this unevenness from firm to firm. The first three experiments on the synthetic economies of Section~\ref{subsec:sim_inflation}, which the main text summarizes in a footnote, vary the incidence of new money, who trades with whom, and the tail of the degree distribution. The fourth lets the number of firms grow.

\subsection{The incidence of new money}
\label{subsec:sim_theta}

\begin{figure}[H]
\centering
\includegraphics[width=0.72\textwidth]{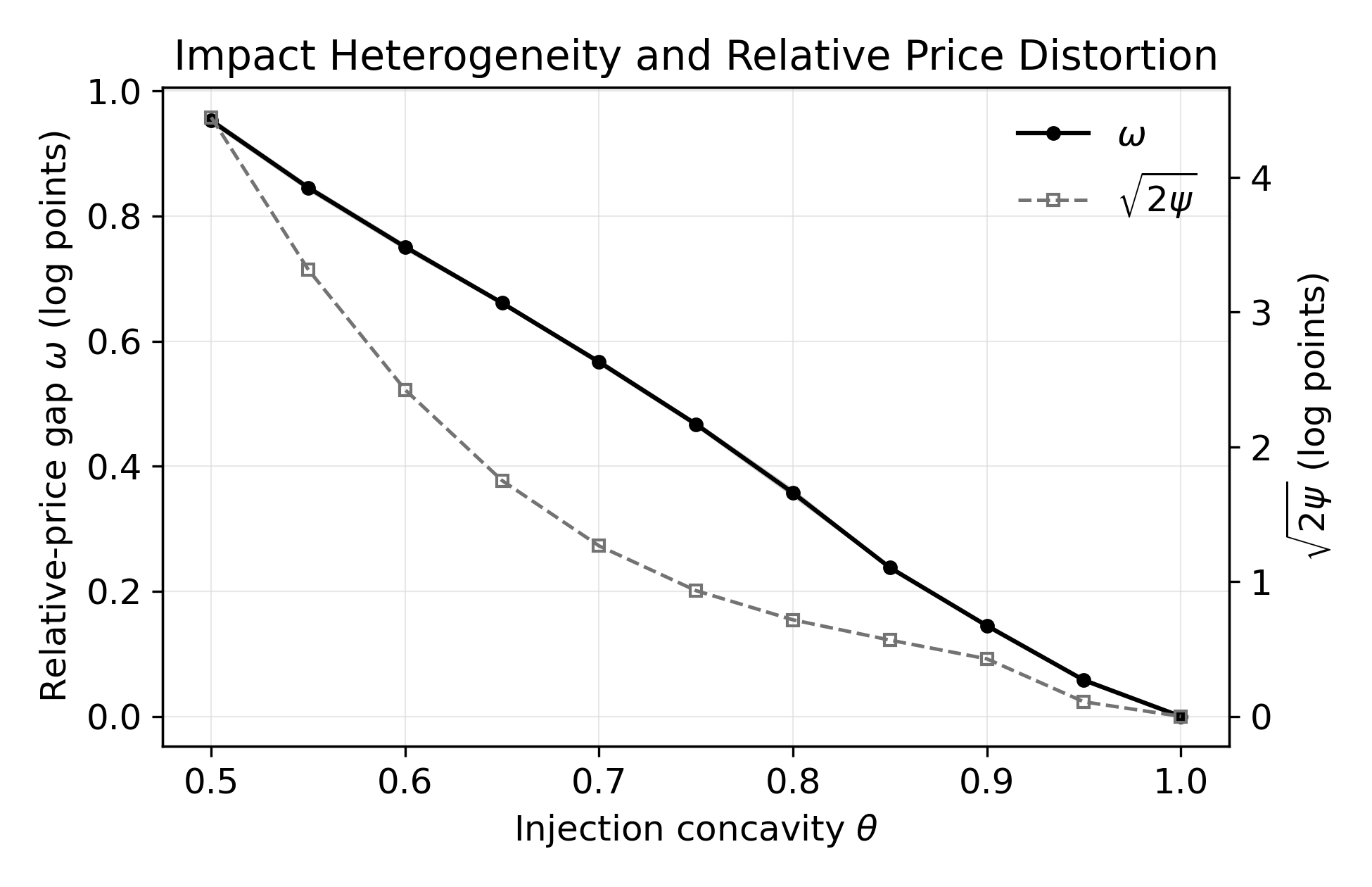}
\caption[The injection exponent and relative-price distortion]{\textbf{The injection exponent and relative-price distortion.}
Relative price gap $\omega$ (solid, left axis) and entropy-equivalent dispersion $\sqrt{2\psi}$ (dashed, right axis), in log points, against the injection exponent $\theta$, from $0.5$ to $1.0$. Curves are means over $20$ network draws, with bands at $\pm1$ standard deviation.}
\label{fig:theta_distortion}
\end{figure}

The injection exponent $\theta$ sets how unevenly new money arrives (Section~\ref{subsec:monetary_process}). At $\theta=1$ each firm receives new money in proportion to its balance, and lower values tilt new money toward small firms. Relative-price distortion therefore rises as $\theta$ falls (Proposition~\ref{prop:comparative_statics}), and it vanishes at $\theta=1$, where the stationary misalignment $\xi_{\mathrm{ub}}$ is zero. We vary $\theta$ from $0.5$ to $1.0$, holding the network fixed, inflation at $\pi=0.5\%$ per period, and the assortativity target at generation at $-0.2$, so that only the incidence of new money changes. Both measures of relative-price distortion rise steeply as the injection exponent falls (Figure~\ref{fig:theta_distortion}), and at $\theta=0.5$ the relative price gap is $2.7$ times its value at $\theta=0.8$. As $\theta$ approaches one, both measures converge to zero, and at $\theta=1$ the relative price gap is zero to the numerical floor of the computation.

\subsection{Buyer--seller assortativity}
\label{subsec:sim_assortativity}

The analysis maintains disassortative mixing (Assumption~\ref{assu:assortativity}), and so it never crosses neutral mixing into assortative networks. The computational economy does. We take one synthetic network of the family of Figure~\ref{fig:inflation_distortion} of the main text, rewire it twenty times at each target from disassortative to assortative mixing, and index each rewired network by the buyer--seller (out--in) Newman coefficient $\rho_{\mathrm{BS}}$ that the rewiring achieves.\footnote{The Newman coefficient $\rho_{\mathrm{BS}}$ shares the sign of the assortativity exponent $\nu$ but not its scale (Section~\ref{oa:app:nu_calibration}).} The rewiring falls short of its targets and saturates near $\rho_{\mathrm{BS}}=-0.045$ on one side and $+0.018$ on the other.

\begin{figure}[H]
\centering
\includegraphics[width=0.72\textwidth]{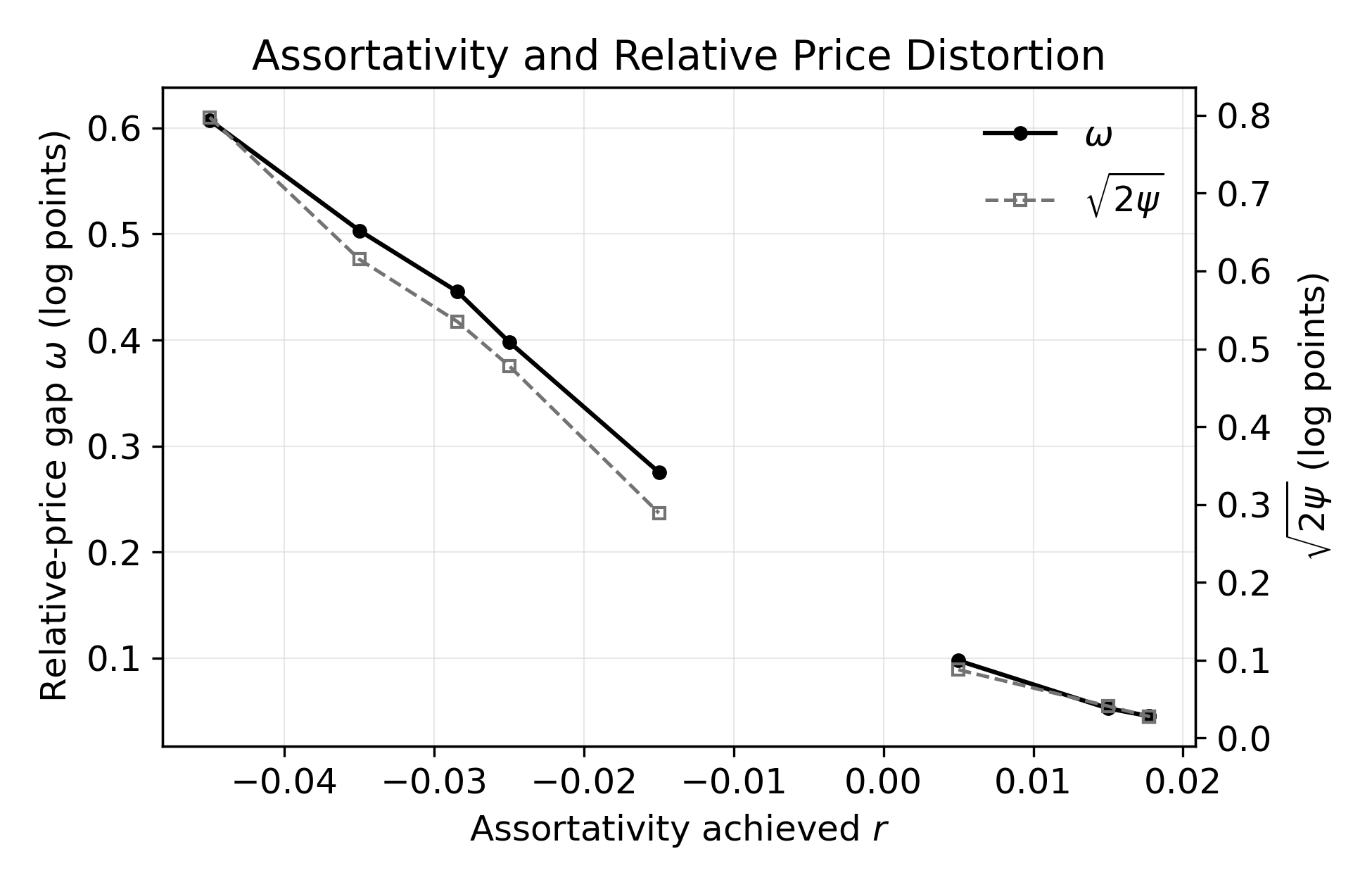}
\caption[Assortativity and relative-price distortion]{\textbf{Assortativity and relative-price distortion.}
Relative price gap $\omega$ (solid, left axis) and entropy-equivalent dispersion $\sqrt{2\psi}$ (dashed, right axis), in log points, against the buyer--seller (out--in) Newman coefficient $\rho_{\mathrm{BS}}$ that degree-preserving rewiring achieves, from $-0.045$ to $+0.018$. The rewiring saturates at the plotted endpoints. The networks are twenty rewirings, at each target, of one synthetic network of the main text, with inflation fixed at $\pi=0.5\%$ per period. Each point is the mean over the twenty rewirings, with bands at $\pm1$ standard deviation.}
\label{fig:assort_distortion}
\end{figure}

Both measures of relative-price distortion fall steadily and steeply as the network is carried from disassortative to assortative mixing (Figure~\ref{fig:assort_distortion}). The relative price gap drops roughly thirteen-fold, from $0.61$ log points at $\rho_{\mathrm{BS}}=-0.045$ to $0.045$ at $\rho_{\mathrm{BS}}=+0.018$, and the entropy-equivalent dispersion $\sqrt{2\psi}$ falls from $0.80$ to $0.03$ log points. No other feature of the network that we vary moves relative-price distortion by as large a factor. On the disassortative side, where the analysis applies, relative-price distortion rises with the strength of disassortative mixing, as Proposition~\ref{prop:comparative_statics} predicts, and it responds most steeply there. But the decline does not stop at neutral mixing. It continues smoothly across $\rho_{\mathrm{BS}}=0$, and both measures remain strictly positive on either side of it.

The rewiring preserves every firm's numbers of suppliers and of customers, so the degree distribution of the economy, however skewed, is the same at every point of the sweep. What changes is who trades with whom (Section~\ref{subsec:comparative_statics}), and with it the differences in size among the firms through which a unit of new money travels. A firm's relative price moves with its nominal demand relative to the nominal demand of the firms around it, its suppliers, its customers, and its customers' other suppliers. And since the injection rule discriminates only by size, new money moves such relative prices only insofar as it treats neighboring firms differently.

And this is precisely why assortative mixing disarms the injection. Like trades with like, so the suppliers, customers, and rivals of each firm come from its own size class, and the injection rule hands all of them nearly the same share of new money per unit of size. Hume's experiment then plays out in miniature within each size class, with all prices rising together and no relative price moving. Disassortative mixing does the reverse. It seats small firms beside large ones, so the outsized receipts of the small firms are delivered one link out to the large firms. The difference in size between trading partners then renews the Cantillon tilt at every round of trade.\footnote{The analytical relative-price distortion vanishes at $\nu=0$, whereas the computational economy measures positive relative-price distortion on either side of a zero Newman coefficient. The two are consistent. This is because the analytical zero is a degeneracy of the degree representation (Appendix~\ref{oa:subsec:eigenvector_proxies}), and because in a heavy-tailed network $\rho_{\mathrm{BS}}=0$ need not mean $\nu=0$ (Section~\ref{oa:app:nu_calibration}).}

The reconstructed United States network of Section~\ref{subsec:sim_us} lies inside the range of these rewired networks, near their neutral end, with a Newman coefficient of $-0.008$ on the full network. But the one direct slope of the mean nearest-neighbor degree on an observed network is much steeper than Newman's coefficient suggests (Section~\ref{oa:app:nu_calibration}), and no identified mapping links the two. We therefore report the relative-price distortion measured on the reconstruction without rescaling it by the rewired networks.

\subsection{The tail of the degree distribution}
\label{subsec:sim_exponent}

The tail of the degree distribution is the one determinant whose effect on relative-price distortion the analysis leaves open (Section~\ref{subsec:comparative_statics}). There are two ways in which the tail can affect relative-price distortion. It shapes the dispersion of the Cantillon exposure, and it can also move the spectral gap, which sets how fast money spreads through the network. The computational economy measures both effects of the tail together, for nothing in it holds the spectral gap fixed. We vary the tail exponent $\alpha$ from $1.1$ to $1.9$, and we hold inflation at $\pi=0.5\%$ per period, the injection exponent at $\theta=0.5$, and the assortativity target at generation at $-0.2$. A lower tail exponent means a heavier tail, in which a few large firms account for a larger share of all buyer--seller links.

\begin{figure}[H]
\centering
\includegraphics[width=0.72\textwidth]{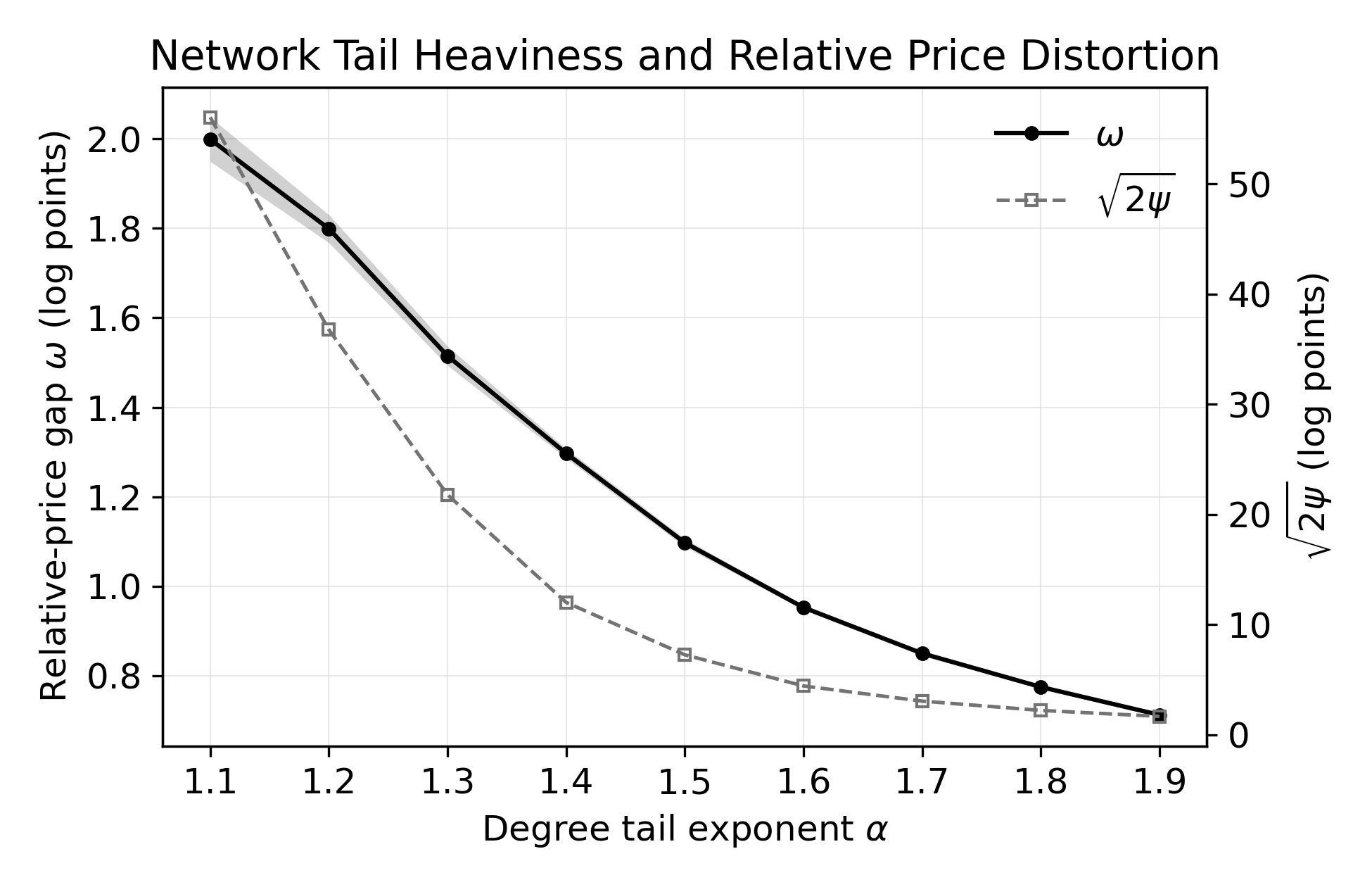}
\caption[The tail exponent and relative-price distortion]{\textbf{The tail exponent and relative-price distortion.}
Relative price gap $\omega$ (solid, left axis) and entropy-equivalent dispersion $\sqrt{2\psi}$ (dashed, right axis), in log points, against the tail exponent $\alpha$ of the out-degree distribution, $\Pr(D>d)\propto d^{-\alpha}$, over $\alpha\in[1.1,1.9]$. A smaller exponent denotes a heavier tail and more dominant hubs. Network density is held fixed across the grid, so the sweep moves the tail alone. Inflation is fixed at $\pi=0.5\%$ per period. Curves are means over $20$ network draws, with bands at $\pm1$ standard deviation.}
\label{fig:exponent_distortion}
\end{figure}

Heavier tails distort relative prices substantially more (Figure~\ref{fig:exponent_distortion}). As the tail exponent falls from $1.9$ to $1.1$, the relative price gap nearly triples, from $0.71$ to $2.00$ log points, and relative-price entropy rises more steeply still. On this generator we can also measure the speed at which money spreads through the network. A heavier tail concentrates the network on a few large firms and narrows its spectral gap, from $1-\lambda_2=0.75$ at $\alpha=1.9$ to $0.52$ at $\alpha=1.1$, steadily across the sweep and with the density of links held fixed.\footnote{The direction in which a heavier tail moves the spectral gap depends on the family of graphs. On degree-symmetric configuration models a heavier tail widens the spectral gap instead, for there the large firms serve as shortcuts along which money spreads. On our directed generator, with independent out-degrees, the spectral gap narrows as the tail becomes heavier. How the spectral gap of a scale-free network depends on its tail exponent is a classical question of random-graph theory \citep{chungluvu2003}.} Monetary disturbances therefore dissipate more slowly as the tail becomes heavier. But the sweep moves the dispersion of the Cantillon exposure at the same time, so it does not apportion the rise in relative-price distortion between the two channels. What the sweep does settle, on this generator, is the sign of the effect of the tail on relative-price distortion.

\subsection{The size of the economy}
\label{subsec:sim_size}

A large enough economy would be immune to relative-price distortion if relative-price distortion were an average of independent disturbances to firms, as aggregate volatility is in the granular theory of fluctuations. Averaging across firms would then drive it toward zero, at the rate $n^{-1/2}$, or at the rate $1/\ln n$ when firm sizes follow Zipf's law \citep{gabaix2011}. But the relative price gap is a different kind of object, for it is nothing but a dispersion of log relative prices around their equilibrium values. Independent disturbances to firms would leave this dispersion at its population value, however many firms the economy contained. The number of firms can therefore change relative-price distortion only by changing the population value itself, through the persistence of the network or the concentration of its largest firms (Section~\ref{subsec:comparative_statics}).

We compare two families of networks, and in both the mean degree and the environment of each firm stay fixed while the number of firms rises from $10^2$ to $10^5$. Erd\H{o}s--R\'enyi graphs, with Poisson degrees and zero assortativity, have no dominant hubs, so their distribution of degrees, and with it the distribution of Cantillon exposures, is the same at every size. In a power-law family with a fixed tail exponent, by contrast, the largest firm grows with the number of firms, at the extreme-value rate $n^{1/\alpha}$. The heterogeneity of firms therefore rises with size, although the law that generates the network does not change.

\begin{figure}[H]
\centering
\includegraphics[width=0.72\textwidth]{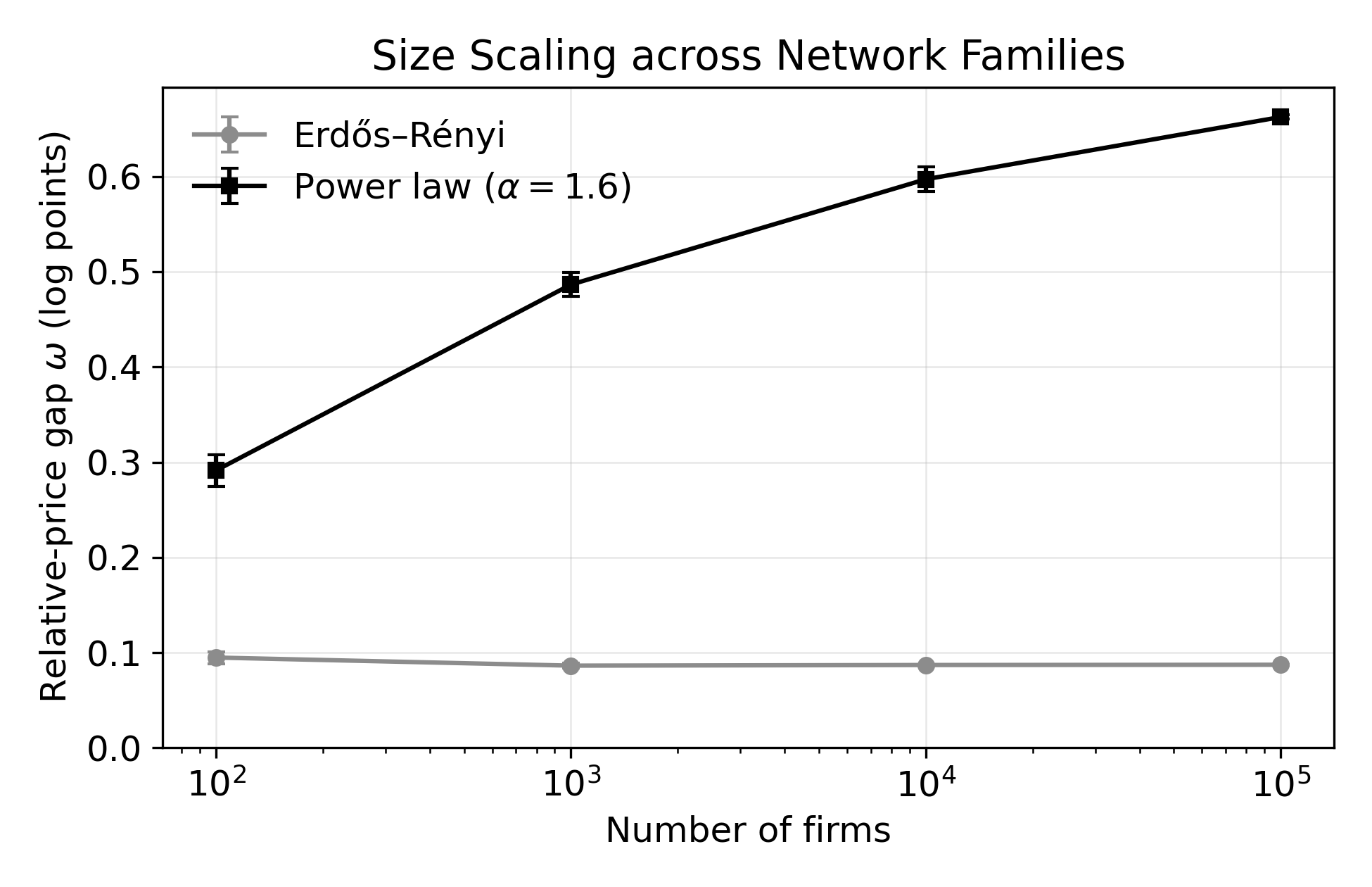}
\caption[Relative-price distortion and the size of the economy]{\textbf{Relative-price distortion and the size of the economy.}
Relative price gap $\omega$, in log points, against the number of firms, $n\in\{10^2,\,10^3,\,10^4,\,10^5\}$, for an Erd\H{o}s--R\'enyi family with Poisson degrees, mean degree $25$, and zero assortativity, and for a scale-free power-law family with tail exponent $\alpha=1.6$ and zero assortativity. Inflation is fixed at $\pi=1\%$ per period. Each point is the mean over $10$ fresh network draws, with error bars at $\pm1$ standard deviation.}
\label{fig:distortion_scaling}
\end{figure}

The number of firms as such leaves relative-price distortion unchanged (Figure~\ref{fig:distortion_scaling}). On the Erd\H{o}s--R\'enyi family the relative price gap is flat, at $0.095$, $0.086$, $0.087$, and $0.087$ log points from $10^2$ to $10^5$ firms, and only the bands across network draws narrow as the number of firms grows. But on the power-law family, whose largest firms grow with the economy, the relative price gap rises from $0.29$ to $0.66$ log points. Relative-price distortion therefore does not average away as the economy grows, and it grows with the economy exactly where growth deepens the heterogeneity of firms.

\section{The Magnitude of Relative-Price Distortion}
\label{oa:sim_magnitude}

Whether the relative-price distortion of the production network economy is large or small depends on the yardstick against which it is set. The natural yardsticks are the workhorse models of price setting, for in them the relative-price distortion that inflation causes is the dispersion of relative prices that staggered price setting creates \citep{woodford2003,yun2005optimal}. There are two such models, the first of which is the Calvo model, in which firms reset their prices at random dates \citep{calvo1983staggered}. And the second is the menu-cost model, in which a firm resets its price once inflation has carried it far enough from its optimal price \citep{sheshinski1977inflation}. Both models measure this dispersion by a standard deviation of log relative prices, the scale on which we read our two measures of relative-price distortion. Table~\ref{tab:distortion_yardsticks} sets them, for the reconstructed United States network at $\pi=1\%$ per period, beside the dispersion that each workhorse model generates at the same rate of inflation and the same frequency of resets. At that rate of inflation half of all firms reset their prices each period (Table~\ref{oa:tab:calibration}). The comparison tells us not only how large our relative-price distortion is, but also whether it is of the Calvo or of the menu-cost kind.

\begin{table}[H]
\centering
\caption{The measured relative-price distortion against its yardsticks}
\label{tab:distortion_yardsticks}
\footnotesize
\setlength{\tabcolsep}{4pt}
\begin{tabularx}{\textwidth}{@{}>{\raggedright\arraybackslash}p{0.18\textwidth}>{\raggedright\arraybackslash}p{0.42\textwidth}>{\raggedright\arraybackslash}p{0.17\textwidth}>{\raggedright\arraybackslash}X@{}}
\toprule
Yardstick & Object and benchmark & Magnitude (log points) & Source\\
\midrule
This paper & SD of log displacements from the network general equilibrium, United States network at $\pi=1\%$ per period, with half of all firms resetting each period & $\approx1.1$ entropy-equivalent ($\omega\approx0.75$ balance weighted) & Figure~\ref{fig:recon_us_inflation} of the main text\\[4pt]
Calvo staggering & $\sqrt{2\psi^{\mathrm{C}}}$, the entropy-equivalent SD of log displacements from the common frictionless benchmark, at $\pi=1\%$ per period and price reset probability $c=\tfrac12$ & $\approx1.4$ ($1.39$ exact, $1.41$ to first order) & Lemma~\ref{oa:lem:workhorse_benchmark}, \citet{AscariSbordone2014}\\[4pt]
Menu cost & $\sqrt{2\psi^{\mathrm{MC}}}$, the entropy-equivalent SD of log displacements from the common frictionless benchmark, at $\pi=1\%$ per period and price spells of two periods & $\approx0.50$ & Lemma~\ref{oa:lem:workhorse_benchmark}, \citet{sheshinski1977inflation,CaplinSpulber1987QJE}\\
\bottomrule
\end{tabularx}
\\[4pt]
\parbox{\textwidth}{\footnotesize \emph{Notes}: Magnitudes are standard deviations or entropy-equivalent standard deviations of log relative prices, in log points ($\times100$). The Calvo and menu-cost rows share the rate of inflation and the frequency of resets of the first row. Only the first row uses the network general equilibrium. The remaining rows carry their own benchmarks, so the table compares orders of magnitude, not identical economic objects.}
\end{table}

The comparisons with the Calvo and the menu-cost models run in opposite directions, and together they locate the mechanism. In the Calvo model, posted prices are vintages of past optimal prices, so the log displacements inherit the geometric distribution of price ages. With a price reset probability $c$, the entropy-equivalent dispersion of the Calvo model is $\sqrt{2\psi^{\mathrm{C}}}=\pi\sqrt{1-c}\,/\,c+O(\pi^2)$ (Lemma~\ref{oa:lem:workhorse_benchmark}). At $c=\tfrac12$ and $\pi=1\%$ the production network economy sustains about four fifths of the exact Calvo dispersion of Table~\ref{tab:distortion_yardsticks} on the entropy-equivalent measure, and about half of it on the relative price gap.\footnote{Each comparison sets the same measure side by side, although the two economies rest on different equilibrium price vectors. In the Calvo model firms differ only in the ages of their prices, so its relative price gap and its entropy-equivalent dispersion coincide to first order.} But in the Calvo model one price reset probability governs both the dispersion of relative prices and the size of price change, so both rise in proportion to inflation. It is this implication that the data of the Great Inflation reject \citep{nakamura2018elusive}. The rejection has no purchase on the production network economy, for in the steady state, across the firms that reset, its log displacements are unrelated to the size of price change (Section~\ref{subsec:sim_us}).

The comparison with the menu-cost model runs the other way. In the menu-cost model, inflation is the only reason a price falls behind its market, so every price falls behind in the same direction. The gap between a firm's price and its optimal price then stays inside the inaction band by construction. At the matched frequency a price spell lasts two periods, so the gaps sit at zero and at one period's inflation in equal measure. Their standard deviation is about $0.5$ log points, less than half the entropy-equivalent dispersion we measure and two thirds of our relative price gap. The relative-price distortion of the production network economy differs from the menu-cost dispersion in kind as well as in size. The telling difference is that closing the inaction band leaves the standing network distortion in place (Theorem~\ref{thm:size_not_measure}), whereas it removes the dispersion of the menu-cost model altogether.

Put together, the two comparisons show the production network economy combining what neither workhorse model combines. Its dispersion of relative prices is of the order that Calvo staggering generates, while its size of price change, as in a menu-cost model, barely moves with inflation. At the matched frequency the three economies share one size of price change, about $2\%$. What sets them apart is how that size responds to inflation, in proportion to it under Calvo pricing and hardly at all under the menu-cost rule and in the production network economy (Figure~\ref{fig:recon_us_size_freq} of the main text). The production network economy can combine the two because its standing network distortion is a property of the injection of money and of the network, and not of the price reset rule.

The comparisons so far set our relative-price distortion beside theoretical yardsticks. A direct measurement of relative-price distortion in the data points the same way. In the micro price data that underlie the consumer price index of the United Kingdom, the effect of inflation on relative-price distortion has been identified \citep{AdamAlexandrovWeber2024}. The cross-sectional dispersion of prices is dominated by movements of flexible prices and does not move with inflation. But the component of that dispersion that measures distortion rises with the deviation of inflation from its optimal rate. The relative-price distortion measured in these data and the standing network distortion of our model are complements. The first is born of the pricing friction and vanishes when prices are flexible. The second is born of the propagation of new money through the network and survives flexible prices. Which means that the dispersion of relative prices that is a cost of staggered price setting \citep{yun2005optimal} has, in the production network economy, a component that staggering does not produce and that resetting prices cannot remove.

\section{Notation}
\label{oa:appendix:notation}

We group the recurring symbols of the main text and of Appendices~B and~C by role. The text defines each object where it first appears.

\small
\setlength{\tabcolsep}{4pt}
\setlength{\LTpre}{6pt}\setlength{\LTpost}{10pt}
\relpenalty=100 \binoppenalty=100
\renewcommand{\arraystretch}{1.15}
\begin{longtable}{@{}>{\raggedright\arraybackslash}p{0.31\textwidth}>{\raggedright\arraybackslash}p{0.625\textwidth}@{}}
\toprule
\endhead
\bottomrule
\endfoot
\multicolumn{2}{@{}l}{\textit{Network and degrees}}\\*
$\mathbf A=(a_{ij})$ & Propagation matrix, column-stochastic, irreducible, and aperiodic, with $a_{ij}$ the share of input $i$ in firm $j$'s expenditure.\\
$\lambda_j,\ \mathbf v_j,\ \mathbf u_j$ & Eigenvalues of $\mathbf A$, with $\lambda_1=1>\lambda_2>|\lambda_3|$, and right and left eigenvectors normalized so that $\mathbf u_j^\top\mathbf v_k=\mathbf 1\{j=k\}$.\\
$1-\lambda_2$ & Spectral gap of $\mathbf A$.\\
$d_i,\ d_{\max}$ & Degree of firm $i$, the common value of its supplier and customer counts, and the upper end of the degree support $[1,d_{\max}]$.\\
$\alpha$ & Tail exponent of the degree distribution, $\Pr(D>d)\propto d^{-\alpha}$.\\
$\nu\in(-1,0)$ & Assortativity exponent, $\bar d_i\propto d_i^{\nu}$, with $\bar d_i$ the mean nearest-neighbor degree of firm $i$.\\
$\rho_{\mathrm{BS}}$ & Newman coefficient of buyer--seller assortativity, which shares the sign of $\nu$ but not its scale.\\
$\E_n[H(d)]$ & Empirical degree moment $n^{-1}\sum_iH(d_i)$ on a realized network, with population limit $\E[H(d)]$.\\
$\zeta,\ \wgt{i}{\zeta}$ & Weight exponent, and degree weight $d_i^{\zeta}/\sum_jd_j^{\zeta}$, with $\wgt{i}{1}=m_i^{*}$ under Assumption~\ref{assu:degree_structure}.\\
$\mathbf d^{\mathrm{sh}}$ & Normalized degree vector $\mathbf d/\mathbf 1^\top\mathbf d$.\\
\addlinespace
\multicolumn{2}{@{}l}{\textit{Production and prices}}\\*
$\varsigma\in(0,1)$ & Returns to scale of the Cobb--Douglas technology.\\
$x_{ij},\ q_i^{(t)}$ & Quantity of good $i$ used by firm $j$, and output of firm $i$.\\
$\mathcal J_i^{(t)},\ \widetilde q_i^{(t)}$ & Inventory in the computational economy, and available supply $q_i^{(t)}+\mathcal J_i^{(t)}$.\\
$p_i^{(t)},\ \bar p_i^{(t)}$ & Market-clearing price $\mathcal D_i^{(t)}/\widetilde q_i^{(t)}$, and posted price $p_i^{(t-u_i^{(t)})}$.\\
$(\mathbf m^*,\mathbf q^*,\mathbf p^*)$ & Zero-injection equilibrium: stationary balances, output, and prices.\\
$r_i^{*},\ \bar r_i^{(t)}$ & Equilibrium and posted relative prices against an arbitrary firm $k$, which cancels from both measures of relative-price distortion.\\
$\widehat r_i^{(t)},\ \widehat r_i^{\,*}$ & Current price shares at market-clearing prices, and equilibrium price shares.\\
$g_i^{(t)},\ \bar g_i^{(t)}$ & Log displacements of the detrended market-clearing and posted prices from their equilibrium values.\\
$\vartheta,\ \vartheta_\star$ & Output--degree exponent in $q_i^*\approx d_i^{\vartheta}$, and the derived exponent $-\varsigma\nu/(1-\varsigma\nu)$.\\
\addlinespace
\multicolumn{2}{@{}l}{\textit{Money and injection}}\\*
$\mathbf m_t,\ M_t$ & Money balances, and the money stock $\mathbf 1^\top\mathbf m_t$, equal to $(1+\pi)^t$ under sustained injection.\\
$\pi$ & Rate of inflation, the fraction of the money stock injected at every date.\\
$\widetilde{\mathbf m}_t$ & Post-injection balances $\mathbf m_t+\pi M_t\boldsymbol\gamma_t$, with $\mathbf m_{t+1}=\mathbf A\widetilde{\mathbf m}_t$.\\
$\boldsymbol{\mathcal D}_t$ & Nominal demand, equal to $\mathbf m_t$.\\
$\widehat{\mathbf m}_t,\ \widehat{\mathbf m}_{\mathrm{ub}}$ & Normalized balances $\mathbf m_t/M_t$ and their balanced-growth limit.\\
$\widehat{\mathbf n}_{\mathrm{ub}}$ & Normalized post-injection balances $(\widehat{\mathbf m}_{\mathrm{ub}}+\pi\boldsymbol\gamma_{\mathrm{ub}})/(1+\pi)$ at the balanced-growth fixed point.\\
$\mathbf q_{\mathrm{ub}},\ \widehat{\mathbf p}_{\mathrm{ub}}$ & Balanced-growth output and detrended prices.\\
$\boldsymbol\gamma_t,\ \boldsymbol\gamma_{\mathrm{ub}}$ & Injection profile in the unit simplex, and the stationary injection profile.\\
$\theta\in(0,1)$ & Injection exponent, $\boldsymbol\gamma_t\propto\mathbf m_t^{\circ\theta}$.\\
$\boldsymbol\tau,\ \iota_i$ & Injection tilt $\boldsymbol\gamma_0-\mathbf m^{*}$, and incidence ratio $\gamma_{0,i}/m_i^{*}$.\\
$\xi_t,\ \xi_0,\ \xi_{\mathrm{ub}}$ & Injection misalignment $\E_{\boldsymbol\gamma_t}[d^{\nu}]-\E_n[d^{1+\nu}]/\E_n[d]$, its value at the injection profile evaluated at the zero-injection equilibrium, and the stationary misalignment.\\
\addlinespace
\multicolumn{2}{@{}l}{\textit{Transmission and the degree representation}}\\*
$\boldsymbol\mu,\ \mathbf y$ & Balance response and output response, the first-order responses of normalized balances and of output to sustained injection, with $\boldsymbol\mu-\mathbf y$ the price response.\\
$\varkappa$ & Projection ratio $\varsigma(1-\lambda_2)/(1-\varsigma)$.\\
$\delta_i,\ \widehat\delta_i$ & Excess degree $d_i^{\nu^{2}}-\E_n[d^{\nu^{2}}]$, and normalized excess degree $\delta_i/d_i$.\\
$\kappa,\ c_m$ & Amplitude of the second-eigenmode response, and money-per-degree constant $1/\sum_jd_j$.\\
$\mathcal C_i$ & Cantillon exposure $\kappa\delta_i/m_i^{*}=(\kappa/c_m)\widehat\delta_i$.\\
$d^{*},\ \hat d$ & Threshold degree $(\E_n[d^{\nu^{2}}])^{1/\nu^{2}}$, at which excess degree and the Cantillon exposure change sign, and degree $d^{*}(1-\nu^{2})^{-1/\nu^{2}}$ of the unconstrained peak of the Cantillon exposure.\\
$P_i^{(T)},\ S_i^{(T)}$ & Permanent trend and transitory wave of nominal demand.\\
$\ell_i^{(T)},\ \ell_i^{\max}$ & Level distortion of firm $i$'s nominal demand, $S_i^{(T)}/P_i^{(T)}\approx\pi\mathcal C_i\mathcal X_T$, and its amplitude $\sup_s|\ell_i^{(s)}|$.\\
$\mathcal X_T,\ \tilde\lambda$ & Propagation kernel $\sum_{t=0}^{T-1}\xi_t\tilde\lambda^{T-t}$, and its decay factor $\lambda_2/(1+\pi)$.\\
$\mathbf h_\perp(\boldsymbol\gamma)$ & Accumulated non-Perron response to a stationary injection profile.\\
$Q(\alpha,\nu),\ \mathcal V(\alpha,\nu;\vartheta)$ & Dispersion functionals of the Cantillon exposure under the Pareto approximation, entering $\digamma_\omega$ and $\digamma_\psi$.\\
\addlinespace
\multicolumn{2}{@{}l}{\textit{Price adjustment}}\\*
$I_i^{(t)},\ u_i^{(t)}$ & Price reset indicator and post-decision price age of firm $i$.\\
$\bar u$ & Maximal price age.\\
$s_i^{(t)}$ & Pre-reset gap $\log(p_i^{(t)}/\bar p_i^{(t-1)})$.\\
$\eta,\ \beta,\ c$ & Hazard function of the pre-reset gap, its state-dependence exponent, and its scale constant, with $\eta(s)\sim c\,s^{\beta}$ as $s\to0$. At $\beta=0$ the scale constant $c$ is the price reset probability of the Calvo benchmark.\\
$u_\pi^\star$ & Natural age scale $\pi^{-\beta/(\beta+1)}$.\\
$\mathcal P_{\bar u},\ s_0,\ \chi$ & Cap-nonbinding range of inflation rates, and the two fixed small constants that define it (Definition~\ref{oa:def:cap_range}).\\
$\mathcal S_i(u)$ & Survival function, the probability that a price spell of firm $i$ survives through age $u$.\\
$\mathfrak x_i,\ \mathfrak x_i^{\mathrm{flex}}$ & Network displacement that a posted price carries from its price reset date, and the network displacement it would carry had it been reset at the current date.\\
$b_t,\ b_0,\ b_w,\ b_\pi$ & Inaction half-width of the $(S,s)$ rule of the computational experiments, and its baseline half-width, responsiveness weight, and band compression.\\
\addlinespace
\multicolumn{2}{@{}l}{\textit{Statistics}}\\*
$\phi_t$ & Average size of price change conditional on adjustment.\\
$\omega,\ \psi$ & Relative price gap and relative-price entropy, with superscripts $\mathrm{flex}$ and $\mathrm{stick}$ for market-clearing and posted prices.\\
$\digamma_\omega,\ \digamma_\psi$ & Leading coefficients of the relative price gap and of relative-price entropy in Theorem~\ref{theorem:price_deviation}.\\
$\phi^{\mathrm C},\ \psi^{\mathrm C},\ \phi^{\mathrm{MC}},\ \psi^{\mathrm{MC}}$ & Size of price change and relative-price entropy in the drift-only Calvo and menu-cost benchmarks.\\
\addlinespace
\multicolumn{2}{@{}l}{\textit{Calibration}}\\*
$z_i,\ z_s,\ z_\ell$ & Firm size, and the mean sizes of a study's small- and large-firm groups.\\
$\mathfrak r$ & Ratio of the per-size fresh-liquidity intake of the small group to that of the large group.\\
\end{longtable}

\end{document}